\documentclass[12pt]{book}
\renewcommand{\baselinestretch}{1.5}

\usepackage{a4wide}
\usepackage{fancyhdr}
\usepackage{graphicx}
\usepackage{amssymbol}
\usepackage{float}
\usepackage{epsfig,latexsym,fancyheadings,amssymb}
\usepackage{rotating}
\usepackage[breaklinks = true]{hyperref}
\usepackage{colortbl}
\usepackage[numbers,sort&compress]{natbib}
\usepackage[inner=4.25cm,outer=2.54cm,top=2.54cm,bottom=2.54cm]{geometry}

\usepackage{sectsty,textcase} 
\allsectionsfont{\sffamily}
\definecolor{bostonuniversityred}{rgb}{0.8, 0.0, 0.0}
 	\definecolor{blue(ryb)}{rgb}{0.01, 0.28, 1.0}
 	\definecolor{burgundy}{rgb}{0.5, 0.0, 0.13}
 	\definecolor{cobalt}{rgb}{0.0, 0.28, 0.67}
 	\definecolor{britishracinggreen}{rgb}{0.0, 0.26, 0.15}
 	\definecolor{bubbles}{rgb}{0.91, 1.0, 1.0}
 	\definecolor{americanrose}{rgb}{1.0, 0.01, 0.24}
 	\definecolor{ao(english)}{rgb}{0.0, 0.5, 0.0}
 	\definecolor{babypink}{rgb}{0.96, 0.76, 0.76}
 	\definecolor{blue(pigment)}{rgb}{0.2, 0.2, 0.6}
 	\definecolor{bondiblue}{rgb}{0.0, 0.58, 0.71}
 	\definecolor{brightube}{rgb}{0.82, 0.62, 0.91}
 	\definecolor{blizzardblue}{rgb}{0.67, 0.9, 0.93}
 	\definecolor{desert}{rgb}{0.76, 0.6, 0.42}
 	\definecolor{electriclavender}{rgb}{0.96, 0.73, 1.0}
 	\definecolor{beige}{rgb}{0.96, 0.96, 0.86}
 		\definecolor{midnightblue}{rgb}{0.1, 0.1, 0.44}
 	\definecolor{richelectricblue}{rgb}{0.03, 0.57, 0.82}

\usepackage[Glenn]{fncychap}
\renewcommand{\chaptermark}[1]%
         {\markboth{\thechapter.\ #1}{}}
\renewcommand{\sectionmark}[1]%
         {\markright{\thesection\ #1}}
\newcommand{\LMUTitle}[9]{

  \newpage
  \thispagestyle{empty}

  \cleardoublepage
  \thispagestyle{empty}
 \begin{center}
    \includegraphics[width=1.50in]{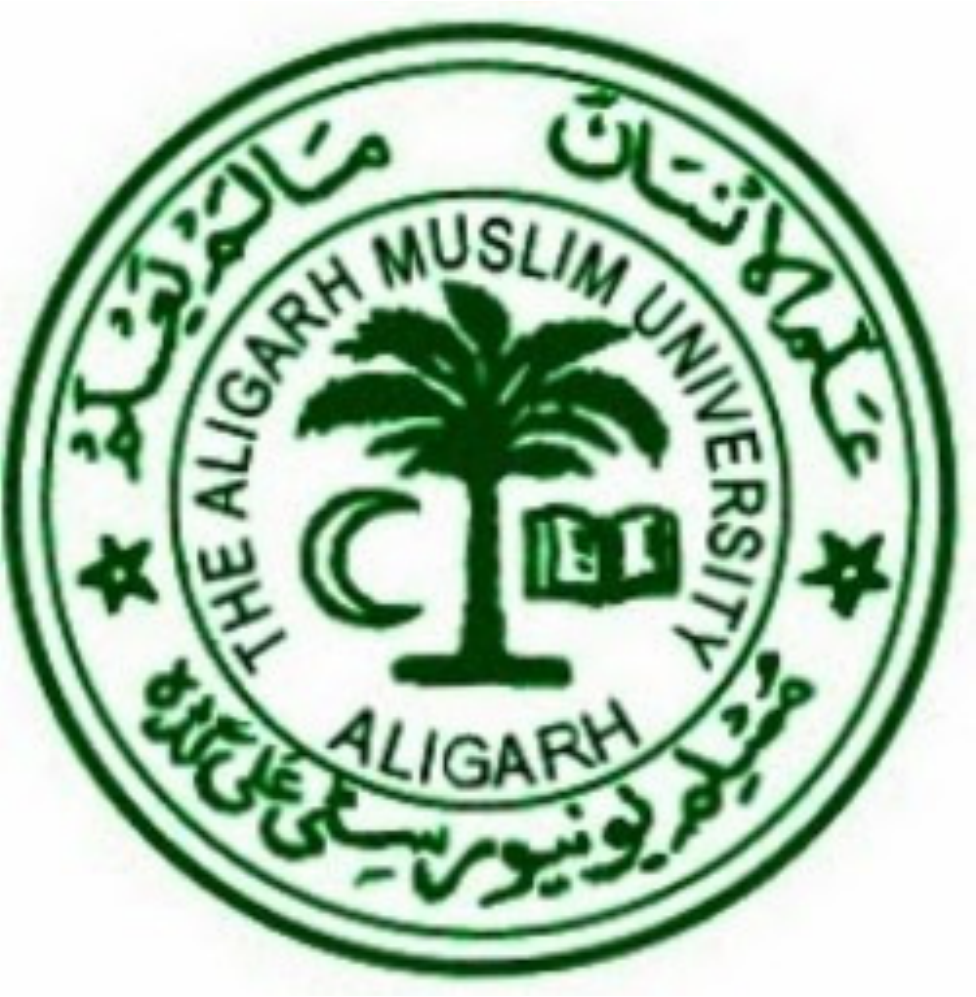}
  \end{center}
\vspace{-0.5cm}
  {\parindent0cm
  \rule{\linewidth}{.5ex}}
  \begin{flushright}
\vspace{-0.5cm}
      \begin{center}\LARGE{{\bf #1}}\end{center}
  \end{flushright}
\vspace{-0.5cm}
  \rule{\linewidth}{.5ex}

\vspace{-0.30cm}
  \begin{center}
    \large  {\bf THESIS}\\[0.15cm]
    \large { SUBMITTED  FOR THE
	   AWARD OF THE DEGREE OF }\\[0.15cm]
	\large {\bf #3}\\
    \large IN\\
    \large {\bf PHYSICS}\\[0.15cm]
    \large BY \\[0.20cm]

    \large {\bf #2}\\[0.20cm]

    \large {Under the Supervision of}\\
    \large {\bf #4}\\

    \large {and Co-supervision of}\\
    \large {\bf #5}\\[0.6cm]

    \large  {\bf #6}\\
	\large {\bf #7}\\
	\large {\bf #8}\\
	\large {\bf 2021}
  \end{center}

  \newpage
  \thispagestyle{empty}
 
  \newpage
  \thispagestyle{empty}
 
  \begin{center}

	\textsc{\large {\bf IN THE NAME OF ALLAH}}\\[1cm]		
	{\large {\bf THE BENEFICENT, THE MERCIFUL}}\\[8cm]
	{\large {\bf DEDICATED TO MY BELOVED PARENTS}}\\[0.5cm]

\end{center}

  \cleardoublepage
}

\begin{document}
  \frontmatter
  \LMUTitle
      {A Systematic Study Of Nuclear Matter: Finite Nuclei To Neutron Star}               
      {ABDUL QUDDUS} 						
      {\textit{DOCTOR OF PHILOSOPHY} }                           		  
      {DR. SHAKEB AHMAD}               			
      {PROF. SURESH KUMAR PATRA}			
      {DEPARTMENT OF PHYSICS}                          
      {ALIGARH MUSLIM UNIVERSITY}                            
      {ALIGARH (INDIA)}                                      
%
%
   \Acknowledgement
 \markboth{Acknowledgements}{Acknowledgements}

\chapter{Acknowledgements}{\label{ack}}

\rule\linewidth{.5ex}
 \vspace*{\stretch{1}}

All praises and thanks are due to {\bf Almighty Allah}, the most beneficent and the most
merciful, for blessing me with wisdom, strength, and required enthusiasm to
overcome all the obstacles in the way of this toilsome journey for knowledge. In utter
gratitude, I bow my head before Him.

This thesis would not have been possible without the support of many people, and this
limited space would not be enough to express all my sincere appreciation and gratitude
to them. I take this opportunity to place on record my heartfelt thanks to all those who
helped me to materialize this arduous task.

First of all, I would like to express my sincere gratitude to my supervisor {\bf Dr. Shakeb Ahmad}, 
Associate Professor, Department of Physics, Aligarh Muslim University (AMU), Aligarh who introduced me to 
the field of research in nuclear theory. His patience, interest, 
enthusiasm, valuable guidance, constant support, and strong motivation helped me to complete the thesis in its present form.
%
I wish to express my immense thanks to my co-supervisor {\bf Prof. Suresh Kumar Patra}, Institute of Physics, Bhubaneswar (IOPB)
for his consistent efforts to bring out of me what I have done here. 
His suggestions and criticisms have been indispensable to the successful completion of this work.
It has been an honor to work with him.

I am immensely grateful to Prof. Bhanu Pratap Singh, Chairperson, Department of Physics, Aligarh Muslim University, Aligarh
for providing me all the necessary facilities, encouragement, and inspiration. I warmly thank
Prof. Tauheed Ahmad \& Prof. Afzal Ansari, former chairpersons, Department of Physics, AMU, Aligarh for their 
valuable and extensive support provided to me to carry out the work during their tenure.
I would like to thank the Director of IOPB for providing lodging, necessary computational and 
library facilities required to carry out the work, and some times to-and-fro fare during my visits to the institute 
to work with my co-supervisor. 

All the faculty members of the department have been very kind enough to extend their
help at various circumstances during my research, whenever I approached them, and I do hereby
acknowledge all of them. 
The thesis would not have come to successful completion without the help of the office and seminar staffs of the Department of Physics, 
AMU, Aligarh. I am highly thankful to them for their co-operation in providing the required official assistance and valuable literature.

I would like to thank our collaborators Prof. B. V. Carlson (Instituto Tecnol\'ogico de Aeron\'autica, S\~ao Jos\'e dos Campos, 
S\~ao Paulo, Brazil), Dr. Grigorios Panotopoulos (Instituto Superior T{\'e}cnico-IST, Universidade de Lisboa-UL, Lisboa, Portugal), 
Dr. Bharat Kumar (National Institute of Technology, Rourkela), Dr. Kabir Chakravarti 
(Inter-University Centre for Astronomy and Astrophysics (IUCAA), Pune), Prof. Sukanta Bose (IUCAA, Pune), Dr. Manpreet Kaur 
(IOPB), Dr. Mrutunjaya Bhuyan (University of Malaya, Malaysia), K. C. Naik (SoA University, Bhubaneswar) 
for their significant role in various published articles. The contribution of Dr. Bharat Kumar is indispensable in discussing  
various physics problems, their coding, and in making possible my visits at IUCAA, Pune and Tsukuba University, Tsukuba, Japan.  
 
I must thank Prof. Mitko K. Gaidarov (Institute for Nuclear Research and Nuclear Energy, Bulgarian Academy of Sciences), 
Prof. Peter Ring (Technical University of Munich), Prof. P. Arumugam (Indian Institute of Technology, Roorkee),
Prof. J. N. De (Saha Institute of Nuclear Physics (SINP), Kolkata), Prof. B. K. Agrawal (SINP, Kolkata), 
Dr. F.~Cappella (Dip. di Fisica, Universit\'a di Roma ``La Sapienza", Rome, Italy) for enlightening discussions and 
correspondence. I am grateful to Prof. Sanjay Reddy (University of Washington, Seattle), Prof. James Lattimer 
(Stony Brook University, Stony Brook), Prof. Takashi Nakatsukasa (University of Tsukuba, Tsukuba), Prof. Javid Ahmad Sheikh 
(Cluster University Srinagar) for sharing their knowledge and giving valuable suggestions 
at various schools/conferences I have attended. I also thank other eminent scientists who delivered wonderful lectures at the CNT Schools of Kolkata, 
SERC School of Srinagar, WE-Heraeus School of MPIK Germany, ERICE School of Italy, OMEG Conference of Kyoto University, DAE symposiums, 
AMU conference, etc. I really enjoyed the lectures of all of them. Whatever I have learned in my Ph.D., the contributions of these 
schools and conferences played a significant role. I must thank the organizers of these schools/conferences for giving me the chance to 
participate in these scientific events, for providing great hospitality, and partial financial assistance.     
I am grateful to Prof. Sukanta Bose and Prof. Takashi Nakatsukasa, again, for hosting me at their respective institute and giving me the 
opportunity to deliver invited talk. 
I am thankful to the referees for enhancing the quality of the published papers, which result in this complied work. 

I am grateful to the Department of Science and Technology (DST) and Council of Scientific and Industrial Research (CSIR) 
for providing full/partial travel airfare to attend the ERICE school 2018 and OMEG15 Conference of Japan, 
respectively. I am also thankful to the European Physical Society (EPS) for providing me the EPS fellowship to bear local expenses incurred in 
ERICE School of 2018 and 2019, the organizer of OMEG15 conference, Kyoto for providing financial assistance for lodging, and the organizers 
of WE-Heraeus school of MPIK, Germany for bearing local expenses there. 
Financial Assistance provided by the DST (India), in the form of INSPIRE Fellowship with order no.
{\bf No./DST/INSPIRE Fellowship/IF160131} is gratefully acknowledged.     

I would like to extend my appreciation to each participant of the schools mentioned above 
for providing their nice company during the 
schools. I am ever indebted to my labmates Dr. Afaque Karim, Mr. Tasleem A. Siddiqui, and Dr. Tabassum Naz 
and admire their distinguished helping nature and contribution to my work. 
I express my gratitude to Dr. Syed Mohammad Faisal (Michigan Medicine School, University of Michigan, Michigan) 
for helping me at various stages of the Ph.D. tenure, his motivation and inspiration. 
I wish to thank all my friends at AMU and IOPB, especially, the following for their friendship and all the good times we had together:
Shan Ahmad (AMU), Zahid Ali (AMU), Syed Bakhtiyar (AMU), Mohammad Shoeb (AMU), Syed Nafisul Hasan (AMU), 
Mohammad Mohsin Nizam Ansari (AMU), Usuf Rahman (AMU), Muqaddar SK (IOPB), Rupam Mondal (IOPB), Avnish Yadav (IOPB), Abhishek (IOPB). 
I thank Ansari, Siddiqui, Rahman, Ishfaq (AMU), and Parveen Bano (Sambalpur University) for reading 
chapter(s) of the thesis carefully and giving their suggestions for its better presentation. Most of their corrections are reflected in this 
thesis. 

Last but not the least, I would like to pay high regards to my parents, Mr. Rafi Ahmad Khan and Mrs. Noor Fatima, 
who encouraged and helped me at every stage of my personal and academic life. I cannot thank enough for the constant 
support (emotional and financial) and unconditional love that they have given without question over the years. 
I owe everything to them. On a personal note, I express my thanks to my beloved brothers, sisters, nieces, and brother-in-laws
 whose blessings, boundless affection, and encouragement helped me achieve this work successfully. 
I wish to express my heartfelt thank to my beloved wife, Mrs. Nazia Muslim for her love, being a source of happiness, 
moral encouragement, and caring. 
Besides this, several people have knowingly and unknowingly helped me in the successful completion of this work. I acknowledge all of them.

\vspace*{\stretch{1}}

\begin{flushright}
\LARGE
 {\scshape {\bf Abdul Quddus}}
\end{flushright}

%
  \markboth{List Of Publications}{List Of Publications}
 \include{zusammenfassung}   
  \tableofcontents
  \markboth{Table of Contents}{Contents}
  \listoffigures
  \markboth{List of Figures}{List of Figures}
  \include{Abbildungsverzeichnis}

  \listoftables
    \include{Tabellenverzeichnis}
 \markboth{List of Tables}{List of Tables}

%
%
%

  \mainmatter\setcounter{page}{1}


\chapter{Introduction}{\label{chap1}}  

\rule\linewidth{.5ex}


The classification of nuclear matter, characterization of its properties, and analysis of various phenomena 
emerging from or happening in nuclear matter remain an area of intense activity. In a broad prospect, the nuclear matter is 
categorized as finite nuclei and infinite nuclear matter. Since the discovery of an atomic nucleus by Rutherford and his collaborators
through the famous gold-foil scattering experiment in 1911, the pursuit of understanding the structure of a nucleus is continued yet. 
An atomic nucleus exhibits a variety of nuclear phenomena, e.g., fission, fusion, and radioactivity.  
The infinite nuclear matter is defined as an ensemble of infinite (practically, very large) number of constitute particles. 
A neutron star is a good example of an asymmetric infinite nuclear matter that have a mysterious nature. The first theoretical 
speculation on the existence of neutron star came in to picture after Landau's paper of 1932, ``density of matter becomes so great 
that atomic nuclei come in contact, forming one gigantic nucleus"~\cite{landau32}.   
Equation of state of nuclear matter is a key ingredient to study the structure of neutron star, the collapse of massive stellar core, 
the dynamics of heavy-ion collisions, and in determining the boundaries of neutron and proton drip line nuclei.    
The Equation of state depends sensitively on the nature of nucleon-nucleon interaction (strong nuclear force), which is being explored by 
nuclear physicists both theoretically and experimentally. 


%
Theoretically, there are several models to comprehend the dynamics of a nuclear many-body problem.
At the most fundamental level are ab-initio theories. For example, No-core shell model~\cite{ag}, Unitary correlator method~\cite{ai}, 
and Coupled cluster calculations~\cite{ah} are used to study the properties of finite nuclei up to A$\sim$20~\cite{aj}. 
At present, the fully microscopic calculation for heavier nuclei is not feasible since these models require heavy computational time. 
There are several models, derived from microscopic theory, that use effective interaction at the intermediate level. Shell model~\cite{ar}, 
deformed shell model (Nilsson model)~\cite{nlsn1,nlsn2}, Triaxial Projected Shell Model (TPSM)~\cite{review,jh99} have been 
well-turned in explaining/interpreting the properties of nuclei.  
On the other hand, the macroscopic models such as the liquid drop model~\cite{ak,al,ama} assorted with shell correction of a nucleus 
constitute the macroscopic-microscopic approach~\cite{moller95}. Over the years, the macroscopic-microscopic models have been well-turned 
in studying nuclear properties with the limitation that its dependence on phenomenological input begets uncertainties in 
the observable parameters when extrapolating for exotic nuclei. 

Moreover, the self-consistent mean-field models that fall between ab-initio and
macroscopic-microscopic approaches have been proved to be good approximations for the phenomenological
description of nuclear matter at nuclear saturation and sub/supra-normal densities. 
Recently, the self-consistent mean-field relativistic and non-relativistic formalisms successfully predict a large number of 
nuclear phenomenon near the nuclear drip lines and at astrophysical sites~\cite{estal01,chabanat98}.
Likewise, the Density Functional Theory (DFT) for electronic systems, mean-field models of nuclei methodically map the 
many-body problem onto a single-body problem~\cite{dft1,dft2}.
Among the effective mean-field models for nuclear matter (finite nuclei to neutron star), the relativistic mean-field (RMF) 
approach is one of the most successful and widely used self-consistent formalisms.
The properties of finite nuclei and neutron star predicted by the RMF models are nearly the same as that predicted by the 
non-relativistic mean-field models such as Skyrme~\cite{aw} and Gogny interactions~\cite{dech80}. 
The advantages of RMF models~\cite{Ring:2012gr,wal74} over its non-relativistic counterparts are:
\begin{itemize}
\item{ Conspicuous treatment of mesonic degrees of freedom. }
\item{ RMF models take into account the spin-orbit force automatically while in the case of non-relativistic descriptions additional terms 
have to be added externally. }
\item {The relativistic effects at higher densities are taken care of appropriately within RMF model~\cite{wal74}.}
\item{ The RMF models well reproduce the isospin dependence of spin-orbit effect like the isotope shifts in the $Pb$ region while 
Skyrme and Gogny models are unable to do so.}
\end{itemize}   

The first RMF theory, known as the linear sigma-omega model, was proposed by Walecka~\cite{wal74,serot86} in which nucleons interact 
through the exchange of $\sigma-$ and $\omega-$ mesons. With time, this model has been improved/upgraded by 
adding other mesons and incorporating higher-order terms of the mesonic degree of freedom to
explain appropriately the nuclear phenomenon of stable nuclei, drip lines nuclei, and infinite nuclear matter. The mediating mesons are 
collectively known as the effective fields and denoted by the classical numbers.  
The detailed formalism of the RMF model, its development to explain the whole nuclear chart, infinite nuclear matter, and neutron star have been 
discussed in the next chapter. We have adopted an extended version of the RMF model, known as the effective field theory motivated relativistic 
mean-field model to study the bulk and surface properties of finite nuclei at zero/finite temperature and to generate an equation of state to 
study the effects of dark matter on neutron star properties. Here, I discuss briefly some of the frontier fields of 
research in the structure of finite nuclei, and neutron stars and what properties of the nuclear matter 
we have studied in this dissertation.  

%


Nuclei lie away from $\beta-$ stability line with large neutron to proton asymmetry are of great
importance. One of the quests among the nuclear physics community is how to synthesis the exotic
and superheavy nuclei and to explore their applications. 
Since the matter at extreme density and temperature is impossible to create
in a laboratory, a study of neutron-rich nuclei is treated as a tool to understand it.
In the quest for the formation of superheavy nuclei,
the last one with $Z=118$ has been synthesized at Dubna which was named Oganesson \cite{ogan06}
and more superheavy nuclei are expected to synthesize. A lot of theoretical predictions are reported
about the stability of superheavy nuclei against the spontaneous fission, $\alpha$- and $\beta$-decays,
and neutron emission \cite{mehta,elena}. The mere existence of exotic nuclei and superheavy nuclei is 
entirely due to the quantal shell effect, which plays its role against the surface tension and the Coulomb
repulsion. Furthermore, one of the very compelling issue in such exotic systems is the appearance
of new magic numbers and the disappearance of others in moving from $\beta$-stable to drip line
region of the nuclear chart \cite{chou95,otsu01}. For example, beyond the proton number $Z=82$ and
neutron number $N=126$, the next predicted magic numbers are Z = 114, 120, and 126 for the proton
and N = 172 or 184 for the neutron \cite{elena}. The neutron-rich/deficient isotopes, and Z = 120
element which is one of the predicted magic numbers represent a challenge for future experimental
synthesis since they are located at the limit of accessibility with available cold fusion reactions
facility. Therefore, an accurate estimation of their characteristics is essential from the theoretical
side to guide future experiments. Various experiments around the globe like Jyav\"askyl\"a
(Finland) \cite{jyav}, FRIB (US) \cite{frib}, FAIR (Germany) \cite{gsi}, RIKEN (Japan) \cite{riken},
GANIL (France) \cite{ganil}, FLNR (Russia) \cite{flnr}, CSR (China) \cite{csr}, and ORNL (US) \cite{ornl} 
provide a possibility of exploring exotic nuclei, and superheavy nuclei under extreme condition of isospin asymmetry.


Furthermore, nuclear fission is one of the most interesting phenomena from the
time of its discovery by Otto Hahn and F. Strassmann in 1938 \cite{otto}. In this process, a huge amount
of energy is produced when heavy elements like Uranium and Thorium are irradiated with slow neutrons. 
In nature, $^{233,235}U$ and $^{239}Pu$ are thermally fissile nuclei that have the tendency 
to undergo fission even when a zero energy neutron (at room temperature) strikes them. 
Of these, $^{235}U$ is naturally available with an isotopic fractional abundance of
$\approx0.7\%$ \cite{nndc}. $^{233}U$ and $^{239}Pu$ are formed by $^{232}Th$ and $^{238}U$,
respectively, through neutron absorption and subsequent $\beta-$ decays. 
It is important to mention that Satpathy {\it et. al.}, have found several neutron-rich uranium and thorium 
isotopes with neutron number 154 $\leq$ N $\leq$ 172 showing thermally fissile behavior \cite{satpathy}. 
These predicted neutron-rich thermally fissile nuclei can produce more energy than that of the naturally available thermally fissile nuclei. 
Due to the great importance of naturally available and neutron-rich thermally fissile nuclei, it is worth studying their properties.


Besides nuclear bulk properties, it is worth knowing the nuclear symmetry energy that is directly connected 
with the isospin asymmetry of the system; either infinite nuclear matter or finite nuclei. 
It is an important quantity having significant role in different areas of
nuclear physics, for example, in structure of ground-state nuclei \cite{Niksic08,Van10,Dalen10}, physics
of giant collective excitation \cite{Rodin07}, dynamics of heavy-ion reactions \cite{Chen08,Colonna09},
and physics of neutron star \cite{Steiner05,james,fattoyev12,dutra12,dutra14}. It determines various neutron
star properties such as mass-radius trajectory, its cooling rate, the thickness of the crust, and the
moment of inertia \cite{lrp}. The astrophysical observations and availability of exotic beam in a
laboratory have raised an interest in symmetry energy \cite{gai11}. In the last three decades,
the density dependence of nuclear symmetry energy has played a great role to understand nuclei near the
drip line \cite{wd80}. For interpreting the neutron-rich
nuclei and the neutron star matter, the characterization of the symmetry energy through experiments is
a crucial step. But, the symmetry energy is not a directly measurable quantity. It is extracted from
the observables related to it. Danielewicz has demonstrated that the ratio of the bulk symmetry energy
to the surface symmetry energy is related to the neutron skin-thickness \cite{daniel03}. 
Even the precise measurement of neutron skin-thickness is difficult, yet \cite{prex1}
it is one of the sensitive probes for nuclear symmetry energy. It is found
that the radius of a neutron star is correlated to the density dependence of the symmetry energy at
a saturation point \cite{lattimer07}. Furthermore, the $L$ coefficient (or say, pressure $P$) is correlated
with the neutron skin-thickness of $^{208}Pb$ \cite{brown,rj02,cent09,rocaprl11} and the radius of a
neutron star. 


Neutron stars (NSs) are fascinating objects that are born from the collapse of massive stars and reach central densities 
that may exceed those found in atomic nuclei by up to an order of magnitude. 
They are considered to be unique cosmic laboratories to explore
properties of ultra-dense matter under extreme conditions of density and neutron-proton asymmetry. The density inside an NS 
span from a few $g cm^{-3}$ at the surface to more than $10^{15} g cm^{-3}$ at its center. Depending upon its density profile and the 
possible nature of particles in a particular density domain, NS matter is categorized as the outer crust, 
inner crust, outer core, and inner core. 
Each part of the NS structure is governed by different forces.  
It is a well-known fact that the mass of an NS is dominated by the core contribution.
The properties of an NS are predicted by its equation of state (EOS), a certain relation between
energy density and pressure. EOSs are used in the Tolman-Oppenheimer-Volkoff (TOV) equations \cite{tov} to study the mass-radius relation 
and other physical quantities of NS like the moment of inertia, tidal deformability, etc.. 
On the other hand, by measuring the mass and radius of NS (pulsar) with the help of Chandra X-ray Observatory, 
Neutron Star Interior Composition Explorer (NICER), and other experimental setups, its EOS can be constrained. 
Moreover, the tidal deformability $\Lambda$ of an NS from the GW170817 data \cite{abbott17}, the historical first detection of gravitational waves from the binary neutron-star (BNS)
merger by the LIGO-Virgo collaboration, provides a new probe to the interior of NS and their nuclear EOS \cite{annala}.
Currently, the most accurate constraint on the high-density behavior of an EOS comes from the observations of a
few massive pulsars (or GW170817) with maximum mass $M_{max}=(2.01 \pm 0.04)M_\odot$ (or $M_{max} \lesssim 2.17M_\odot$)
\cite{marg17,antoni13}, respectively. However, Ref. \cite{rezzo18} has inferred the maximum mass $M \lesssim 2.16^{+0.16}_{-0.15} M_\odot$
using the quasi-universal relation between the maximum mass and the mass-shedding limit.
The mass and radius of the isolated 205.53 Hz millisecond pulsar PSR J0030+0451 are estimated to be $M=1.44^{+0.15}_{-0.14}M_\odot$ 
(68\% C.L.), $R=13.02^{+1.24}_{-1.06}$km \cite{miller19} and $M=1.34^{+0.15}_{-0.16}M_\odot$ (84\% C.L.), 
$R=12.71^{+1.14}_{-1.19}$km \cite{riley19} using the NICER data.  
Future measurments of NS properties can finely constrain its exact EOS and can tell confidently what type of matter is present 
inside the inner core of an NS. Till then, there is a possibility of non-standard particles too such as dark matter 
besides hadron, strange matter, and quark-gluon plasma.  

One of the most chaotic and enthralling conundra in physics is the problem of dark matter in the Universe. Several Cosmological and Astrophysical observations suggest that at least $90\%$ mass of the Universe is due to some non-luminous matter, yet to be discovered, the so-called “dark matter” (DM). The term ``dark matter" was coined by Zwicky in 1933 when he found some evidence about the missing mass in studying a cluster of galaxies known as ``Coma" \cite{zwicky}. He found that the expansion of the space (red-shift)
in the Coma cluster could not be explained in terms of the known luminous mass. He applied the virial theorem and concluded that a large amount of DM must be present to keep these galaxies bound together. The measurements of the cosmic microwave background (CMB), too, suggest that DM is necessary to explain structure formation \cite{carlos}. Structure formation implies that clumps of neutral particles arose through
the gravitational attraction, form neutral atoms that were attracted gravitationally by DM to form the galaxies. 
Another evidence of DM is the high temperature of the gas detected in clusters through its X-ray emission \cite{carlos,berg06}. 
Currently, there are a plethora of modern observations that support and confirm the existence of DM on a wide range of scales.
More than 20 experiments worldwide for DM direct detection searches are either running or in preparation, and some of them are 
the following: the DArk MAtter (DAMA) experiment \cite{dama1,dama2},
Cryogenic Dark Matter Search (CDMS) experiment \cite{cdms}, EDELWEISS experiment \cite{edel}, IGEX \cite{igex}, ZEPLIN \cite{zeplin},
 GErmanium DEtectors in ONe cryostat (GEDEON) \cite{morales}, CRESST \cite{cresst}, GErmanium in liquid NItrogen Underground Setup
(GENIUS) \cite{genius}, and LHC. Furthermore, Fermi-LAT, GAMMA-400, IceCube, Kamiokande, and AMS-02 are some of the indirect DM detection experiments \cite{jennifer}.
Besides, a lot of theoretical studies have been done to find the nature of DM particles or to constrain the DM variables 
by applying different assumptions. In this thesis, we have considered WIMP as a DM candidate and have tried to constrain its parameters 
with the help of gravitational-waves data, emitted from binary neutron star merger. 
%

The main aim of the thesis is to study the properties of nuclear matter, i.e., finite nuclei to infinite nuclear matter, at zero and finite 
temperature within effective field theory motived relativistic mean-field model by using some of the recent parameter sets. For this, 
we have chosen exotic, superheavy, and natural/neutron-rich thermally fissile nuclei and studied ground as well as 
excited-state bulk and surface properties of nuclei. We have also tried to figure out the possible constraints on the DM 
variables by considering WIMP, a DM candidate, inside the NS core and using gravitational-wave data GW170817.  

The organization of the thesis is the following: After a brief introduction, here, in this chapter, 
I outline E-RMF formalism at zero and finite temperature for finite nuclei in Chapter \ref{chap2}. 
The equations of motion are derived for the $\sigma -$, $\omega -$, $\rho -$, $\delta -$, and electromagnetic field 
by applying the variational principle on Lagrangian density and using mean-field approximations. 
The formalism for a temperature-dependent equation of state for infinite nuclear matter and 
its key parameters are discussed in the same chapter. 
The BCS pairing at finite temperature and the Quasi-BCS pairing correlation for the open-shell nuclei are also discussed in the chapter.
At the end of Chapter \ref{chap2}, I have briefed about the development of force parameters within the RMF model and 
what interactions we have used in this study. 

In Chapter \ref{chap3},  
we have studied the fission parameters of hot natural thermally fissile $^{234,236}$U and $^{240}$Pu nuclei, and neutron-rich 
thermally fissile $^{244-262}Th$ and $^{246-264}U$ nuclei within the temperature-dependent 
axially-deformed effective field theory motivated relativistic mean-field (E-TRMF) formalism by using the recently developed FSUGarnet, 
and IOPB-I parameter sets.
The results obtained by these two forces are compared with the results of
the well known and widely accepted NL3 parameter set. The excitation
energy $E^*$, shell correction energy $\delta E_{shell}$, single-particle energy for neutrons and protons
$\epsilon_{n,p}$, level density parameter $a$, neutron skin-thickness $\Delta R$, quadrupole and hexadecapole deformation parameters, 
two neutron separation energy $S_{2n}$, and asymmetry energy coefficient $a_{sym}$ of these natural/neutron-rich thermally
fissile nuclei are calculated at finite temperature.
The dependency of level density parameter and other observables on the temperature and the force parameters (interaction Lagrangian) are
discussed there.

In Chapter \ref{chap4}, 
we present effective surface properties such as the symmetry energy, neutron pressure, and symmetry energy curvature that are 
calculated using the coherent density fluctuation model. The isotopic chains of $O$, $Ca$, $Ni$, $Zr$, $Sn$, $Pb$, and $Z =$ 120 
are considered in the present analysis, which covers nuclei over the whole nuclear chart. The matter density distributions
of these nuclei along with the ground-state bulk properties are calculated within the spherically
symmetric E-RMF model by using the recently developed IOPB-I, FSUGarnet, and G3 parameter sets and compared with the results of the NL3 set. 

In Chapter \ref{chap5},  
we show the temperature-dependent symmetry energy and its relevant quantities for $^{234,236,250}U$, and $^{240}Pu$ nuclei. 
The temperature-dependent relativistic mean-field (TRMF) model with FSUGarnet, IOPB-I, and NL3 parameter sets is used to 
obtain the ground and excited-state bulk properties of finite nuclei and the energy density, pressure, and the symmetry
energy of infinite nuclear matter. 
The nuclear matter observables at the local density
of the nuclei serve as an input to the local density approximation to obtain the effective symmetry energy coefficient, neutron
pressure, and the symmetry energy curvature of $^{234,236,250}U$ and $^{240}Pu$ nuclei.

In Chapter \ref{chap6},  
we have investigated for the first time the effects of DM inside an NS adopting the $\sim 10~GeV$ WIMP hypothesis as suggested by the results of the DAMA/LIBRA collaboration, which can be realized e.g. in the framework of the
Next-to-Minimal Supersymmetric Standard Model (NMSSM).
The dark matter particles interact with the baryonic matter of a neutron star through the Higgs bosons. 
The dark matter variables are essentially fixed using the results of the DAMA/LIBRA experiment, which are then used to build the Lagrangian density for the WIMP-nucleon interaction inside a neutron star.
We have used the E-RMF model to study the equations of state in the presence of dark matter. The predicted
equations of state are used in the Tolman-Oppenheimer-Volkoff equations to obtain the mass-radius relations, the moment of inertia, and effects of the tidal field on a neutron star. The calculated properties are compared with the corresponding data of the GW170817 event.


Finally, the thesis work is summarized and concluded in Chapter \ref{chap8}. 

\chapter{Relativistic Mean-Field Theory for Finite Nuclei and Nuclear Matter}{\label{chap2}}  

\rule\linewidth{.5ex}

From the last five decades, the relativistic mean-field (RMF) formalism is one of the most successful and widely used
theories for both finite nuclei and infinite nuclear matter including the study of a neutron star. It is
nothing but the relativistic generalization of the non-relativistic Hartree or Hartree-Fock Bogoliubov theory with the
advantages that it takes into account the spin-orbit interaction automatically and works better in high-density
region. In this theory, nucleons (treated as point particles) are considered to oscillate independently in a harmonic oscillator motion in the
mean-fields generated by the exchange of mesons and photons, by strictly obeying relativity and causality. 
These exchanged mesons are neither quantized nor play the role of particles. Thus, all types of mesons can be taken in RMF models. 
Since pions have negative parity, thus they do not contribute to the RMF model for a real nucleus. 
In general, mesons with the lowest quantum numbers and as few as possible to explain experimental data of a nucleus are used in RMF models. 
The list of used mesons with their quantum numbers is given below. 
\begin{itemize}
\item  $\sigma-$mesons:($J^\pi$,T = $0^{+}$,0), it contributes to a long- and mid-range attractive interaction.
\item  $\omega-$mesons:($1^{-}$,0), it provides to a short-range repulsive interaction.
\item  $\rho-$mesons:($1^{-}$,1), takes into account the isospin-dependent part of the nuclear interaction.
\end{itemize}
	 

In this chapter, we will discuss the Effective field theory motivated relativistic mean-field formalism, BCS and Quasi BCS pairings, 
equation of state for nuclear matter and its key parameters, and parameters used in this thesis. 

%

\section{Effective Field Theory Motivated Relativistic Mean-Field (E-RMF) Formalism}
\label{ermf}

In this section, the effective field theory motivated relativistic mean-field (E-RMF)
theory \cite{furnstahl97} is outlined briefly. 
In principle, the E-RMF Lagrangian has an infinite number of terms with all type of self-
and cross-couplings. Thus, it is necessary to develop a truncation scheme to handle the model numerically for the
calculations of finite and infinite nuclear matter properties.
The meson fields included in the Lagrangian are smaller than the
mass of nucleon. Their ratios are used as a truncation scheme as it
is done in Refs. \cite{furnstahl97,muller96,serot97,estal01}. This means $\Phi/M$, $W/M$ and $R/M$
are the expansion parameters. The constraint of naturalness is also introduced in the truncation
scheme to avoid ambiguities in the expansion. In other words, the coupling constants written with a
suitable dimensionless form should be $\sim 1$. Imposing these conditions, one can then estimate the
contributions coming from different terms of the Lagrangian by counting powers in the expansion up to
a certain order of accuracy, and the coupling constants should not be truncated arbitrarily.
It is shown that the Lagrangian up to $4^{th}$ order of dimension is a good approximation to predict
finite nuclei and nuclear matter observables up to considerable satisfaction \cite{furnstahl97,muller96,serot97,estal01}.
Thus, in the present calculations, we have considered the contribution of the terms in the E-RMF Lagrangian up to
$4^{th}$ order of expansion. The nucleon-meson E-RMF Lagrangian density
with  $\sigma-$, $\omega-$, $\rho-$mesons and photon $A^{\mu}$ fields is given as
\cite{chai15,iopb1}:
\begin{eqnarray}
{\cal L}({r_{\perp},z}) & = & \bar\varphi({r_{\perp},z}) \left(i\gamma^\mu\partial\mu -M + g_s \sigma - g_\omega \gamma^\mu \omega_\mu - 
g_\rho \gamma^\mu \tau \vec{\rho}_\mu 
-e \gamma^\mu \frac{1+\tau_3}{2} A_\mu \right) \cdot
\nonumber \\[3mm]
& &
\varphi({r_{\perp},z})
+ \frac{1}{2}\left(\partial^\mu\sigma\partial_\mu\sigma-m_s^2\sigma^2 \right)
-\frac{1}{4}V^{\mu\upsilon}V_{\mu\upsilon} + \frac{1}{2}m_\omega^2 \omega^\mu \omega_\mu
\nonumber \\[3mm]
& &
-\frac{1}{4} \vec{R}^{\mu\upsilon} \vec{R}_{\mu\upsilon} 
+ \frac{1}{2}m_\rho^2\vec{\rho}^\mu\vec{\rho}_\mu-\frac{1}{4}F^{\mu\upsilon} F_{\mu\upsilon}
- m_s^2 \sigma^2 \left(\frac{k_3}{3!}\frac{g_s\sigma}{M} + \frac{k_4}{4!}\frac{g_s^2\sigma^2}{M^2} \right ) 
\nonumber \\[3mm]
& &
+\frac{1}{4!} \zeta_0 g_\omega^2 (\omega^\mu \omega_\mu)^2
- \Lambda_\omega g_\omega^2 g_\rho^2(\omega^\mu \omega_\mu)(\vec{\rho}^\mu \vec{\rho}_\mu)
\;,
\end{eqnarray}
with
\begin{eqnarray}
V^{\mu\upsilon} & = & \partial^\mu \omega^\upsilon -\partial^\upsilon \omega^\mu
\;,
\end{eqnarray}
\begin{eqnarray}
\vec{R}^{\mu\upsilon} & = & \partial^\mu \vec{\rho}^\upsilon -\partial^\upsilon\vec{\rho}^\mu
\;,
\end{eqnarray}
\begin{eqnarray}
F^{\mu\upsilon} & = & \partial^\mu A^\upsilon -\partial^\upsilon A^\mu
\;.
\end{eqnarray}
Here $\sigma$, $\omega$, and $\rho$ are the mesonic fields having masses $m_s$, $m_\omega$ and $m_\rho$
with coupling constants $g_s$, $g_\omega$ and $g_\rho$ for $\sigma-$, $\omega-$ and $\rho-$mesons, respectively.
$A^\mu$ is the photon field that is responsible for electromagnetic interaction with coupling strength
$\frac{e^2}{4\pi}$. By using the variational principle and applying mean-field approximations, the equations
of motion for the nucleon and Boson fields are obtained. Redefining fields as $\Phi = g_s\sigma $,
$W = g_\omega \omega^0$, $R = g_\rho \vec{\rho}^0$ and $A = eA^0 $, the Dirac equation corresponding to
the above Lagrangian density is
\begin{eqnarray}
\Bigg\{-i \mbox{\boldmath$\alpha$} \!\cdot\! \mbox{\boldmath$\nabla$}
& + & \beta [M - \Phi(r_{\perp},z)] + W(r_{\perp},z)
  +\frac{1}{2} \tau_3 R(r_{\perp},z) 
\nonumber \\[3mm] 
& &
+ \frac{1 +\tau_3}{2}A(r_{\perp},z)
     \bigg\}  
\varphi_i (r_{\perp},z) =
     \varepsilon_i \, \varphi_i (r_{\perp},z) \,.
     \label{eq5}
     \end{eqnarray}
The mean-field equations for $\Phi$, $W$, $R$, and $A$ are given as
\begin{eqnarray}
-\Delta \Phi(r_{\perp},z) + m_s^2 \Phi(r_{\perp},z)  & = &
g_s^2 \rho_s(r_{\perp},z)
-{m_s^2\over M}\Phi^2 (r_{\perp},z)
\left({\kappa_3\over 2}+{\kappa_4\over 3!}{\Phi(r_{\perp},z)\over M}
\right )
 \,,
 \label{eq6}  \\[3mm]
-\Delta W(r_{\perp},z) +  m_\omega^2 W(r_{\perp},z)  & = &
g_\omega^2  \rho(r_{\perp},z) 
-{1\over 3!}\zeta_0 W^3(r_{\perp},z)
-2\;\Lambda_{\omega} {R^{2}(r_{\perp},z)} W(r_{\perp},z) \,,
\label{eq7}  \\[3mm]
-\Delta R(r_{\perp},z) +  m_{\rho}^2 R(r_{\perp},z)  & = &
{1 \over 2 }g_{\rho}^2 \rho_{3}(r_{\perp},z) 
-2\;\Lambda_{\omega} R(r_{\perp},z) {W^{2}(r_{\perp},z)} \,,
\label{eq8}  \\[3mm]
-\Delta A(r_{\perp},z)   & = &
e^2 \rho_{\rm p}(r_{\perp},z)    \,.
\label{eq9}
\end{eqnarray}
The baryon, scalar, isovector, and proton densities used in the above equations are defined as
\begin{eqnarray}
\rho(r_{\perp},z) & = &
\sum_i n_i \varphi_i^\dagger(r_{\perp},z) \varphi_i(r_{\perp},z) 
\nonumber\\
        &=&\rho_{p}(r_{\perp},z)+\rho_{n}(r_{\perp},z) \,,
\label{den_nuc}
\end{eqnarray}
\begin{eqnarray}
\rho_s(r_{\perp},z) & = &
\sum_i n_i \varphi_i^\dagger(r_{\perp},z) \beta \varphi_i(r_{\perp},z)
\nonumber \\
        &=&\rho_{s p}(r_{\perp},z) +  \rho_{s n}(r_{\perp},z) \,,
\label{scalar_density}
\end{eqnarray}
\begin{eqnarray}
\rho_3 (r_{\perp},z) & = &
\sum_i n_i \varphi_i^\dagger(r_{\perp},z) \tau_3 \varphi_i(r_{\perp},z) \nonumber \\
&=& \rho_{p} (r_{\perp},z) -  \rho_{n} (r_{\perp},z) \,,
\label{isovect_density}
\end{eqnarray}
\begin{eqnarray}
\rho_{\rm p}(r_{\perp},z) & = &
\sum_i n_i \varphi_i^\dagger(r_{\perp},z) \left (\frac{1 +\tau_3}{2}
\right)  \varphi_i(r_{\perp},z).
\label{charge_density}
\end{eqnarray}
Where $\beta$ and $\tau_3$ have their usual meanings. We have taken summation
 over $i$, where $i$ stands for all nucleon. The factor $n_i$,
used in the density expressions is nothing but occupation probability which is
described in the next section. The effective mass of
nucleon due to its motion in the mean-field potential is given as
$M^{\ast}=M- \Phi (r_{\perp},z)$ and the
vector potential is
$V(r_{\perp},z)=W(r_{\perp},z)+\frac{1}{2} \tau_{3}R(r_{\perp},z)
+\frac{(1+\tau_3)}{2}A(r_{\perp},z).$
The energy densities for nucleonic, and mesonic fields corresponding to the Lagrangian density are
\begin{eqnarray}
{\cal E}_{nucl.}({r_{\perp},z}) & = &  \sum_i \varphi_i^\dagger({r_{\perp},z})
\Bigg\{ -i \mbox{\boldmath$\alpha$} \!\cdot\! \mbox{\boldmath$\nabla$}
+ \beta \left[M - \Phi (r_{\perp},z) \right] + W({r_{\perp},z})
\nonumber \\[3mm]
& &
+ \frac{1}{2}\tau_3 R({r_{\perp},z})
+ \frac{1+\tau_3}{2} A ({r_{\perp},z})\bigg\}\varphi_i (r_{\perp},z)
\;,
\label{eq11}
\end{eqnarray}
and
\begin{eqnarray}
{\cal E}_{mes.}({r_{\perp},z}) & = & \frac{1}{2g_s^2}\left [\left(\mbox{\boldmath $\nabla$}\Phi({r_{\perp},z})\right)^2
+ m_s^2 \Phi^2({r_{\perp},z})\right]
  + \left (  \frac{\kappa_3}{3!}\frac{\Phi({r_{\perp},z})}{M}
  + \frac{\kappa_4}{4!}\frac{\Phi^2({r_{\perp},z})}{M^2}\right )
\nonumber \\[3mm]
& &
\cdot   \frac{m_s^2}{g_s^2} \Phi^2({r_{\perp},z})
- \frac{1}{2g_\omega^2}\left[\left(\mbox{\boldmath $\nabla$}W({r_{\perp},z})\right)^2+{m_\omega^2} W^2 ({r_{\perp},z})\right]
\nonumber \\[3mm]
& &
- \frac{1}{2g_\rho^2}\left[\left(\mbox{\boldmath $\nabla$}R({r_{\perp},z})\right)^2+m_\rho^2 R^2 ({r_{\perp},z})\right]
-  \frac{\zeta_0}{4!} \frac{1}{ g_\omega^2 } W^4 ({r_{\perp},z})
\nonumber \\[3mm]
& & 
-\Lambda_{\omega}R^{2}(r_{\perp},z)\!\cdot\! W^{2}(r_{\perp},z)\;.
\end{eqnarray}
To solve the set of coupled differential equations \ref{eq5}$-$\ref{eq9}, 
we expand the Boson and Fermion fields in an axially-deformed
harmonic oscillator basis with $\beta_0$ as the initial deformation.
The set of equations are solved self-iteratively till the convergence is
achieved.
The center of mass correction is subtracted within the non-relativistic approximation \cite{negele}. The calculation
is extended to finite temperature T through the occupation number $n_i$ in the BCS pairing formalism.
The quadrupole deformation parameter
$\beta_2$ is estimated from the resulting quadrupole moments of the
protons and neutrons as $Q = Q_n + Q_p = \sqrt{\frac{16\pi}5} (\frac3{4\pi} AR^2\beta_2)$, (where $R=1.2A^{1/3}$fm ).
The total energy at finite temperature T is given by \cite{blunden87,reinhard89,gam90},
\begin{eqnarray}
E(T) = \sum_i \epsilon_i n_i + E_{mes.} +E_{pair} + E_{c.m.} - AM,
\end{eqnarray}
with
\begin{equation}
E_{mes.}  =  \int d^3r {\cal E}_{mes.}({r_{\perp},z}) ,
\end{equation}
\vspace{-3mm}
\begin{eqnarray}
E_{pair} = - \triangle\sum_{i>0}u_i v_i = -\frac{\triangle^2}{G},
\end{eqnarray}
\vspace{-3mm}
\begin{equation}
E_{c.m.}= -\frac{3}{4}\times 41A^{-1/3}.
\end{equation}
Here, $\epsilon_i$ is the single-particle energy, $n_i$ is the occupation
probability and $E_{pair}$ is the pairing energy obtained from the BCS formalism. The $u_i^2$ and
$v_i^2$ are the probabilities of unoccupied and occupied states, respectively.

\section{Pairing and Temperature Dependent E-RMF Formalism}\label{sec:bcs}

There are experimental evidence that even-even nuclei are more stable than even-odd or odd-odd isotopes.
The pairing correlation plays a distinct role in open-shell nuclei. The total binding energy of open-shell
nuclei deviates slightly from the experimental value when pairing correlation is not considered. To explain
this effect, Aage Bohr, Ben Mottelson, and Pines suggested BCS pairing in nuclei \cite{bohr} just after the
formulation of BCS theory for electrons in metals.  The BCS pairing in nuclei is analogous to the pairing
of electrons (Cooper pair) in super-conductors. It is used to explain the energy gap in single-particle spectrum.
The detail formalism is given in Refs. \cite{bharat17,seinthil17}, but for completeness, we are briefly highlighting
some essential part of the formalism.  The BCS pairing state is defined as:
\begin{equation}
\mid \Psi_0 >^{BCS} = \prod_{j,m>0}(u_j + 
v_j\varphi_{j,m}^{\dagger}\varphi_{j,-m}^{\dagger})\mid 0>,
\end{equation}
where j and m are the quantum numbers of the state.
In the mean-field formalism, the violation of particle number is seen due to the pairing correlation, i.e.,
the appearance of terms like $\varphi^{\dagger}\varphi^{\dagger}$ or $\varphi\varphi$, which are responsible for pairing
correlations. Thus, we neglect such type of interaction at the RMF level and taking externally the pairing
effect through the constant gap BCS pairing.
The  pairing interaction energy in terms of occupation
probabilities $v_i^2$ and $u_i^2=1-v_i^2$ (where $i$ stands for nucleon) is written
as~\cite{patra93,pres82}:
\begin{equation}
E_{pair}=-G\left[\sum_{i>0}u_i v_i\right]^2,
\end{equation}
with $G$ is the pairing force constant.
The variational approach with respect to the occupation number $v_i^2$ gives the BCS equation
\cite{pres82}:
\begin{equation}
2\epsilon_iu_iv_i-\triangle(u_i^2-v_i^2)=0,
\label{eqn:bcs}
\end{equation}
with the pairing gap $\triangle=G\sum_{i>0}u_iv_i$. The pairing gap ($\triangle$) of proton
and neutron is taken from the empirical formula \cite{gam90,va73}:
\begin{equation}
\triangle = 12 \times A^{-1/2}.
\end{equation}
The temperature introduced in the partial occupancies through the BCS approximation (here) is given by,
\begin{equation}
n_i=v_i^2=\frac{1}{2}\left[1-\frac{\epsilon_i-\lambda}{\tilde{\epsilon_i}}[1-2 f(\tilde{\epsilon_i},T)]\right],
\end{equation}
with
\begin{eqnarray}
f(\tilde{\epsilon_i},T) = \frac{1}{(1+exp[{\tilde{\epsilon_i}/T}])} ,  &  \nonumber \\[3mm]
\tilde{\epsilon_i} = \sqrt{(\epsilon_i-\lambda)^2+\triangle^2}.&
\end{eqnarray}
The function $f(\tilde{\epsilon_i},T)$ represents the Fermi Dirac distribution for quasi-particle
energy $\tilde{\epsilon_i}$. The chemical potential $\lambda_p (\lambda_n)$
for protons (neutrons) is obtained from the constraints of particle number equations
\begin{eqnarray}
\sum_i n_i^{Z}  = Z, \nonumber \\
\sum_i n_i^{N} =  N.
\end{eqnarray}
The sum is taken over all the protons and neutrons states. The entropy is obtained by,
\begin{equation}
S = - \sum_i \left[n_i\, lnn_i + (1 - n_i)\, ln (1- n_i)\right].
\end{equation}
The total energy and the gap parameter are obtained by minimizing the free energy,
\begin{equation}
F = E - TS.
\end{equation}
In constant pairing gap calculations, for a particular value of pairing gap $\triangle$
and force constant $G$, the pairing energy $E_{pair}$ diverges, if
it is extended to infinite configuration space. In fact, in all
realistic calculations with finite range forces, $\triangle$ is not
constant but decreases with large angular momenta states above the Fermi
surface. Therefore, a pairing window in all the equations is extended
up to the level $|\epsilon_i-\lambda|\leq 2(41A^{-1/3})$ as a function
of the single-particle energy. The factor of 2 has been determined to reproduce the pairing correlation energy for neutrons in $^{118}$Sn
using Gogny force \cite{gam90,patra93,dech80}.

However, for nuclei far from the line of stability, nuclei having high spin and nuclei at high temperature
or nuclei under fission, the BCS pairing is a poor approximation.
In general, the BCS approximation with a constant gap/force scheme is adopted on top of the RMF/E-RMF formalism to take care of
the pairing correlation. However, this prescription does not hold good for drip line nuclei, 
nuclei having high spin, and nuclei at high temperature or nuclei under fission as the seniority
pairing recipe fails for such exotic nuclei. This is because the coupling to the continuum in the normal BCS
approximation is not taken correctly. This deficiency can be removed to some extent by including a few quasi-bound
states owing to their centrifugal barrier \cite{chabanat98}. These quasi-bound states mock up the influence of
the continuum.  Here, we follow the procedure of Refs. \cite{patra01}. For drip line nuclei, there are no bound
single-particle levels above the Fermi surface. In surface properties calculations (Chapter \ref{chap4}), 
we take the bound-state contributions and the levels coming
from the quasi-bound states at positive energies \cite{chabanat98} and the expressions for $E_{pair}$ is written as:
\begin{eqnarray}
E_{pair} &=& -\frac{\Delta_i^2}{G_i} , \\ \nonumber
\end{eqnarray}
where $\Delta_i$ and $G_i (= C_i/A)$ are, respectively, the pairing gap and strength with $i = n, p$.
The $C_i$ are chosen in a way to reproduce the binding energy of a nucleus with mass number $A$.
For G2 set, $C_n = 21$ and $C_p = 22.5$ MeV, and for FSUGarnet, G3 and IOPB-I sets, $C_n = 19$ and $C_p = 21$ MeV.

\section{Temperature-Dependent Equation of State of Nuclear Matter and its Key Parameters}
{\label{nuclear-matter}}
In this thesis, we have used the most recent parameter sets within the E-RMF model, having contributions 
of $\delta -$ meson and photon up to 2$^{nd}$ order exponent and the rest up to 4$^{th}$ order of
exponents, which is a reasonably good approximation to predict the finite nuclei
and the nuclear matter observables up to considerable satisfaction \cite{G3}. The energy density at finite temperature,
obtained within the E-RMF Lagrangian by applying mean-field approximation, is given as:
\begin{eqnarray}
{\cal E}({r,T}) & = &  \sum_i n_i(T) \varphi_i^\dagger({r})
\Bigg\{ -i \mbox{\boldmath$\alpha$} \!\cdot\! \mbox{\boldmath$\nabla$}
+ \beta \left[M - \Phi (r) - \tau_3 D(r)\right] + W({r})
+ \frac{1}{2}\tau_3 R({r})
\nonumber \\[3mm]
& &
+ \frac{1+\tau_3}{2} A ({r})
- \frac{i \beta\mbox{\boldmath$\alpha$}}{2M}\!\cdot\!
  \left (f_\omega \mbox{\boldmath$\nabla$} W({r})
  + \frac{1}{2}f_\rho\tau_3 \mbox{\boldmath$\nabla$} R({r}) \right)
  \Bigg\} \varphi_i (r)
\nonumber \\[3mm]
& & \null
  + \left ( \frac{1}{2}
  + \frac{\kappa_3}{3!}\frac{\Phi({r})}{M}
  + \frac{\kappa_4}{4!}\frac{\Phi^2({r})}{M^2}\right )
   \frac{m_s^2}{g_s^2} \Phi^2({r})
- \frac{\zeta_0}{4!} \frac{1}{ g_\omega^2 } W^4 ({r})
\nonumber \\[3mm]
& & \null 
+ \frac{1}{2g_s^2}\left( 1 +
\alpha_1\frac{\Phi({r})}{M}\right) \left(
\mbox{\boldmath $\nabla$}\Phi({r})\right)^2
 - \frac{1}{2g_\omega^2}\left( 1 +\alpha_2\frac{\Phi({r})}{M}\right)
 \left( \mbox{\boldmath $\nabla$} W({r})  \right)^2
 \nonumber \\[3mm]
 & &  \null 
 - \frac{1}{2}\left(1 + \eta_1 \frac{\Phi({r})}{M} +
 \frac{\eta_2}{2} \frac{\Phi^2 ({r})}{M^2} \right)
  \frac{m_\omega^2}{g_\omega^2} W^2 ({r})
   - \frac{1}{2e^2} \left( \mbox{\boldmath $\nabla$} A({r})\right)^2
    \nonumber \\[3mm]
  & & \null
   - \frac{1}{2g_\rho^2} \left( \mbox{\boldmath $\nabla$} R({r})\right)^2
   - \frac{1}{2} \left( 1 + \eta_\rho \frac{\Phi({r})}{M} \right)
   \frac{m_\rho^2}{g_\rho^2} R^2({r})
   -\Lambda_{\omega}\left(R^{2}(r)\times W^{2}(r)\right)
    \nonumber \\[3mm]
  & & \null
   +\frac{1}{2 g_{\delta}^{2}}\left( \mbox{\boldmath $\nabla$} D({r})\right)^2
   +\frac{1}{2}\frac{ {  m_{\delta}}^2}{g_{\delta}^{2}}\left(D^{2}(r)\right)\;,
\label{eq1}
\end{eqnarray}
where $D$ is the delta field redefined as $D=g_\delta \delta$ with mass $m_\delta$, and coupling constant $g_\delta$. The rest 
of the symbols have the same meaning as defined in Sec \ref{ermf}. 
Using the Euler-Lagrangian equation of motion to Eq. \ref{eq1}, we get two first-order coupled differential
equations for nucleons and four second-order differential equations for the four types of the meson fields \cite{iopb1} 
(as expressed in Sec \ref{ermf}). We transformed the Dirac equation into a Shr$\ddot{o}$dinger-like
form as it is done in Ref. \cite{iopb1} by eliminating the smaller component of the Dirac spinor. Further, the equation
is solved by following the procedure of Vautherin and Brink \cite{aw} with a fourth-order Runge-Kutta
algorithm and the meson fields are solved by the Newton-Raphson method \cite{patra01,iopb1, G3}.
A brief way out to solve these equations of motion for nucleons and mesons fields is presented in Appendix \ref{append1}. 

{\bf  Energy and pressure density of infinite nuclear matter:} The temporal ($<T_{00}>$) and spatial
($<T_{ii}>$) components of the energy-momentum tensor $<T_{\mu\nu}>$ give the energy density and
pressure of the system \cite{singh14}. For static uniform infinite nuclear matter, the derivatives
of the meson field and electromagnetic interaction vanish. The expressions for energy density and
pressure at finite temperature without pairing are given by \cite{iopb1,alamT,dutraT}:
\begin{eqnarray}\label{eqn:eos1}
{\cal{E}}(T) & = &  \frac{2}{(2\pi)^{3}}\int d^{3}k \epsilon_{i}^\ast (k) (f_{i+}+f_{i-})+\rho  W+
\frac{ m_{s}^2\Phi^{2}}{g_{s}^2}\Bigg(\frac{1}{2}+\frac{\kappa_{3}}{3!}
\frac{\Phi }{M} + \frac{\kappa_4}{4!}\frac{\Phi^2}{M^2}\Bigg)
\nonumber\\
&&
 -\frac{1}{2}m_{\omega}^2\frac{W^{2}}{g_{\omega}^2}\Bigg(1+\eta_{1}\frac{\Phi}{M}+\frac{\eta_{2}}{2}\frac{\Phi ^2}{M^2}\Bigg)-\frac{1}{4!}\frac{\zeta_{0}W^{4}}
        {g_{\omega}^2}+\frac{1}{2}\rho_{3} R
 \nonumber\\
 &&
-\frac{1}{2}\Bigg(1+\frac{\eta_{\rho}\Phi}{M}\Bigg)\frac{m_{\rho}^2}{g_{\rho}^2}R^{2}-\Lambda_{\omega}  (R^{2}\times W^{2})
+\frac{1}{2}\frac{m_{\delta}^2}{g_{\delta}^{2}}\left(D^{2} \right),
        \label{eq20}
\end{eqnarray}
\begin{eqnarray}\label{eqn:eos2}
P(T) & = &  \frac{2}{3 (2\pi)^{3}}\int d^{3}k \frac{k^2}{\epsilon_{i}^\ast (k)} (f_{i+}+f_{i-})-
\frac{ m_{s}^2\Phi^{2}}{g_{s}^2}\Bigg(\frac{1}{2}+\frac{\kappa_{3}}{3!}
\frac{\Phi }{M}+ \frac{\kappa_4}{4!}\frac{\Phi^2}{M^2}  \Bigg)
\nonumber\\
& &
 +\frac{1}{2}m_{\omega}^2\frac{W^{2}}{g_{\omega}^2}\Bigg(1+\eta_{1}\frac{\Phi}{M}+\frac{\eta_{2}}{2}\frac{\Phi ^2}{M^2}\Bigg)+\frac{1}{4!}\frac{\zeta_{0}W^{4}}{g_{\omega}^2}
  \nonumber\\
& &
+\frac{1}{2}\Bigg(1+\frac{\eta_{\rho}\Phi}{M}\Bigg)\frac{m_{\rho}^2}{g_{\rho}^2}R^{2}+\Lambda_{\omega} (R^{2}\times W^{2})
-\frac{1}{2}\frac{m_{\delta}^2}{g_{\delta}^{2}}\left(D^{2}\right),
        \label{eq21}
\end{eqnarray}
with the equilibrium distribution functions defined as
\begin{eqnarray}
f_{i_\pm}=\frac{1}{1+exp[(\epsilon_i^\ast \mp \nu_i)/T};,
\end{eqnarray}
where $\epsilon_i^\ast=(k^2+M_i^{\ast 2})^{1/2}$ $(i= p,n)$, $M_{p,n}^\ast=M_{p,n}-\Phi\mp D$, $k$
is the momentum of nucleon, and the nucleon effective chemical potential is given by
\begin{eqnarray}
\nu_i=\mu_i-W-\frac{1}{2}\tau_3R ;,
\end{eqnarray}
where $\tau_3$ is the third component of the isospin operator.

{\bf Symmetry energy and incompressibility coefficient:} The binding energy per nucleon ${\cal E}/A$
= $e(\rho, \alpha)$ (where $\rho$ is the baryon density) can be expanded through Taylor series expansion method in terms
of the isospin asymmetry parameter $\alpha\left(=\frac{\rho_n-\rho_p}{\rho_n+\rho_p}\right)$:
\begin{eqnarray}
e^{NM}(\rho, \alpha) &=& \frac{{\cal E}}{\rho_{B}} - M ={e}^{NM}(\rho, \alpha = 0) \nonumber\\ 
&& + S^{NM}(\rho) \alpha^2 + {\cal O}^{NM}(\alpha^4),
{\cal O}^{NM}(\alpha^4) ,
\end{eqnarray}
where ${e}(\rho,\alpha=0)$ is the energy density of symmetric nuclear matter (SNM) and $S(\rho)$, is
the symmetry energy of the system, as defined below. Odd powers of $\alpha$ are forbidden by the
isospin symmetry. The terms proportional to $\alpha^4$ and higher-order are found to be negligible,
and
\begin{eqnarray}
S^{NM}(\rho)&=&\frac{1}{2}\left[\frac{\partial^2 {e}^{NM}(\rho, \alpha)} 
{\partial \alpha^2}\right]_{\alpha=0}.
\label{ssym}
\end{eqnarray}
Near the saturation density $\rho_0$, the symmetry energy can be expanded in a Taylor series as:
\begin{eqnarray}
S^{NM}(\rho)&=&J^{NM} + L^{NM}{\cal Y} + \frac{1}{2}K^{NM}_{sym}{\cal Y}^2 \nonumber \\
&& +\frac{1}{6}Q^{NM}_{sym}{\cal Y}^3 + {\cal O}^{NM}[{\cal Y}^4],
        \label{slda}
\end{eqnarray}
where $J^{NM}=S^{NM}(\rho_0)$ is the symmetry energy at saturation and
${\cal Y} = \frac{\rho-\rho_0}{3\rho_0}$. The coefficients $L^{NM}(\rho_0)$, $K^{NM}_{sym}(\rho_0)$,
and $Q^{NM}_{sym}$ are defined as:
\begin{eqnarray}
L^{NM}=3\rho\frac{\partial S^{NM}(\rho)}{\partial {\rho}}\bigg{|}_{ \rho=\rho_0}= \frac{3 P^{NM}}{\rho} 
\bigg{|}_{ \rho=\rho_0},\;
\label{lsym}
\end{eqnarray}
\begin{eqnarray}
K^{NM}_{sym}=9\rho^2\frac{\partial^2 S^{NM}(\rho)}{\partial {\rho}^2}\bigg{|}_{ {\rho=\rho_0}},\;
\label{ksym}
\end{eqnarray}
\begin{eqnarray}
Q^{NM}_{sym}=27\rho^3\frac{\partial^3 S^{NM}(\rho)}{\partial {\rho}^3}\bigg{|}_{ {\rho=\rho_0}}.\;
\label{qsym}
\end{eqnarray}
Here, $L^{NM}(\rho_0)$, $P^{NM}(\rho_0)$, $K^{NM}_{sym}(\rho_0)$, and $Q^{NM}_{sym}(\rho_0)$) represent 
the slope parameter of the symmetry energy, neutron pressure, symmetry energy curvature, and skew-ness coefficient 
at the saturation density, respectively. 
It is to note that these nuclear matter properties at saturation (i.e., $\rho_0$, $S(\rho_0)$) 
are model dependent and vary with certain
uncertainties. More details of these quantities and their values along with the
allowed ranges for the non-relativistic and relativistic mean-field models with various force
parameters can be found in Refs. \cite{dutra12,dutra14}.

\section{Parameter Chosen}

The solution of RMF equations depends on the parameters of the Lagrangian density. 
The parameters (meson masses, the coupling constants and the
non-linear $\sigma$ couplings) are obtained by fitting experimental data of finite nuclei and nuclear matter.
There are large number of parameter sets $\sim 265$ available in the literature \cite{iopb1,G3,dutra12}. All the forces are
designed with an aim to explain certain nuclear phenomena either in normal or in extreme conditions.
In a relativistic mean-field Lagrangian, every coupling term has its own effect to explain some
physical quantities. For example, the self-coupling terms in the $\sigma-$meson take care of the
3-body interaction which helps to explain the Coester band problem and the incompressibility coefficient
$K_{\infty}$ of nuclear matter at saturation \cite{bogu77,fujita,pie}. In the absence of these non-linear $\sigma-$couplings,
the earlier force parameters predict a large value of $K_{\infty}\sim 540$ MeV \cite{serot86,wal74}. The inclusion of non-linear 
mesonic terms and vector-isovector mesons in the Lagrangian not only explain the incompressibility $K_{\infty}$ of nuclear matter 
to a considerable range but also reproduce the nuclear bulk properties like binding energy BE, root mean square (RMS) 
charge radius $R_{ch}$, neutron skin-thickness $\Delta R$ and quadrupole deformation parameter $\beta_2$ etc. 
remarkably well for stable as well as drip-lines nuclei. The non-linear
term of the isoscalar-vector meson plays a crucial role to soften the nuclear equation of state (EOS) \cite{toki}.
By adjusting the coupling constant $\zeta_0$, one can reproduce the experimental data of the sub-saturation density \cite{toki,G3}. The
finite nuclear system is in the region of sub-saturation density and this coupling could be important
to describe the phenomena of finite nuclei.

For quite some time, the cross-coupling of the isovector-vector $\rho-$meson and the
isoscalar-vector $\omega-$meson is ignored in the calculations. Even the effective field theory motivated
relativistic mean-field (E-RMF) Lagrangian \cite{furnstahl97} does not include this term in its original
formalism. For the first time, Todd-Rutel and Piekarewicz \cite{pika05} realized the effect of this coupling
in the correlation of neutron skin-thickness and the radius of a neutron star.  This coupling constant
$\Lambda_{\omega}$ influences the neutrons' distribution radius $R_n$ without affecting many other properties
like protons' distribution radius $R_p$ or binding energy of a nucleus. The RMF parameter set without
$\Lambda_{\omega}$ coupling predicts a larger incompressibility coefficient $K_{\infty}$ than the non-relativistic
Skyrme/Gogny interactions or empirical data. However, this value of $K_{\infty}$ agree with such predictions
when the $\Lambda_{\omega}$ coupling present in the Lagrangian. Thus, the parameter $\Lambda_{\omega}$
can be used as a bridge between the non-relativistic and relativistic mean-field models. Although the
contribution of this coupling is marginal for the calculation of bulk properties of finite
nuclei, the inclusion of $\Lambda_{\omega}$ in the E-RMF formalism is important
for its softening nature to the nuclear equation of states. The inclusion of non-linear term of $\omega$ field and
cross-coupling of vector fields ($\Lambda_{\omega}$) reproduce
experimental values of GMR and IVGDR well comparative to those of NL3, and hence, they are needed for reproducing
a few nuclear collective modes \cite{pika05}. These two terms also soften both EOS of symmetric nuclear matter
and symmetry energy.

In this thesis, calculation is carried out by considering different RMF sets,
namely G3 \cite{G3}, FSUGarnet \cite{chai15}, IOPB-I \cite{iopb1} and NL3 \cite{lala97}. 
The G3 \cite{G3} is one of the latest parameter sets in the series having contribution of $\delta-$ meson too and describing 
properties of finite nuclei and infinite nuclear matter well. The NL3 is the oldest among them and one of the most successful forces
for finite nuclei all over the mass table. It produces excellent results for binding energy, charge
radius, and quadrupole deformation parameter not only for $\beta-$stability nuclei but also for nuclei away
from the valley of stability. However, NL3, being a stiffer EOS, is not a good choice to deal with the nuclear matter and is ruled out by 
gravitational-waves data on binary neutron star merger. On the other hand, the FSUGarnet is a recent parameter set \cite{chai15}. It is
seen in Ref. \cite{iopb1} that this set reproduces the neutron skin-thickness $\Delta R=R_n-R_p$ with the
recent data up to a satisfactory level along with other bulk properties. The IOPB-I is the latest in this series and
reproduces the results with an excellent agreement with the data. It is to be noted that the FSUGarnet reproduces
the neutron star mass in the lower limit, i.e., $M = 2.06 M_{\odot}$ and the IOPB-I gives the upper limit of
neutron star mass $M = 2.15 M_{\odot}$ \cite{iopb1}. These FSUGarnet and IOPB-I parameters have the additional
non-linear term of isoscalar vector meson and cross-coupling of vector mesons $\omega-, \rho-$ over NL3 set.
To our knowledge, for the first time, the
FSUGarnet and IOPB-I are used for the calculations of deform nuclei. Also, for the first time these two
sets are applied to finite temperature calculations for nuclei. The values of the parameters and their nuclear
matter properties are depicted in Table \ref{force1}. 

\renewcommand{\baselinestretch}{1.0}
\begin{table}
\caption{The FSUGarnet \cite{chai15}, IOPB-I \cite{iopb1}, G3 \cite{G3}, and NL3 \cite{lala97} parameter sets
are listed. The nucleon mass $M$ is 939.0 MeV in all the sets.  All the coupling constants are dimensionless,
except $k_3$ which is in fm$^{-1}$. The lower panel of the table shows the nuclear matter properties at
saturation density $\rho_{0}$ (fm$^{-3}$).}
\renewcommand{\tabcolsep}{0.25cm}
\renewcommand{\arraystretch}{1.25}
\begin{tabular}{cccccccccc}
\hline
\hline
\multicolumn{1}{c}{}
&\multicolumn{1}{c}{NL3}
&\multicolumn{1}{c}{FSUGarnet}
&\multicolumn{1}{c}{G3}
&\multicolumn{1}{c}{IOPB-I}\\
\hline
$m_{s}/M$  &  0.541  &  0.529&  0.559&0.533  \\
$m_{\omega}/M$  &  0.833  & 0.833 &  0.832&0.833  \\
$m_{\rho}/M$  &  0.812 & 0.812 &  0.820&0.812  \\
$m_{\delta}/M$   & 0.0  &  0.0&   1.043&0.0  \\
$g_{s}/4 \pi$  &  0.813  &  0.837 &  0.782 &0.827 \\
$g_{\omega}/4 \pi$  &  1.024  & 1.091 &  0.923&1.062 \\
$g_{\rho}/4 \pi$  &  0.712  & 1.105&  0.962 &0.885  \\
$g_{\delta}/4 \pi$  &  0.0  &  0.0&  0.160& 0.0 \\
$k_{3} $   &  1.465  & 1.368&    2.606 &1.496 \\
$k_{4}$  &  -5.688  &  -1.397& 1.694 &-2.932  \\
$\zeta_{0}$  &  0.0  &4.410&  1.010  &3.103  \\
$\eta_{1}$  &  0.0  & 0.0&  0.424 &0.0  \\
$\eta_{2}$  &  0.0  & 0.0&  0.114 &0.0  \\
$\eta_{\rho}$  &  0.0  & 0.0&  0.645& 0.0  \\
$\Lambda_{\omega}$  &  0.0  &0.043 &  0.038&0.024   \\
$\alpha_{1}$  &  0.0  & 0.0&   2.000&0.0  \\
$\alpha_{2}$  &  0.0  & 0.0&  -1.468&0.0  \\
$f_\omega/4$  &  0.0  & 0.0&  0.220&0.0 \\
$f_\rho/4$  &  0.0  & 0.0&    1.239&0.0 \\
$f_\rho/4$  &  0.0  & 0.0&    1.239&0.0 \\
$\beta_\sigma$  &  0.0  & 0.0& -0.087& 0.0  \\
$\beta_\omega$  &  0.0  & 0.0& -0.484& 0.0  \\
\hline
\hline
$\rho_{0}$ (fm$^{-3})$ &  0.148  &  0.153&  0.148&0.149  \\
$\mathcal{E}_{0}$(MeV)  &  -16.29  & -16.23 &  -16.02&-16.10  \\
$M^{*}/M$  &  0.595 & 0.578 &  0.699&0.593  \\
$J$(MeV)   & 37.43  &  30.95&   31.84&33.30  \\
$K_{\infty}$(MeV)&271.38&229.5&243.96&222.65\\
$L$(MeV)  &  118.65  &  51.04 &  49.31&63.58 \\
$K_{sym}$(MeV)  &  101.34  & 59.36 & -106.07&-37.09 \\
$Q_{sym}$(MeV)  &  177.90  & 130.93&  915.47 &862.70  \\
\hline
\hline
\end{tabular}
\label{force1}
\end{table}
\renewcommand{\baselinestretch}{1.5}

\chapter{Temperature-Dependent Bulk Properties of Finite Nuclei}{\label{chap3}} 

\rule\linewidth{.5ex}

\section{\label{sec:level1} Introduction}

Out of $\sim$ 300 known stable nuclei in nature, the bottom part of the periodic table, known as the actinide series,
encompasses the elements from Z = 89 to 103 which have applications in smoke detectors, gas mantles, as
a fuel in nuclear reactors and nuclear weapon, etc. Among the actinide, Thorium ($Th$) and Uranium ($U$) are 
the most abundant elements in nature with their isotopic fraction: 100\%\ of $^{232}Th$ and 0.0054\%\ $^{234}U$, 
0.7204\%\ $^{235}U$ and 99.2742\%\ $^{238}U$ \cite{nndc}. The isotopes $^{233}U$ and $^{239}Pu$ are 
synthesized from $^{232}Th$ and $^{238}U$, respectively by the bombardment of neutron and subsequent 
$\beta-$decay processes. 
The isotopes $^{233,235}U$ and $^{239}Pu$ are quite stable in general as long as it is not disturbed by
an almost zero energy external agent, such as a thermal neutron. Hence, these types of isotopes are called the
thermally fissile nuclei. These thermally fissile nuclei have great importance for controlled energy production
in nuclear reactors.
Due to the limited availability of these nuclei, 
we concern about whether any other heavier isotopes possessing thermally fissile characteristics exist or 
can be synthesized in a laboratory. If such isotopes exist, then it is important to analyze their fission parameters, 
which is a challenging task on both theoretical and experimental grounds. 
It is important to mention here that, Satpathy et. al. \cite{satpathy} have reported a series of $Th$ and $U$
isotopes with $N = 154-172$ to be thermally fissile on the basis of fission barrier and neutron separation energy 
\cite{satpathy,Howard}. Neutron-rich thermally fissile isotopes are an effective alternative for the production of fission energy 
more than that of naturally available thermally fissile nuclei due to the excess magnitude of neutrons. 

The neutron-rich thermally fissile nuclei undergo exotic decay mode, which is called multi-fragmentation 
fission \cite{satpathy}. In this process, along with the two heavy fission fragments, several prompt scissions 
neutrons are supposed to be released. Such properties of  
 nuclei may have a significant role in stellar evolution and $r-$process nucleosynthesis. The estimation of various 
nuclear cross-section $i.e.$, the measure of production probability of these neutron-rich thermally fissile 
$Th$ and $U$ isotopes may be useful for energy generation in future \cite{Panda}. 
Due to the significance of neutron-rich thermally fissile nuclei for energy production, 
it is important to investigate their structural and reaction properties. 
Moreover, before occurring fission in naturally available thermally fissile $^{233,235}U$ and $^{239}Pu$ nuclei, they form 
compound nuclei $^{234,236}U$ and $^{240}Pu$ after absorbing slow neutrons.
The compound nuclei ($^{234,236}U$ and $^{240}Pu$) oscillate in different modes (quadrupole, hexadecapole) 
of vibrations and finally
reaches to the scission point. In the process, the compound nucleus exhibits various stages including an increase 
in temperature (T). To understand the fission dynamics, it is important to study the nuclear properties, 
like nuclear excitation energy $E^*$, change in shapes and sizes of a nucleus, the variation of specific heat $C$, 
effect of shell structure, change of single-particle energy, and inverse level density parameter. All these 
observables are crucial quantities to understand the fission phenomena and we aim to analyze these properties 
with temperature.   

Recently, the relative mass distribution of thermally fissile nuclei for binary \cite{bharat17} and 
ternary \cite{seinthil17} fission processes are reported. Here, it is shown that the relative yield of 
fission fragments depends very much on the temperature of the system. The level density parameter is also 
influenced a lot by temperature, which is a key quantity in fission study. The neutron skin-thickness 
has a direct correlation with the equation of state (EOS) of nuclear matter
\cite{brown,horo86}. It is to be noted that the neutron star EOS is the main ingredient that is used to predict 
the mass and radius of the star. 
The asymmetry energy coefficient $a_{sym}$ is an important quantity for various nuclear properties, such as
to establish proper boundaries for neutron and proton drip lines, a study of heavy ion collision, 
physics of supernovae explosions and neutron star \cite{baran05,lattimer07,baron85}. Thus, it attracts the attention for the 
analysis of neutron skin-thickness and asymmetry energy coefficient as a function of temperature.

To study the fission process, a large number of models have been proposed
\cite{fong71,chenya88,lest04,fong56,maran09,toke,bord,rama85,swiat,diehl74,wilkins}.
The liquid drop model successfully explains the fission of a nucleus \cite{diehl74,wilkins,pauli}
and the semi-empirical mass formula is  the simple oldest tool to get a rough 
estimation of the energy released in a binary fission process.
Most of the time, the liquid drop concept is applied to study the fission 
phenomenon, where the shell effect of the nucleus generally ignored. But, the shell effect plays
a vital role in the stability of the nucleus not only at T=0 but also at finite temperature. This shell 
structure is considered to be responsible for the formation of superheavy nuclei in the {\it superheavy island}.
Thus, the microscopic model could be a better framework for such types of studies, where the shell structure 
of the nucleus is included automatically. In this aspect the Hartree or Hartree-Fock approach of non-relativistic 
mean-field \cite{bender} or relativistic mean-field (RMF) \cite{bogu77} formalisms could be some of the ideal theories. 

The pioneering work of Vautherin and Brink \cite{vat70}, who has applied the Skyrme interaction in a self-consistent method
for the calculation of ground-state properties of finite nuclei opened a new dimension in the quantitative estimation of 
nuclear properties. Subsequently, the Hartree-Fock and time-dependent Hartree-Fock
formalisms \cite{pal} are also implemented to study the properties of fission. Most recently, the microscopic
relativistic mean-field approximation, which is another successful theory in nuclear physics is
used for the study of nuclear fission \cite{skp10}. 
The detailed description of RMF formalism and the advantages of using it are given in Chapter \ref{chap2}. 
In this work, we have applied the recently developed FSUGarnet \cite{chai15}, and IOPB-I \cite{iopb1} parameter sets and 
NL3 set~\cite{lala97} (for the sake of comparison) within the framework of temperature-dependent relativistic mean-field (TRMF) formalism.
In this chapter, we have investigated the ground as well as the excited state properties of $^{234}U$, $^{236}U$, $^{240}Pu$, and 
neutron-rich thermally fissile $^{244-262}Th$ and $^{246-264}U$ nuclei. These properties are the following: shell correction to energy $\delta 
E_{shell}$, excitation energy $E^*$, variation of specific heat $C$, level density parameter $a$, deformation parameters, and asymmetry energy 
coefficient $a_{sym}$. Along with these nuclei, we have taken $^{208}Pb$ as a
representative case to examine our calculation for spherical nuclei.

\section{Ground-State Properties of Nuclei}

\renewcommand{\baselinestretch}{1.0}
\begin{table}
\caption{The ground-state binding energy (MeV), charge radius (fm), and neutron skin-thickness (fm)
for $^{208}Pb$, $^{227,228,229,230,232}Th$, $^{233,234,235,236,238}U$, and $^{240}Pu$ corresponding to NL3 \cite{lala97}, 
FSUGarnet \cite{chai15}, and IOPB-I \cite{iopb1} sets are compared with the experimental data \cite{nndc,Angeli2013}.}
\scalebox{1.0}{
\begin{tabular}{cccccccccc}
\hline
\hline
                        \multicolumn{1}{c}{Nucl.}
                        &\multicolumn{1}{c}{Obs.}
                        &\multicolumn{1}{c}{NL3}
                        &\multicolumn{1}{c}{FSUGarnet}
                        &\multicolumn{1}{c}{IOPB-I}
                        &\multicolumn{1}{c}{Exp.}\\
                        \hline
				  & BE          &1639.04&1639.04&1639.04&1636.96\\
                        $^{208}Pb$& $R_{\rm c}$&5.520  &5.550  &5.580  &5.500\\
 				  &$\Delta{R}$  &0.278  &0.162  &0.225  &-\\\hline
                        	  & BE		&1739.9 &1739.5 &1741.9 &1735.9\\
                        $^{227}Th$& $R_{\rm c}$&5.749 &5.757 &5.788 &5.740\\
                        	  &$\Delta{R}$  &0.272 &0.160 &0.207 &-\\
                        	  &BE        	&1745.5&1744.7&1747.2&1743.1\\
                        $^{228}Th$& $R_{\rm c}$&5.749&5.757&5.796&5.749\\
                        &$\Delta{R}$&0.280&0.155&0.211&-\\
                        &BE&1751.0&1749.8&1752.4&1748.2\\
                        $^{229}Th$& $R_{\rm c}$&5.757&5.764&5.804&5.756\\
                        &$\Delta{R}$&0.287&0.161&0.217&-\\
                        &BE&1756.3&1754.9&1757.5&1754.9\\
                        $^{230}Th$& $R_{\rm c}$&5.764&5.772&5.811&5.7670\\
                        &$\Delta{R}$&0.295&0.165&0.223&-\\
                        &BE&1766.7&1765.0&1767.3&1766.7\\
                        $^{232}Th$& $R_{\rm c}$&5.778&5.798&5.836&5.785\\
                        &$\Delta{R}$&0310&0.178&0.235&-\\\hline
                        &BE&1771.768&1772.130&1774.213&1771.499\\
                        $^{233}U$&$R_{\rm c}$&5.836&5.848&0266&5.820\\
                        &$\Delta{R}$&0.266&0.153&0.205&-\\
                        &BE&1777.400&1778.218&1780.487&1778.400\\
                        $^{234}U$&$R_{\rm c}$&5.841&5.839&5.878&5.829\\
                        &$\Delta{R}$&0.273&0.157&0.159&-\\
                        &BE&1783.450&1784.020&1783.450&1783.650\\
                        $^{235}U$&$R_{\rm c}$&5.848&5.850&5.848&5.834\\
                        &$\Delta{R}$&0.281&0.219&0.281&-\\
                        &BE&1789.080&1789.627&1792.880&1790.296\\
                        $^{236}U$&$R_{\rm c}$&5.855&5.860&5.901&5.843\\
                        &$\Delta{R}$&0.288&0.166&0.225&-\\
                        &BE&1799.973&1800.344&1803.103&1801.660\\
                        $^{238}U$&$R_{c}$&5.869&5.878&5.919&5.857\\
                        &$\Delta{R}$&0.303&0.176&0.236&-\\\hline
				  & BE		&1812.4&1814.5&1816.8&1814.7\\
			$^{240}Pu$& $R_{\rm c}$&5.902 &5.911 &5.953 &5.874\\
				  &$\Delta{R}$  &0.272  &0.160  &0.207  &-\\
\hline
\hline
\end{tabular}}
\label{table2}
\end{table}
\renewcommand{\baselinestretch}{1.5}

Before proceeding to the calculation at finite temperature, first, we have calculated the ground-state 
properties of $^{208}Pb$, known $U$ and $Th$ isotopes, and $^{240}Pu$ using NL3, FSUGarnet, and IOPB-I parameter sets. The 
calculated results are compared with the experimental data wherever available \cite{nndc,Angeli2013}. 
The binding energy (BE), charge radius ($R_c$), and neutron skin-thickness ($\Delta R$) for the nuclei 
$^{208}Pb$, $^{227,228,229,230,232}Th$, $^{233,234,235,236,238}U$, and $^{240}Pu$ are shown in Table \ref{table2}. The neutron skin-thickness 
is defined as $\Delta R=R_n-R_p$, where $R_n$ and $R_p$ are the roots mean square radius of neutrons and 
protons distribution, respectively. From the table, it is clear that all the parameter sets reproduce the 
experimental data remarkably well. After getting a good agreement with the data, we have further 
proceeded to use our selected sets to calculate the bulk properties of experimentally unknown neutron-rich 
thermally fissile nuclei. The calculated $BE$, $R_{c}$, and $\Delta{R}$ for the neutron-rich thermally 
fissile $^{244-262}Th$ and $^{246-264}U$ nuclei are shown in Table \ref{table3}. It is clear from the table that 
all the parameter sets predict almost similar results for the binding energy, charge radius $R_{c}$, 
and the neutron skin-thickness $\Delta{R}$. A further inspection of Table \ref{table3} shows that except $^{244}Th$ 
and $^{246-250}U$, the $BE$ predicted by the NL3 set are larger than the results obtained 
by the other two sets. As expected, the $\Delta{R}$ values are larger for the NL3 set. 
Furthermore, the values of $R_{c}$ with IOPB-I are found to be larger than the prediction
of FSUGarnet and NL3 sets. In general, all the observables obtained by these three forces are in  
excellent agreement with each other. After getting success for these sets in predicting the ground-state 
properties of neutron-rich thermally fissile nuclei, we extend our investigations at finite T which are
presented in the following sections. 
\renewcommand{\baselinestretch}{1.0}
\begin{table*}
\caption{The ground-state binding energy (MeV), charge radius (fm), and the neutron skin-thickness (fm) 
for $^{244-262}Th$ and $^{246-264}U$ with NL3 \cite{lala97}, FSUGarnet \cite{chai15}, and IOPB-I \cite{iopb1}  
force parameters.}
\scalebox{0.7}{
\begin{tabular}{|ccccc|ccccc|}
\hline
Nucleus&Observables&NL3&FSUGarnet&IOPB-I&Nucleus&Observables&NL3&FSUGarnet&IOPB-I\\\hline
		&BE&-1823.7&-1819.3&-1824.0&
		&BE&-1838.6&-1837.2&-1841.1\\
		$^{244}Th$&$R_c$&5.860&5.903&5.946&
		$^{246}U$&$R_c$&5.917&5.931&5.975\\
		&$\Delta{R}$&0.404&0.234&0.308&&
		$\Delta{R}$&0.363&0.214&0.284\\
		& BE&-1827.9&-1823.0&-1827.9&
		&BE&-1843.0&-1841.2&-1845.3\\
		$^{245}Th$&$R_c$&5.867&5.908&5.957&
		$^{247}U$&$R_c$&5.923&5.936&5.980\\
		&$\Delta{R}$&0.411&0.240&0.314&&
		$\Delta{R}$&0.370&0.219&0.290\\
		& BE&-1832.1&-1826.6&-1831.7&
		&BE&-1847.3&-1845.1&-1849.4\\
		$^{246}Th$&$R_c$&5.873&5.913&5.956&
		$^{248}U$&$R_c$&5.928& 5.941&5.985\\
		&$\Delta{R}$&0.418&0.245&0.320&&
		$\Delta{R}$&0.377&0.225&0.296\\
		& BE&-1836.2&-1830.1&-1835.4&
		&BE&-1851.4&-1848.9&-1853.4\\
		$^{247}Th$&$R_c$&5.880&5.917&5.961&
		$^{249}U$&$R_c$&5.933& 5.945&5.989\\
		&$\Delta{R}$&0.425&0.250&0.327&&
		$\Delta{R}$&0.384&0.230&0.302\\
		& BE&-1840.2&-1833.5&-1839.1&
		&BE&-1855.5&-1852.7&-1857.4\\
		$^{248}Th$&$R_c$&5.886&5.921&5.965&
		$^{250}U$&$R_c$&5.937&5.949&5.993\\
		&$\Delta{R}$&0.431&0.256&0.333&&
		$\Delta{R}$&0.391&0.235&0.309\\
		& BE&-1844.2&-1836.9&-1842.6&
		&BE&-1861.3&-1856.3&-1861.2\\
		$^{249}Th$&$R_c$&5.892&5.925&5.969&
		$^{251}U$&$R_c$&5.913&5.952&5.997\\
		&$\Delta{R}$&0.438&0.261&0.339&&
		$\Delta{R}$&0.406&0.240&0.315\\
		& BE&-1848.1&-1840.2&-1846.1&
		&BE&-1865.5&-1859.9&-1865.0\\
		$^{250}Th$&$R_c$&5.898&5.928&5.972&
		$^{252}U$&$R_c$&5.919&5.956& 5.600\\
		&$\Delta{R}$&0.444&0.266&0.346&&
		$\Delta{R}$&0.413&0.245&0.321\\
		& BE&-1852.0&-1843.4&-1849.6&
		&BE&-1869.7&-1863.4&-1868.8\\
		$^{251}Th$&$R_c$&5.904&5.932&5.975&
		$^{253}U$&$R_c$&5.925&5.958&6.003\\
		&$\Delta{R}$&0.451&0.271&0.352&&
		$\Delta{R}$&0.419&0.250&0.327\\
		& BE&-1855.7&-1846.5&-1853.0&
		&BE&-1873.8&-1866.9&-1872.4\\
		$^{252}Th$&$R_c$&5.910&5.935&5.978&
		$^{254}U$&$R_c$&5.931&5.961&6.005\\
		&$\Delta{R}$&0.457&0.276&0.359&&
		$\Delta{R}$& 0.426&0.255&0.334\\
		& BE&-1859.4&-1849.6&-1856.3&
		&BE&-1877.8&-1870.3&-1876.1\\
		$^{253}Th$&$R_c$&5.916&5.938&5.981&
		$^{255}U$&$R_c$&5.936&5.964&6.008\\
		&$\Delta{R}$&0.467&0.282&0.365&&
		$\Delta{R}$&0.432&0.260&0.340\\
		& BE&-1863.1&-1852.7&-1859.7&
		&BE&-1881.8&-1873.7&-1879.7\\
		$^{254}Th$&$R_c$&5.922&5.941&5.984&
		$^{256}U$&$R_c$&5.942&5.967&6.010\\
		&$\Delta{R}$&0.470&0.287&0.371&&
		$\Delta{R}$&0.438&0.265&0.346\\
		& BE&-1866.7&-1855.7&-1862.9&
		&BE&-1885.8&-1877.0&-1883.3\\
		$^{255}Th$&$R_c$&5.927&5.944&5.987&
		$^{257}U$&$R_c$&5.948&5.970&6.013\\
		&$\Delta{R}$&0.476&0.292&0.378&&
		$\Delta{R}$&0.445&0.270&0.352\\
		& BE&-1870.2&-1858.6&-1866.1&
		&BE&-1889.7&-1880.3&-1886.8\\
		$^{256}Th$&$R_c$&5.932&5.947&5.990&
		$^{258}U$&$R_c$&5.953&5.973&6.016\\
		&$\Delta{R}$&0.483&0.297&0.384&&
		$\Delta{R}$&0.451&0.276&0.358\\
		& BE&-1873.6&-1861.5&-1869.3&
		&BE&-1893.5&-1883.5&-1890.3\\
		$^{257}Th$&$R_c$&5.938&5.951&5.993&
		$^{259}U$&$R_c$&5.959&5.976&6.018\\
		&$\Delta{R}$&0.489&0.303&0.390&&
		$\Delta{R}$&0.457&0.281&0.365\\
		& BE&-1877.0&-1864.3&-1872.4&
		&BE&-1897.3&-1886.6&-1893.7\\
		$^{258}Th$&$R_c$&5.943&5.954&5.996&
		$^{260}U$&$R_c$&5.964&5.979&6.021\\
		&$\Delta{R}$&0.496&0.308&0.397&&
		$\Delta{R}$&0.463&0.286&0.371\\
		& BE&-1880.3&-1867.1&-1875.5&
		&BE&-1901.0&-1889.7&-1897.1\\
		$^{259}Th$&$R_c$&5.948&5.957&5.999&
		$^{261}U$&$R_c$&5.970&5.982&6.024\\
		&$\Delta{R}$&0.502&0.313&0.404&&
		$\Delta{R}$&0.469&0.291&0.377\\
		& BE&-1883.5&-1869.8&-1878.5&
		&BE&-1904.7&-1892.8&-1900.4\\
		$^{260}Th$&$R_c$&5.953&5.960&6.002&
		$^{262}U$&$R_c$&5.980&5.985&6.027\\
		&$\Delta{R}$&0.508&0.319&0.410&&
		$\Delta{R}$&0.482&0.296&0.384\\
		& BE&-1885.6&-1872.5&-1881.5&
		&BE&-1908.3&-1895.8&-1903.7\\
		$^{261}Th$&$R_c$& 5.954&5.963&6.005&
		$^{263}U$&$R_c$&5.980&5.987&6.029\\
		&$\Delta{R}$&0.516&0.324332&0.417&&
		$\Delta{R}$&0.482&0.301&0.390\\
		& BE&-1888.8&-1875.1&-1884.4&
		&BE&-1911.8&-1898.7&-1907.1\\
		$^{262}Th$&$R_c$&5.959&5.966&6.008&
		$^{264}U$&$R_c$&5.985&5.989&6.030\\
		&$\Delta{R}$&0.523&0.330&0.423&&
		$\Delta{R}$&0.488&0.307&0.396\\\hline
\end{tabular}}
\label{table3}
\end{table*}
\renewcommand{\baselinestretch}{1.5}
%
%
\section{The Properties of Nuclei at Finite Temperature}
\subsection{Shell Melting Point and Strutinsky Shell Correction Energy}
\begin{figure}
	\includegraphics[width=0.9\columnwidth,height=8.0cm]{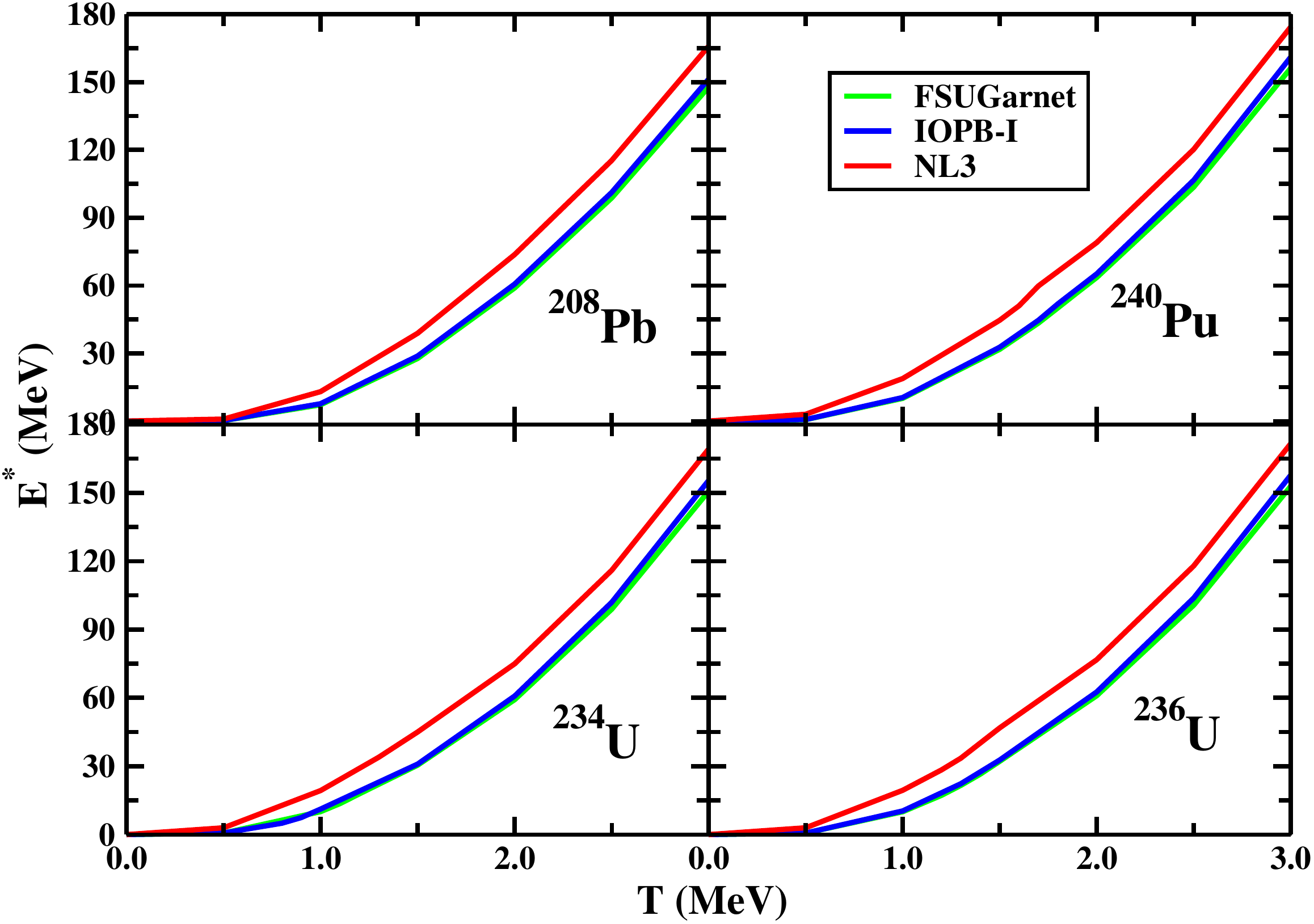}
	\caption{The excitation energy $E^*$ as a function of 
        temperature T for $^{208}Pb$, $^{234}U$, $^{236}U$, and $^{240}Pu$ with FSUGarnet, IOPB-I, and NL3 parameter
        sets.}
\label{exc0}
\end{figure}
\begin{figure}
\includegraphics[width=0.9\columnwidth,height=9.0cm]{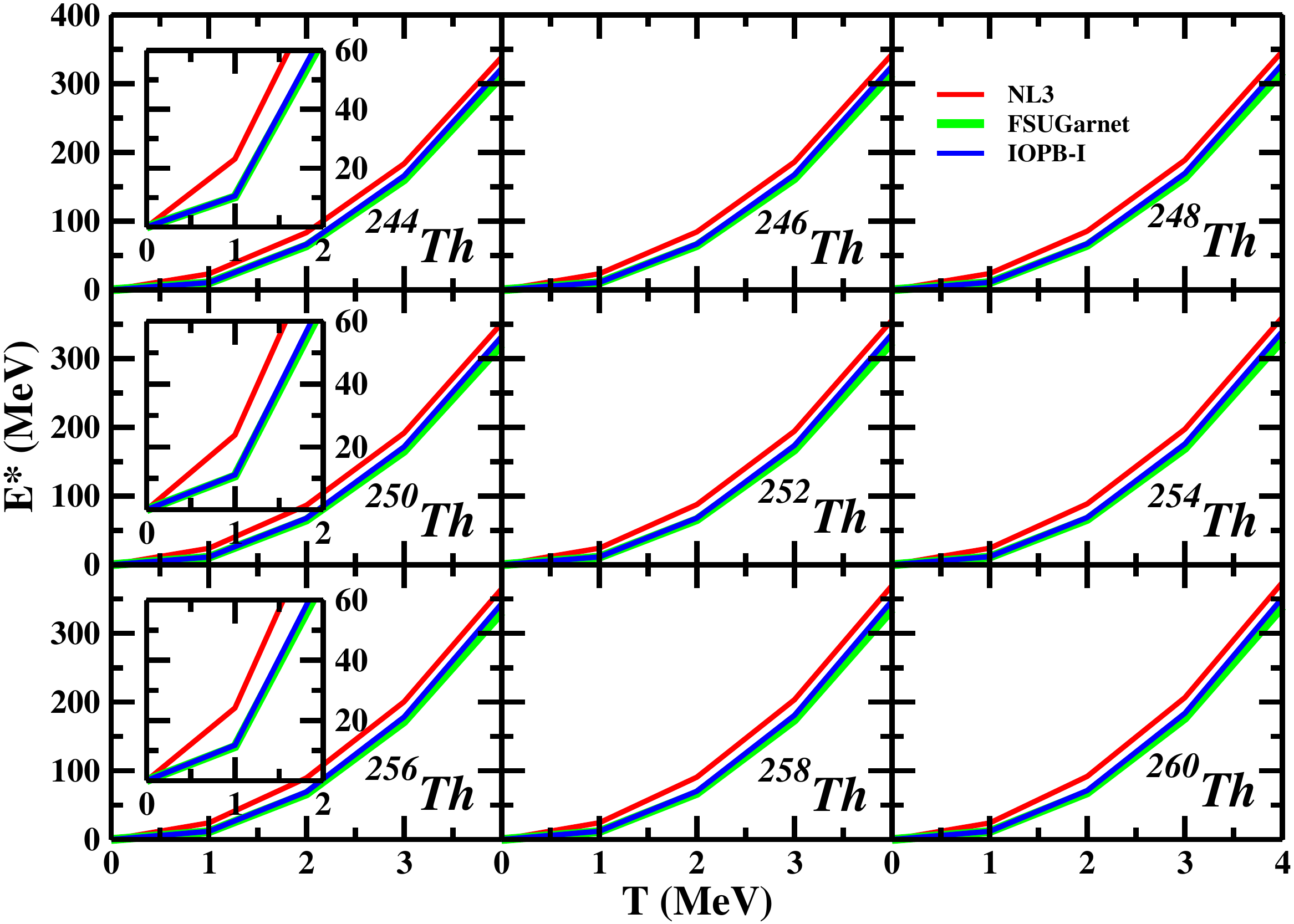}
\caption{Same as Figure \ref{exc0} but for 
$^{244,246,248,250,252,254,256,258,260}Th$.}
\label{exc1}
\end{figure}

\begin{figure}
\includegraphics[width=0.9\columnwidth,height=9.0cm]{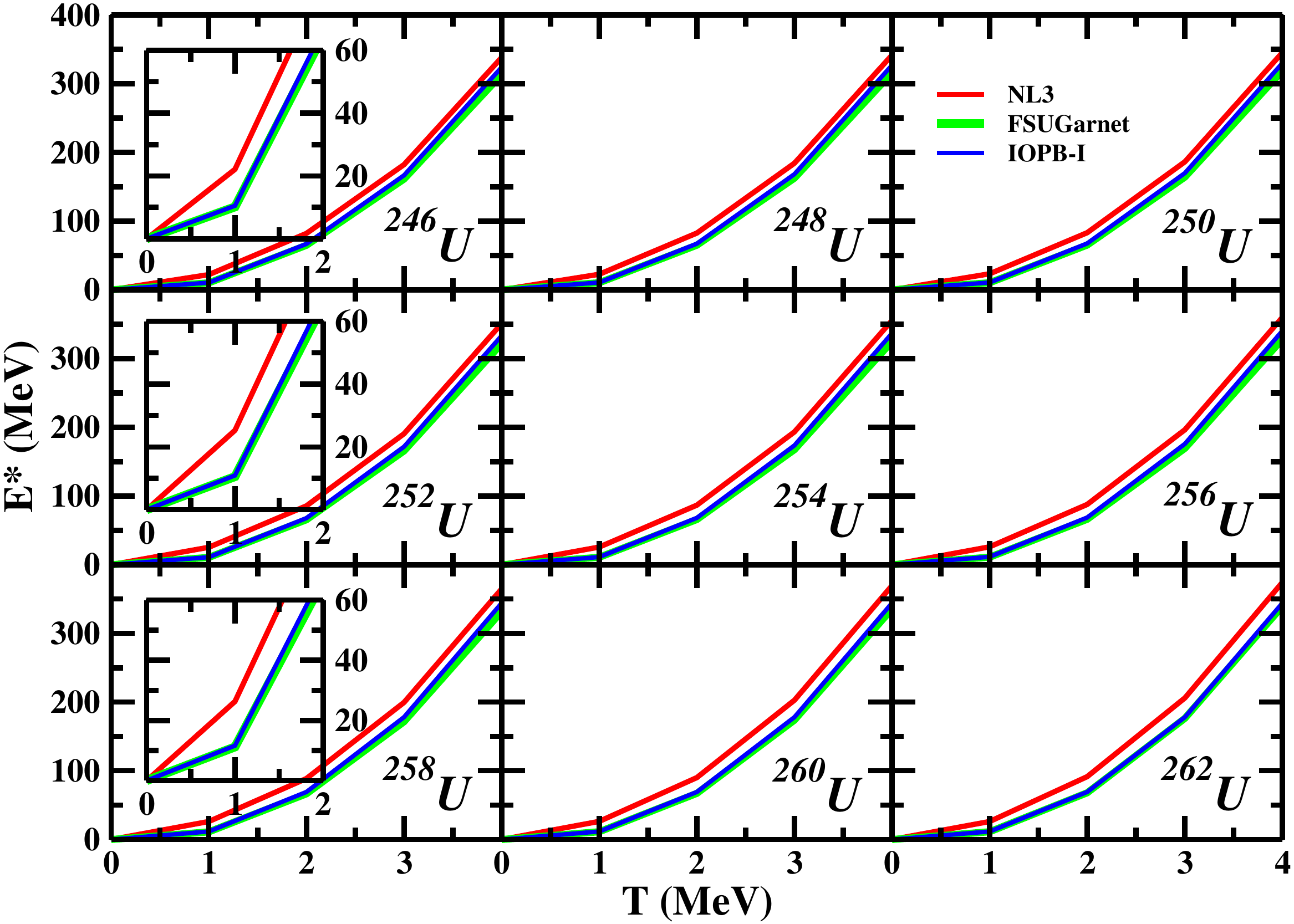}
\caption{Same as Figure \ref{exc0} but for $^{246,248,250,252,254,256,258,260,262}U$.}
\label{exc2}
\end{figure}

The nuclear excitation energy $E^*$ is one of the key quantities in fission dynamics. The excitation energy
very much depends on the state of the nucleus. It is defined as how far a nucleus is excited from its ground-state 
and can be measured from the relation $E^*=E(T)-E(T=0)$, where $E(T)$ is the binding energy of a 
nucleus at finite T and E(T=0) is the ground-state binding energy. The variation of $E^*$ as a function of T is 
shown in Figure~\ref{exc0} for $^{208}Pb$, $^{234,236}U$, and $^{240}Pu$ with NL3, FSUGarnet, and IOPB-I sets.  
The same is plotted for $^{244,246,248,250,252,254,256,258,260}Th$, and $^{246,248,250,252,254,256,258,260,262}U$ in 
Figures \ref{exc1}, and \ref{exc2}, respectively. 
One can see from the figures that the variation of excitation energy is almost quadratic satisfying the relation $E^* = aT^2$. 
similar to the binding energy, the results of $E^*$ for FSUGarnet and IOPB-I 
coincide with each other, but smaller than the predicted values of the NL3 set.  

\begin{figure}
        \includegraphics[width=0.9\columnwidth,height=9.0cm]{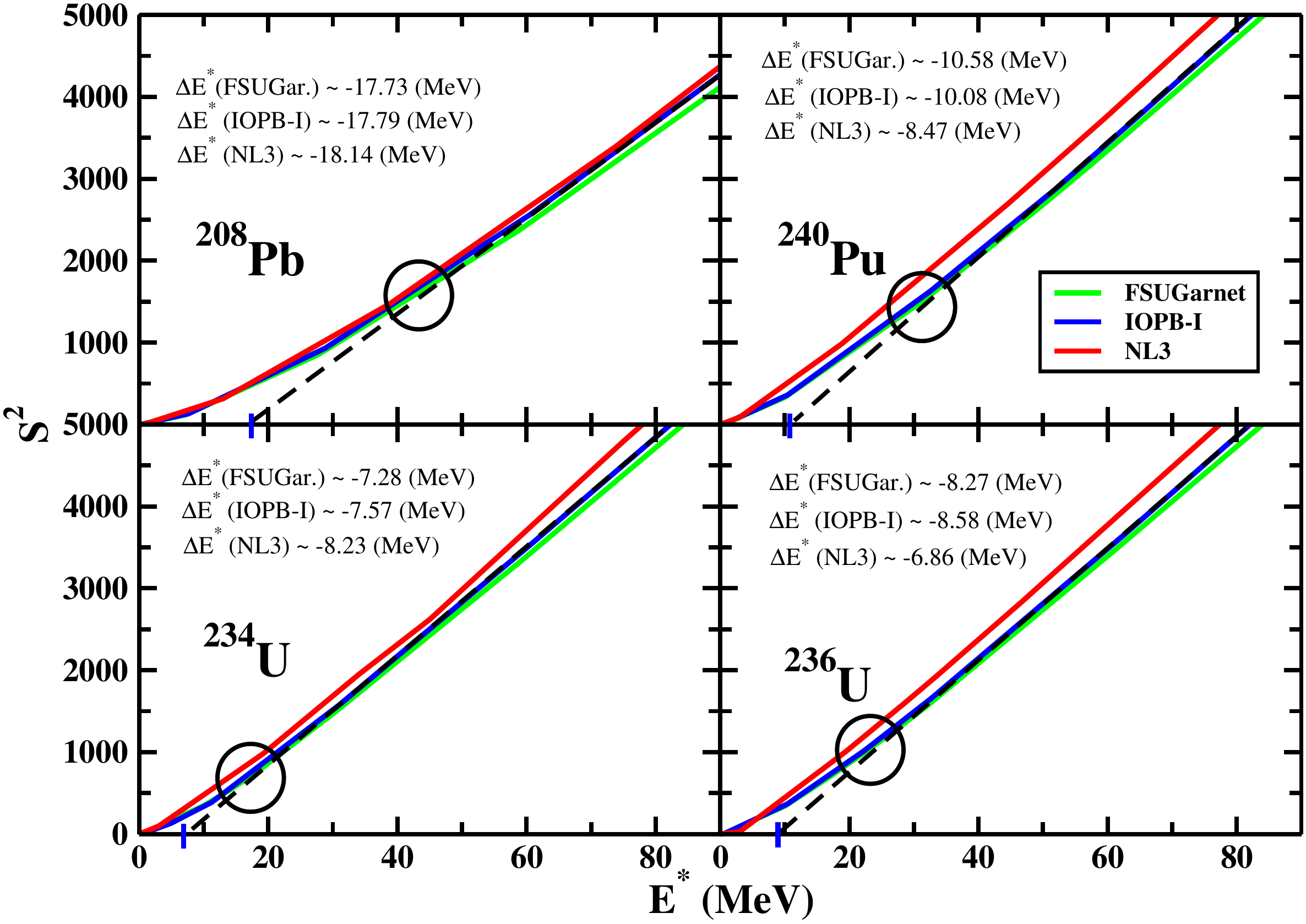}
        \caption{The square of the entropy $S^2$ versus excitation
        energy $E^*$ for $^{208}Pb$, $^{234}U$, $^{236}U$, and $^{240}Pu$ with FSUGarnet, IOPB-I, and NL3 parameter
        sets. The circle marked in the curve shows the shell melting point.}
\label{s2e0}
\end{figure}
\begin{figure}
\includegraphics[width=0.9\columnwidth,height=9.0cm]{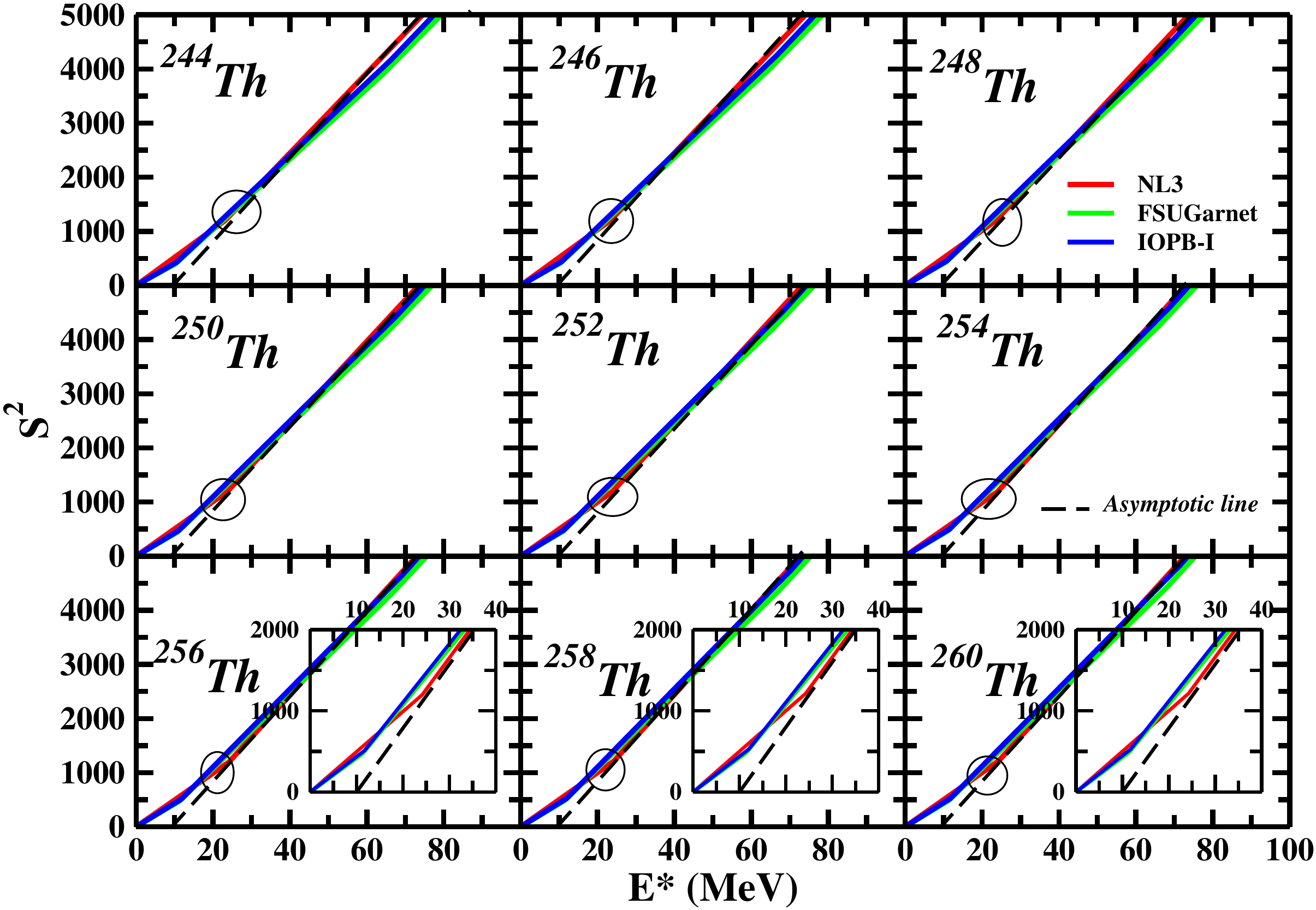}
\caption{Same as Figure \ref{s2e0}, but for
$^{244,246,248,250,252,254,256,258,260}Th$.}
\label{s2e1}
\end{figure}
\begin{figure}
\includegraphics[width=0.9\columnwidth,height=9.0cm]{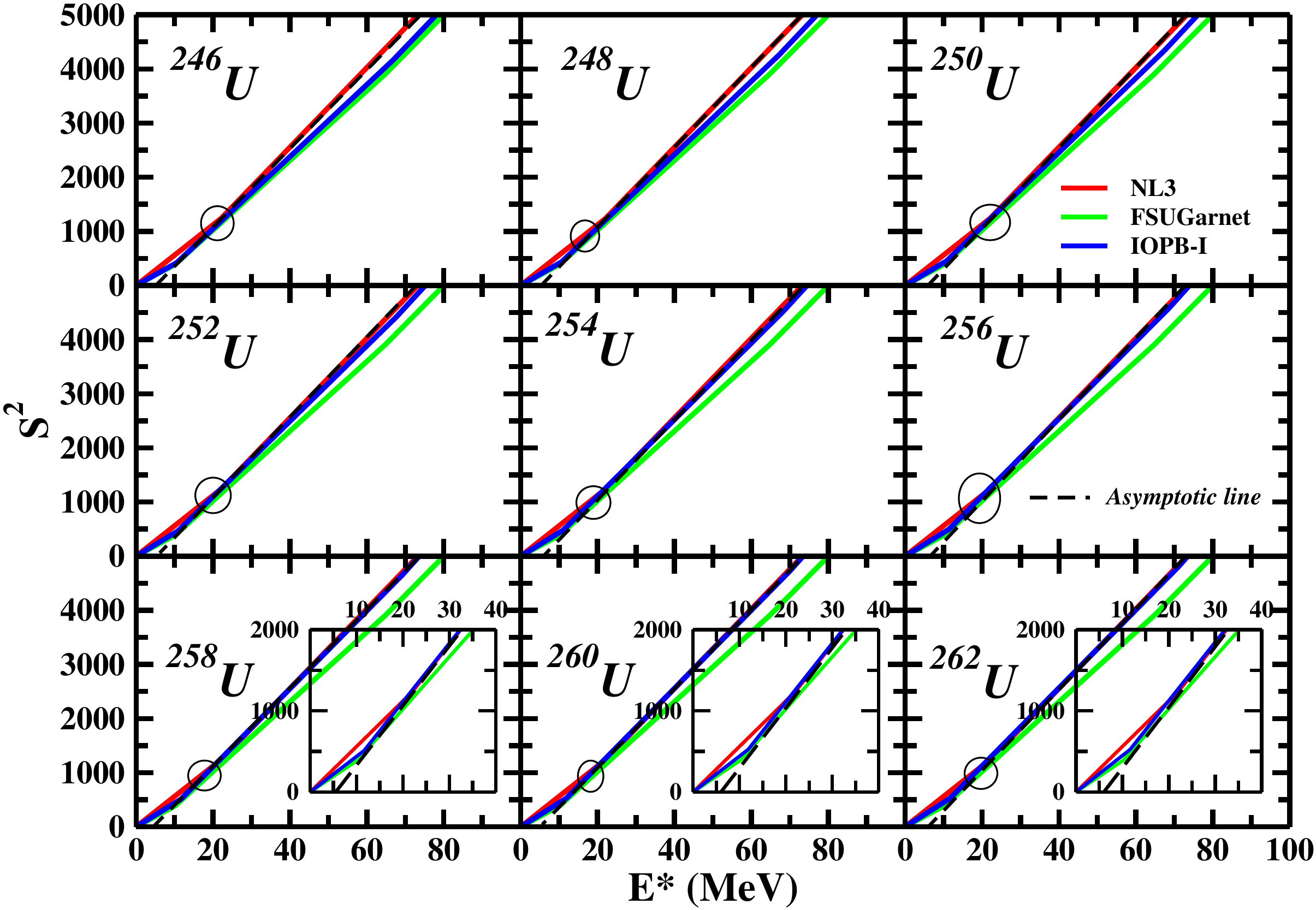}
\caption{Same as Figure \ref{s2e0}, but for $^{246,248,250,252,254,256,258,260,262}U$.}
\label{s2e2}
\end{figure}

The excitation energy $E^*$ has a direct relation
with the entropy S, i.e., the disorderliness of the system. The expected relation of S with $E^*$ and the level 
density parameter $a$ from Fermi gas model is $S^2=4aE^*$ \cite{rama70}. 
However, this straight-line relation of $S^2$ versus $E^*$
deviates at low excitation energy due to the shell structure of nucleus \cite{rama70}. The values of $S^2$
as a function of $E^*$ are shown in Figure~\ref{s2e0}-\ref{s2e2} for some of the nuclei as a representative case. The intercept of the curve 
on the $E^*-$axis is a measure of shell correction energy to a nucleus. Thus, the actual relation of $E^*$ with $S^2$ 
can be written as $S^2=4a(E^*\pm\triangle{E^*})$, where $\pm\triangle{E^*}=$ shell correction energy. Beyond these ``slope
points" the $S^2$ versus $E^*$ curve increase in a straight line as shown in the figures (the slope point is marked by a circle). 
Thus, one can interpret that beyond this particular
excitation energy a nucleus as a whole does not have a shell structure, and this point can be considered as the melting
point of the shell in the nucleus. The value of this point depends on the ground-state shell structure of a
nucleus. The experimental evidence of washing out of the shell effects at around 40 MeV excitation energy has also been pointed out in Ref 
\cite{Chaudhuri15}. Shell correction for some of the considered nuclei, obtained from the intercepts on the $ E^*-$ axis, are depicted 
in Figures~\ref{s2e0}-\ref{s2e2} corresponding to parameters set considered here. 
For example, shell correction energies are $\sim -17.79$, $-7.57$, $-8.58$, and $-10.08$ MeV 
for $^{208}Pb$, $^{234,236}U$, and $^{240}Pu$ respectively, corresponding to IOPB-I parameter set. 
The values for all three parameter sets are almost same with a little difference in NL3 model (see Figure~\ref{s2e0}-\ref{s2e2}).  
The shell correction energies of the nuclei are further calculated within the Strutinsky averaging method.

\begin{figure}
	\includegraphics[width=0.9\columnwidth,height=8.0cm]{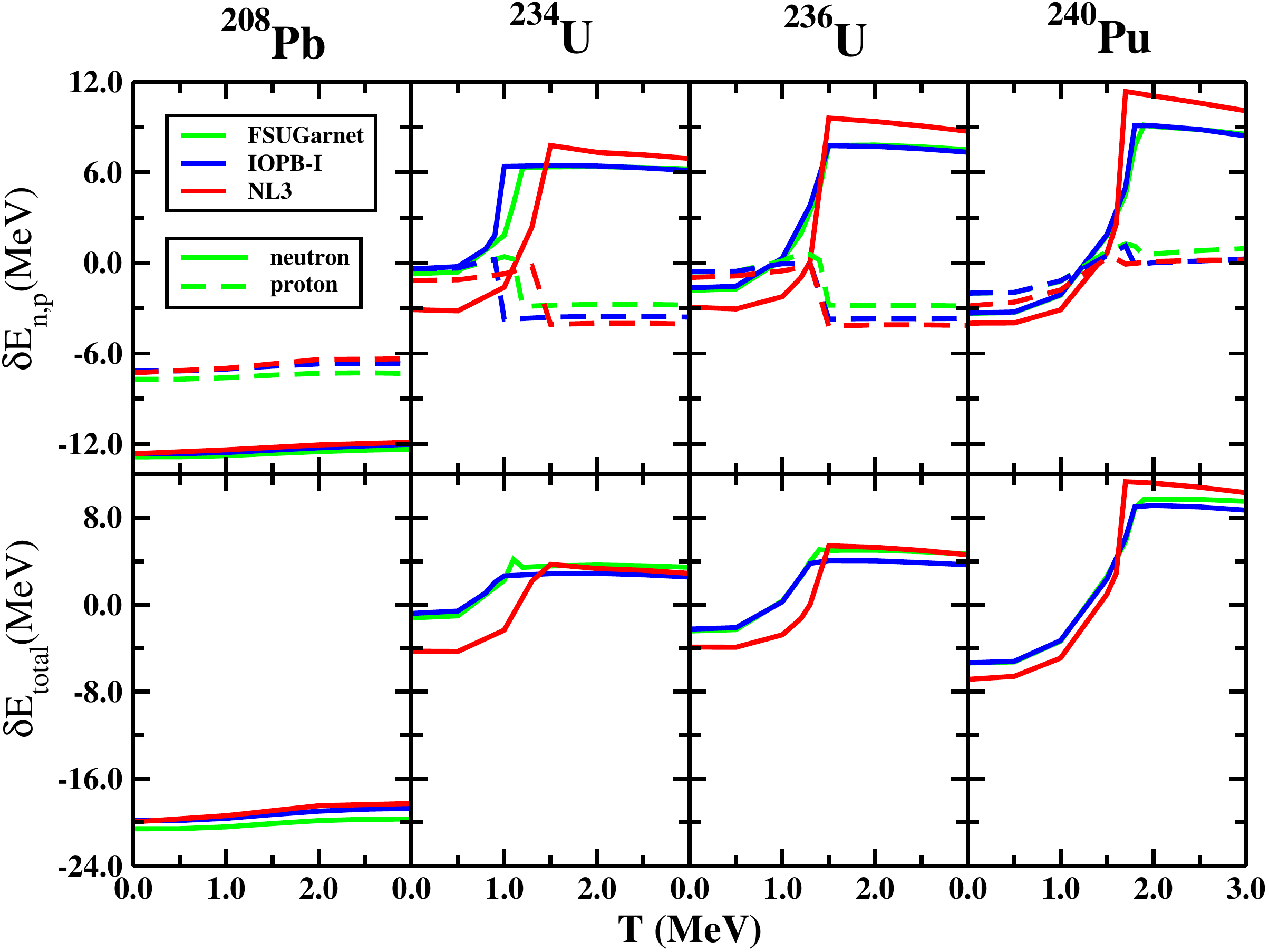}
	\caption{The nuclear shell correction to the energy $\delta{E_{shell}}$ 
as a function of temperature T for the nuclei $^{208}Pb$, $^{234}U$, $^{236}U$, and $^{240}Pu$ with FSUGarnet, IOPB-I, and NL3
        parameter sets. The bold and dot-dashed lines in the upper row represent the shell correction for neutrons and protons, respectively. 
The lower row represents the total $\delta{E_{shell}}$.}
\label{shc0}
\end{figure}
\begin{figure}
\includegraphics[width=0.9\columnwidth,height=9.0cm]{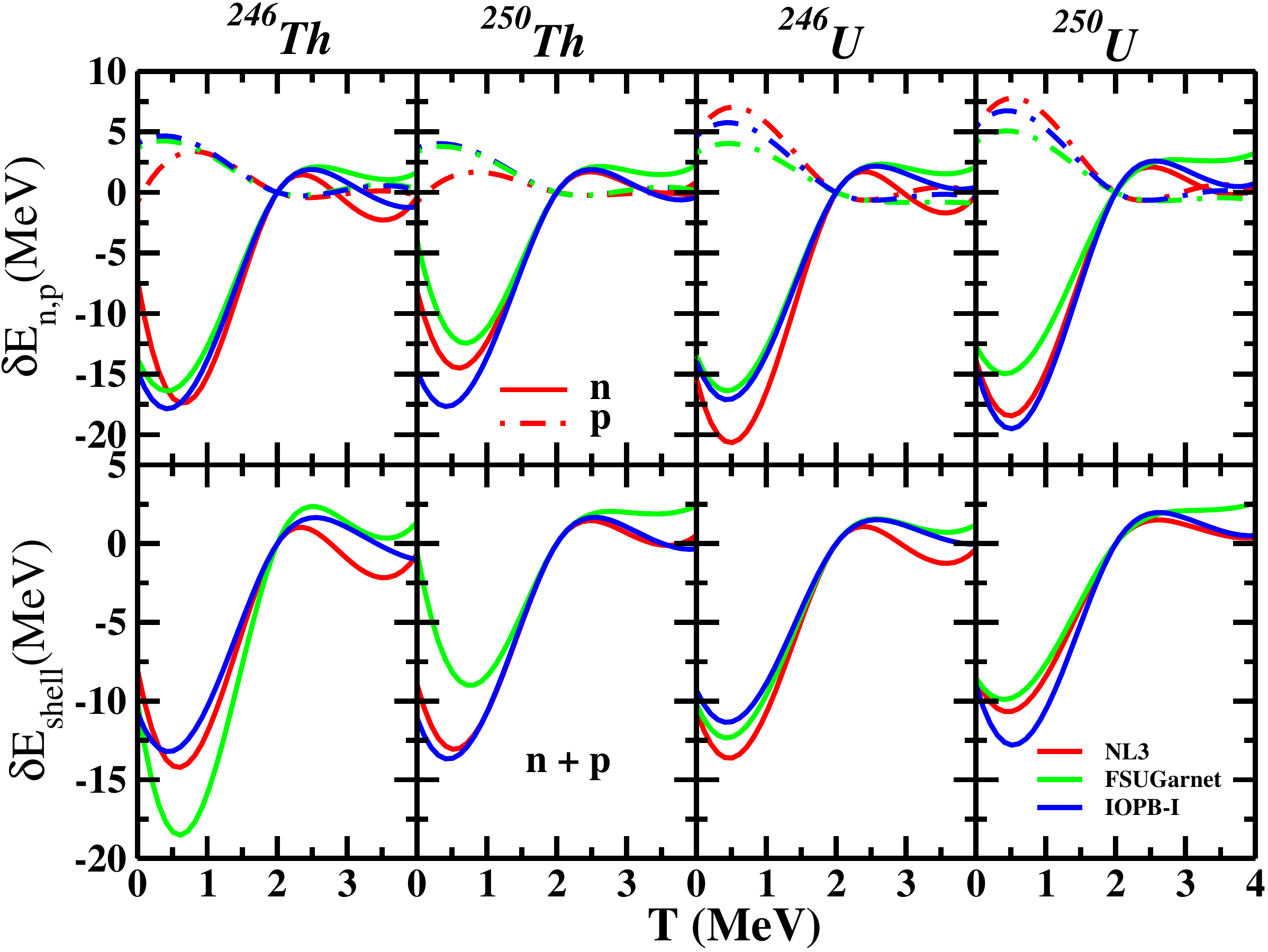}
\caption{Same as Figure \ref{shc0}, but for the nuclei $^{246,250}Th$ and $^{246,250}U$. } 
\label{shc}
\end{figure}

Strutinsky was the first to calculate the fluctuating part of the total energy within the shell model 
$i.e.$, the oscillatory part $E_{osc}$ by averaging the level density of a nucleus, and hence this method 
is known as Strutinsky averaging method or Strutinsky shell correction scheme \cite{stru}.  Some of the 
steps of averaging the level density of a nucleus are the following. The shell model energy is divided into 
an oscillating part, $E_{osc}$, and a smoothly varying part, $\tilde{E}_{sh}$ 
$i.e.$ $E_{sh}=\sum_{i=1}^{A}\epsilon_i=E_{osc}+\tilde{E}_{sh}$. To calculate the average part of the 
energy, it is quite useful to introduce the concept of the level density $g(\epsilon)$ by defining 
$g(\epsilon) d\epsilon$ as the number of levels in the energy interval between $\epsilon$ and 
$\epsilon+d\epsilon$. In the shell model, the level density is given by  
$g(\epsilon)=\sum_{i}\delta(\epsilon-\epsilon_i)$. The shell model energy can be written in terms of 
level density as: $E_{sh}=\int_{-\infty}^{\lambda}\epsilon g(\epsilon)d\epsilon$. The shell model levels 
are grouped into bunches with an average distance of $\hbar\omega_o\simeq41A^{-1/3}$ (MeV). So the level 
density $g$ shows oscillations with nearly this frequency. Since the fluctuations in the shell model energy 
$E_{sh}$ are due to these oscillations, one can calculate the smooth part $\tilde{E}_{sh}$ of the energy by 
introducing a continuous function  $\tilde{g}(\epsilon)$, which represents the smooth part of the level 
density $g(\epsilon)$. The corresponding Fermi energy $\tilde{\lambda}$ can be calculated by the condition 
$A=\int_{-\infty}^{\tilde{\lambda}}\tilde{g}(\epsilon)d\epsilon$. For the smooth part of the energy we 
finally get $\tilde{E}_{sh}=\int_{-\infty}^{\tilde{\lambda}}\tilde{g}(\epsilon)d\epsilon$. Hence, the total 
energy $E$ of the system is given by $E=E_{LDM}+E_{osc}=E_{LDM}+E_{sh}-\tilde{E}_{sh}$. 

The temperature-dependent single-particle energy for protons and neutrons calculated with NL3, 
FUSGarnet, and IOPB-I sets within the TRMF model are used as the inputs to the Strutinsky scheme to evaluate the shell 
correction energy.  The results are presented in Figures \ref{shc0} and \ref{shc}. The upper panel is the shell correction 
energies for both the neutrons (bold) and protons (dot-dashed) and the lower panel shows the total shell 
correction energy. Figure \ref{shc0} shows the contributions of all microscopic corrections while Figure \ref{shc} have the 
corrections from the shell closure only (discussed below). As expected, the shell corrections (both protons and neutrons) for
$^{208}Pb$, almost remains constant with temperature. Contrary to the behavior of $^{208}Pb$, the
$\delta{E_{shell}}$ for $^{234,236}U$ and $^{240}Pu$, initially increase with T for neutron up to the transition
point and then suddenly decrease monotonously. On the other hand, we get an abrupt change of $\delta{E_{shell}}$
for proton at the transition point and remains a constant value as shown in Figure~\ref{shc0}. This transition point, i.e., the
shell melting point coincides with the results obtained from the $S^2\sim E^*$ curve (Figure~\ref{s2e0}).
After a certain temperature, the value of $\delta{E_n}$ for $Th$ and $U$ starts 
increasing with T up to the transition point and then remains an almost constant. The value of $\delta{E_p}$ 
decreases with T till the transition point reached and then it follows almost constant nature. From the figure, 
it is clear that $\delta{E_{shell}}$ also follows the same trend as of neutron. It is worth mentioning that 
the total energy of a nucleus is not just the sum of the average part ($E_{av.}$ or $E_{LDM}$) and the correction 
due to shell effect ($\delta E_{shell}$). It has also the contributions of deformation energy, the degeneracy of 
the levels, correction due to pairing ($\delta E_{pair}$), and some microscopic fluctuation in the levels 
(or say, $\delta E_{fluc})$. The correction part of the energy $\delta E_{total}$ contains the contribution 
from all the effects mentioned above 
$i.e.$ $\delta{E_{total}}=\delta{E_{shell}}+\delta{E_{def}}+\delta{E_{pair}}+\delta{E_{fluc}}$. Thus, 
the $\delta{E_{total}}$ has some definite non zero value even at finite T. In order to find the sole contribution 
of the shell correction energy ($\delta{E_{shell}}$), we normalize the $\delta E_{total}$ and taken out all other 
contributions (such as $\delta{E_{def}}$, $\delta{E_{pair}}$, and $\delta{E_{fluc}}$). For example, in the case of 
$^{246}U$ with FSUGarnet, we have normalized the $\delta{E_n}$ by a factor of -12.04 MeV and that of $\delta{E_p}$ 
by +2.31 MeV (The idea to make $\delta E_{shell} = \delta E_n + \delta E_p = 0$ at the transition temperature). 
By doing this, the rest of the correction parts (as discussed above) have been removed and the remaining is the 
shell correction energy ($\delta E_{shell}$) only. It is clear from the figures that the $E_{shell}$ approaches
to zero with T and maintains its nature beyond $T_c$ (Fig. \ref{shc}). This type of normalization is performed 
accordingly to all the other considered nuclei with different force parameters.

\begin{figure}
	\includegraphics[width=0.9\columnwidth,height=9.0cm]{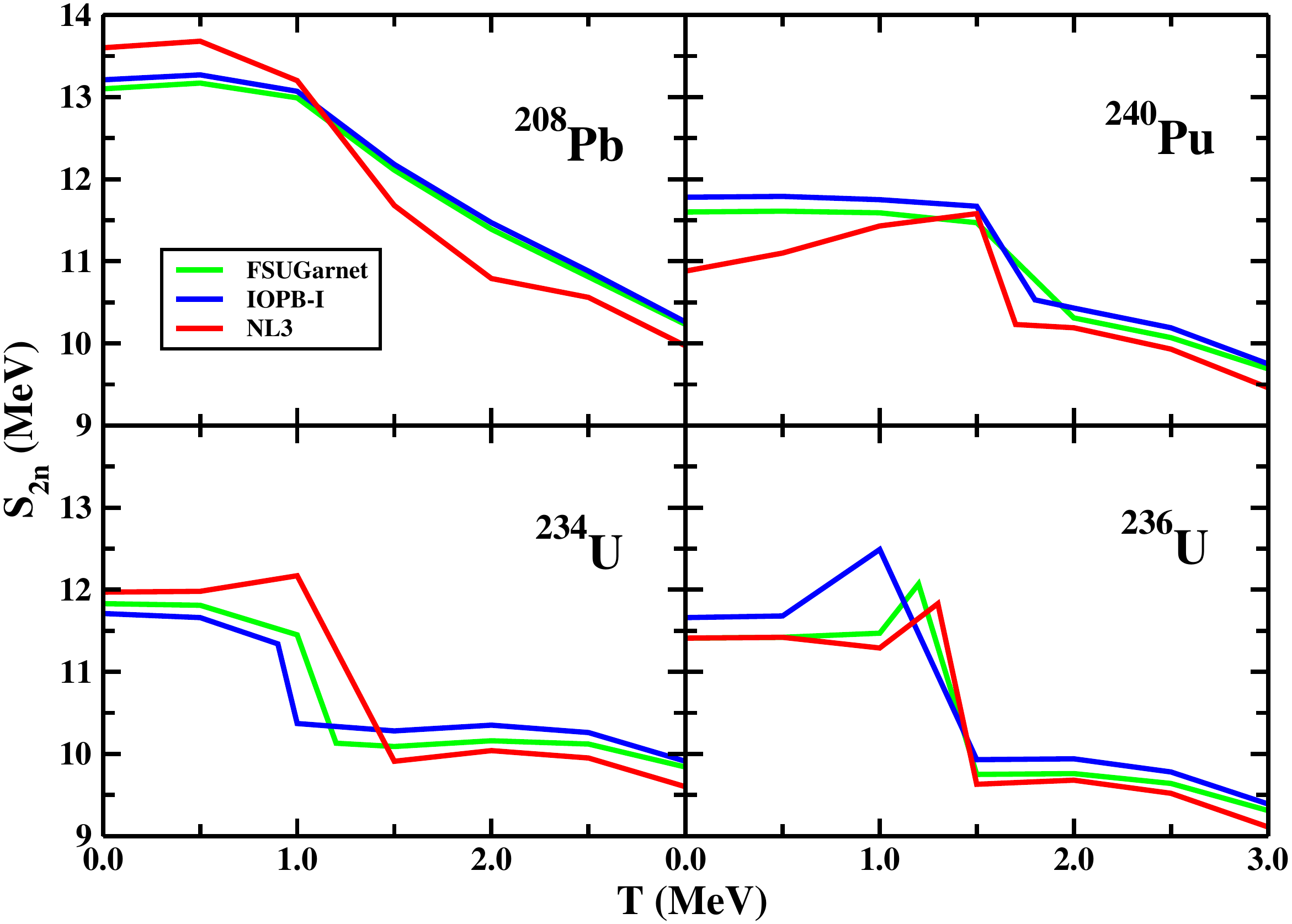}
	\caption{The $S_{2n}-$separation energy as a function 
        of temperature T for the nuclei $^{208}Pb$, $^{234}U$, $^{236}U$, and $^{240}Pu$ with FSUGarnet, IOPB-I, and NL3
        parameter sets. } 
\label{s2n0}
\end{figure}
\begin{figure}
\includegraphics[width=0.9\columnwidth,height=9.0cm]{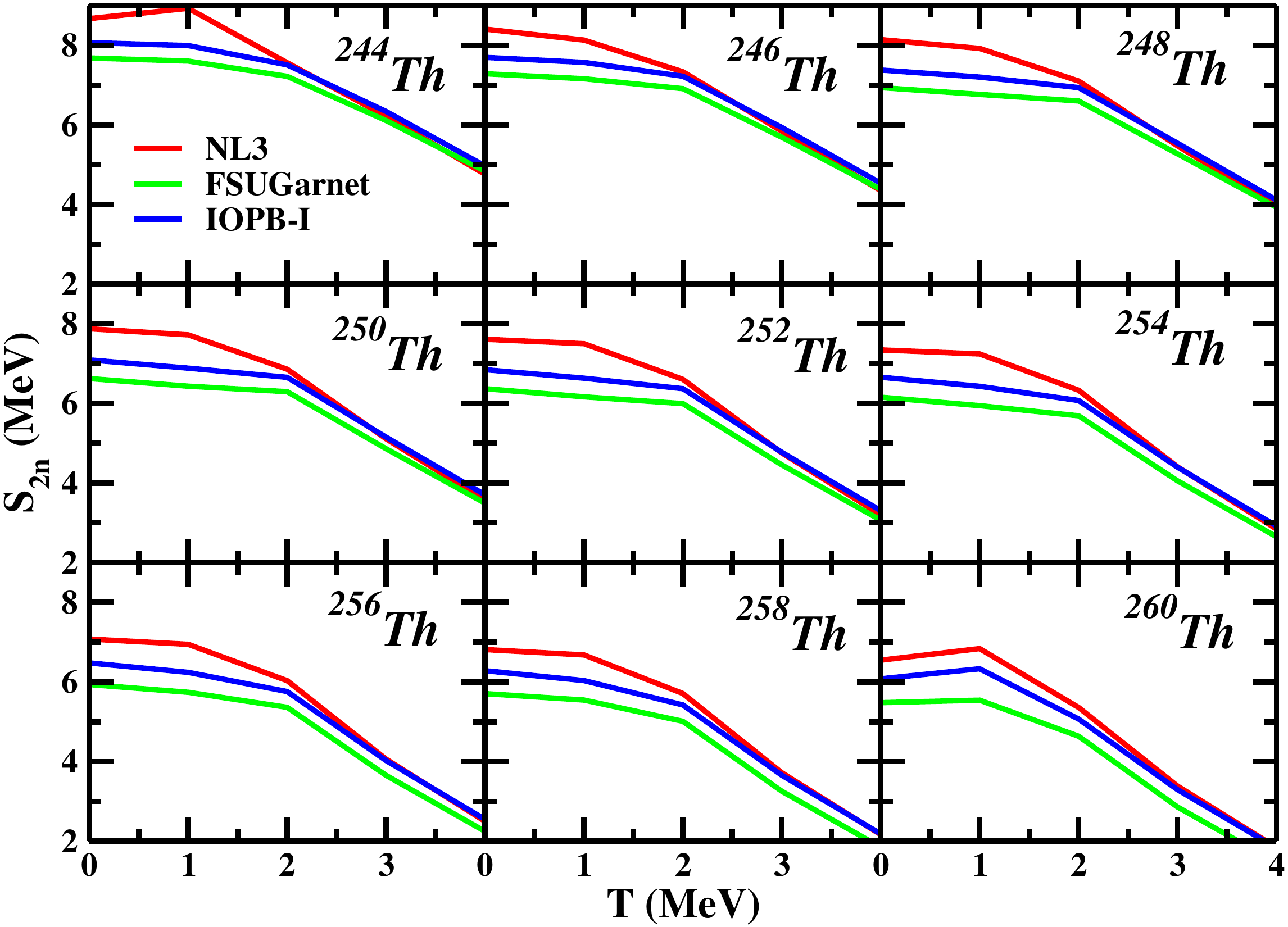}
\caption{Same as Figure~\ref{s2n0}, but for the nuclei $^{244,246,248,250,252,254,256,258,260}Th$. }
\label{s2n1}
\end{figure}

\begin{figure}
\includegraphics[width=0.9\columnwidth,height=9.0cm]{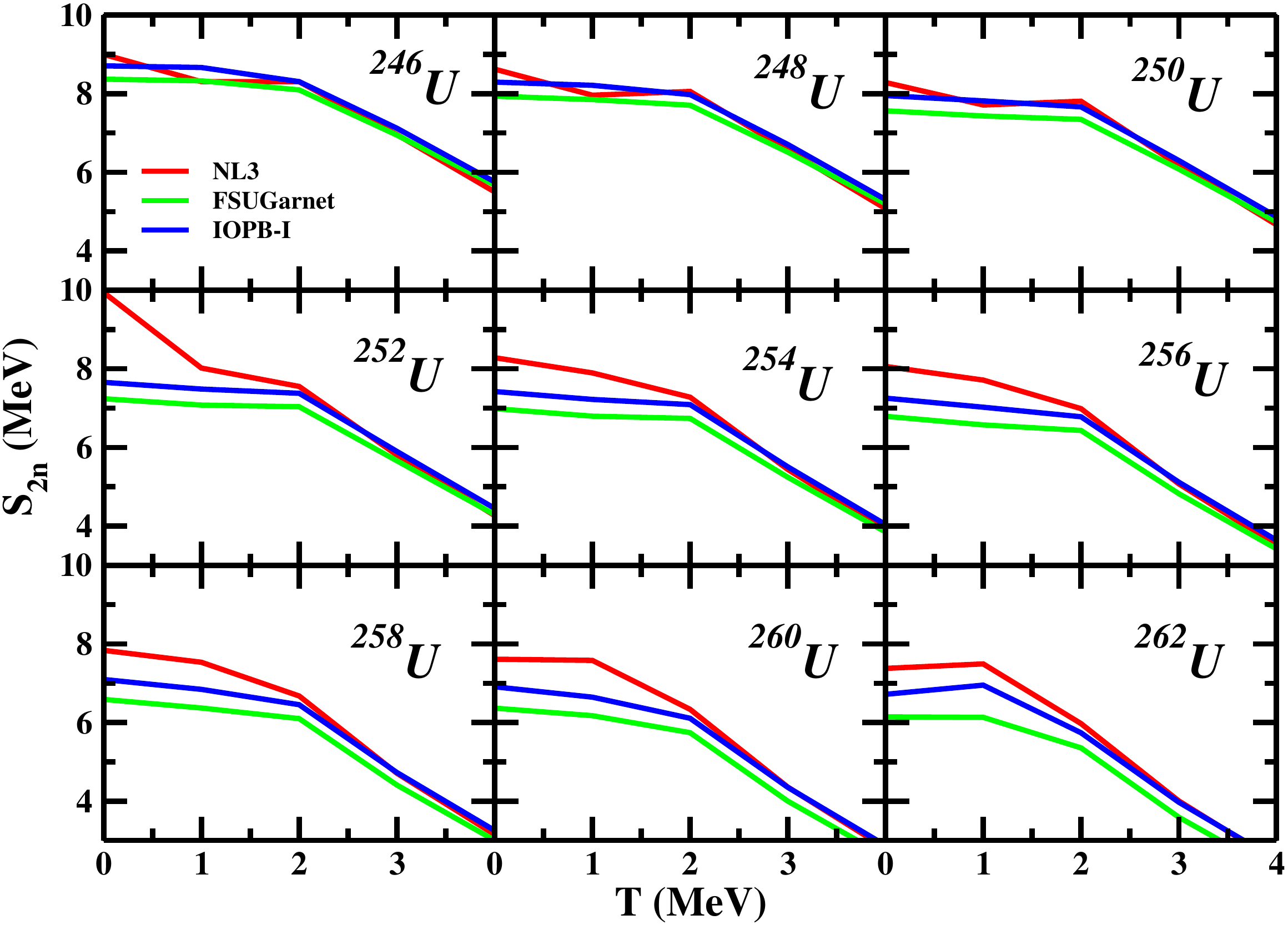}
\caption{Same as Figure~\ref{s2n0}, but for $^{246,248,250,252,254,256,258,260,262}U$.}
\label{s2n2}
\end{figure}

The two neutron separation energy ($S_{2n}$) provides information regarding the structure of a nucleus. 
More specifically, it gives an idea about how easily a pair of neutron can be emitted from a nucleus. It is calculated as 
the difference of ground-state $BE$ of two isotopes, i.e. $S_{2n}(Z,N)=BE(Z,N-2)-BE(Z,N)$. This is also used as a probe to show the magicity 
of a nucleus. In that case, it gives a kink (more $S_{2n}$ values) corresponding to a closed shell nucleus, 
when it is investigated for the whole isotopic series. The variation of $S_{2n}$ as a function of T for $^{208}Pb$, 
$^{234}U$, $^{236}U$, $^{240}Pu$, and for neutron-rich thermally fissile nuclei 
$^{244,246,248,250,252,254,256,258,260}Th$, and $^{246,248,250,252,254,256,258,260,262}U$ with NL3, FSUGarnet, 
and IOPB-I are shown in Figures \ref{s2n0}, \ref{s2n1}, and \ref{s2n2}. 
For $^{208}Pb$, there is no definite transition point in the 2n-separation energy.
In case of $^{236}U$, all forces give the same transition point (i.e., $T=1.5$ Mev), whereas in $^{234}U$ and $^{240}Pu$
it is parameter-dependent as shown in Figure~\ref{s2n0}.
 For example, in the case of $^{234}U$ we noticed the transition point
for NL3 a bit later than IOPB-I and FSUGarnet but for $^{240}Pu$, the transition point is opposite in nature (see Figure~\ref{s2n0}).
Moreover, It can be noticed from Figures \ref{s2n1} and \ref{s2n2} that all the nuclei with the three-parameter sets give the 
same transition points ($\thicksim2$ MeV). It is also noted from Figures \ref{s2n1} and \ref{s2n2} that the $S_{2n}$ values 
calculated with NL3 set are larger than 
those predicted by the other two sets except for the nuclei $^{246}U$, $^{248}U$, and $^{250}U$. The 
IOPB-I predicts a moderate $S_{2n}$ for all the nuclei in comparison to the other two sets. 
The $S_{2n}$ value decreases gradually for $^{208}Pb$. But, for other nuclei this behavior is different. There is a sudden fall at the
transition temperature. Beyond these temperatures, the $S_{2n}$ value decreases
almost smoothly. Thus, It can be concluded from the graph that the probability of emission of neutrons grows as temperature
increases. This probability becomes very high at and beyond transition points.

\subsection{Single-Particle Energy, Deformation Parameters and Shape Transition}

\begin{figure}
	\includegraphics[width=0.5\columnwidth,height=14.0cm]{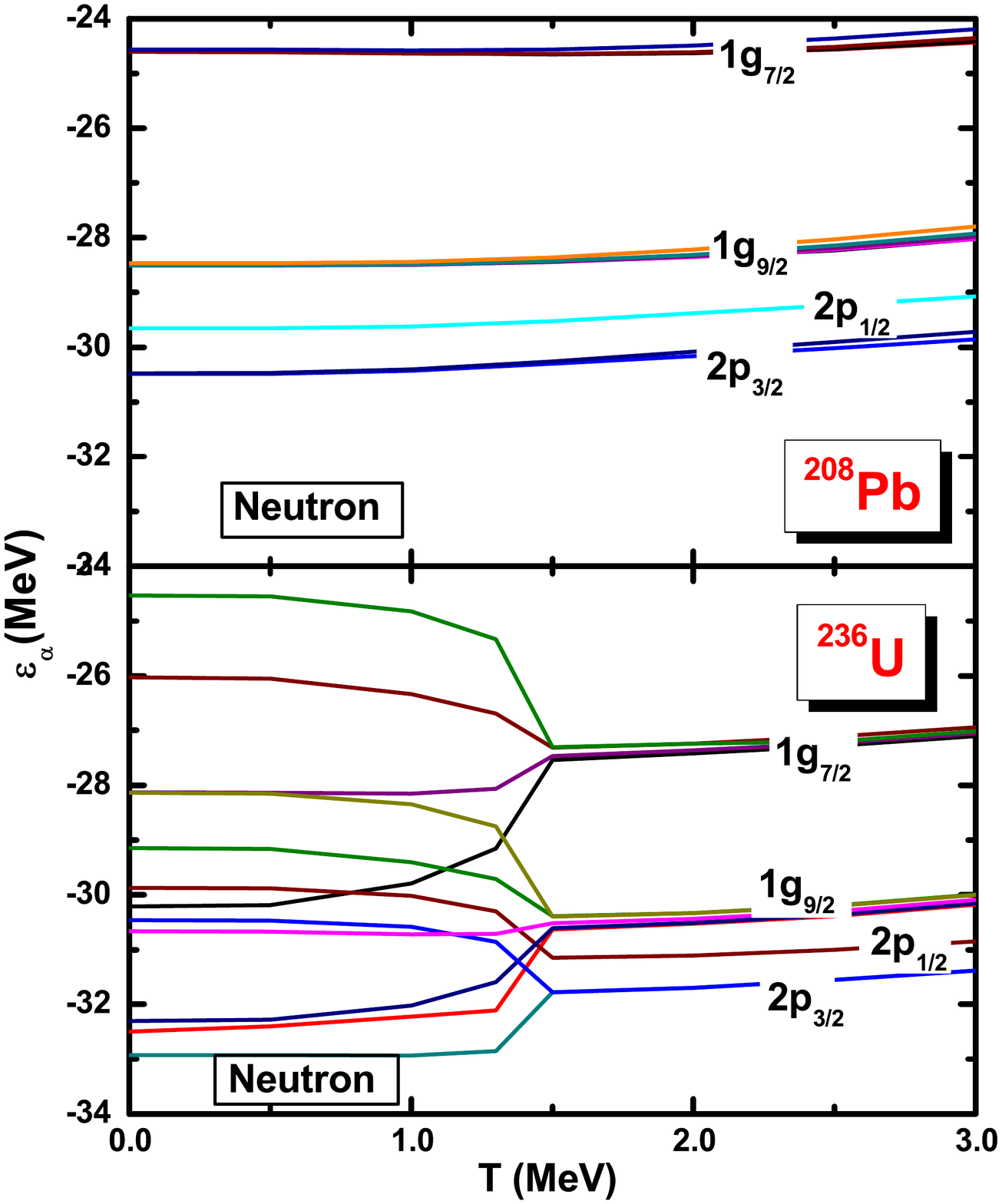}
	\includegraphics[width=0.5\columnwidth,height=14.0cm]{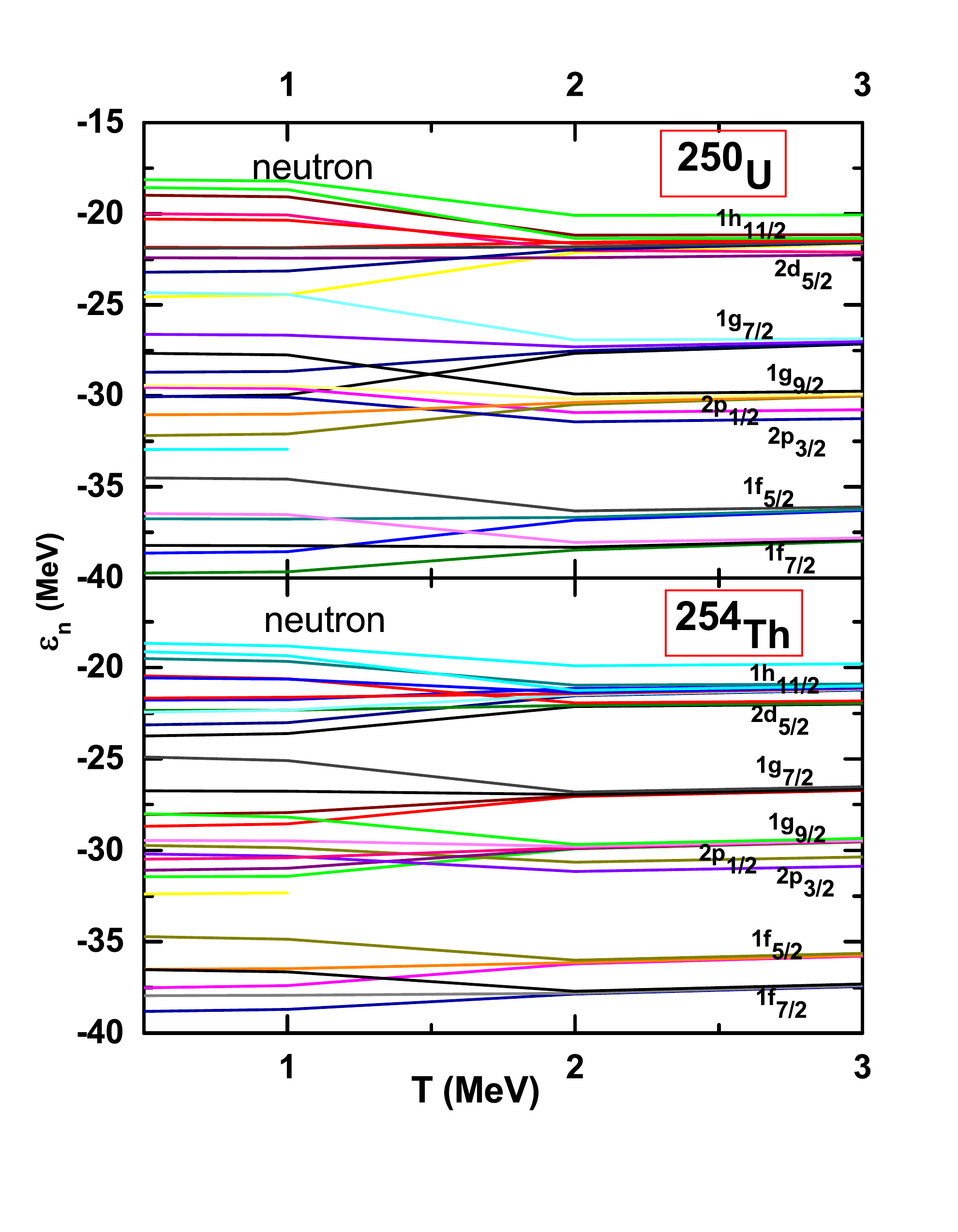}
        \caption{The change of neutron single-particle spectrum $\epsilon_{\alpha}$ for some selected levels
        as a function of temperature for the nuclei (a) $^{208}Pb$ and $^{236}U$, (b) $^{250}U$, and $^{254}Th$, with IOPB-I parameter set.}
\label{spe0}
\end{figure} 
%

Figure~\ref{spe0} represents a few single-particle levels $\epsilon_{\alpha}$ of neutrons for the 
nuclei $^{208}Pb$ and $^{236}U$ (left panel), and $^{254}Th$ and $^{250}U$ (right panel) as a representing cases with IOPB-I set. 
From the figure, it is clear that the single-particle energy 
of various $m$ states arising from different orbitals merge to a single degenerate level with an increase of temperature 
after a particular T except for $^{208}Pb$ nucleus. We call this temperature $T=T_c$ as the shell melting or transition
temperature because this is the temperature at which the nucleus changes from deformed to spherical state and all the
$m-$ levels converged to a single degenerate one \cite{Gambhir62,Agarwal64}. The same value of critical temperature is obtained 
in the case of the single-particle spectrum of protons for the nuclei, i.e., non-degenerate levels become degenerate. 
The same behavior is seen for the other remaining natural/neutron-rich thermally fissile nuclei considered here. 
When we compare this temperature with the shell melting
point (i.e. slope point of the $S^2$ vs $E^*$ curve), change in $\delta{E_{shell}}$ (Figure~\ref{shc0}-\ref{shc}) 
and $S_{2n}$ values (Figure~\ref{s2n0}-\ref{s2n2}), the
merging point in $\epsilon_{n,p}$ matches perfectly with each other. We have increased the temperature further
and analyzed the single-particle levels at higher T, but we have not noticed the re-appearance of 
non-degenerate states. In the case of $^{208}Pb$, the single-particle energy for proton and neutron do not
change with temperature. This is because, there is no change in the shape of this nucleus as it is 
spherical throughout all the T. Although, the shell correction energy changes from negative 
to positive or vice versa, we 
have not found the disappearance of shell correction completely. This means, whatever be the temperature of the
nucleus, the shell nature remains there, of course, the nucleons are in a degenerate state. So, the disappearance of shell effects
implies the redistribution of shells at transition point i.e., from non-degenerate to degenerate. 
In other words, whatever be the temperature of the nucleus, there will be a 
finite value of shell correction energy (as shown in Figure~\ref{shc0}) due to the random motion of nucleons. 
On inspecting the single-particle energy
for the entire spectrum (not shown in the figure), one can find that 
low lying states raise slightly and high lying states decrease slightly. This is due to an increase in the effective mass and 
RMS radius \cite{Gambhir62}. 
We have repeated the calculations for the other two parameter sets (NL3 and FSUGarnet) and find the same scenario (not shown here).  


\begin{figure}
	\hspace{0.8cm}
	\includegraphics[width=0.9\columnwidth,height=9.0cm]{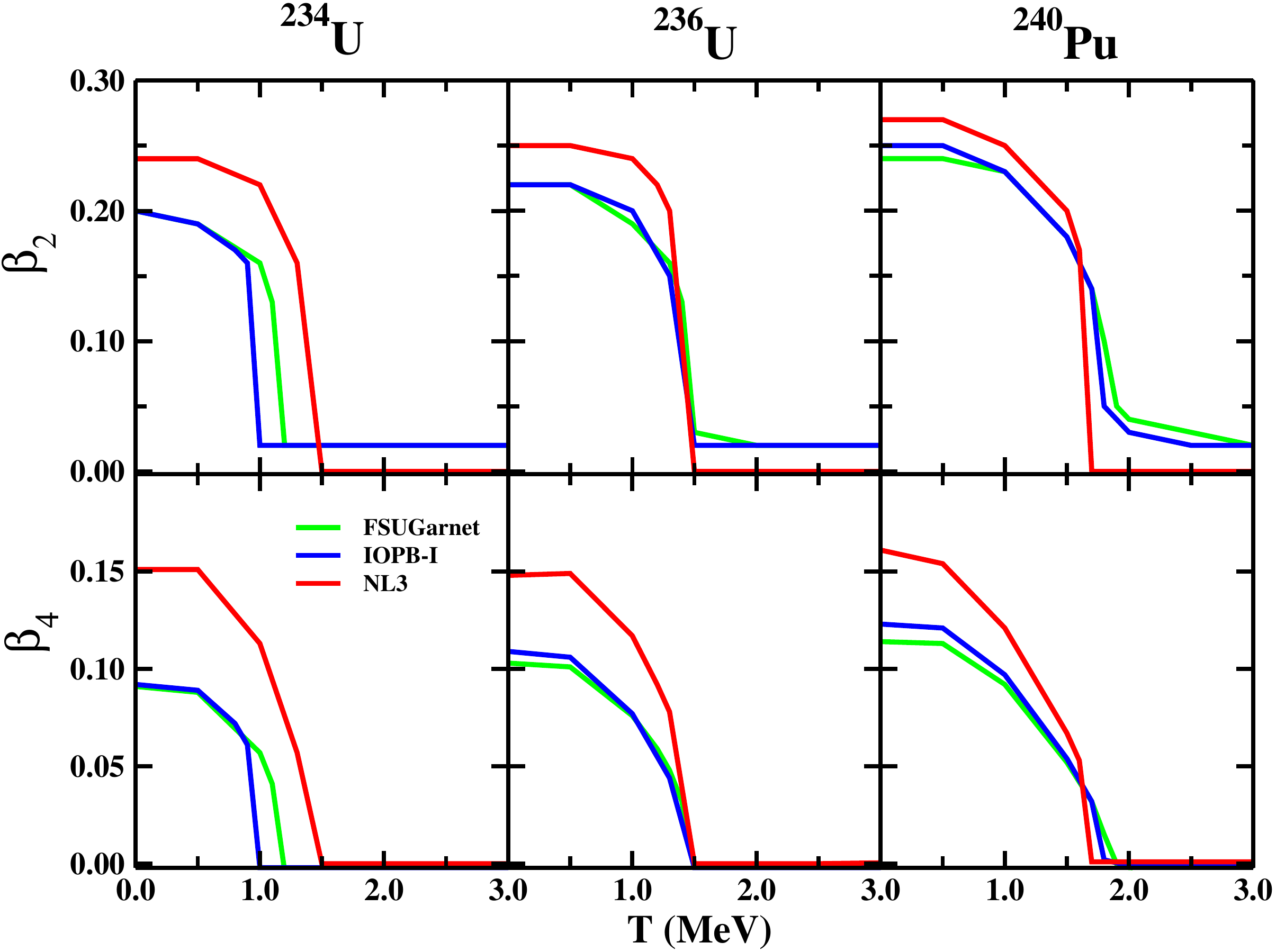}
	\caption{The quadrupole and hexadecapole deformations parameters
        ($\beta_2$ and $\beta_4$,respectively) as a function of temperature T for $^{234}U$, $^{236}U$i, and $^{240}Pu$ with 
        FSUGarnet, IOPB-I, and NL3 parameter sets.}
\label{Fig.7}
\end{figure}
\begin{figure}
\includegraphics[width=0.9\columnwidth,height=9.0cm]{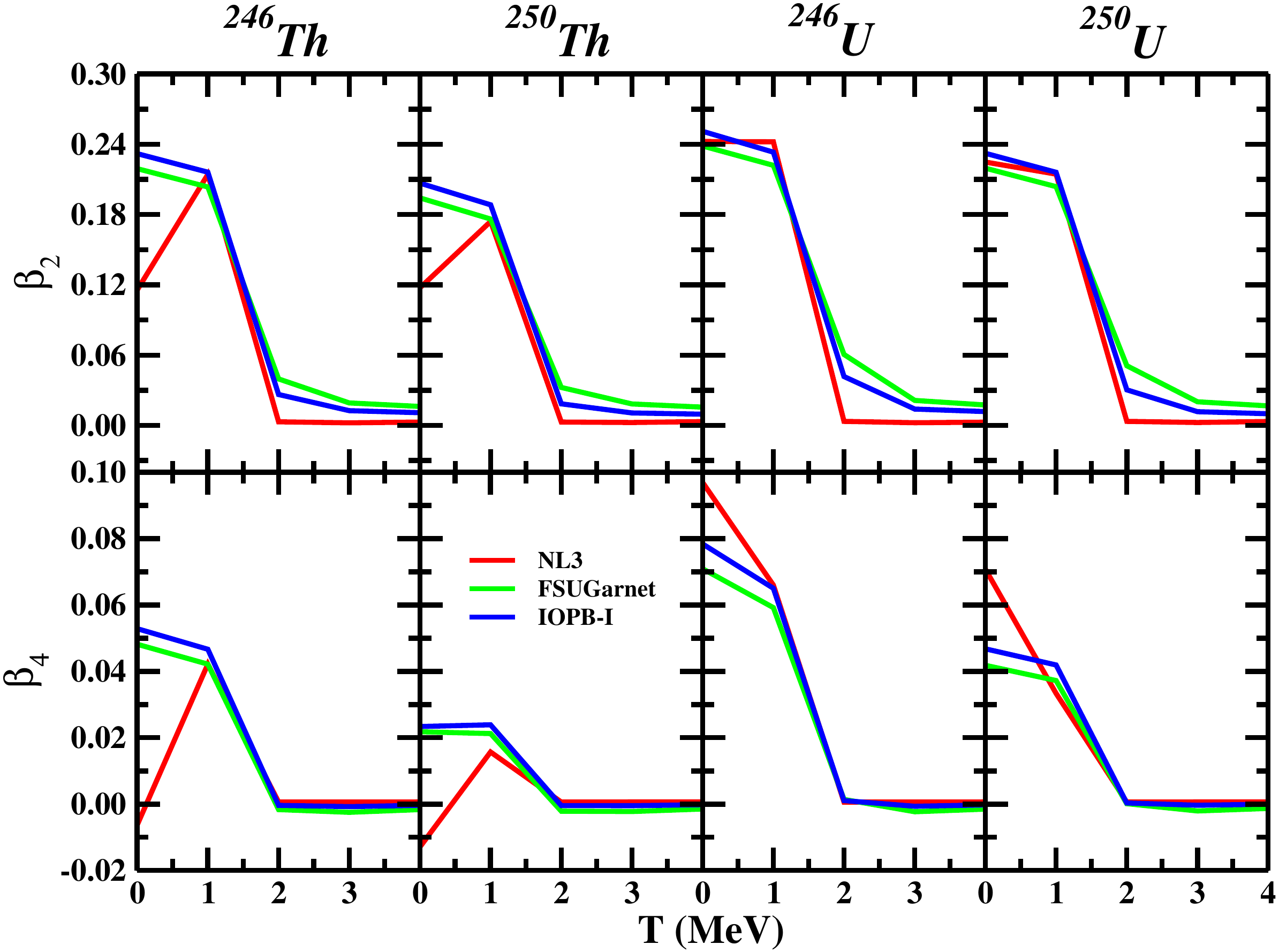}
\caption{Same as Figure~\ref{Fig.7}, but for $^{246,250}Th$, and $^{246,250}U$ nuclei.}
\label{b2b4}
\end{figure}

The quadrupole $\beta_2$ and hexadecapole $\beta_4$ deformation parameters for 
the nuclei $^{234}U$, $^{236}U$, $^{240}Pu$, $^{246,250}Th$, and $^{246,256}U$ are shown in Figures~\ref{Fig.7},\ref{b2b4}. 
The upper panels in the figures are the $\beta_2$ and the lower panels are $\beta_4$ as a function of temperature T. 
Both the deformation parameters drastically decrease with T at $T=T_c$. However, due to more number of
neutrons in the neutron-rich thermally fissile nuclei, the value of $T_c$ is larger as compared to that of
normal thermally fissile nuclei. These results are qualitatively consistent with the previous studies 
for different nuclei \cite{Gambhir62,Agarwal62}. The almost zero value of $\beta_2$ at and beyond the critical temperature $T_c$ implies 
that the shape of a nucleus changes from deformed to spherical. The non-smoothness on the surface 
of a nucleus irrespective of its shape is defined by the hexadecapole deformation parameter $\beta_4$. 
The same behavior for $\beta_4$ implies that on increasing temperature not only the shape of a nucleus changes but also 
its surface becomes smooth. Hence nucleus becomes a perfectly degenerate sphere at temperature $T_c$ and beyond. 
We have checked that even on the further increasing temperature the state of a nucleus does not change again and remains 
spherical as mentioned earlier. While, in the fission process, a nucleus undergoes 
scission point where it is highly deformed and hence, breaks into fragments.
This observation concludes that the nucleus never undergoes fission only by temperature. 
It is necessary to disturb the nucleus physically for fission reaction bombarding thermal neutron.
In other words, $^{233}U$, $^{235}U$ and $^{239}Pu$ have half-lives $T_{1/2}=$
$1.59 \times 10^5$ years, $7.04 \times10^8$ years and $2.4 \times10^4$ years, respectively \cite{nndc} and never undergo 
spontaneous fission. The fission reaction takes place whenever a zero
energy (thermal) neutron hits externally. For example, the thermally fissile nucleus $^{233}U$ is synthesized
through $^{232}Th$ after the absorption of neutron and subsequent $\beta-$decays. The synthesized $^{233}U$
lies in the excited state (high T) but does not undergo spontaneous fission. In other words, its fission
occurs after absorbing a thermal neutron. 


\begin{figure}
        \includegraphics[width=0.9\columnwidth,height=9.0cm]{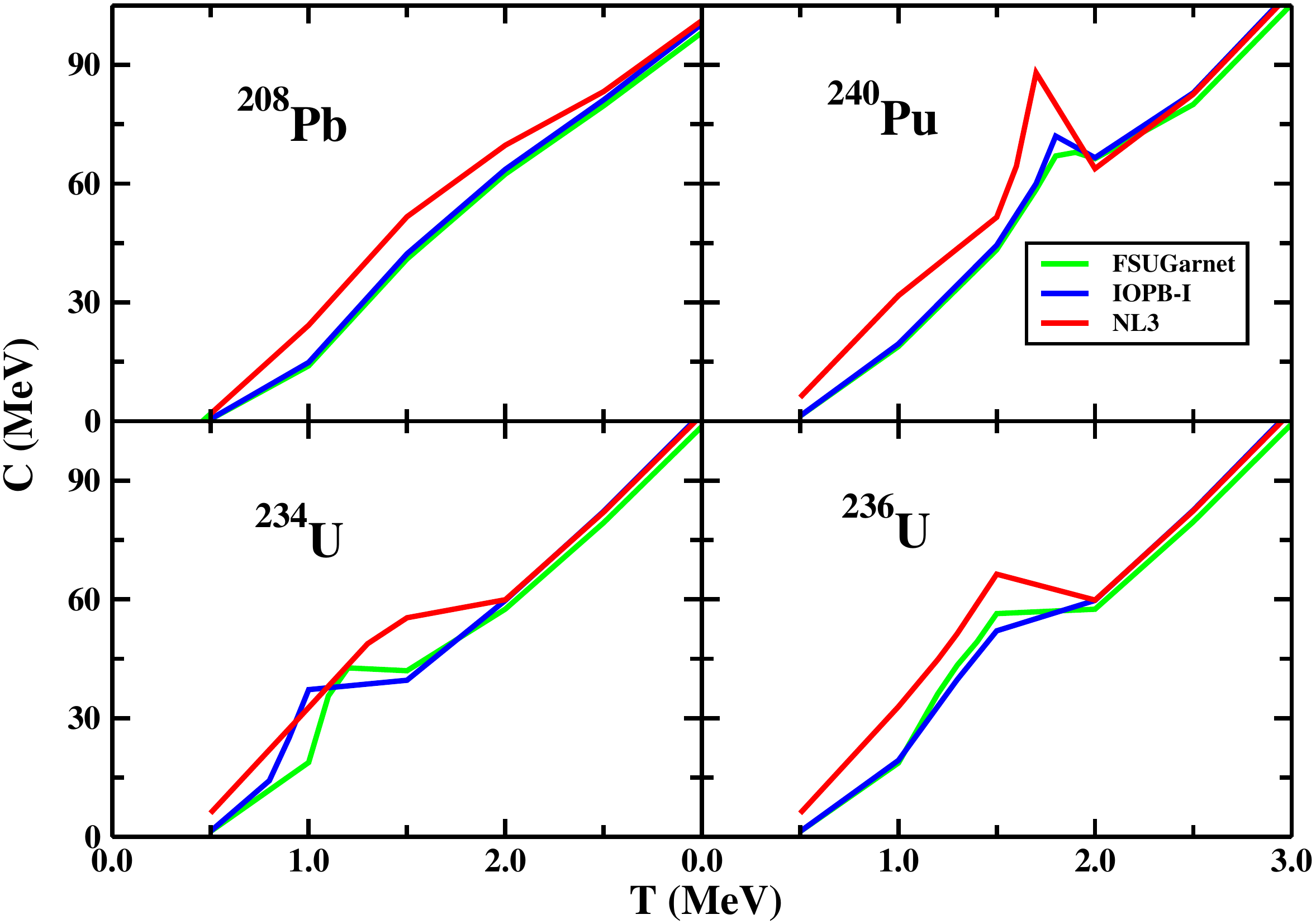}
        \caption{The specific heat for the nuclei $^{208}Pb$,
        $^{234}U$, $^{236}U$, and $^{240}Pu$ as a function of temperature T corresponding to the FSUGarnet, IOPB-I, and NL3
        parameter sets.}
\label{Fig.8}
\end{figure}

For further study and clarification of the phase transition at critical temperatures, we have calculated specific heat for 
the nuclei $^{208}Pb$, $^{234}U$, $^{236}U$, and $^{240}Pu$. The specific heat of a nucleus 
 is defined as:

\begin{eqnarray}
C(T) & = & \frac{\partial E^*}{\partial T}\;.
\end{eqnarray}
The variation of specific heat with temperature is shown in Figure~\ref{Fig.8}. The kinks in the curves 
for different nuclei correspond to the critical temperatures where shape transition takes place. 
This value differs with the nucleus and in agreement with the obtained results shown in the previous graphs (Figures \ref{s2e0}-\ref{Fig.7}). 
But, it can be seen that for $^{208}Pb$, there is no such kink which signifies that 
it remains spherical at all temperatures. The pattern of the curves is the same as studied earlier 
for $^{166}Er$ and $^{170}Er$ \cite{Agarwal62}. 
At low temperature, FSUGarnet, and IOPB-I results match and underestimate those of NL3 but at higher temperature i.e., 
beyond critical temperatures, all the curves overlap with each other. 
The values of transition temperature are different for different parameter sets.

\subsection{Root Mean Square Radius and Neutron Skin-thickness}

\begin{figure}
        \includegraphics[width=0.9\columnwidth,height=9.0cm]{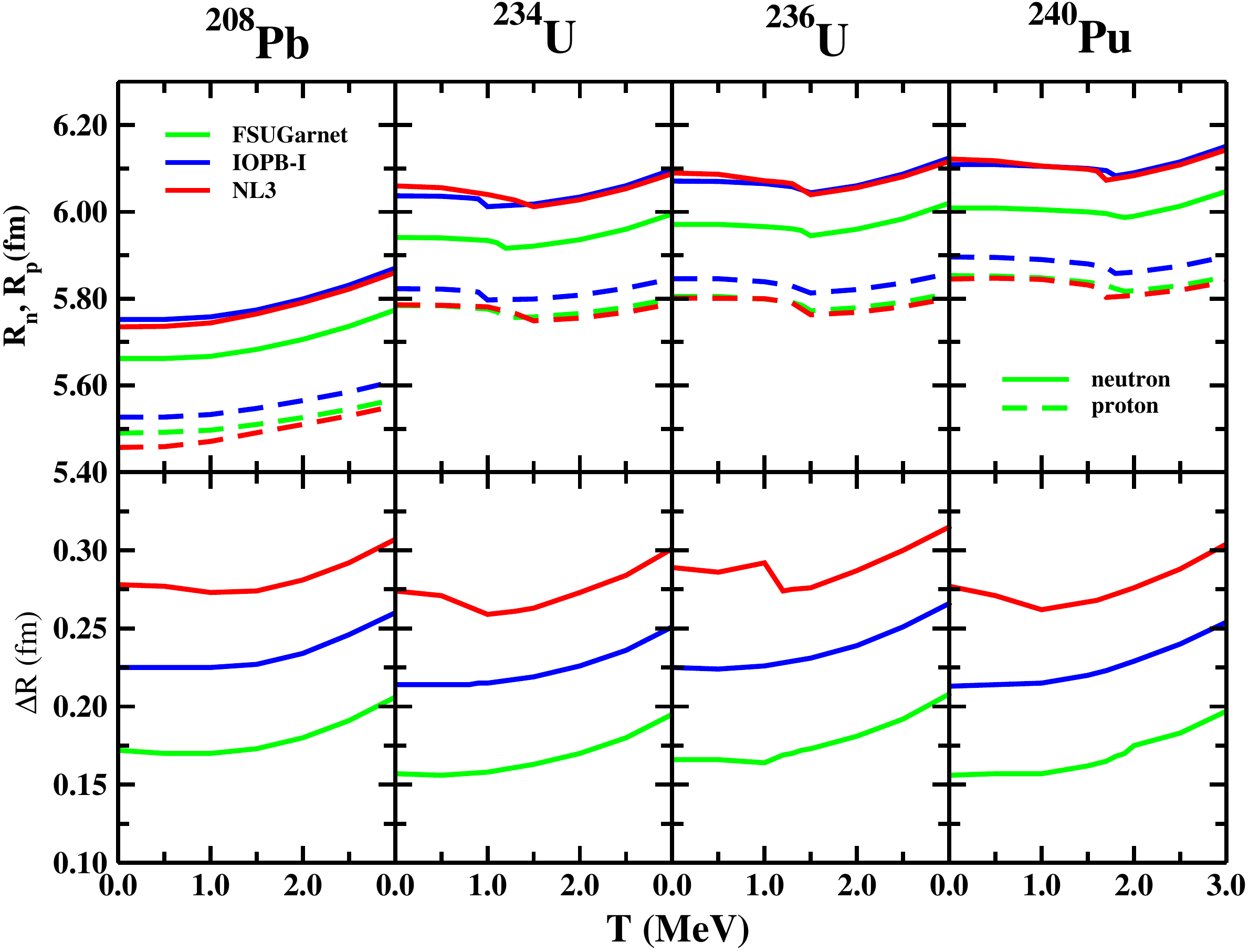}
        \caption{(a) The variation of root mean square (RMS) neutron $R_n$ and proton $R_p$ radii,
and (b) the neutron skin-thickness $\Delta {R}$ as a function of temperature T for the nuclei $^{208}Pb$,
        $^{234}U$, $^{236}U$, and $^{240}Pu$ with FSUGarnet, IOPB-I, and NL3
        parameter sets.}
\label{rnrp0}
\end{figure}
\begin{figure}
\includegraphics[width=0.9\columnwidth,height=9.0cm]{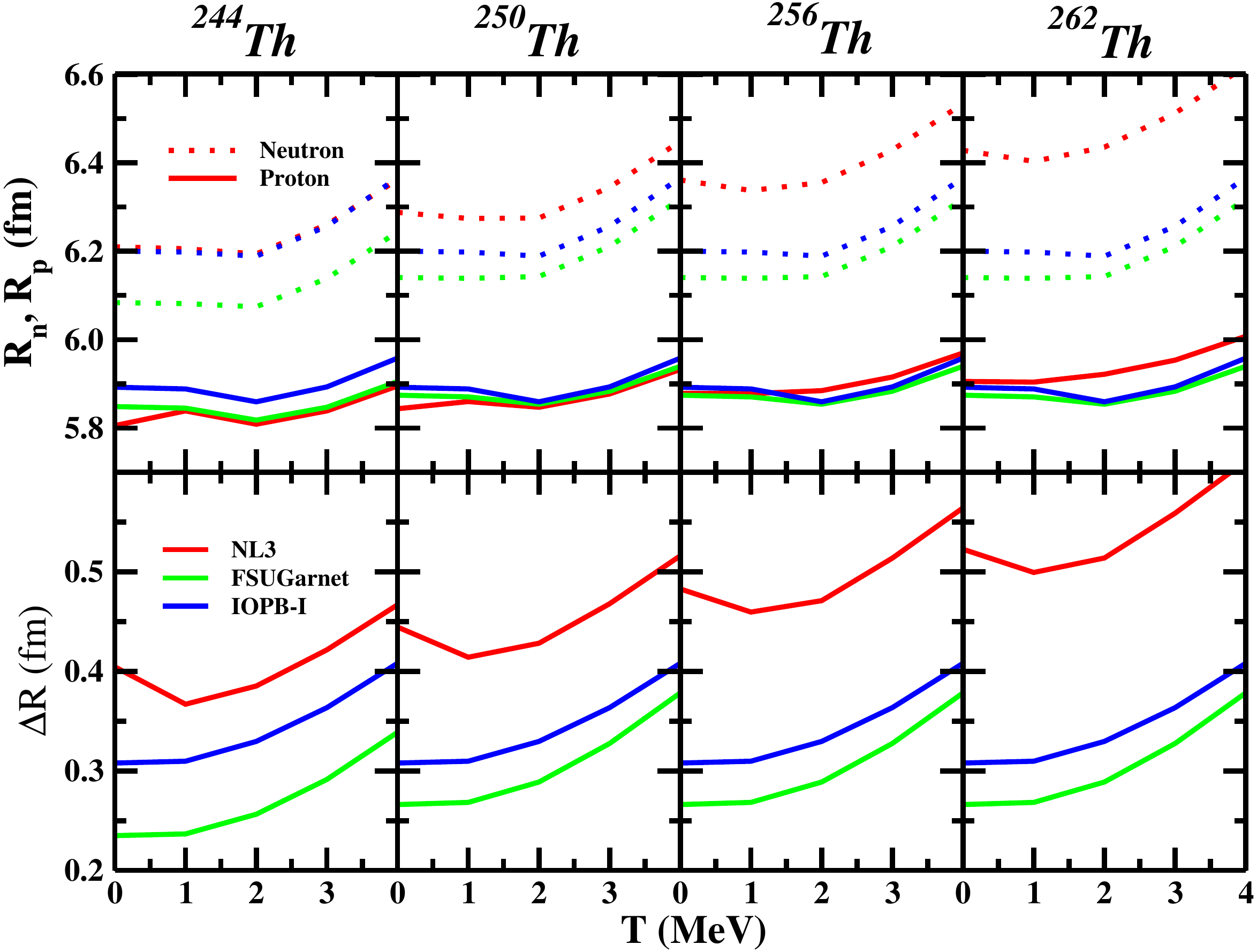}
\caption{Same as Figure \ref{rnrp0}, but for the nuclei $^{244,250,256,262}Th$. }	
\label{rnrp1}
\end{figure}
\begin{figure}
\includegraphics[width=0.9\columnwidth,height=9.0cm]{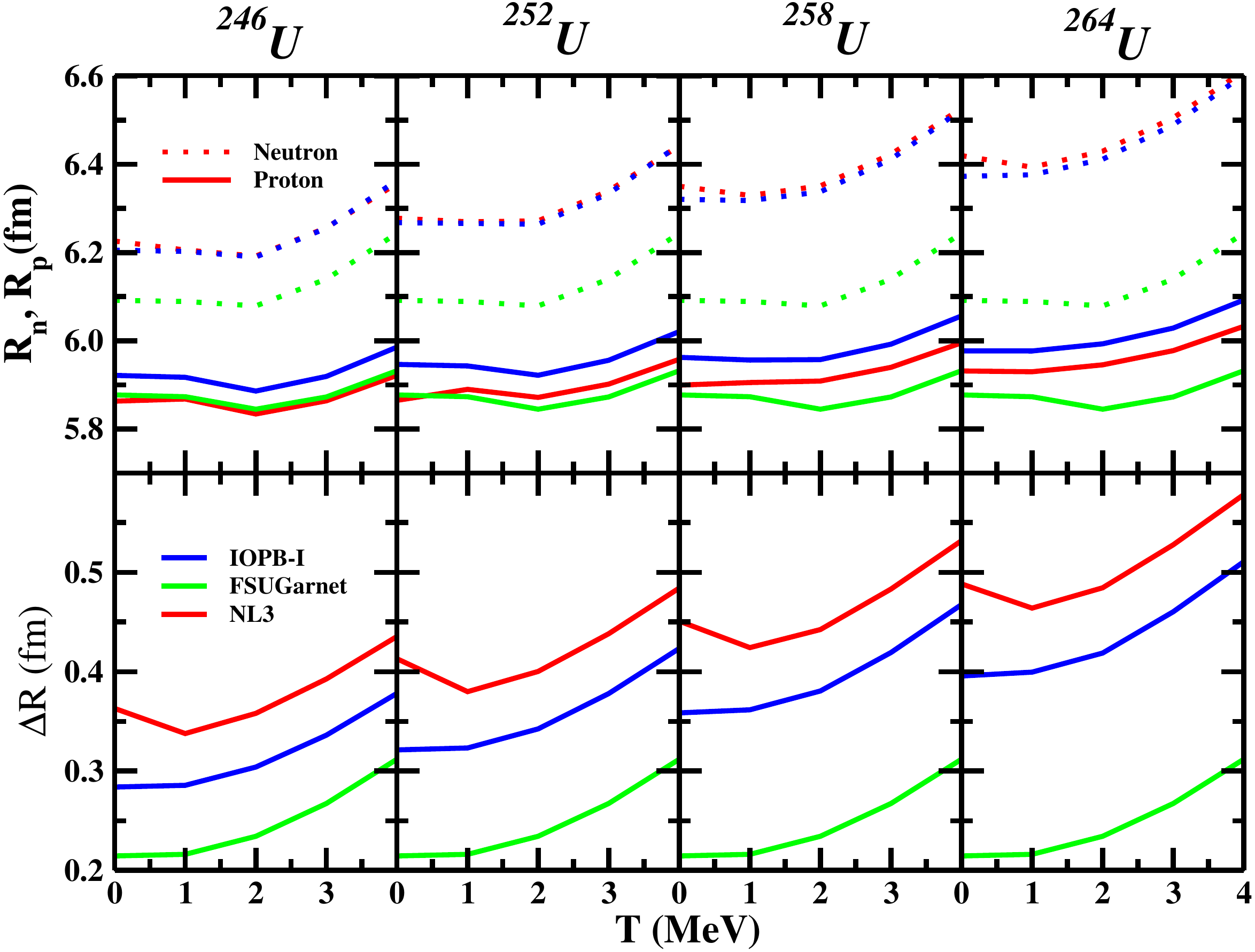}
\caption{Same as Figure \ref{rnrp0}, but for $^{246,252,258,264}U$.}	
\label{rnrp2}
\end{figure}

The root mean square radii for $^{208}Pb$, $^{234}U$, $^{236}U$, and $^{240}Pu$ are shown in Figure~\ref{rnrp0}. 
The proton and neutron radii are presented in the upper panel whereas the lower panel 
has the skin-thickness for the nuclei. The RMS radius first increases slowly at a lower temperature, and at a higher 
temperature, it increases rapidly for $^{208}Pb$. This behavior is consistent with that discussed in Refs. \cite{Gambhir62,ant17}. 
For the remaining nuclei, the RMS radius first decreases with temperature up to a point ($T_c$) and beyond that point, 
these values increase rapidly with temperature. This is because the deformed nucleus first undergoes 
a phase transition to a spherical shape and hence, radius decreases slightly, and beyond the said point these increase rapidly. 
Once the nucleus becomes perfectly spherical ($\beta_2=0$) and again the 
temperature is supplied to the nucleus, it expands monotonously, like the normal expansion of a metal. The
values of temperature where the RMS radii are minimum correspond to the same transition temperature
obtained in the previous graphs.
The same behavior of RMS radii is observed for $^{244,250,256,262}Th$ and $^{246,252,258,264}U$ nuclei as shown 
in Figures \ref{rnrp1} and \ref{rnrp2}.
It is also observed from the figures that 
neutron radius calculated by IOPB-I and NL3 force parameters coincide well with each other for $^{208}Pb$, $^{234}U$, 
$^{236}U$, $^{240}Pu$, $^{244}Th$, and $^{246,252}U$, but overestimate to that of FSUGarnet
set for all the considered nuclei.
 For proton radius, FSUGarnet and NL3 results coincide with each other except for $^{252,258,264}U$ and underestimate 
that of IOPB-I results.  

As discussed, the neutron skin-thickness ($\Delta R$) plays a crucial role in constraining the equation of states (EOS) of 
nuclear matter and to predict the radius of a neutron star. Because of the neutral nature of neutrons, 
the neutrons' distribution radius $R_n$ can not be measured as precisely as that of protons' distributions 
radius ($R_p$).  
Although there is large uncertainty in the determination of $R_n$, some of the precise measurements are done
 \cite{expskin} and the PREX-II experimental results are the most awaited data in this regard \cite{prex1}.
Recently, it is reported \cite{fatto,abbott17} using the Gravitational-wave observation data GW170817
that the upper limit of $\Delta R$ should be $\leq$ 0.25 fm for $^{208}Pb$. The calculated values of $\Delta R$ 
for $^{208}Pb$ corresponding to the NL3, FSUGarnet, and IOPB-I are 0.28, 0.16, and 0.22 fm, respectively, 
which can be seen from Figure~\ref{rnrp0}.
The behavior of neutron skin-thickness is also almost the same as of nucleus radius. The kinks here correspond to the same transition temperatures.
The value of neutron skin-thickness obtained by IOPB-I is preferred over NL3 and FSUGarnet. 
As we know, NL3 predicts a larger $\Delta R$
value which gives a larger neutron star radius \cite{iopb1}. Similarly, FSUGarnet predicts a smaller $\Delta R$ and expected
a smaller neutron star radius.

Although, there has not been established a direct relation of the radius ($R_n$, $R_p$ or $R_m$) 
of a nucleus with the fission yields, the neutrons distribution radius $R_n$ and the neutron skin-thickness 
$\Delta R$ is quite informative for the fragments' production during the fission 
process. The $R_n$ is quite important for the fission mass distribution. In Ref. \cite{skp10} Patra 
et al.  have shown that the larger the $\Delta R$ (neutron-rich nuclei), the more 
neutron-rich fragments evolve from the neck in the multi-fragmentation fission process. Due to an 
increase in the temperature, the neutrons in a nucleus are excited more than that of the proton 
(see Figs. \ref{rnrp1} and \ref{rnrp2}) and manifest a greater yield as compared to the lower temperature fission process. 
As mentioned above, it is a known fact that the NL3 set predicts a larger neutrons' distribution radius 
and produces a larger neutron star radius compared to the other successful RMF forces. In our present 
calculations, though the NL3 and IOPB-I predict similar $R_n$, the prediction of $\Delta R$ is in the order 
$\Delta R (NL3) > \Delta R (IOPB-I) > \Delta R (FSUGarnet)$. From this observation, it 
may be outlined that the fission yield with asymmetric fragments will be in the same order as that of $\Delta R$,
i.e. more in the case of the NL3 set and less for FSUGarnet.

\subsection{Inverse Level Density Parameter}

\begin{figure}
        \includegraphics[width=0.9\columnwidth,height=9.0cm]{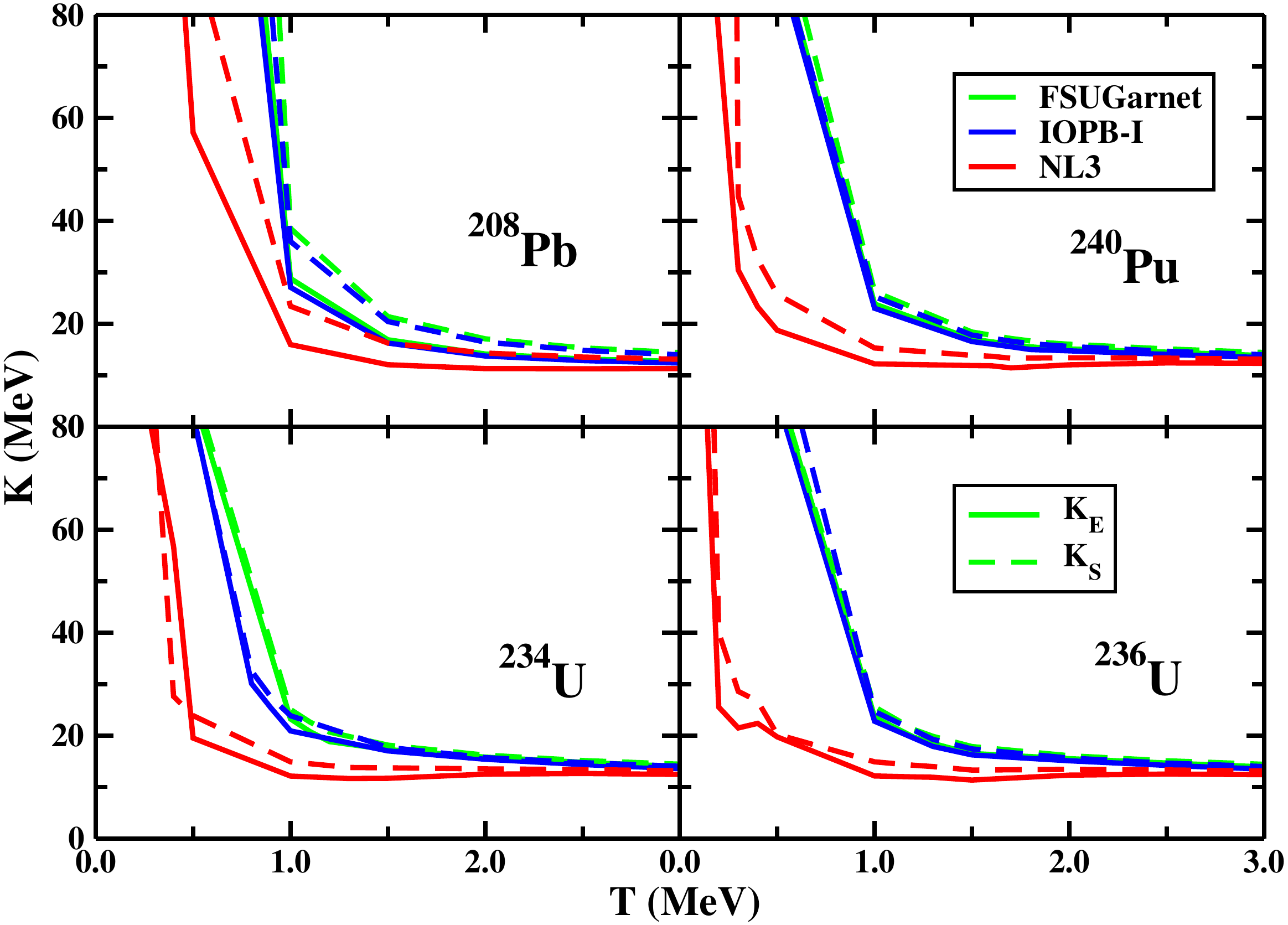}
        \caption{The inverse level density parameter (K)
        as a function of temperature T for the nuclei $^{208}Pb$,
        $^{234}U$, $^{236}U$, and $^{240}Pu$ with FSUGarnet, IOPB-I, and NL3
        parameter sets.}
\label{ilp}
\end{figure}
\begin{figure}
\includegraphics[width=0.9\columnwidth,height=9.0cm]{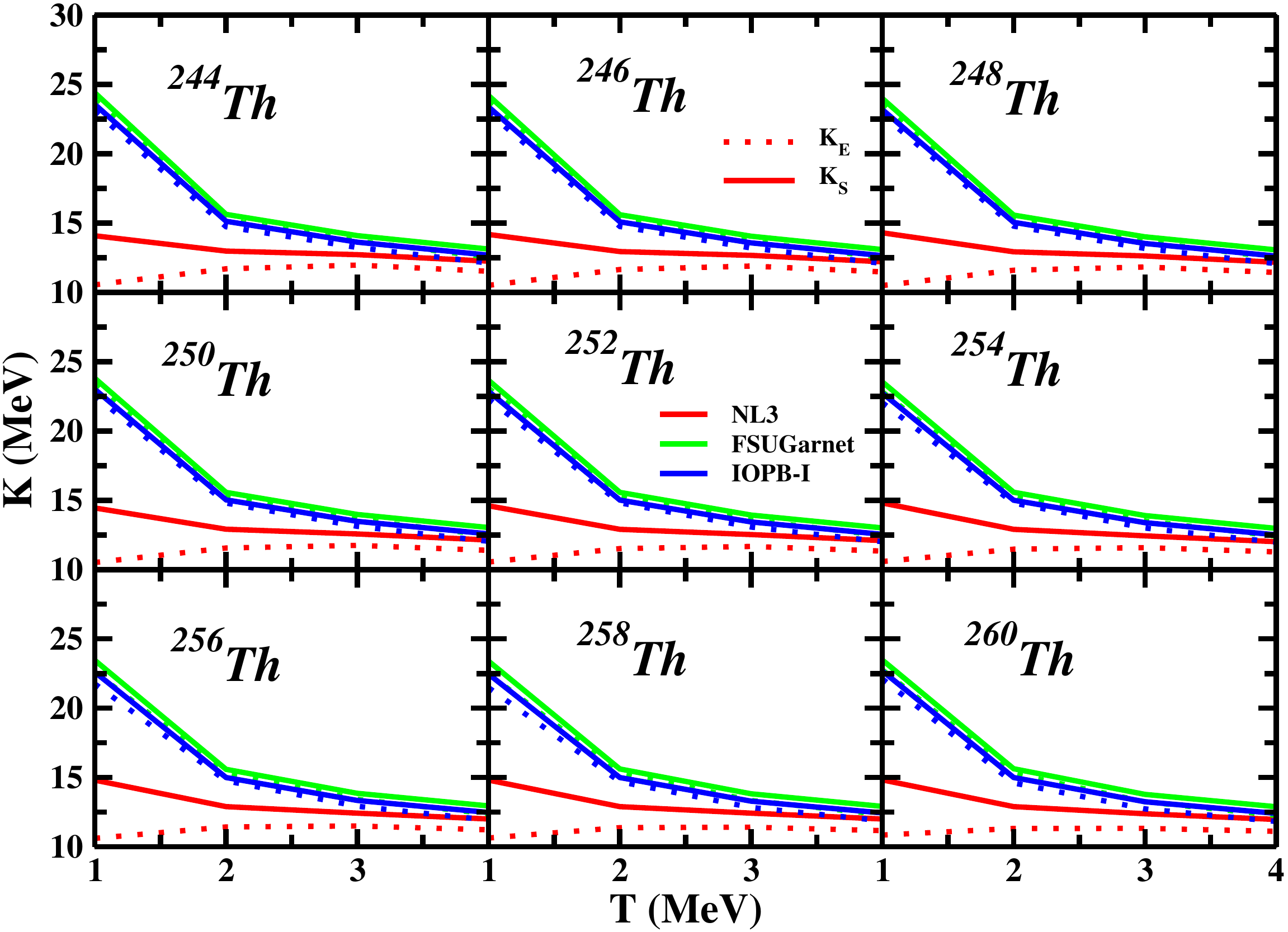}
\caption{Same as Figure~\ref{ilp}, but for the nuclei $^{244,246,248,250,252,254,256,258,260}Th$. }
\label{ilp1}
\end{figure}
\begin{figure}
\includegraphics[width=0.9\columnwidth,height=9.0cm]{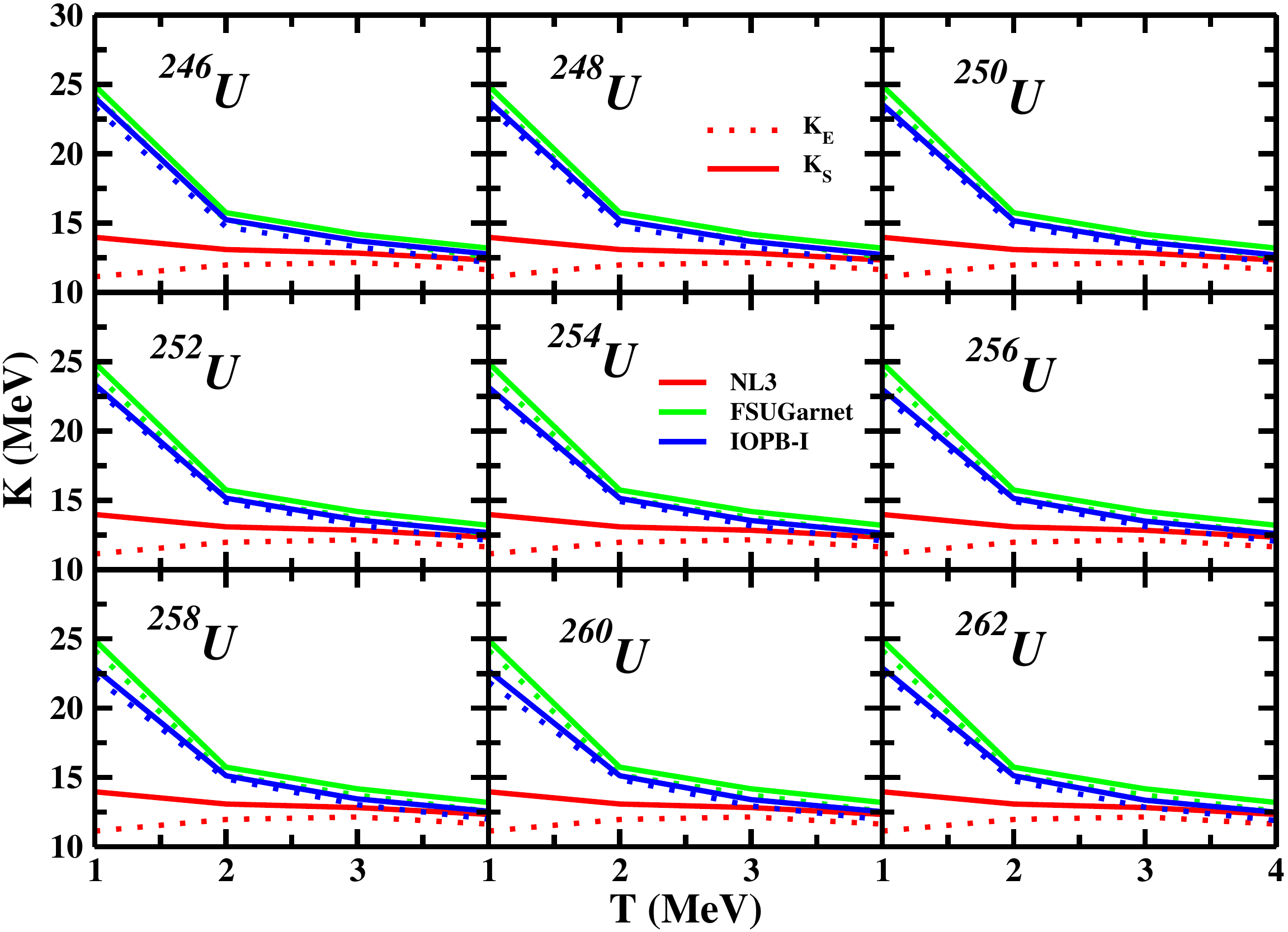}
\caption{Same as Figure \ref{ilp}, but for $^{246,248,250,252,254,256,258,260,262}U$.}
\label{ilp2}
\end{figure}

The level density parameter is a key quantity in the study of nuclear fission \cite{bethe}. 
It is used to predict the probability of fission fragments, 
the yield of fragments, and ultimately angular distribution of fragments \cite{seinthil17,bharat17}. The motivation 
behind choosing natural/neutron-rich thermally fissile nuclei as discussed above is to study fission parameters 
at the finite temperature which may be useful for theoretical and experimental fission study. 
The dependence of the level density parameter on temperature and its sensitivity to parameters chosen are discussed below. 
It can be obtained using the excitation energy and entropy as follows:

\begin{eqnarray}
E^* & = & aT^2 \:,
\label{eq34}
\end{eqnarray}
\begin{eqnarray}
S & = & 2aT \:.
\label{eq35}
\end{eqnarray}

The parameter $a$ obtained from equations (\ref{eq34}) and (\ref{eq35}) are equal, when it is independent of 
temperature \cite{Agarwal98}. This is true for higher temperatures i.e., beyond the critical point. But, at low temperatures, 
it is quite sensitive to T. The inverse level density parameter 
defined as $ K = A/a$ (where A is the mass of the nucleus) is presented in Figures~\ref{ilp}-\ref{ilp2}. The bold and dashed lines 
correspond to the K values 
obtained by using the above two equations (Eqs. \ref{eq34}, \ref{eq35}) and are represented by the symbols $ K_E $ and $ K_S$. 
At low temperature, values of K shoot up, and then it becomes smooth with the variation of temperature. The pattern of $ K_E$ and $K_S $ 
are same and there is no appreciable change as temperature rises except the kinks which again correspond to critical points. 
For the nuclei $^{248,250,252,254,256,258,260,262}U$, 
the values of $K_E$ and $K_S$ are distinct for all the parameter sets at low temperatures and become almost constant with increasing T.
The constant K shows the vanishing of the shell structure of the nucleus at the higher temperature (See Fig.~\ref{ilp}). 
These excitation energies are consistent with those shown in Figure~\ref{s2e0}-\ref{s2e2} 
wherein the slope of $ S^2$ versus $ E^*$ curve representing level 
density parameter. It is clear from the figure that FSUGarnet and IOPB-I curves overlap with each other and overestimate 
the values obtained by the NL3 model. This sensitivity of the level density parameter on the choice of Lagrangian density 
will affect the fragment distributions in the fission process.

\subsection{Asymmetry Energy Coefficient}

%

The properties of hot nuclei play a vital role in both nuclear physics and astrophysics \cite{shlomo05}. These properties are 
the excitation energy, entropy, symmetry energy, and density distribution $etc.$. Among these properties, symmetry energy 
and its dependence on density and temperature have a crucial role in understanding various phenomena in heavy-ion collision, 
supernovae explosions, and neutron star \cite{baran05,lattimer07}. As defined in Chapter \ref{chap1}, it is a measure of energy gain in converting isospin 
symmetric nuclear matter to asymmetric one. The temperature-dependence of symmetry energy plays its role in changing the location 
of the neutron drip line. It has key importance for the liquid-gas phase transition of asymmetric nuclear matter, 
the dynamical evolution of mechanism of a massive star, and the supernovae explosion \cite{baron85}. 

Experimentally, nuclear symmetry energy is not a directly measurable quantity. It is extracted indirectly from the observables 
that are related to it \cite{Shetty,kowalski}. Theoretically, there are the different definitions for different systems. 
For infinite nuclear matter, symmetry energy is the expansion of  binding energy per nucleon in terms of isospin asymmetry 
parameter, i.e., $ \alpha = (\rho_n - \rho_p)/\rho $ as \cite{DeA12}:

\begin{eqnarray}
\epsilon (\rho,\alpha) & = & \epsilon (\rho,0) + a^{\upsilon}_{sym} (\rho)\alpha^2 + O(\alpha^4) \:, 
\end{eqnarray}
where $\rho = \rho_n + \rho_p $ is the nucleon number density. The coefficient $a^{\upsilon}_{sym}$ in the second term is the asymmetry 
energy coefficient of nuclear matter at density $\rho$.   
For finite nuclei, symmetry energy is defined as one of the contributing terms due to asymmetry in the system 
in the Bethe-Weizsacker mass formula. The coefficient of symmetry energy is defined as \cite{daniel03};

\begin{eqnarray}
a_{sym}(A) & = & \frac{a^{\upsilon}_{sym}}{1 + (a^{\upsilon}_{sym}/a^s_{sym})A^{-1/3}}
\;,
\end{eqnarray}
where $ a^{\upsilon}_{sym}$ and $ a^s_{sym}$ are the volume and surface asymmetry energy coefficients. The 
$a^{\upsilon}_{sym}$ is considered as the asymmetry energy coefficient of infinite nuclear matter at saturation density.  
The empirical value of $a_{sym}$ for nuclear matter 
at saturation density $\rho_0$ and temperature T=0 is $\sim{30-34}$ MeV \cite{Jiang,Myers,daniel03,DeA12,Agarwal14,hard06}.
Here, we have calculated the asymmetry energy coefficient of the nucleus of mass number A at finite temperature by using 
the method \cite{dean02,De12}:

\begin{eqnarray}
a_{sym}(A,T) = [\epsilon_b(A,X_1,T) - \epsilon_b(A,X_2,T)]/(X^2_1-X^2_2)\;,
\label{asymeqn}
\end{eqnarray}
where $X_1$ and $X_2$ are the neutron excess ($X=\frac{N-Z}{A}$) of a pair 
of nuclei having the same mass number A but different proton number Z. 
We have taken $Z_2 = Z_1-2$ for calculating $a_{sym}$ where, 
$Z_1$ is the atomic number of the considered nucleus. 
The $\epsilon_b$ 
is the energy per particle obtained by subtracting the Coulomb part, i.e., Coulomb 
energy due to an exchange of photon in the interaction of nucleons is subtracted
from the total energy of the nucleus ($\epsilon_b = \epsilon - \epsilon_c$).  
For example, to study the asymmetry energy coefficient $a_{sym}$ for $^{236}U$ at temperature T
using Eq. (34), we estimate the binding energy $\epsilon_b(A,X_1,T)$ for $^{236}U$
at finite temperature T without considering Coulomb contribution. Then the binding
energy of $^{236}Th$ $\epsilon_b(A,X_2,T)$ is measured in similar conditions, i.e.
without Coulomb energy at temperature T. Here $Z_1=92$ for Uranium and $Z_2=90$ for Thorium with
same mass number A=236 are chosen. Note that the value of $Z_2$ need not be $Z_1-2$, but
generally a pair of even-even nuclei are considered whose atomic number differ by 2.
The temperature dependence $a_{sym}$ coefficient  for different isotopic chains had 
been studied by using various definition as mentioned above and with relativistic 
and non-relativistic Extended Thomas Fermi Model \cite{ant17,De12,Zhang14}.
In Ref. \cite{De12}, the effect of choosing different pairs of nuclei on $a_{sym}$ coefficient is discussed.


\begin{figure}
\includegraphics[width=0.9\columnwidth,height=9.0cm]{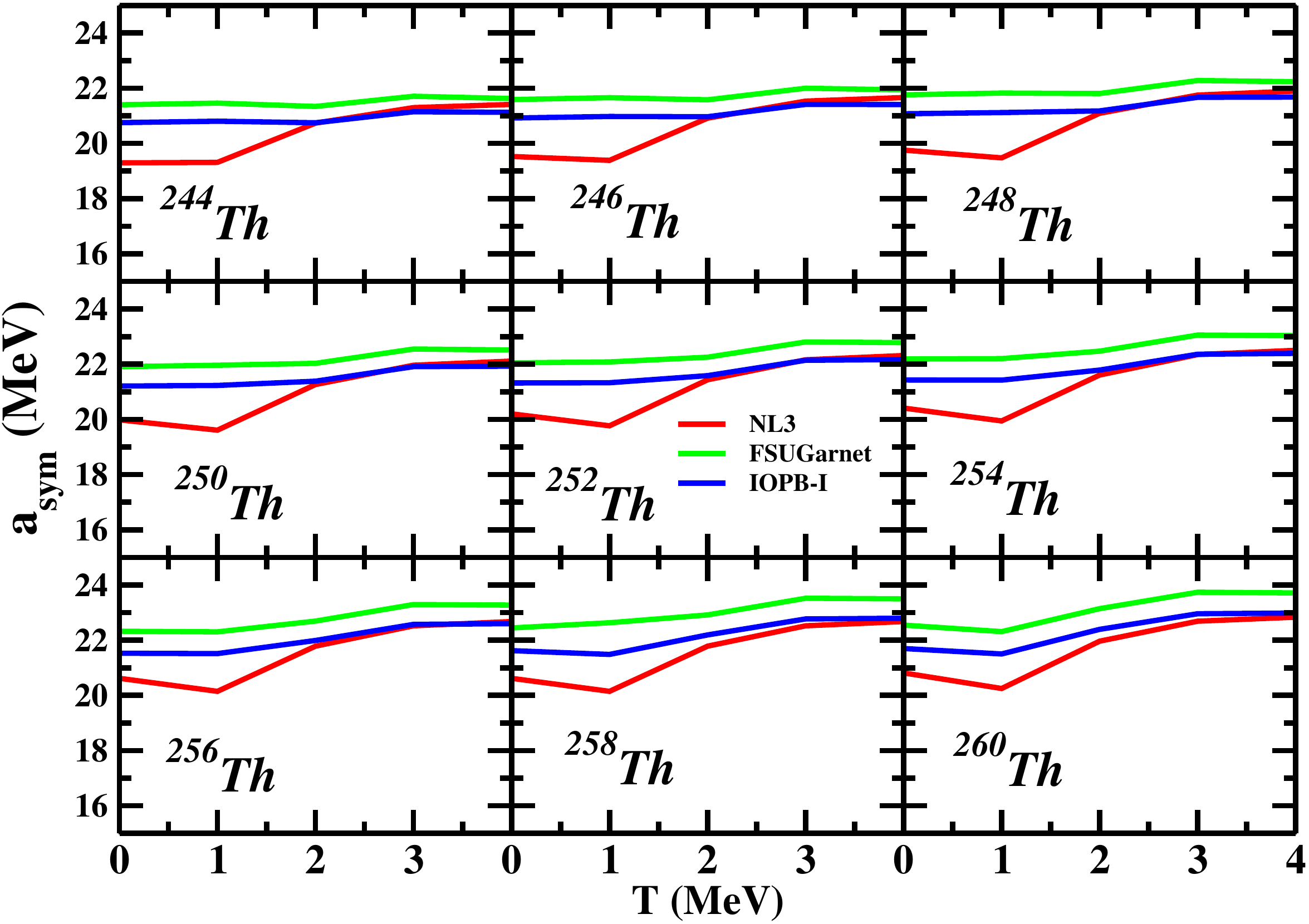} \caption{The asymmetry energy coefficient $a_{sym}$ as a function
        of temperature T for the nuclei $^{244-260}Th$ with FSUGarnet, IOPB-I, and NL3
        parameter sets.nuclei. }
\label{asymt1}
\end{figure}
\begin{figure}
\includegraphics[width=0.9\columnwidth,height=9.0cm]{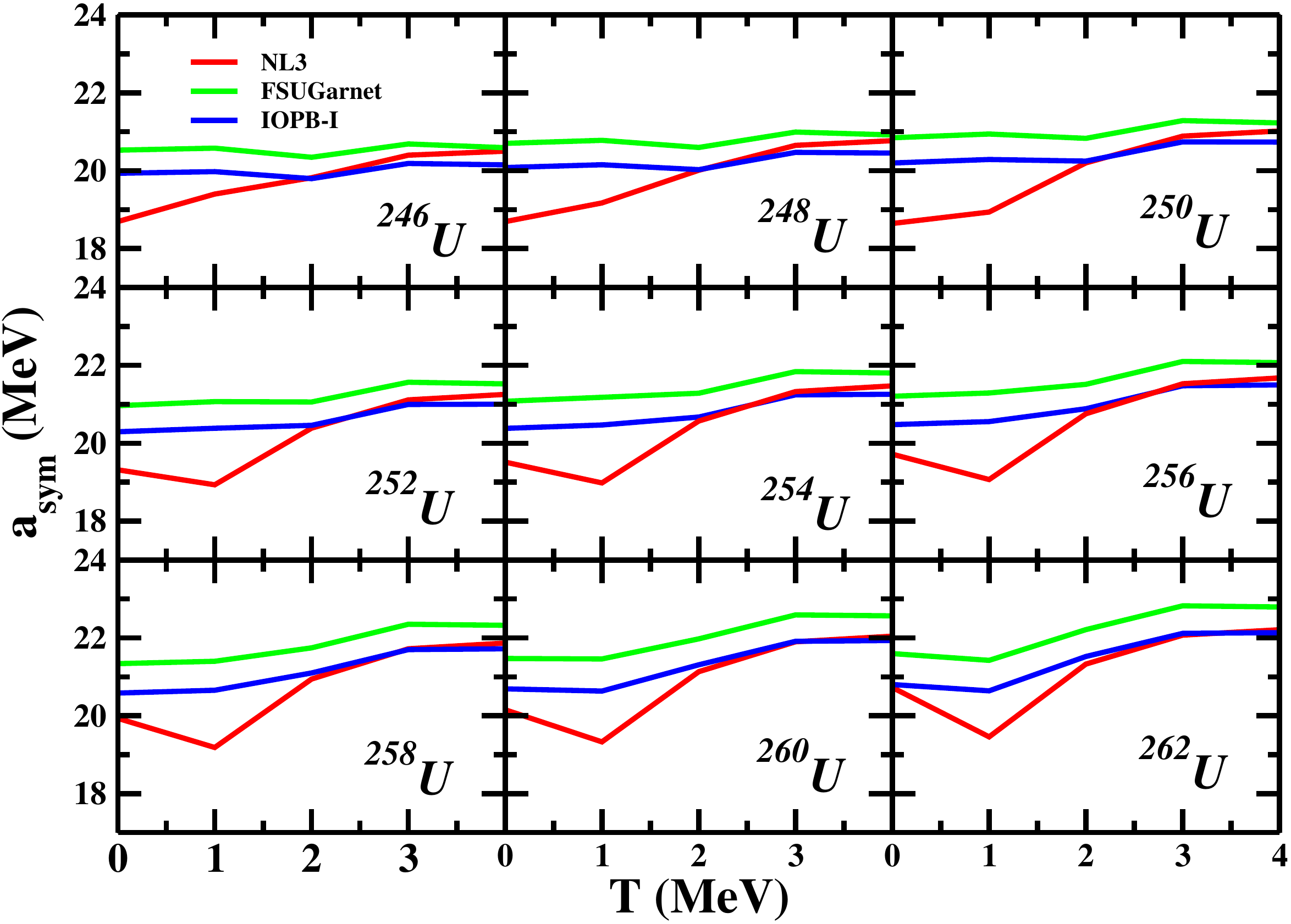} \caption{Same as Figure \ref{asymt1}, but for $^{246-262}U$ nuclei.}
\label{asymt2}
\end{figure}
\begin{figure}
\includegraphics[width=0.9\columnwidth,height=9.0cm]{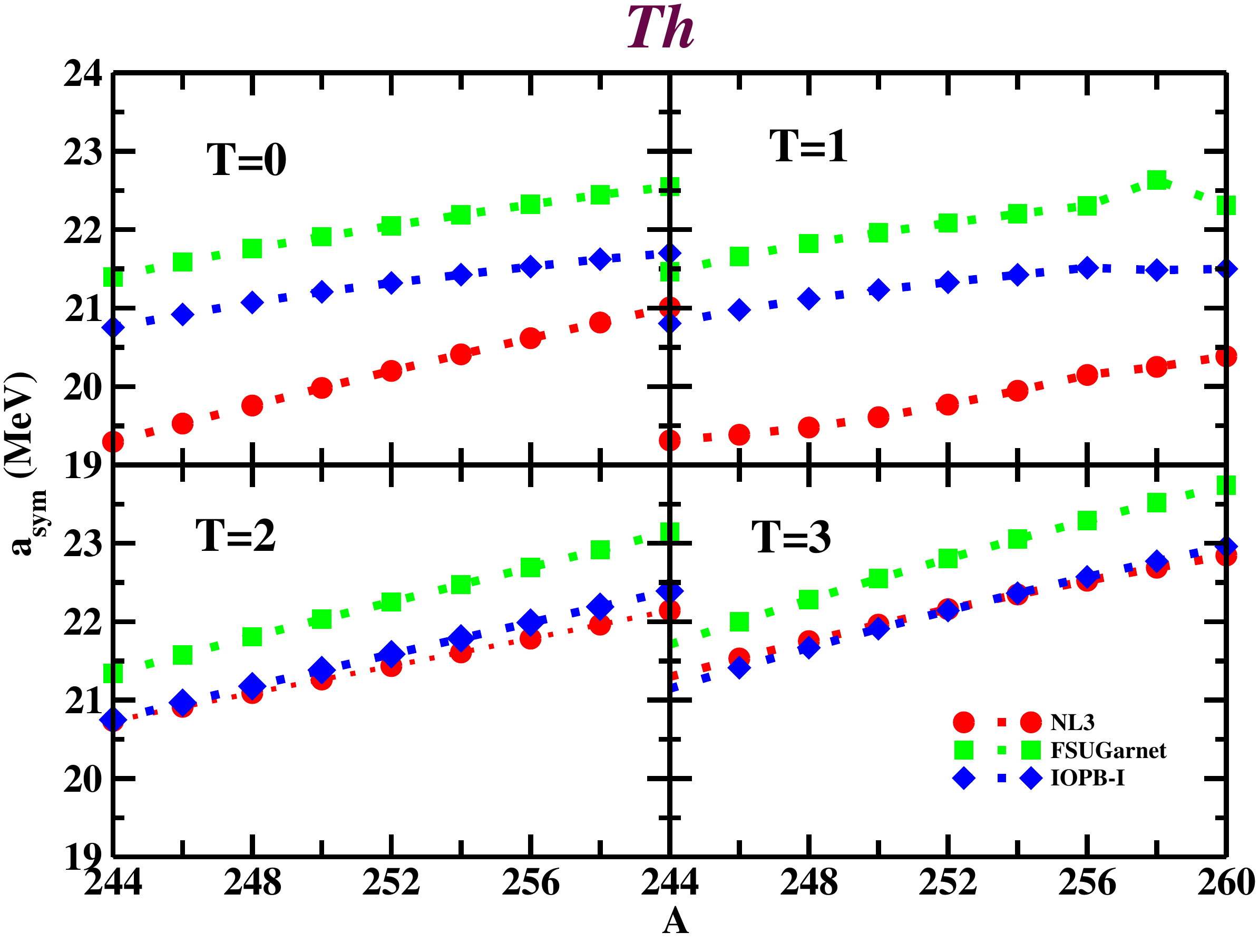} \caption{The variation of asymmetry energy
coefficient $a_{sym}$ with mass number A for $Th$ isotopes at different temperatures.}
\label{asyma1}
\end{figure}
\begin{figure}
\includegraphics[width=0.9\columnwidth,height=9.0cm]{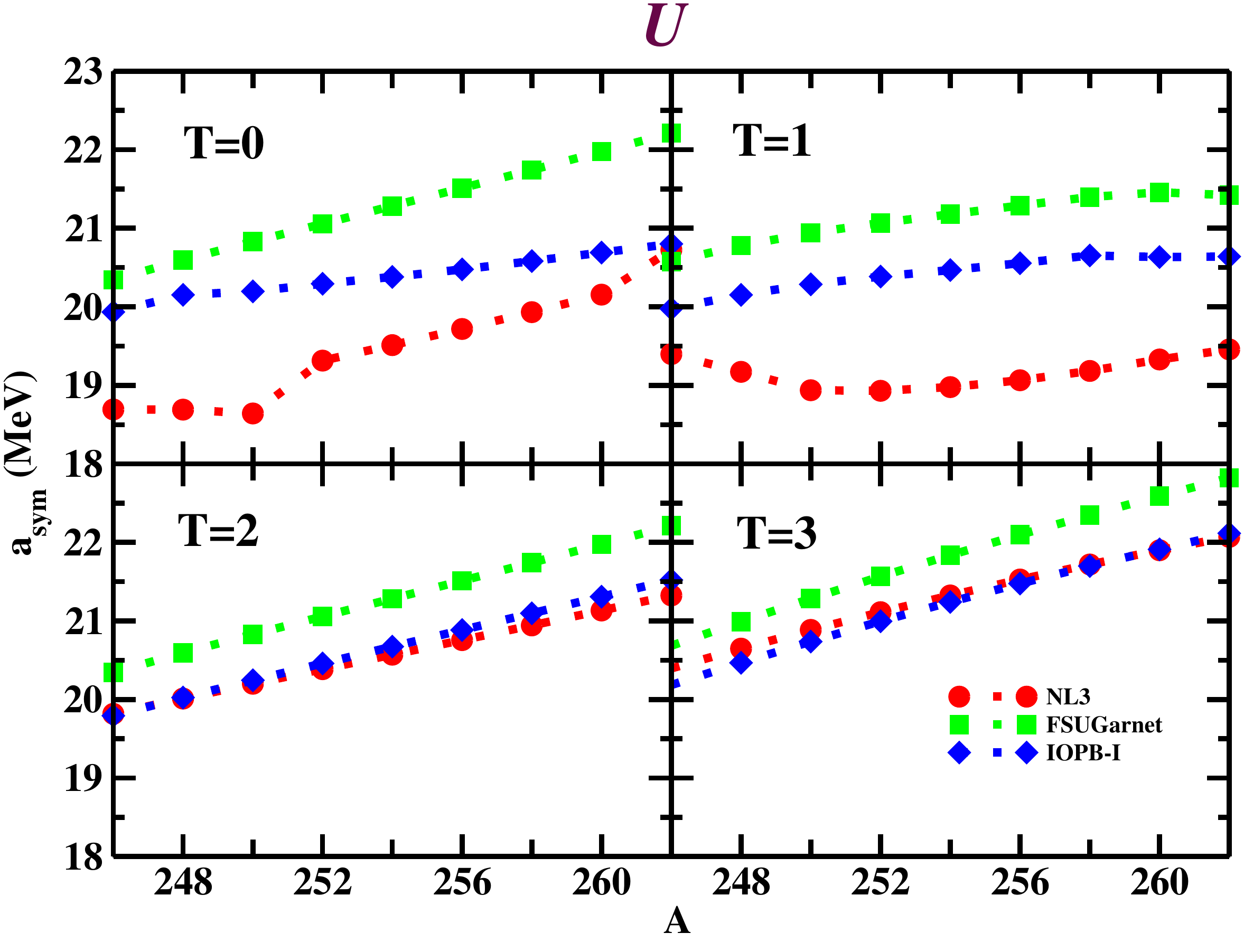} \caption{Same as Fig. \ref{asyma1}, but for $U$ isotopes.}
\label{asyma2}
\end{figure}

The variation of the asymmetry energy coefficient $a_{sym}$ with T is shown in Figures \ref{asymt1}, and \ref{asymt2} 
for some of the considered nuclei as mentioned in the figures. It can be observed from the figures that 
the behavior of asymmetry energy coefficient is force-dependent and vary with different pair of nuclei. 
For instance, FSUGarnet consistently predicts larger 
$a_{sym}$ than those of NL3 and IOPB-I. At lower temperatures, all the considered forces are distinct from one 
another but after $T\approx2$ MeV, results of NL3 and IOPB-I coincide in most of the cases. 
The asymmetry energy coefficient increases with T as shown in Figs. \ref{asymt1}, \ref{asymt2}. However, it is 
shown to be decreased with T in Refs. \cite{sam07,De12}. This disagreement may be due to the choice of pair of 
nuclei used in Eq. \ref{asymeqn} to calculate $a_{sym}$. Hence, a more consistent way is required to study the 
temperature-dependence of symmetry energy within microscopic models, e.g., TRMF formalism. In this approach, we have worked
 by considering the symmetric nuclear matter properties in local density approximation to calculate the corresponding 
properties of finite nuclei at finite T (see Chapter \ref{chap5}), and temperature-independent symmetry energy for finite nuclei 
with the application of the Coherent Density Fluctuation Model in RMF formalism (Chapter \ref{chap4}). 
The variation of $a_{sym}$ with the mass number at different 
temperatures for $Th$ and $U$ nuclei are shown in Figs. \ref{asyma1} and \ref{asyma2}. In both the figures, one 
can observe the gradual increase of $a_{sym}$ with the increase of mass number A almost in a straight line. There 
is a significant change of $a_{sym}$ in the isotopic series of both $Th$ and $U$. This variation of asymmetry 
energy coefficient can help in estimating the various phenomena as mentioned earlier.

It can be noticed from the figures that at all the temperatures, the values of $a_{sym}$ in the isotopic 
chains are maximum for FSUGarnet force as compared NL3 and IOPB-I. The calculated results of $a_{sym}$ 
almost coincide at temperature T=2 and 3 MeV for NL3 and IOPB-I. It is clear that one can not establish 
a direct relation of asymmetry energy parameter $a_{sym}$ of the nucleus with the fission yield, 
but it has a great influence on the equation of states. The effect of $a_{sym}$ on fission properties is not 
yet settled as $a_{sym}$ depends on many factors, such as parameter sets and choice of nuclear pairs, etc. 
Recently, a relation of $a_{sym}$ with the finite and infinite nuclear matter has been established 
\cite{bhu18,ant18,gai11,antonov}, which may be able to eradicate these limitations.

\section{Conclusions}

For the first time, we have applied the recently proposed FSUGarnet and IOPB-I parameter sets
of RMF formalism to deform nuclei at finite temperature. The calculated results are compared 
with those of the NL3 force. We have investigated the temperature-dependent properties of 
$^{234,236}U$, $^{240}Pu$, $^{244-262}Th$, and $^{246-264}U$ nuclei. 
 The quadrupole and hexadecapole deformation parameters $\beta_2$ and $\beta_4$ 
approach to zero at the critical temperature $T_c$, which is about 2 MeV for the neutron-rich thermally fissile nuclei 
(and less for remaining nuclei) irrespective 
of the force parameters. All the analysis, such as the $S_{2n}$ trends, $S^2\sim E^*$ relation, $\epsilon\sim T$ 
(single-particle energy with temperature), $\beta_2$ and $\beta_4$ shape transition and the vanishing of shell 
correction are consistently agree with the transition temperature of $T_c\sim 2$ MeV.

The inverse level density parameter is found to be quite sensitive to the temperature and interactions at low T. 
Because of the similarity in the interactions of FSUGarnet and IOPB-I, the predicted results are also found to
be similar in maximum cases. In spite of similar predictions of FSUGarnet and IOPB-I models, the IOPB-I predicts 
moderate neutron skin-thickness as compared to the rest of the sets. The NL3 overestimates and the FSUGarnet 
underestimates the neutron skin-thickness with respect to IOPB-I. This gives a direct relation of the force 
parameter sets with the EOS for neutron star matter. As a result, the neutron star radius estimated by NL3 has 
a larger value and IOPB-I has a moderate estimation. In the present calculations, though the NL3 and IOPB-I 
predict similar $R_n$, the prediction of $\Delta R$ is in the order $\Delta R (NL3) > \Delta R (IOPB-I) 
> \Delta R (FSUGarnet)$. This result conclude that the fission yield with asymmetric fragments will be more 
in the case of the NL3 set than the FSUGarnet and then with IOPB-I.

We have also studied the symmetry energy coefficient with temperature as well as mass number.  The $a_{sym}$
increases with both T and A. The rate of increase of $a_{sym}$ with T for an isotope is slower as compared
to the rate of increase of asymmetry energy coefficient in an isotopic chain with mass number A. Although, a 
relation of $a_{sym}$ with fission parameters could not be established definitely, we can draw a general conclusion
that with the increase of $a_{sym}$, the generation of mass fragments enhanced in a fission process. This is because 
with an increase in T both $E^*$ and $a_{sym}$ increase.
It is worthy to mention that the value of $a_{sym}$ at the normal nuclear matter density has a large variation 
(with a range of $a_{sym}=-138.96$ to 51.01 MeV) with parameter sets \cite{dutra12} and also it shows a diverse 
trend as a function of nuclear density for different parameter sets \cite{goud}. The latest empirical values of 
$a_{sym}$ at the saturation vary around 32.5 MeV \cite{moller}. The recent works on $a_{sym}$ for finite nuclei 
and infinite nuclear matter may settled down the issue \cite{bhu18,gai11,ant18,antonov}.
The present results may give some guideline for the study of hot thermally fissile nuclei in nuclear fission 
dynamics and nuclear astrophysics. The properties of the neutron-rich thermally fissile nuclei may be useful 
for further theoretical and experimental fission studies for energy production in the future.


\chapter{Temperature-Independent Symmetry Energy of Finite Nuclei and Nuclear Matter}{\label{chap4}}

\rule\linewidth{.5ex}

\section{Introduction} 
\label{intro4}

As discussed in Chapter \ref{chap1}, the nuclear symmetry energy is an important quantity having significant role in different areas of
nuclear physics, for example, in structure of ground-state nuclei \cite{Niksic08,Van10,Dalen10}, physics
of giant collective excitation \cite{Rodin07}, dynamics of heavy-ion reactions \cite{Chen08,Colonna09},
and physics of neutron star \cite{Steiner05,james,fattoyev12,dutra12,dutra14}.
The characterization of the symmetry energy is an important step to interpret neutron-rich nuclei and the neutron star.   
In Refs. \cite{gai11,bhu18}, it has been shown that 
the symmetry energy of finite nuclei can be used as an observable to indicate/determine magic nuclei within the 
Coherent Density Fluctuation Model (CDFM). The neutron pressure of finite nuclei is related to the slope parameter ($L$) of 
symmetry energy at saturation, which is an essential quantity in determining the equation of state (EoS) 
of nuclear matter \cite{gai11,bhu18,latt14,aqsubm}. Furthermore, in finite nuclei, the pressure depends 
on the strength of interaction among nucleons and their distributions. 
The symmetry energy and neutron pressure are collectively termed as effective surface properties, which are extensively defined in Refs. 
\cite{gai11,bhu18} and also illustrated in Sub-sec. \ref{denwef}. 

The importance of the surface properties and their sensitivity to density have motivated us to pursue their systematic study over 
all regions of the nuclear chart. Here, we have investigated the effective surface properties for the 
isotopic series of $O$, $Ca$, $Ni$, $Zr$, $Sn$, $Pb$, and $Z = 120$ nuclei, which cover all the region of the mass 
table from light to superheavy. Recently, the symmetry energy of finite nuclei at a local 
density has been studied by using various formulae of the liquid drop model \cite{wd66,moller95,kp03}, 
the energy density functional of Skyrme force \cite{chen05,yosh06,chen10}, the random phase 
approximation based on the Hartree-Fock (HF) approach \cite{carbone10}, the relativistic 
nucleon-nucleon interaction \cite{lee98,bka10}, and the effective relativistic Lagrangian with 
density-dependent meson-nucleon vertex function \cite{dv03}. In Refs. 
\cite{gai11,bhu18,anto,gai12} the surface properties of the nuclei have been studied by 
folding the nuclear matter properties, within the BEDF  
\cite{brueck68,brueck69}, with the weight functions of the nuclei in the CDFM \cite{anto,antob}. 
The advantages of CDFM over other methods are that this method takes 
care of (i) the fluctuation that arises in the nuclear density distribution via weight function $\vert f (x) 
\vert^2$, and (ii) the momentum distributions through the mixed density matrix (i.e., the Wigner 
distribution function) \cite{bhu18,gai11,gai12}. In other words, the CDFM approach is 
adept to comprise the variation arise from the density and momentum distributions at the surface of 
finite nuclei. The present work covers systematic studies of the effective surface properties 
of several nuclei over the nuclear chart by finding the bulk properties along with densities of the 
nuclei within the effective field theory motivated relativistic mean-field (E-RMF) approach. The 
calculated densities of the nuclei are served as the input to the CDFM to investigate the surface 
properties.


%

\section{The Coherent Density Fluctuation Model (CDFM)}
\label{cdfm}
The CDFM was suggested and developed in Refs. \cite{anto,antob}. In the CDFM, the one-body density 
matrix $\rho$ ({\bf r, r$'$}) of a nucleus can be written as a coherent superposition of infinite 
number of one-body density matrices $\rho_x$ ({\bf r}, {\bf r$'$}) for spherical pieces of the nuclear 
matter called as {\it Fluctons} \cite{gai11,bhu18,anto},
\begin{equation}
\rho_x ({\bf r}) = \rho_0 (x) \Theta (x - \vert {\bf r} \vert), 
\label{denx} 
\end{equation}
with $\rho_o (x) = \frac{3A}{4 \pi x^3}$. The generator coordinate $x$ is the spherical radius of 
the nucleus contained in a uniformly distributed spherical Fermi gas. In finite nuclear system, the 
one body density matrix can be given as \cite{gai11,bhu18,anto,gai12},
\begin{equation}
\rho ({\bf r}, {\bf r'}) = \int_0^{\infty} dx \vert f(x) \vert^2 \rho_x 
({\bf r}, {\bf r'}), 
\label{denr} 
\end{equation}
where, $\vert f(x) \vert^2 $ is the weight function (Eq. (\ref{wfn})). The term $\rho_x ({\bf r}, 
{\bf r'})$ is the coherent superposition of the one body density matrix and defined as,
\begin{eqnarray}
\rho_x ({\bf r}, {\bf r'}) &=& 3 \rho_0 (x) \frac{J_1 \left( k_f (x) \vert 
{\bf r} - {\bf r'} \vert \right)}{\left( k_f (x) \vert {\bf r} - {\bf r'} 
\vert \right)} 
\times \Theta \left(x-\frac{ \vert {\bf r} + {\bf r'} \vert }{2} \right). 
\label{denrr}
\end{eqnarray}
Here, J$_1$ is the first order spherical Bessel function. The Fermi momentum of nucleons in the 
Fluctons with radius $x$ is expressed as $k_f (x)=(3\pi^2/2\rho_0(x))^{1/3} =\gamma/x$, where 
$\gamma=(9\pi A/8)^{1/3}\approx 1.52A^{1/3}$. The Wigner distribution function of the one body 
density matrices in Eq. (\ref{denrr}) is,
\begin{eqnarray}
W ({\bf r}, {\bf k}) =  \int_0^{\infty} dx \vert f(x) \vert^2 W_x ({\bf r}, {\bf k}). 
\label{wing}
\end{eqnarray}
Here, $W_x ({\bf r}, {\bf k})=\frac{4}{8\pi^3}\Theta (x-\vert {\bf r} \vert)\Theta (k_F(x)-\vert 
{\bf k} \vert)$. 
Similarly, the density $\rho$ (r) within CDFM can express in terms of the same weight function as,
\begin{eqnarray}
\rho (r) &=& \int d{\bf k} W ({\bf r}, {\bf k}) 
 = \int_0^{\infty} dx \vert f(x) \vert^2 \frac{3A}{4\pi x^3} \Theta(x-\vert 
{\bf r} \vert)
\label{rhor}
\end{eqnarray}
and it is normalized to the nucleon numbers of the nucleus, $\int \rho ({\bf r})d{\bf r} = A$. By 
taking the $\delta$-function approximation to the Hill-Wheeler integral equation, we can obtain the 
differential equation for the weight function in the generator coordinate \cite{anto,bhu18}. The 
weight function for a given density distribution $\rho$ (r) can be expressed as,
\begin{equation}
|f(x)|^2 = - \left (\frac{1}{\rho_0 (x)} \frac{d\rho (r)}{dr}\right )_{r=x}, 
\label{wfn}
\end{equation}
with $\int_0^{\infty} dx \vert f(x) \vert^2 =1$. For a detailed analytical derivation, one can follows 
Refs. \cite{bhu18,anto,fuch95}. The symmetry energy, neutron pressure, and symmetry energy 
curvature of a finite nucleus are defined below by weighting the corresponding quantities of 
the infinite nuclear matter within the CDFM. The CDFM allows us to make a transition from the properties 
of nuclear matter to those of finite nuclei. Following the CDFM approach, the expression for the 
effective symmetry energy $S$, pressure $P$, and curvature $\Delta K$ for a nucleus can be written 
as \cite{gai11,bhu18,gai12,anto,fuch95,ant17},
\begin{eqnarray}
S = \int_0^{\infty} dx \vert f(x) \vert^2 S_0^{NM} (\rho (x)) ,
\label{s0}
\end{eqnarray}
\begin{eqnarray}
P =  \int_0^{\infty} dx \vert f(x) \vert^2 P_0^{NM} (\rho (x)), 
\label{p0}
\end{eqnarray}
\begin{eqnarray}
\Delta K =  \int_0^{\infty} dx \vert f(x) \vert^2 \Delta K_0^{NM} (\rho (x)).  
\label{k0}
\end{eqnarray}
Here, the quantities on the left-hand-side of Eqs. (\ref{s0}-\ref{k0}) are the surface weighted average 
of the corresponding nuclear matter quantities with local density approximation, which have been 
determined within the method of Brueckner, et al., \cite{brueck68,brueck69}.

In the present work considering the pieces of nuclear matter with density $\rho_0$(x), we have used 
the matrix element V(x) of the nuclear Hamiltonian the corresponding energy of nuclear matter from 
the method of Brueckner {\it et. al.}, \cite{brueck68,brueck69}. In the Brueckner Energy Density Functional (BEDF) 
method, the V(x) is given by:
\begin{eqnarray}
V(x)=A V_0(x) + V_C + V_{CO},
\label{vx}
\end{eqnarray}
where
\begin{eqnarray}
V_0(x)=37.53 [(1+\delta)^{5/3}+(1-\delta)^{5/3}]\rho_0^{2/3}(x) 
+b_1 \rho_0(x) +b_2 \rho_0^{4/3}(x)  \nonumber \\[3mm]
+b_3\rho_0^{5/3}(x) + \delta^2[b_4 \rho_0 (x)+b_5 \rho_0 ^{4/3}(x) + b_6 \rho_0^{5/3}], 
\label{bruc}
\end{eqnarray}
with
$b_1=-741.28$, $b_2=1179.89$, $b_3=-467.54$, $b_4=148.26$, $b_5=372.84$, and $b_6=-769.57$. The 
$V_0$(x) in Eq. \ref{vx} is the energy per particle of nuclear matter (in MeV) which accounts for 
the neutron-proton asymmetry. $V_C$ is the coulomb energy of charge particle (proton) in a flucton,
\begin{eqnarray}
V_C=\frac{3}{5} \frac{Z^2 e^2}{x},
\end{eqnarray}
and $V_{CO}$ is the coulomb exchange energy given by,
\begin{eqnarray}
V_{CO}=0.7386 Z e^2 (3 Z /4 \pi x^3)^{1/3}.
\end{eqnarray}

On substituting $V_0$(x) in Eq. \ref{ssym} and taking its second order derivative, the symmetry energy 
$S_0^{NM}$(x) of nuclear matter with density $\rho_0$(x) is obtained:
\begin{eqnarray}
S_0^{NM}(x) = 41.7 \rho_0^{2/3}(x) + b_4 \rho_0(x) + b_5 \rho_0^{4/3}(x) + b_6 \rho_0^{5/3} (x). 
\end{eqnarray}
The corresponding parameterized expressions for the pressure $P_0^{NM}$(x) and the symmetry energy 
curvature $\Delta K_0^{NM}$(x) for such a system within the BEDF method 
have the forms
\begin{eqnarray}
P_0^{NM}(x) = 27.8 \rho_0^{5/3}(x) + b_4 \rho_0^2(x) + \frac{4}{3} b_5 \rho_0^{7/3}(x)+ \frac{5}{3} 
b_6\rho_0^{8/3}(x), 
\end{eqnarray}
and 
\begin{eqnarray}
\Delta K_0^{NM}(x) = -83.4 \rho_0^{2/3}(x) + 4 b_5 \rho_0^{4/3}(x)+ 10 b_6 \rho_0^{5/3}(x), 
\end{eqnarray}
respectively. These quantities are folded in the Eqs. (\ref{s0}-\ref{k0}) with the weight function to 
find the corresponding quantities of finite nuclei within the CDFM.

\renewcommand{\baselinestretch}{1.0}
\begin{table}
\caption{The calculated binding energy per particle (B/A), and charge radius ($R_{c}$) are compared with the 
available experimental data \cite{audi12,Angeli2013}. The predicted neutron skin-thickness $\Delta R $ 
is also depicted with all the four models and compared with the available experimental data \cite{expskin}.}
\begin{tabular}{cccccccccc}
\hline
\hline
\multicolumn{1}{c}{Nucleus}&
\multicolumn{1}{c}{Obs.}&
\multicolumn{1}{c}{Expt.}&
\multicolumn{1}{c}{NL3}&
\multicolumn{1}{c}{FSUGarnet}&
\multicolumn{1}{c}{G3}&
\multicolumn{1}{c}{IOPB-I}\\
\hline
\hline
$^{16}O$  & B/A             & 7.976 & 7.917 & 7.876 & 8.037 & 7.977  \\
          & R$_{c}$        & 2.699 & 2.714 & 2.690 & 2.707 & 2.705  \\
          & R$_{n}$-R$_{p}$ & -     &-0.026 &-0.029 &-0.028 &-0.027  \\
\\
$^{28}O$  & B/A             & 5.988 & 6.379 & 5.933 & 6.215 & 6.220 \\
          & R$_{c}$        & -     & 2.800 & 2.804 & 2.791 & 2.805 \\
          & R$_{n}$-R$_{p}$ & -     & 0.809 & 0.796 & 0.741 & 0.809 \\
\\
$^{40}Ca$ & B/A             & 8.551 & 8.540 & 8.528 & 8.561 & 8.577  \\
          & R$_{c}$        & 3.478 & 3.466 & 3.438 & 3.459 & 3.458  \\
          & R$_{n}$-R$_{p}$ &-0.08     &-0.046 &-0.051 &-0.049 &-0.049  \\
\\
$^{48}Ca$ & B/A             & 8.666 & 8.636 & 8.609 & 8.671 & 8.638 \\
          & R$_{c}$        & 3.477 & 3.443 & 3.426 & 3.466 & 3.446 \\
          & R$_{n}$-R$_{p}$ & 0.16  & 0.229 & 0.169 & 0.174 & 0.202  \\
\\
$^{68}Ni$ & B/A             & 8.682 & 8.698 & 8.692 & 8.690 & 8.707  \\
          & R$_{c}$        & -     & 3.870 & 3.861 & 3.892 & 3.873  \\
          & R$_{n}$-R$_{p}$ & -     & 0.262 & 0.184 & 0.190 & 0.223  \\
\\
$^{90}Zr$ & B/A             & 8.709 & 8.695 & 8.693 & 8.699 & 8.691  \\
          & R$_{c}$        & 4.269 & 4.253 & 4.231 & 4.276 & 4.253 \\
          & R$_{n}$-R$_{p}$ & 0.09  & 0.115 & 0.065 & 0.068 & 0.091 \\
\\
$^{100}Sn$& B/A             & 8.253 & 8.301 & 8.298 & 8.266 & 8.284 \\
          & R$_{c}$        & -     & 4.469 & 4.426 & 4.497 & 4.464  \\
          & R$_{n}$-R$_{p}$ & -     &-0.073 &-0.078 &-0.079 &-0.077 \\
\\
$^{132}Sn$& B/A             & 8.355 & 8.371 & 8.372 & 8.359 & 8.352 \\
          & R$_{c}$        & 4.709 & 4.697 & 4.687 & 4.732 & 4.706 \\
          & R$_{n}$-R$_{p}$ & -     & 0.349 & 0.224 & 0.243 & 0.287 \\
\\
$^{208}Pb$& B/A             & 7.867 & 7.885 & 7.902 & 7.863 & 7.870  \\
          & R$_{c}$        & 5.501 & 5.509 & 5.496 & 5.541 & 5.521 \\
          & R$_{n}$-R$_{p}$ & 0.17  & 0.283 & 0.162 & 0.180 & 0.221\\
\hline
\hline
\end{tabular}
\label{table4.3}
\end{table}
\renewcommand{\baselinestretch}{1.5}


\section{Ground-State Properties of the Nuclei} {\label {gsp4}}
The main aim of this work is to study the effective surface properties like the symmetry energy $S$,
neutron pressure $P$, and symmetry energy curvature $\Delta K$ for the isotopes (from neutron-deficient
to the neutron-rich side of the nuclear landscape) of the light, heavy, and superheavy nuclei. Before
proceeding to the effective surface properties, we have calculated their ground-state bulk properties
within the E-RMF model. Within the E-RMF formalism, we have used some of the recent force parameters such as the FSUGarnet
\cite{chai15}, IOPB-I \cite{iopb1}, and the G3 \cite{G3}. The FSUGarnet \cite{chai15}, IOPB-I \cite{iopb1},
and G3 \cite{G3} parameters have the advantage that their EoSs are softer compared to the NL3 parameter.
The motive behind choosing these parameter sets has been illustrated in Chapter \ref{chap2}.
From Table \ref{force1}, we notice that the symmetry energy (its then representation is $J$) and its
coefficients $L$, $K_{sym}$, and Q$_{sym}$ are consistent with the allowed empirical and/or experimental
ranges for three force parameters, namely the FSUGarnet, G3, and IOPB-I. The allowed empirical and/or experimental
ranges along with the non-relativistic and relativistic constraints for a large number of force parameter sets
can be found in Refs. \cite{dutra12,dutra14}. In the case of the NL3 force parameter, these values are overestimated
to the allowed constraint ranges. These overestimations are due to the well-known stiffer nature of EoS. Furthermore,
one can find that these ranges get broad with respect to density as well as higher-order derivatives (i.e,
the derived quantities from symmetry energy), which are the accepted behavior of the mean-field models.
The binding energy per particle (B/A), charge radius ($R_{c}$), and neutron skin-thickness
($\Delta R=R_n - R_p $) of some of the double magic nuclei for the FSUGarnet \cite{chai15}, IOPB-I \cite{iopb1},
G3, and NL3 \cite{lala97} parameter sets are listed in Table \ref{table4.3} with the available experimental
data \cite{audi12,Angeli2013}. The calculated properties of finite nuclei corresponding to all chosen parameter
sets are in good agreement with each other. These results are comparable to the available experimental data.
On inspecting the table, it is found that in some cases, the binding energy corresponding to the IOPB-I
force parameter set overestimates the experimental data.


\begin{figure*}[!b]
        \includegraphics[width=0.32\columnwidth,height=5.0cm]{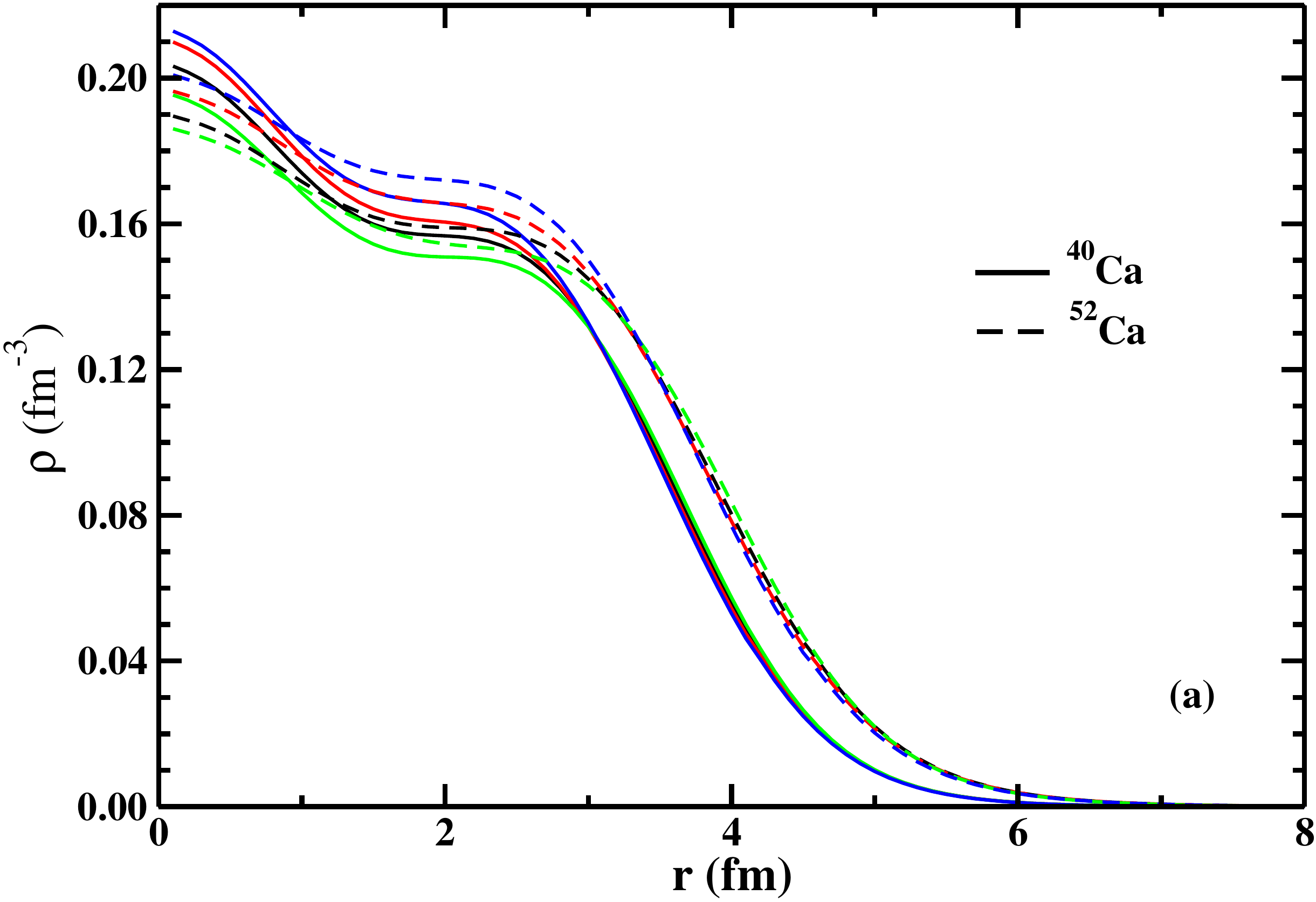}
        \includegraphics[width=0.32\columnwidth,height=5.0cm]{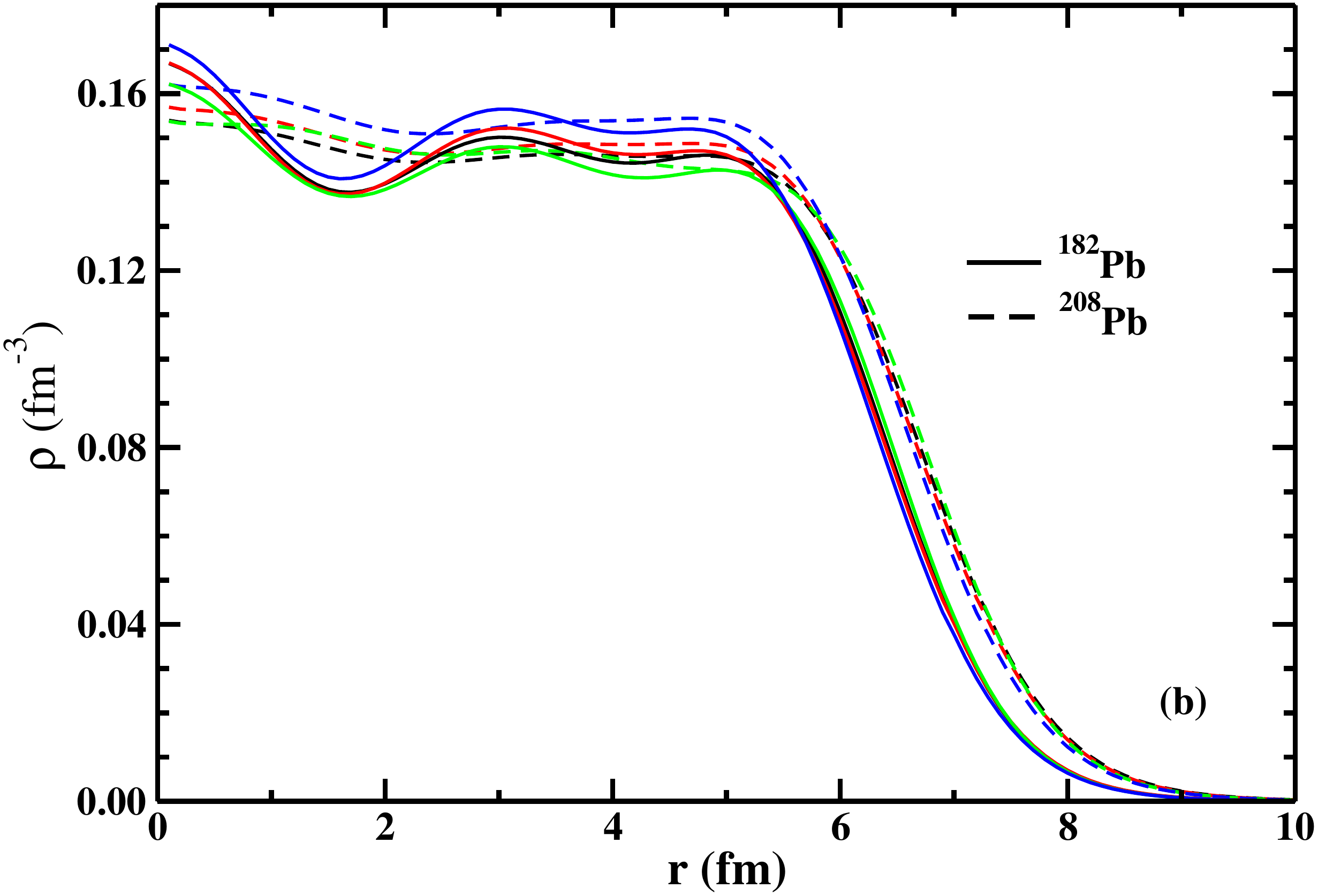}
        \includegraphics[width=0.32\columnwidth,height=5.0cm]{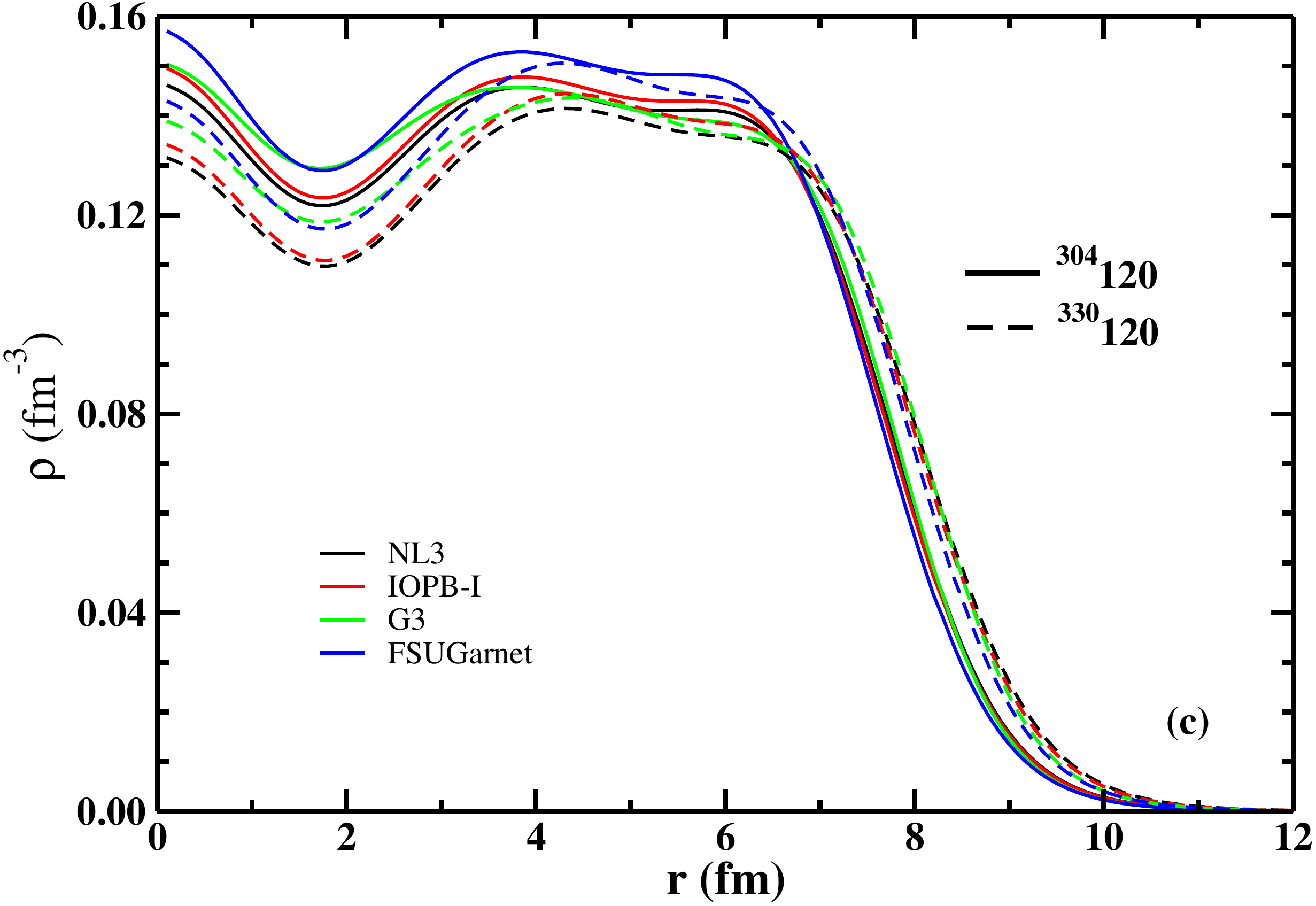}
\caption{(a) The total density profiles for (a) $^{40,52}Ca$, (b) $^{182,208}Pb$, and (c) 
$^{304,330}120$ as the representative cases corresponding to the FSUGarnet \cite{chai15}, IOPB-I \cite{iopb1}, 
G3 \cite{G3}, and NL3 \cite{lala97} parameter sets.}
\label{dens}
\end{figure*}
\begin{figure*}[!b]
        \includegraphics[width=0.35\columnwidth,height=5.05cm]{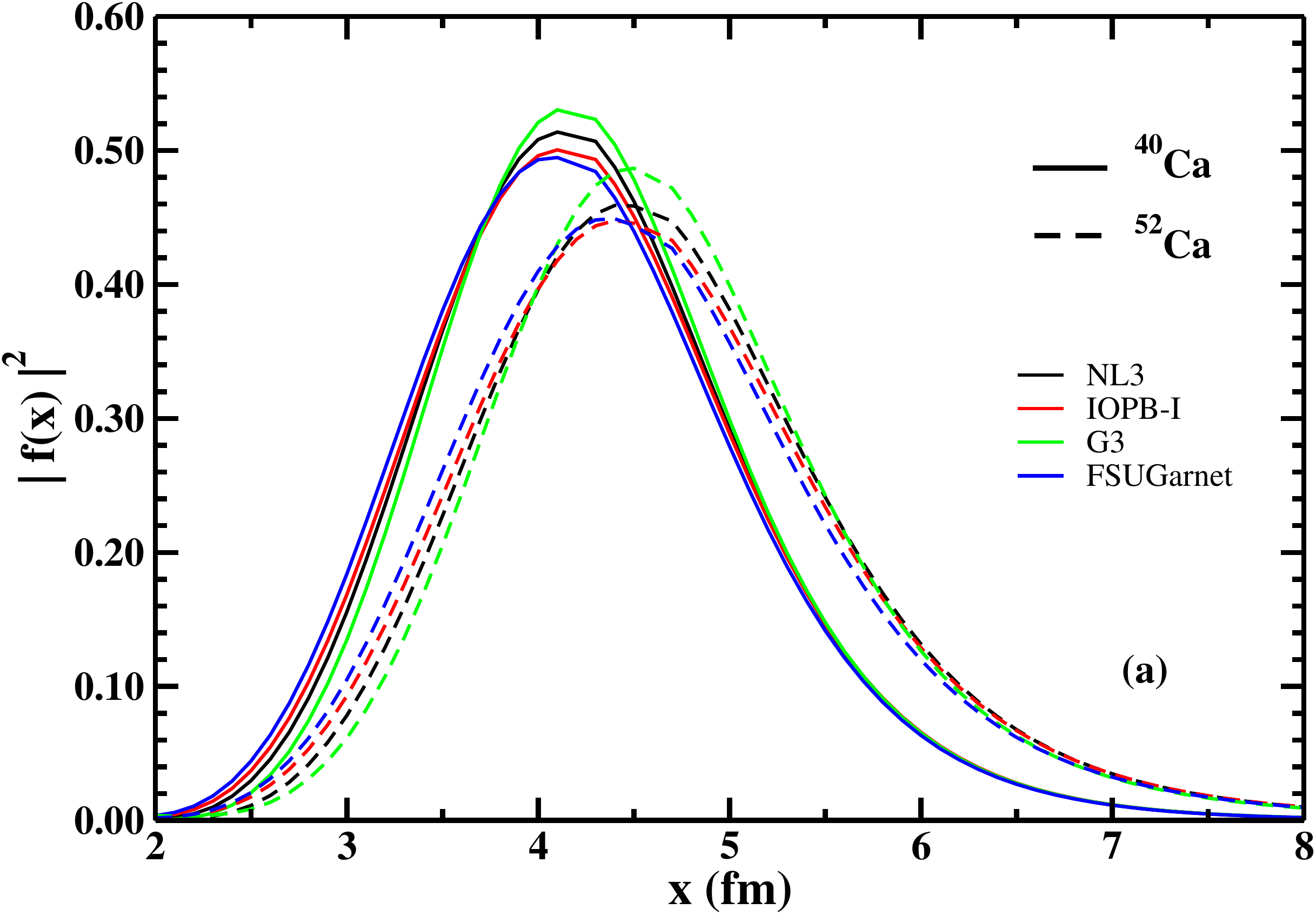}
        \includegraphics[width=0.31\columnwidth,height=5.0cm]{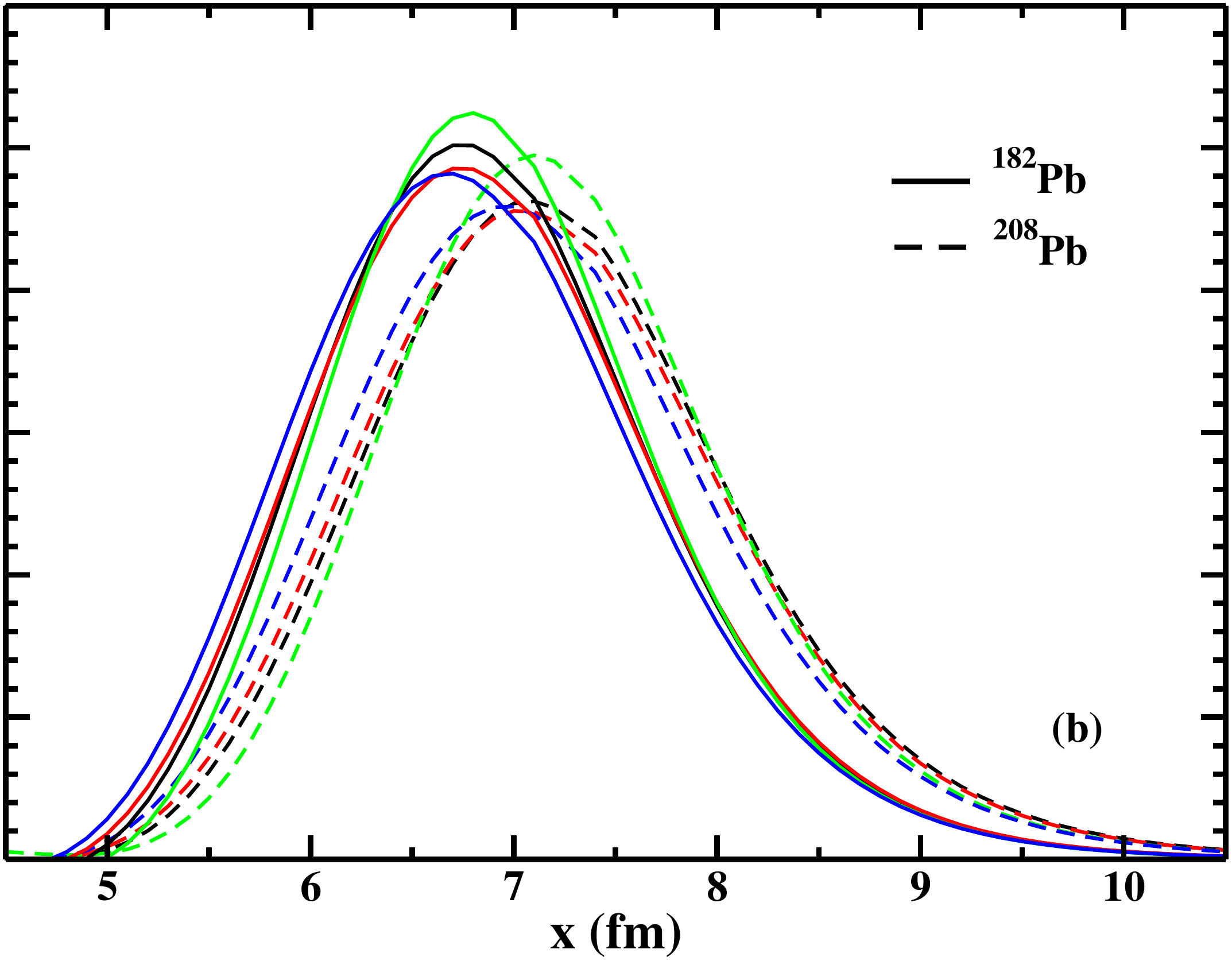}
        \includegraphics[width=0.31\columnwidth,height=5.0cm]{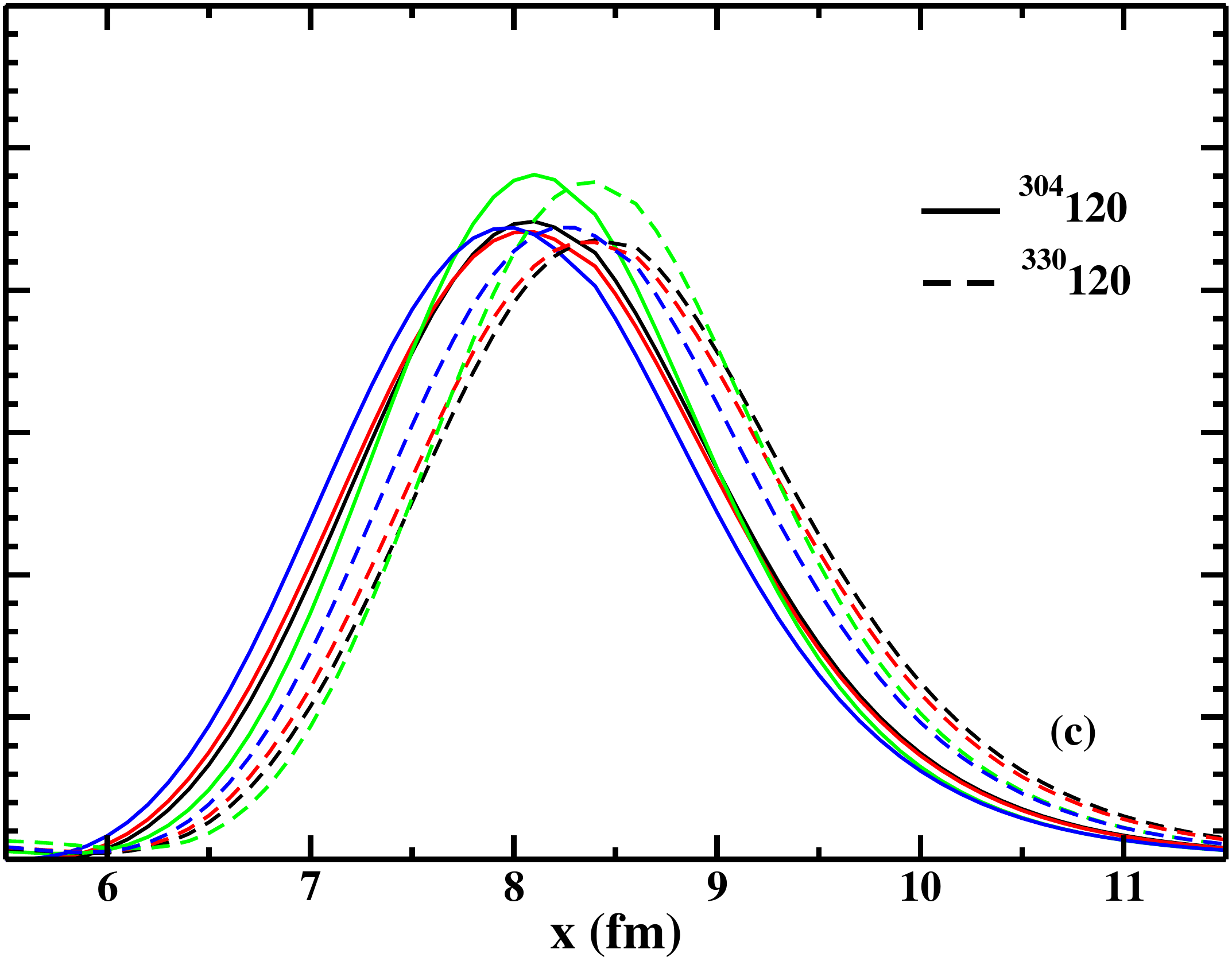}
\caption{The weight function for (a) $^{40,52}Ca$, (b) $^{182,208}Pb$, and (c) $^{304,330}120$ 
as the representative cases corresponding to the FSUGarnet \cite{chai15}, IOPB-I \cite{iopb1}, G3 \cite{G3}, and 
NL3 \cite{lala97} parameter sets.}
        \label{weight}
\end{figure*}


\begin{figure*}[!b]
        \includegraphics[width=0.35\columnwidth,height=5.05cm]{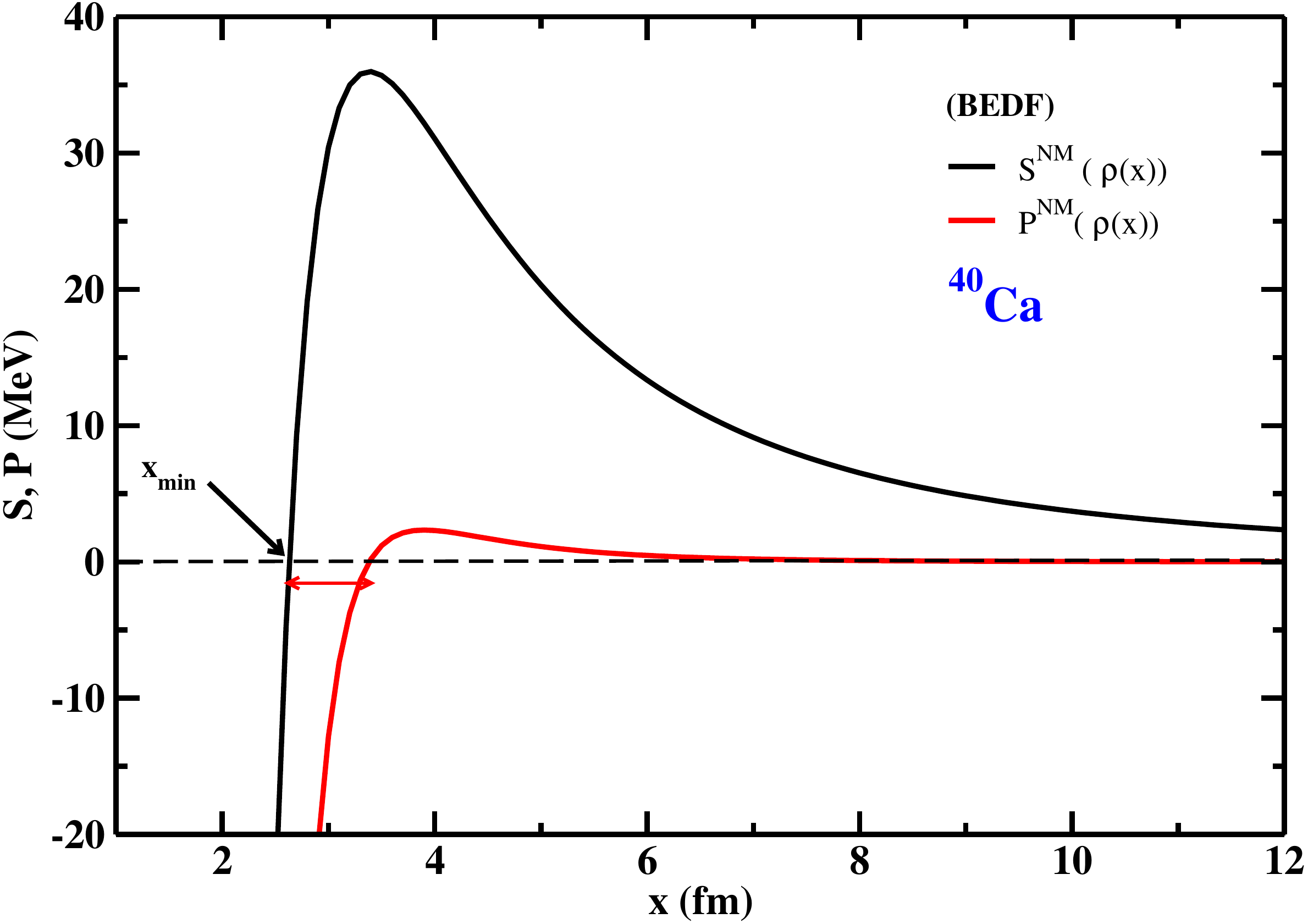}  
        \includegraphics[width=0.31\columnwidth,height=5.0cm]{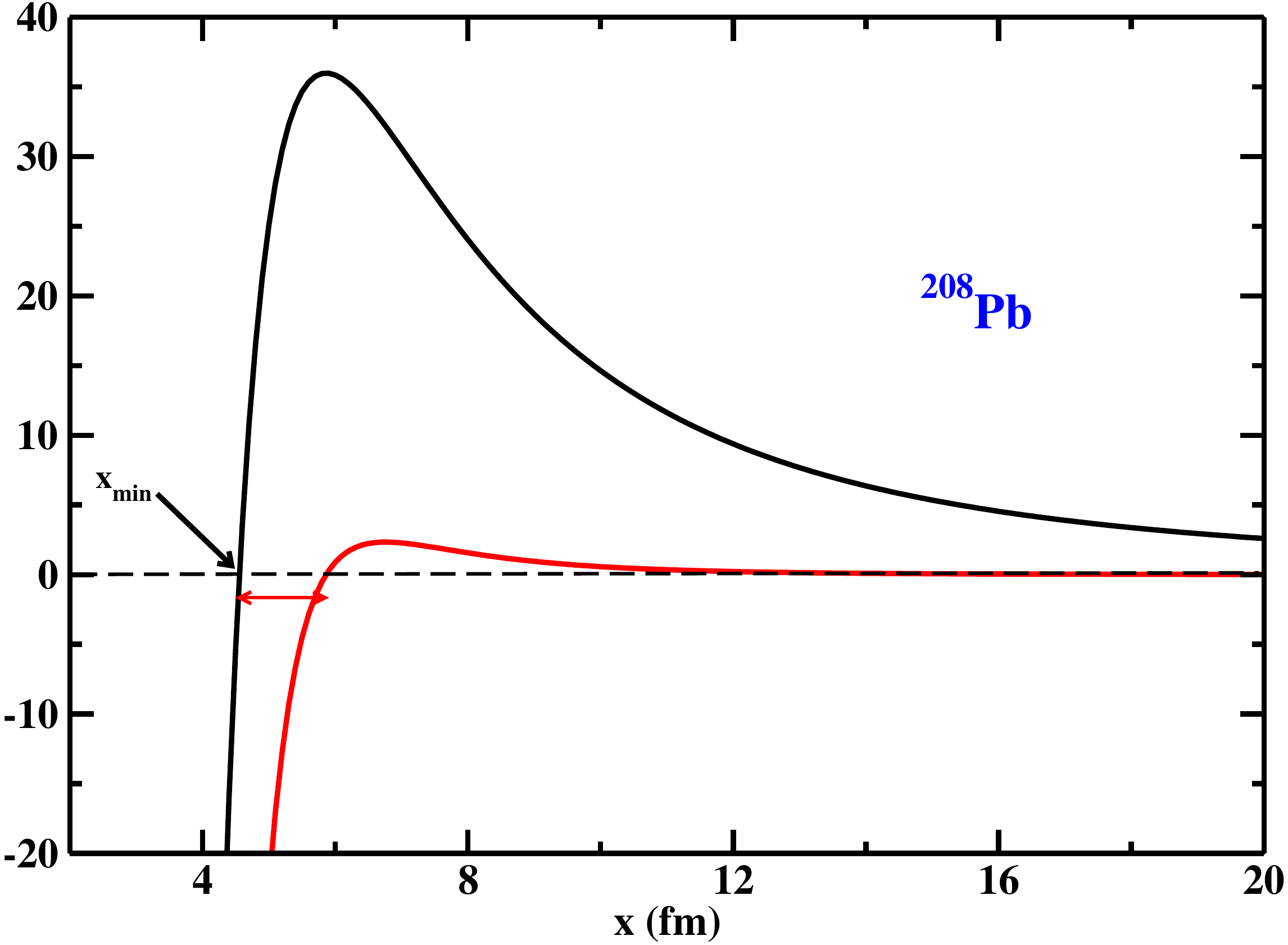}
        \includegraphics[width=0.31\columnwidth,height=5.0cm]{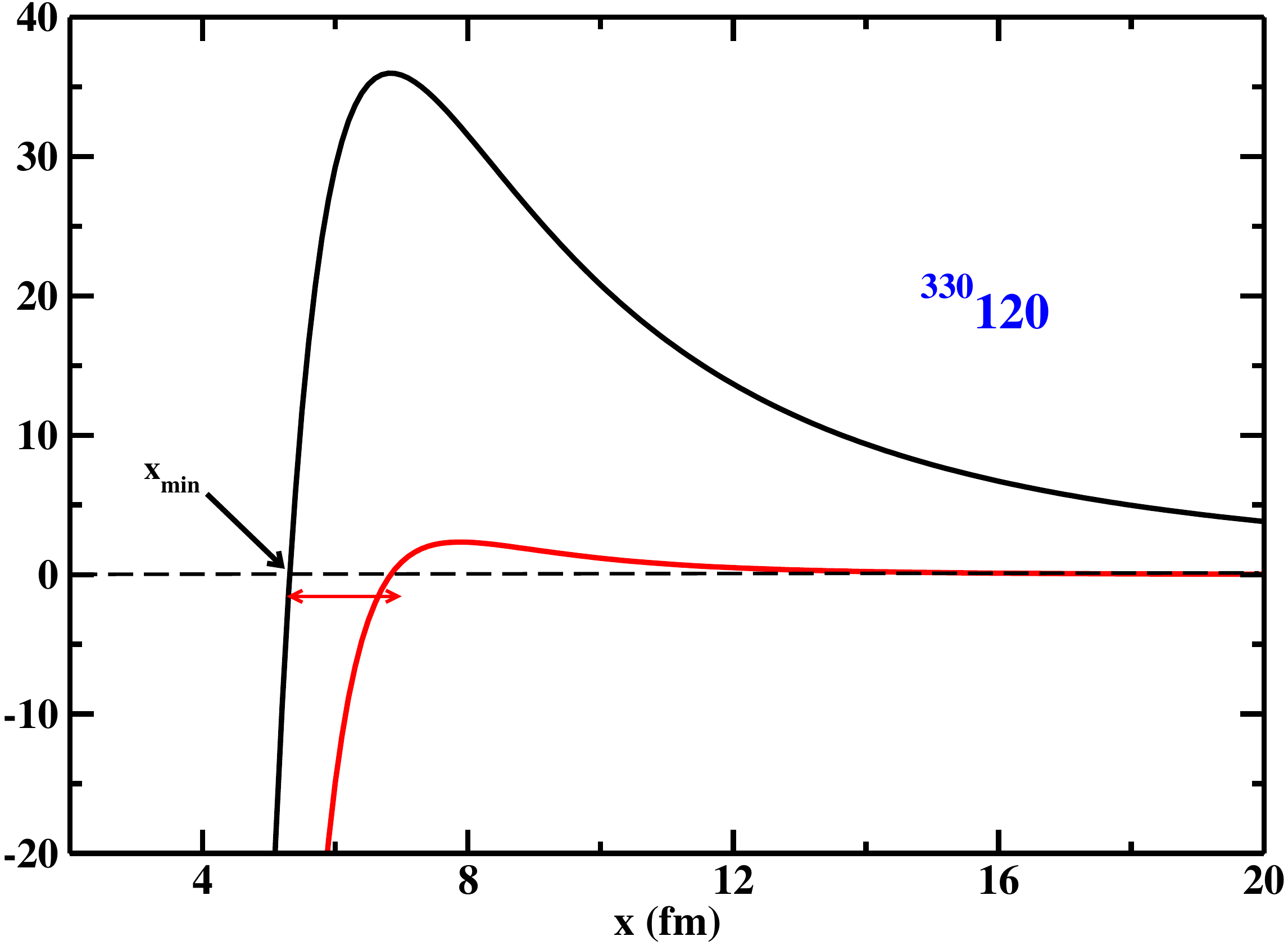}
\caption{The symmetry energy $S^{NM}$, and the neutron presser $P^{NM}$ of nuclear matter at the Flucton's 
density of (a) $Ca$, (b) $Pb$, and (c) $Z = 120$ within the BEDF as a function 
of local coordinate x. The $x_{min}$ represent the lower limit of the integrations (Eqs. \ref{s0}, 
\ref{p0}, and \ref{k0}). The negative values before the $x_{min}$ points are the unphysical values. The dashed lines 
in the figure mark zero to differentiate between positive and negative values of $S^{NM}$, and $P^{NM}$.}
        \label{nms}
\end{figure*}

\begin{figure}[!b]
        \includegraphics[width=0.50\columnwidth,height=5.50cm]{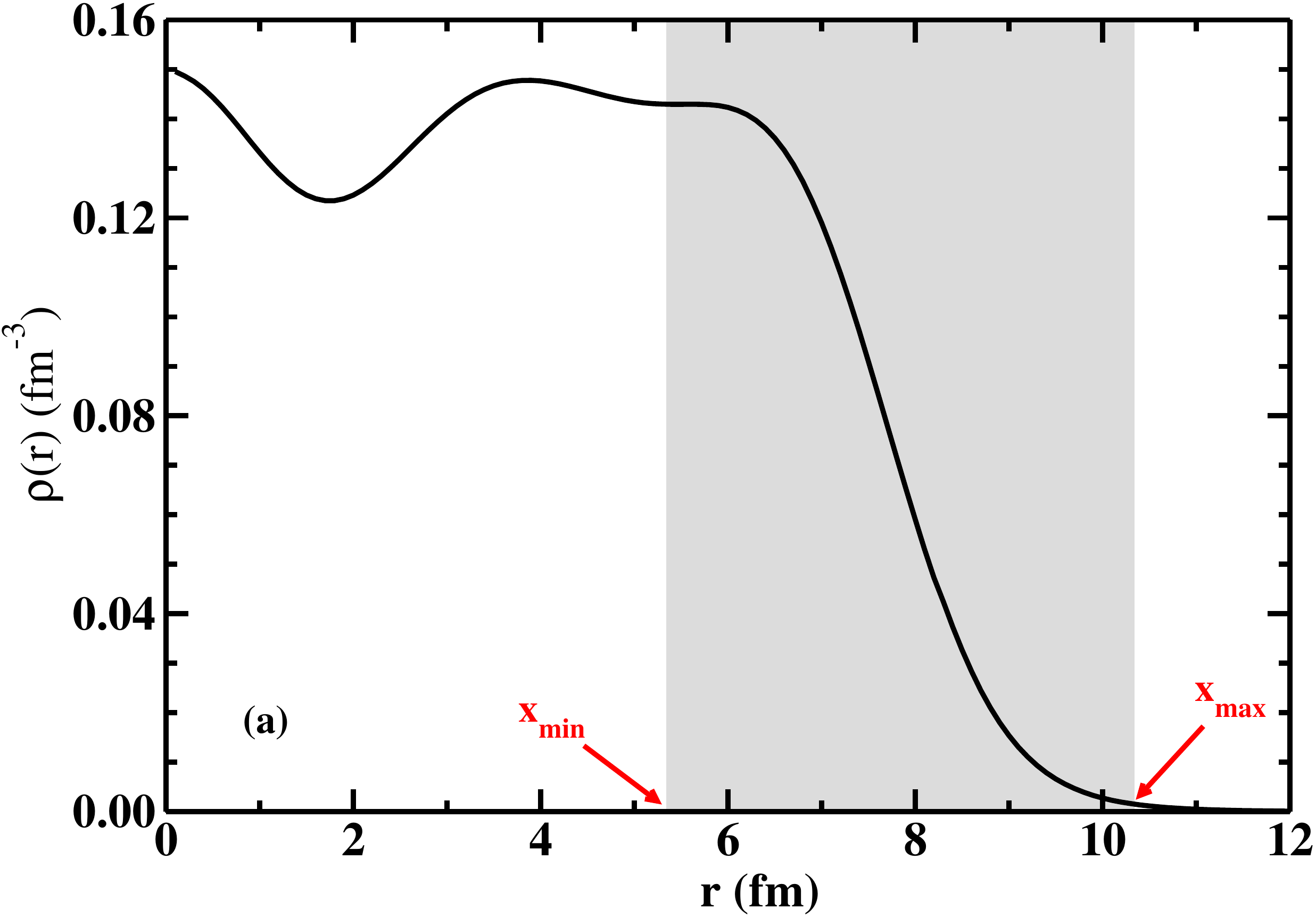}
        \includegraphics[width=0.50\columnwidth,height=5.50cm]{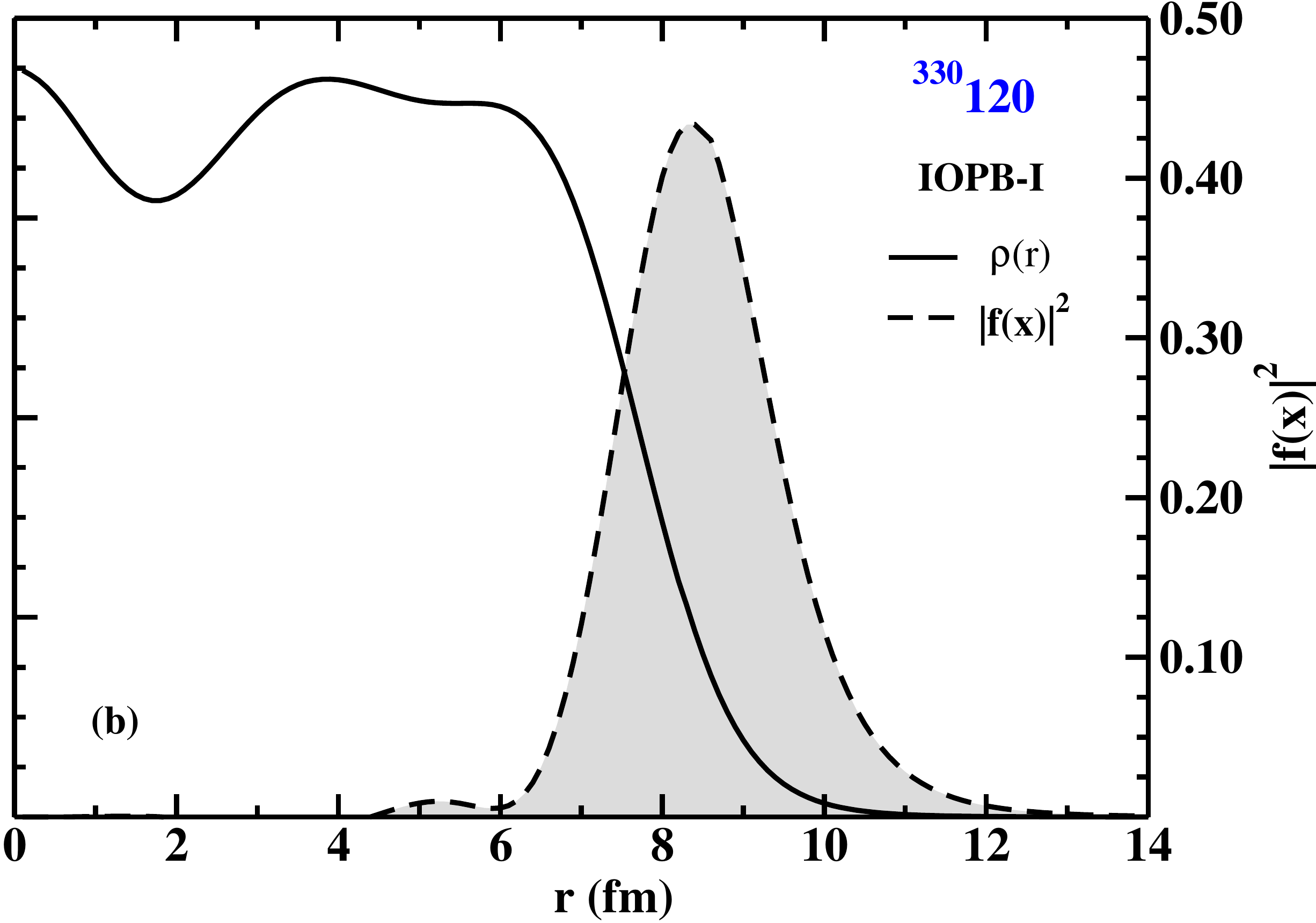}
        \caption{(a) The density for $^{330}120$ with the limits of integrations (Eqs. \ref{s0}, 
\ref{p0}, and \ref{k0}) $i.e.,$ $x_{min}$ and $x_{max}$, (b) the weight function $\vert f(x) \vert^2$ ($x=r$) 
for $^{330}120$ along with its density,  corresponding to IOPB-I parameter set. 
}
        \label{x1x2}
\end{figure}


\section{Densities and Weight Functions for the Nuclei}
\label{denwef}
It is important to note that the measurement of neutrons' density distribution in a nucleus is a difficult 
task because of its neutral nature. As a result, the determination of neutrons' distribution radius R$_n$ has 
been poorly done. On the other hand, the protons' distribution radius R$_p$ has been measured with high accuracy. 
To solve this problem, the recently proposed parity violation experiments PREX-II at JLab \cite{prex1} and the Bates 
Laboratory at MIT were done with polarized beams and targets \cite{bate} which have given better results for the neutron 
distribution \cite{prex1,bate}. The anti-proton experiment at CERN gives the neutron skin-thickness for 26 stable nuclei 
starting from $^{40}Ca$ to $^{238}U$ \cite{expskin}. The recently reported G3 and IOPB-I force parameters in the framework of 
the E-RMF formalism reproduce the neutron skin-thickness $\Delta R$ quite well \cite{iopb1}. This gives us 
confidence that although the measurement of neutron distribution inside the nucleus is not as general as the proton 
even then our chosen forces should be capable enough to reproduce the protons and neutrons' distribution and 
hence the total density distribution of a nucleus.

The total densities of the nuclei, calculated within the spherically symmetric E-RMF formalism corresponding to the 
NL3, IOPB-I, G3, and FSUGarnet parameter sets, are shown in Fig. \ref{dens}. The color code is represented 
in the legends. The bold lines represent the total densities for neutron-deficient nuclei while dashed lines represent 
the densities for neutron-rich isotopes. The panels (a), (b), 
and (c) of the figure show the densities of $^{40,52}Ca$, $^{182,208}Pb$, and $^{304,330}120$, respectively,
as the representative cases. It can be noticed from the figure that the central part of the density is larger 
for lighter isotopes than those of heavier isotopes of that particular series. On the other hand, the surface 
densities are enhanced a bit as a function of radius for heavier isotopes than lighter ones. 

The calculated densities from the E-RMF model are further used in Eq. \ref{wfn} to obtain the weight 
functions for the corresponding nuclei. The weight functions for the nuclei $^{40,52}Ca$, $^{182,208}Pb$, 
and $^{304,330}120$ as the representative cases are shown in the panels (a), (b), and (c) of Fig. 
\ref{weight}, respectively. From the figure, one can notice that the trend of the density profile (Fig. \ref{dens}) is reflected 
in weight functions. In other words, the lower value of the central 
density gives the lesser height of the weight function for an isotope of the particular nucleus. Further, 
it can be noticed in the figure that the maxima of the weight functions shift towards the right (larger $r$) 
with the size of a nucleus increases. The G3 parameter set predicts the larger weight function for all the 
nuclei while the lower one corresponds to the FSUGarnet parameter set. The symmetry energy, neutron pressure,  
and symmetry energy curvature of nuclear matter at local coordinate are folded with the calculated weight 
function of a nucleus which results in the corresponding effective surface properties of the finite nucleus. 
It would be worth illustrating the point that why these quantities are termed as the surface properties 
and how to find the limits of integration (Eqs. \ref{s0}-\ref{k0}).

In principle, the limits of integration in Eqs. \ref{s0}-\ref{k0} are set from $0$ to $\infty$. But, the 
symmetry energy of infinite symmetric nuclear matter within the BEDF method has 
some negative values (unphysical points) in certain regions. To avoid the unphysical points of the 
symmetry energy of nuclear matter, the limits of integration $x_{min}$ and $x_{max}$ are put other than what 
mentioned above. In general, $x_{min}$ and $x_{max}$ are the points where the symmetry energy of nuclear 
matter changes from negative to positive and from positive to negative, respectively \cite{gai12}. For the 
better understanding of the concept of finding $x_{min}$ and $x_{max}$, we present the symmetry energy of 
nuclear matter $S^{NM}(x)$ within BEDF (used in Eq. \ref{s0}) for $^{40}Ca$, 
$^{208}Pb$, and $^{330}120$ in the panels (a), (b), and (c) of Fig. \ref{nms}. The nature of the curves for 
the symmetry energy of nuclear matter is the same for the nuclei shown in the figure with almost the same maximum 
value. However, the curves shift towards the right (larger values of $r$) with nuclei having a large mass 
number. It can easily be noticed from the figure that at $x=2.5$fm, $x=4.3$fm, and $x=5.2$ fm, the 
$S^{NM}(x)$ of nuclear matter changes from negative to positive at Flucton density in $^{40}Ca$, 
$^{208}Pb$, and $^{330}120$, respectively. Thus, these points are considered as $x_{min}$ for the respective 
nuclei. While no point at large $x$ seems to be such that where the $S^{NM}(x)$ changes from positive to 
negative (in this case). Rather, $S^{NM}(x)$ tends to zero at large values of $x$. Thus, the value of 
$x_{max}$ can not be fixed in this way. On the other hand, the densities of the nuclei become almost zero 
at $r = 6.0$fm, $r = 9.4$fm, $r = 10.5$fm for $^{40}Ca$, $^{208}Pb$, and $^{330}120$, respectively. 
Therefore, these points are considered as the $x_{max}$ (upper limit of integrations (Eqs. \ref{s0} - 
\ref{k0})). Figure \ref{x1x2} represents the density of $^{330}120$ with the IOPB-I parameter set as the 
representative case, showing the limits of integration $x_{min}$ and $x_{max}$. The limits are used in 
Eqs. (\ref{s0}-\ref{k0}) to find the symmetry energy, neutron pressure, and symmetry energy curvature, 
respectively. It can be noticed from Fig. \ref{x1x2} that the values of limits do not have any central part 
of the density and lie in the surface region. Hence, the quantities S, P, and $\Delta K$ are known as  
surface properties. For further illustrating the concept of naming these quantities the surface 
properties, the panel (b) of Figure \ref{x1x2} is presented, here, showing the density and weight function 
altogether. It has already been mentioned that the properties of infinite nuclear matter are folded with the 
weight function to obtain the corresponding quantities of finite nuclei. The significant values of weight 
functions (its peak value) lie in the range which corresponds to the surface part of the density. This is 
also one of the reasons to call these quantities the surface properties. 

\begin{figure*}[!b]
        \includegraphics[width=0.5\columnwidth]{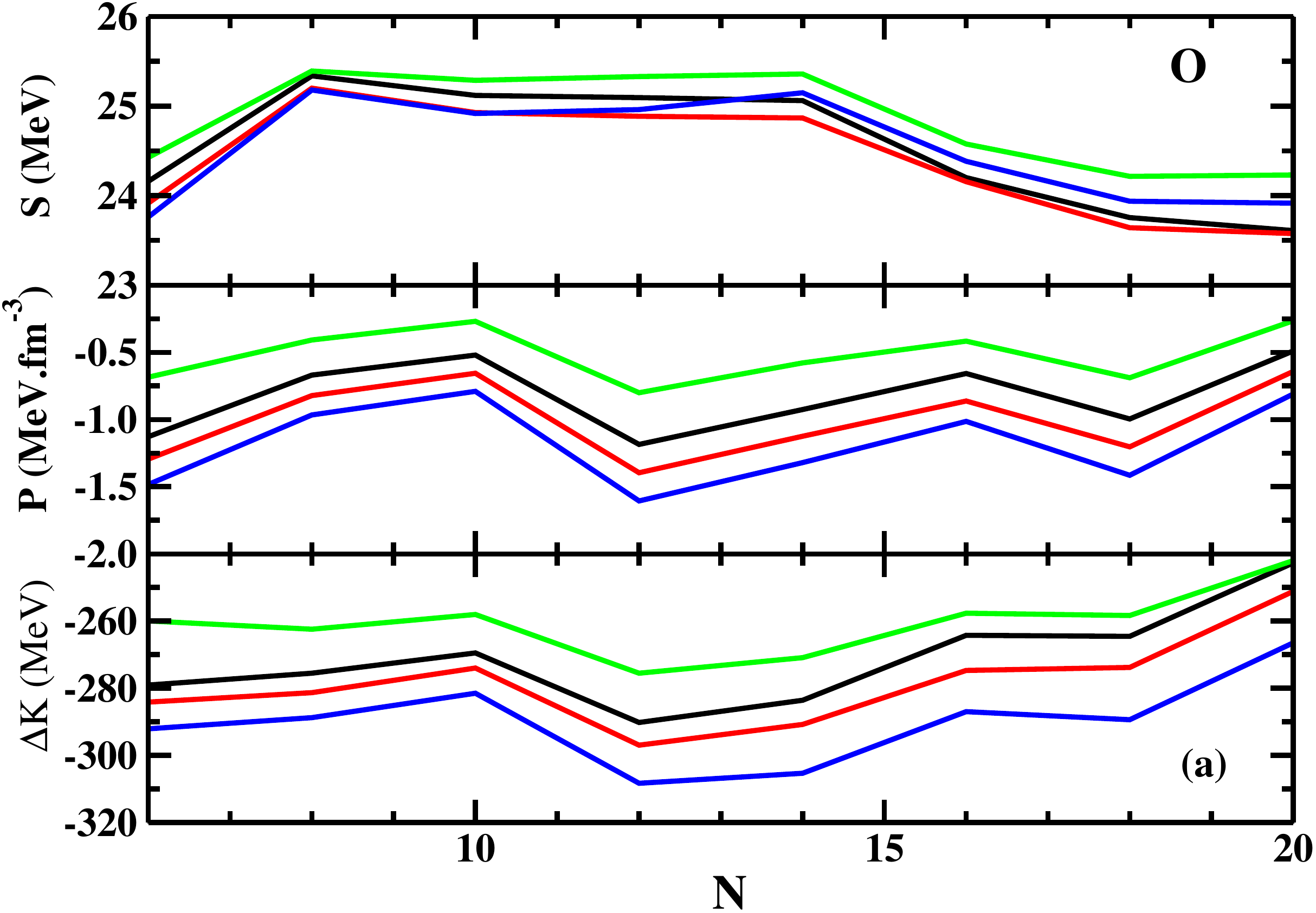}
        \includegraphics[width=0.5\columnwidth]{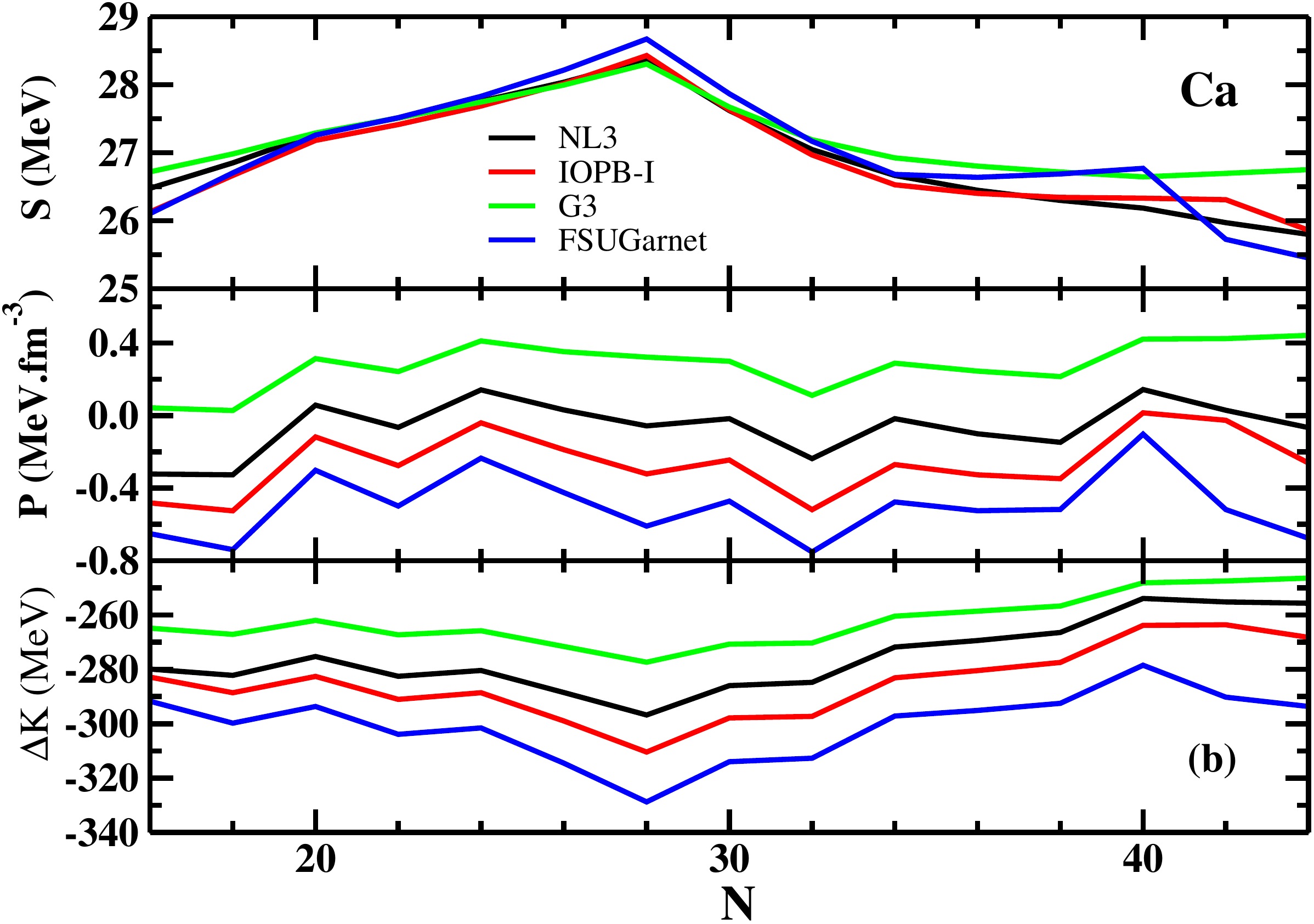}
        \includegraphics[width=0.5\columnwidth]{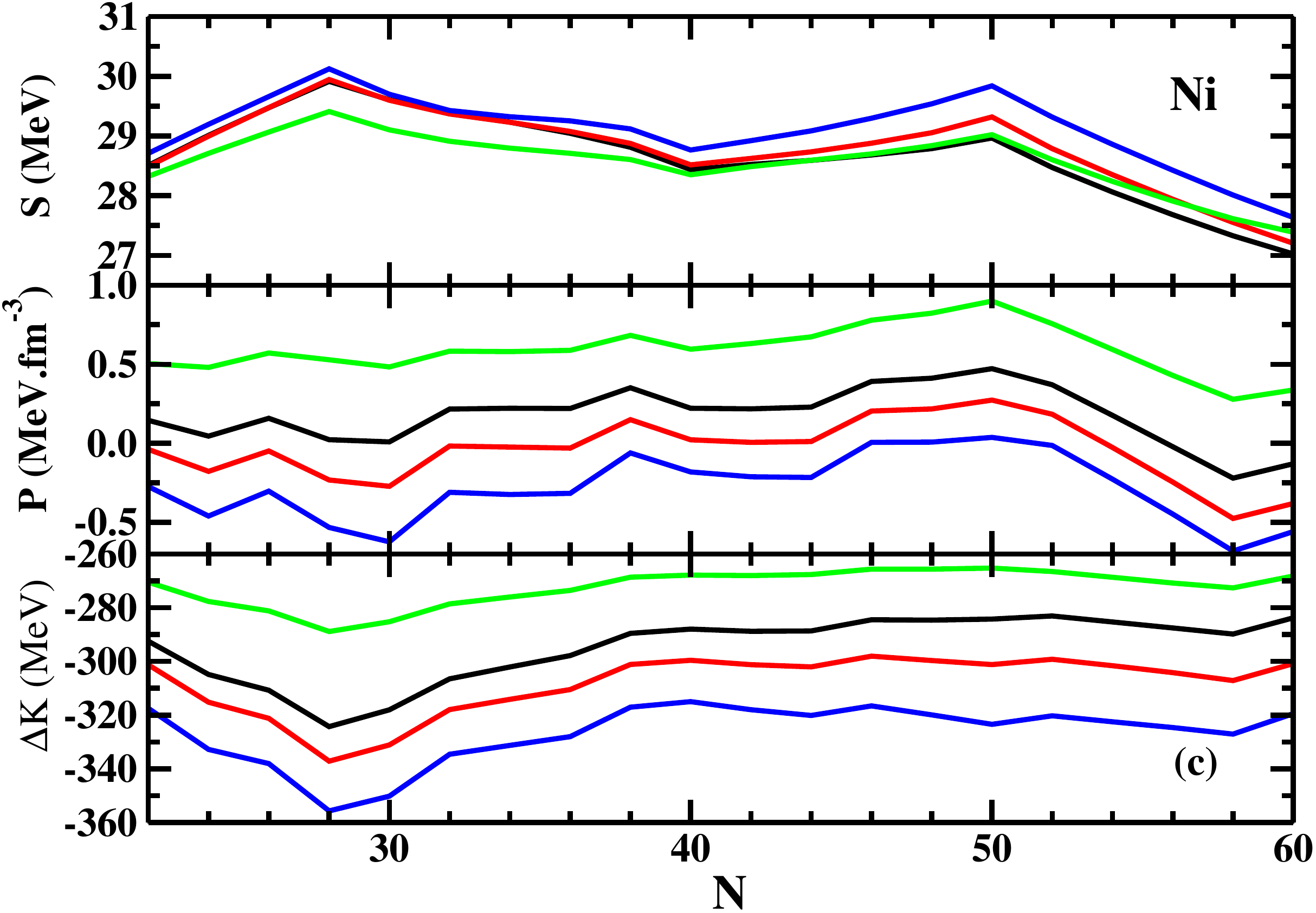}
        \includegraphics[width=0.5\columnwidth]{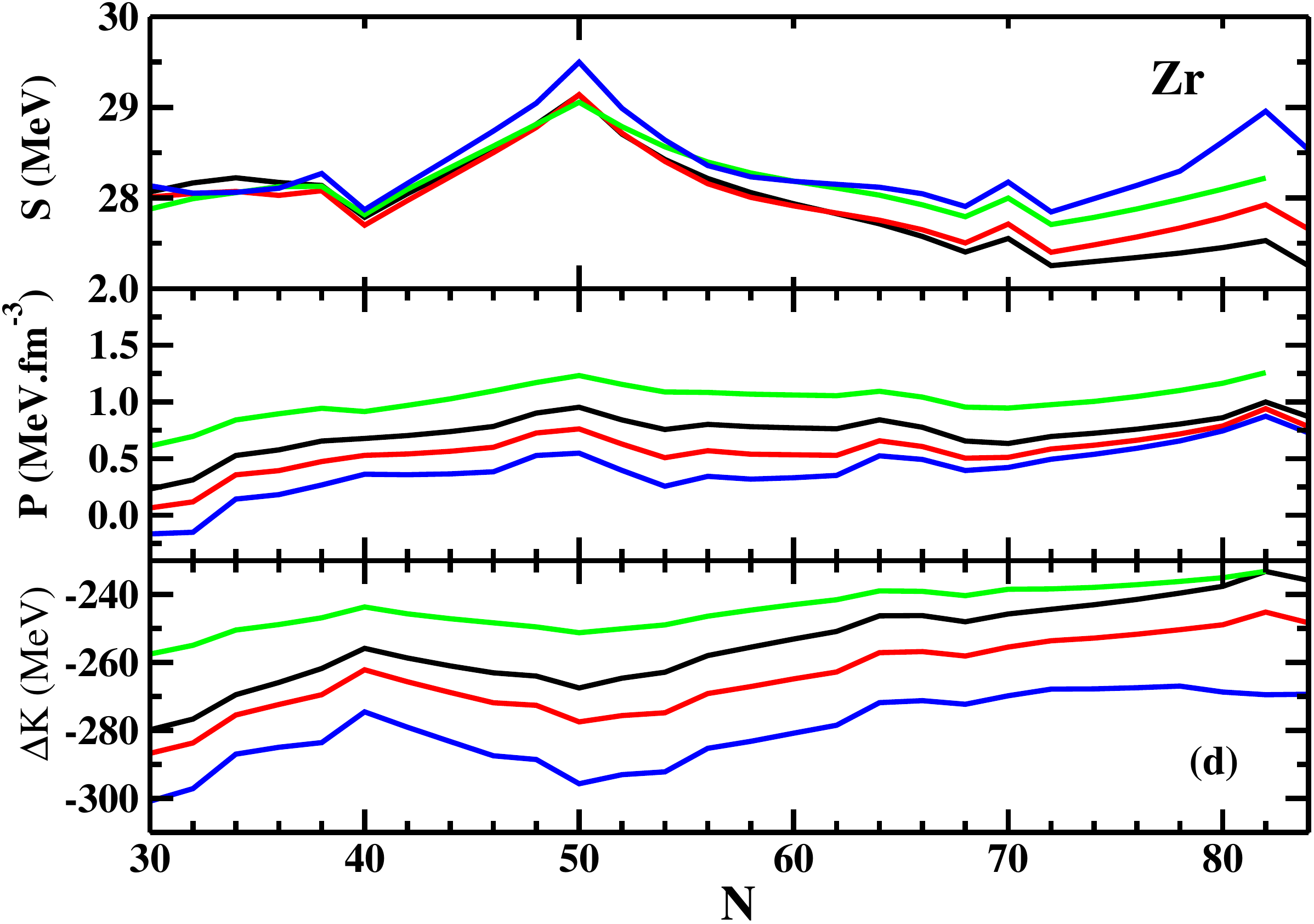}
        \includegraphics[width=0.5\columnwidth]{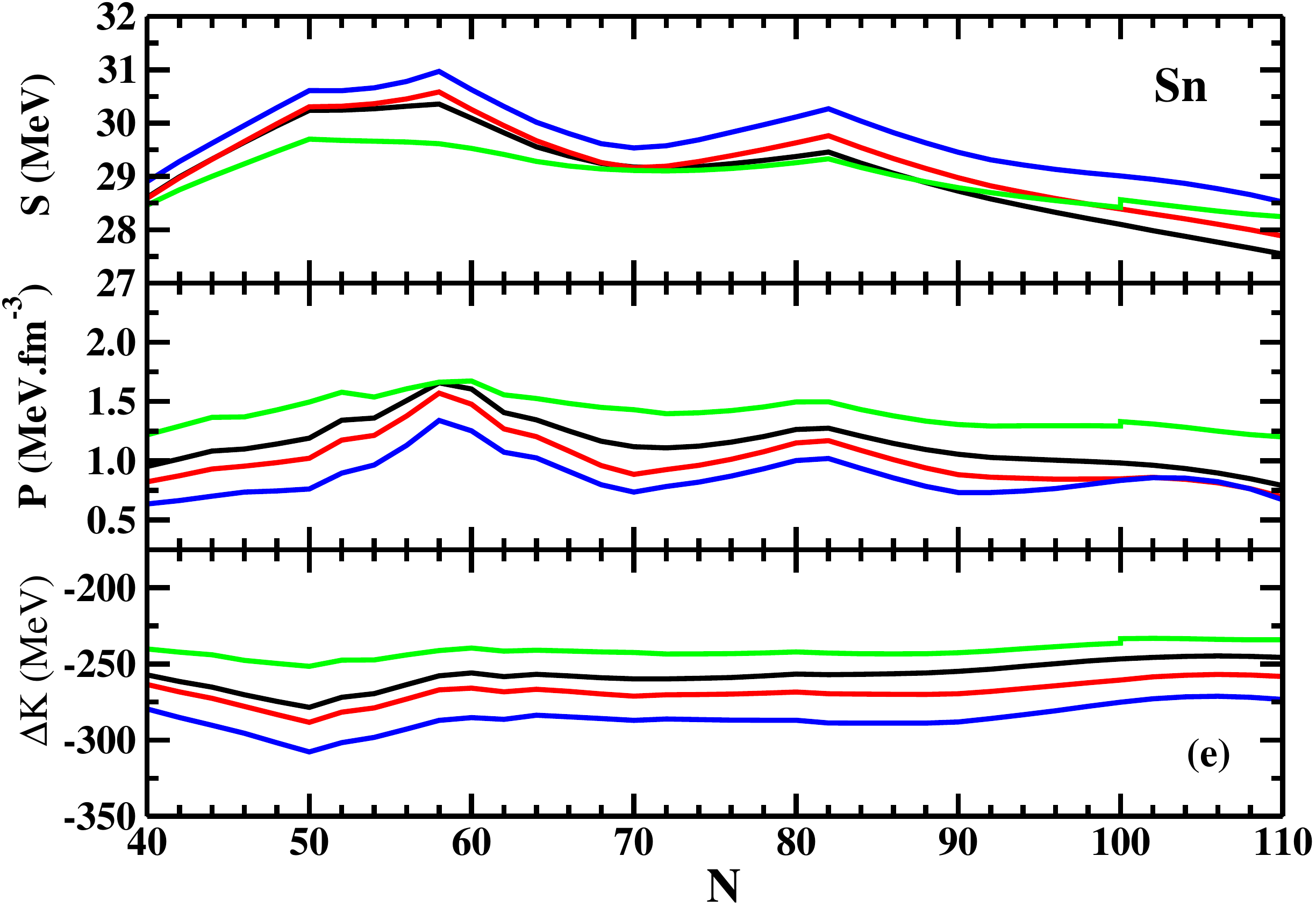}
        \includegraphics[width=0.5\columnwidth]{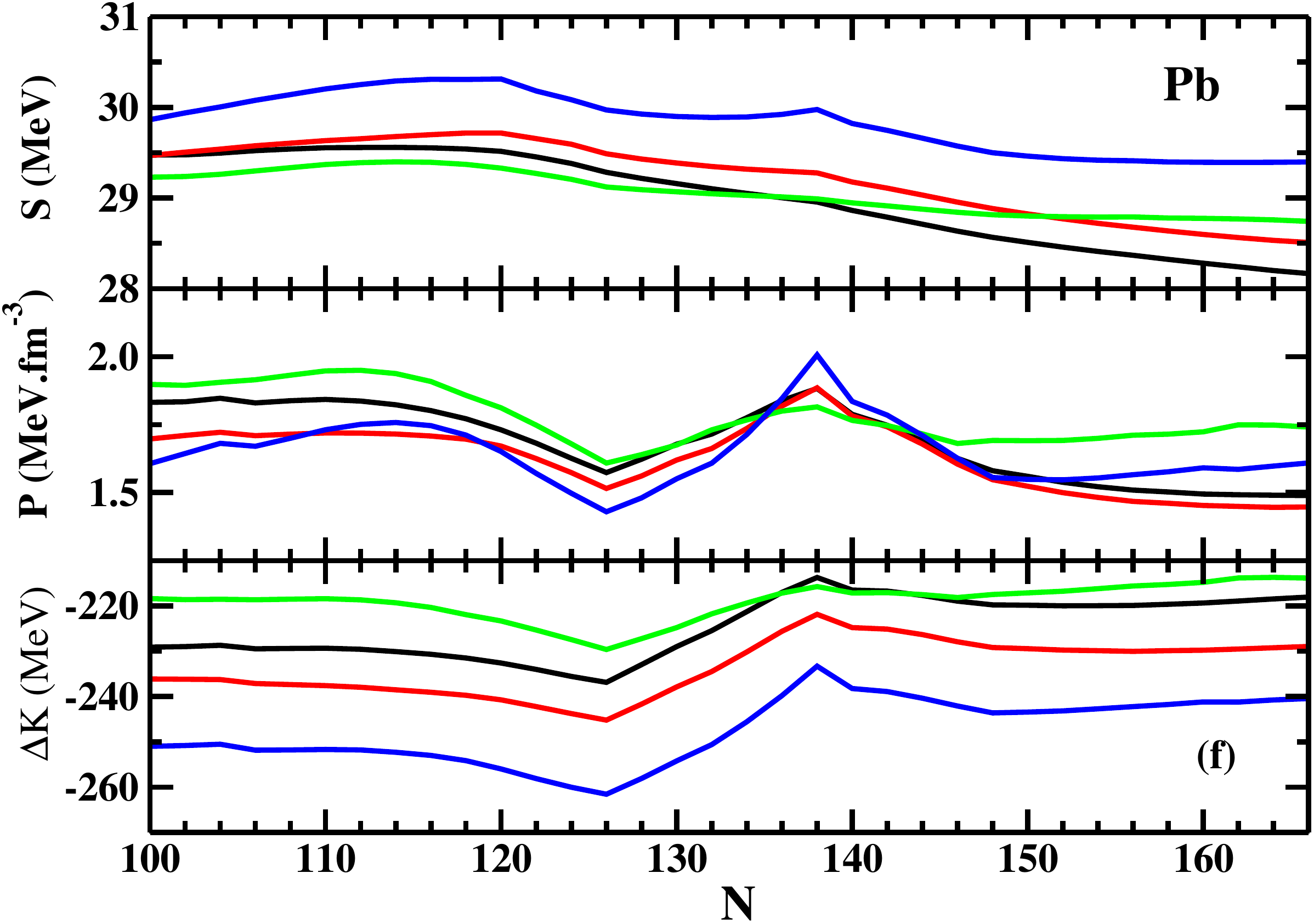}
        \caption{The symmetry energy ($S$), pressure ($P$), and symmetry energy curvature 
($\Delta K$) for the isotopic series of $O$, $Ca$, $Ni$, $Zr$, $Sn$, and $Pb$ nuclei corresponding to NL3, IOPB-I, 
G3, and FSUGarnet parameters sets.}
        \label{spkn}
\end{figure*}

\begin{figure}[!b]
	\centering
        \includegraphics[width=0.8\columnwidth]{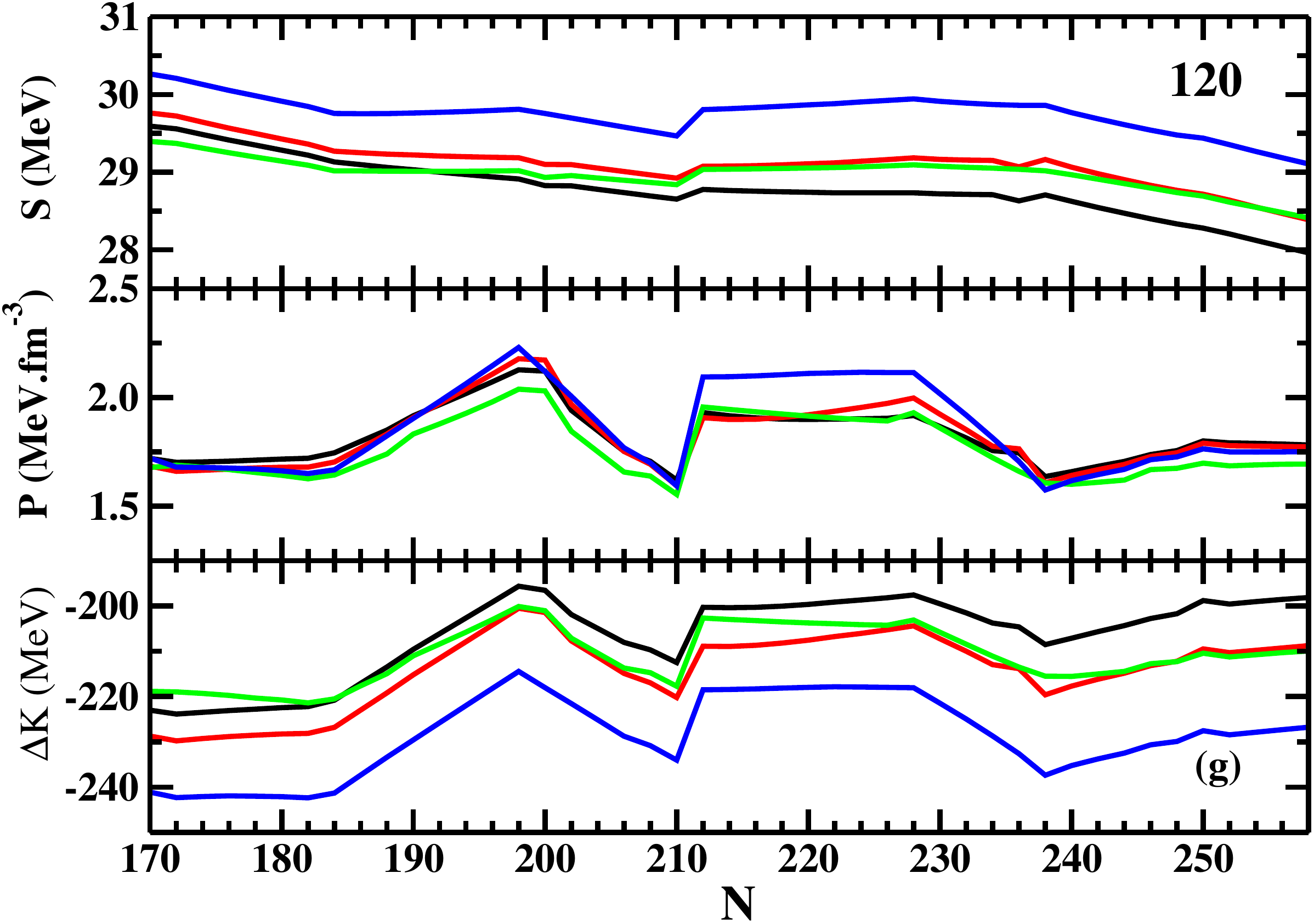}
        \caption{The description is same as in Fig. \ref{spkn} but for $Z = 120$ nuclei. }
        \label{120spkn}
\end{figure}


\section{The Effective Surface Properties of the Nuclei} 
\label{surface}
Before discussing the results of the symmetry energy and its derivatives, it is important to brief why we have 
considered spherical densities of the isotopes despite the fact that some of the isotopes, considered here, might be deformed. 
Generally, the symmetry energy coefficient of a finite nucleus is a bulk property, which mainly depends on the 
isospin asymmetry of the nucleus. Further, the symmetry energy coefficient of a nucleus can be written in terms 
of the volume and surface-symmetry energy coefficients (see Eq. 2 of Ref. \cite{bkaprl}). It is remarked in \cite{nikolov} that 
the volume-symmetry energy is shape-independent, so the surface effects do not play their role in the volume-symmetry energy. 
On the other hand, the surface effects become negligible for heavy and superheavy nuclei since the surface symmetry 
energy coefficient is proportional to $A^{-1/3}$ , where $A$ is the mass number \cite{bkaprl,nikolov}. It is clear from the previous 
sentence that surface effects are important for lighter nuclei and have negligible effects for heavier nuclei. 
The inclusion of the deformation may slightly improve the results but the effects of deformation are very small. 
In some of the recent works \cite{hong15}, authors have shown the effect of deformation on the symmetry energy of finite nuclei 
using Thomas-Fermi approximation over Skyrme energy density functional. 
Following the work, one can find the relative change in the symmetry energy with very large deformation ($\beta_2 \sim 0.6$) 
is around 0.4 MeV. It is mentioned therein \cite{hong15} that the effects of deformation decrease with respect to the mass number. 
So for the sake of computational ease, we have considered the spherical densities of the isotopes.

The effective surface properties for the isotopic series of $O$, $Ca$, $Ni$, $Zr$, $Sn$, and $Pb$ nuclei are shown in 
Fig. \ref{spkn}. The first, second, and third row of each panel of the figure represent the symmetry energy
$S$, neutron pressure $P$, and symmetry energy curvature $\Delta K$, respectively. The value of the 
symmetry energy for finite nuclei lie in the range of 24-31 MeV. It is observed from the figure that the 
symmetry energy is larger for the FSUGarnet parameter set for all the cases except the isotopes of $O$ and few isotopes of $Ca$. 
While G3 set predicts less symmetry energy in most of the cases in Fig. \ref{spkn}. The nature of 
parameter sets get reversed in the cases of neutron pressure $P$ and symmetry energy curvature $\Delta K$. 
For example, the G3 parameter set predicts the larger value of $P$ and $\Delta K$ for all the isotopic series except 
$N = 138$ isotope of $Pb$. 
Furthermore, we find several peaks at neutron numbers, which correspond to the magic numbers and/or shell/sub-shell 
closures for each isotopic chain. These peaks in the symmetry energy curve imply that the 
stability of the nuclei at the magic neutron number is more as compared to the neighboring isotopes. 
The peaks in the symmetry energy curve imply that 
more energy would be required to convert one neutron to a proton or vice versa. Apart from the peaks of the 
symmetry energy at the magic neutron number, a few small peaks are also evolved which may arise due to the 
shell structure on the density distribution of the nuclei. The present investigation predicts a few neutron 
magic numbers beyond the known magic numbers based on the well-known feature of symmetry energy over an 
isotopic chain, which will be studied systematically in the near future.  

The neutron pressure $P$ and symmetry energy curvature $\Delta K$ have the opposite nature to that of the 
symmetry energy with respect to the force parameter sets. It is meant by the opposite nature that the higher 
the symmetry energy of nuclei corresponding to the particular interaction, lower the neutron pressure, and 
symmetry energy curvature values are for the same parameter set and vice versa. Further, we find negative 
neutron pressure for the isotopes of Oxygen nuclei for all parameter sets. Also negative values of $P$ 
are obtained for the isotopic series of $Ca$ and $Ni$ corresponding to the FSUGarnet parameter set and a few of 
them corresponding to IOPB-I set. 
However, the NL3 set predicts negative values of $P$ for some of the isotopes of $Ca$ and neutron-rich isotopes 
of $Ni$. It is to note that the negative value of $P$ arises due to the significant value of weight function 
(the reflection of the behavior of density distribution) in the 
range of local coordinate x (fm), where the pressure of nuclear matter is negative. For example, in Fig. 
\ref{nms} (a), the red arrow bar represents the range of x wherein the pressure is negative. In this range 
of x, the weight function has a non-zero definite value (see Fig. \ref{weight} (a)), which is when multiplied 
by the pressure of nuclear matter (in Eq. \ref{p0}), results in the negative pressure of a nucleus. The non-zero 
definite values of the weight functions in the lower range of x are obtained for lighter nuclei due to their 
small size. On the other hand, the weight functions have negligible values for heavier nuclei in the range 
of x wherein the pressure of nuclear matter is negative for the corresponding nuclei. In general, the 
pressure and symmetry energy curvature values increase with neutron numbers for an isotopic series while the 
symmetry energy decreases with the increase of neutron numbers.

Observing the behavior of effective symmetry energy, neutron pressure, and 
symmetry energy curvature, we find a kink and/or a fall at the magic (or shell closures) neutron numbers. In the case 
of the isotopic chain of $Pb$ nucleus, we did not get a kink in the $S$ curve at N=126. In spite of that, there is a 
fall of the $S$ curve at N=126, beyond which it is almost constant for the few isotopes. This 
signifies that N = 126 is a weak magic number.
Moreover at N=126, the sharp fall of the $P$ and $\Delta K$ curves can be noticed from the figure as they are related 
with the first and second-order derivatives of the symmetry energy. 
Even the sharp kink is not observed in the $S$ curve at N=126 for the isotopic series of $Pb$ likewise at the 
other neutron magic numbers in different isotopic series, the sharp fall in the $P$ and $\Delta K$ curves support 
N=126 as a magic number. Furthermore, in Fig. 5 of Ref. \cite{gai11}, one can find that there are small kinks/deviations 
in the $S$ and $P$ curves at $^{208}Pb$ while sharp kinks are observed at the other neutron magic numbers in 
the isotopic series of $Ni$ (Fig. 1) and $Sn$ (Fig. 6) \cite{gai11}. 

The effective surface properties for the isotopic series of $O$, $Ca$, $Ni$, $Zr$, $Sn$, $Pb$ nuclei motivate us to 
pursue the said calculations for isotopes of experimentally unknown superheavy nuclei. Nowadays, the analysis 
of the superheavy element is a frontier topic in nuclear physics. The discovery of transuranic elements from 
Z = 93$-$118 with Oganesson ($_{118}Og$) is the heaviest element known so far that completes the $7p$ 
orbitals. Hence, the next element Z = 119 will occupy a new row in the Periodic Table. A large number of 
models predict different neutron and proton combinations for the next double close magic nuclei in the 
superheavy stability valley \cite{bhuyan12}. Among them, $Z = 120$ attracts much attention with neutron number 
N = 184 as the next double magic isotope, and near to be synthesized. Therefore, we have calculated the 
effective surface properties for the isotopic chain of Z = 120 nuclei, shown in Fig. \ref{120spkn}. 
We found almost a smooth fall in the symmetry energy up to N = 210, and further, a very miniature growth 
appears up to N= 240 following the previous trends. In other words, there is a moderate decrease in the 
symmetry energy over the isotopic chain of $Z = 120$ with some exception for the neutron number 
212 $\leq$ N $\leq$ 238 (see Fig. \ref{120spkn}). Here, 
we also got peaks in the neutron-rich side at N = 212 and 238 in the symmetry energy curve as similar to 
those in Fig. \ref{spkn}. These neutron numbers can be attributed to the magic neutron numbers. Here also, the symmetry 
energy predicted by the FSUGarnet parameter set is larger compared to the rest of the parameter sets. The 
neutron pressure and symmetry energy curvature are shown, respectively, in the second and third rows of Fig. \ref{120spkn}. 
Similar conclusions can be drawn for the $P$ and $\Delta K$ of $Z = 120$ isotopes, as for the other 
isotopic chains. In general, we get a bit larger values of the effective surface properties for $Z = 120$ 
isotopes as compared to the isotopes of the rest of the nuclei except a few isotopes of $Sn$ and $Pb$. 
Here, we did not find any transparent signature of shell closures or magicity in the symmetry energy over the isotopic 
chain of $Z =120$. Following Ref. \cite{bhu12}, the ground-state configuration of $Z = 120$ isotopes are super-deformed prolate and/or 
oblate shapes followed by a spherical intrinsic excited-state. Here, our calculation is limited to the spherical 
co-ordinate, which may cause for weaken the signature over the isotopic chain. Hence, a self-consistent 
microscopic calculation is required in the deformed basis.

\begin{figure*}[!b]
        \includegraphics[width=0.5\columnwidth]{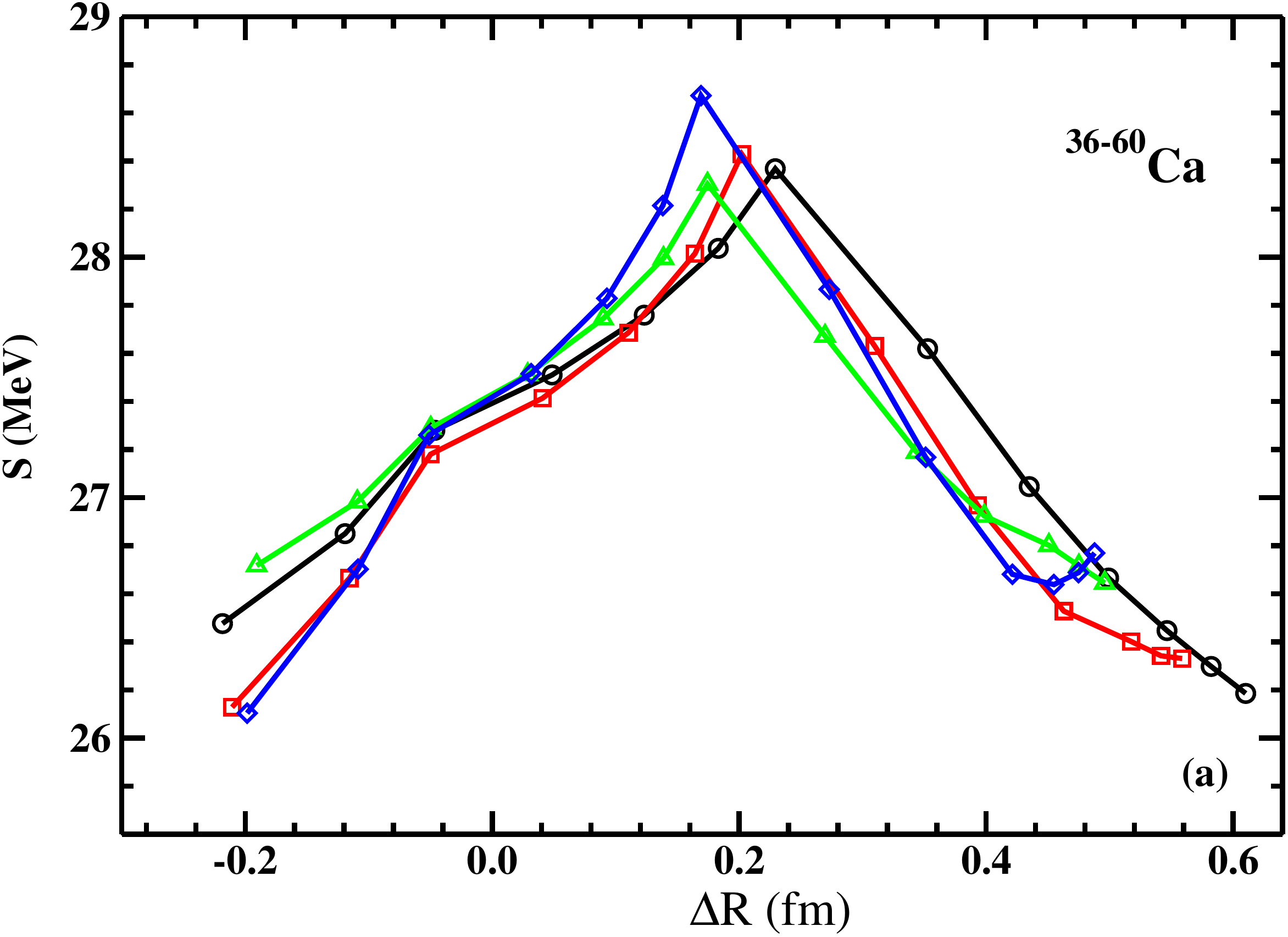}
        \includegraphics[width=0.5\columnwidth]{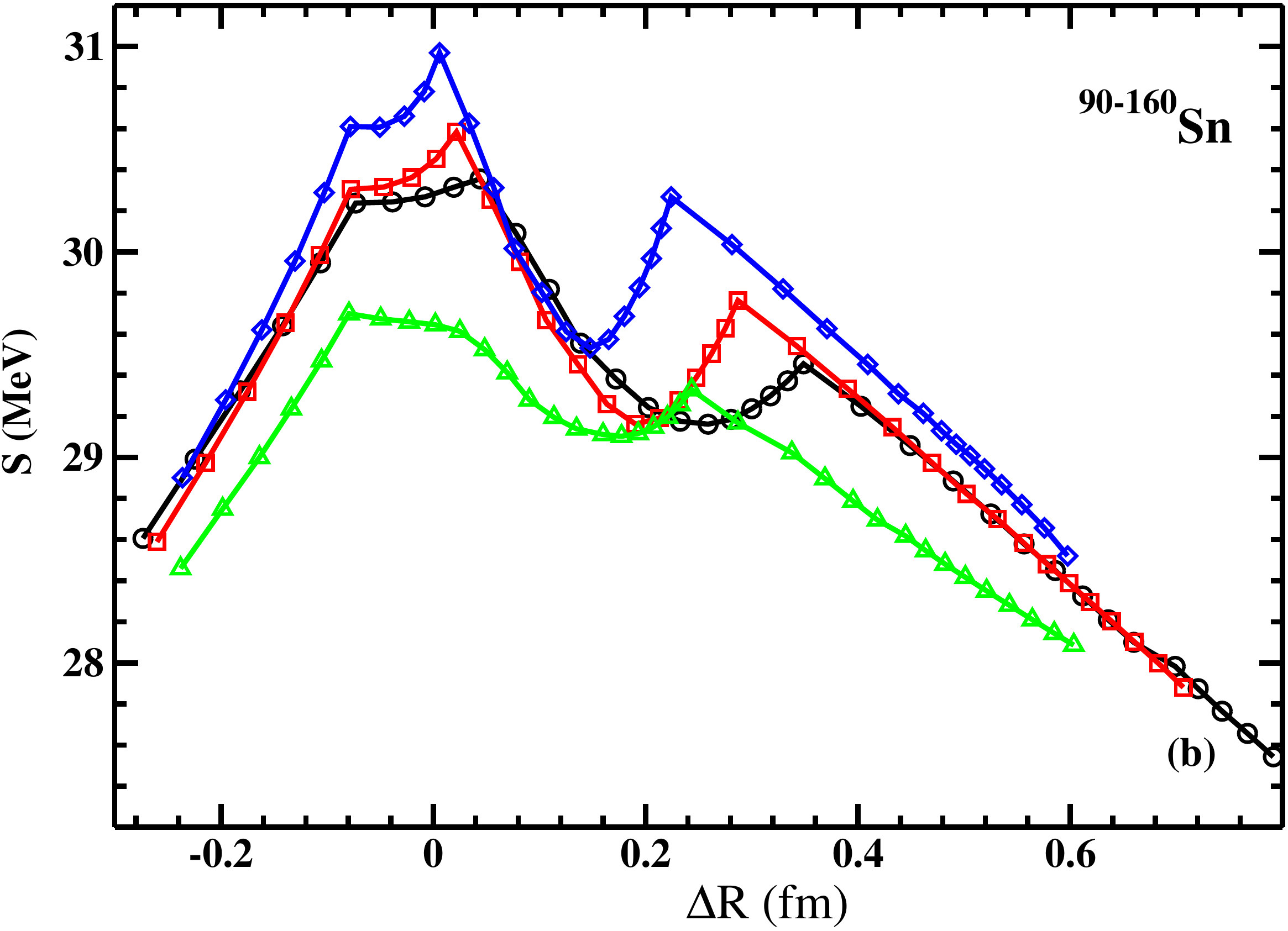}
        \includegraphics[width=0.5\columnwidth]{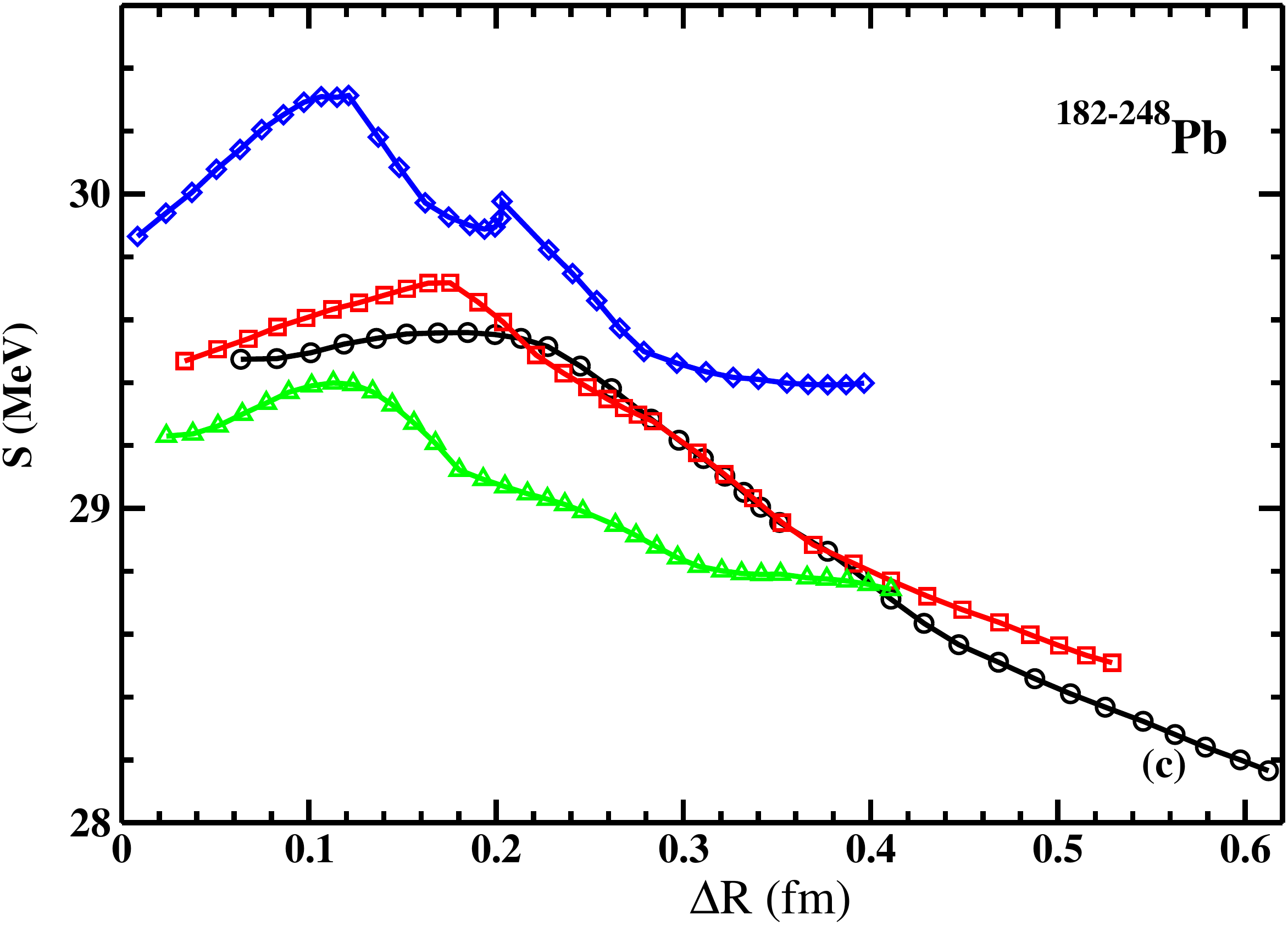}
        \includegraphics[width=0.5\columnwidth]{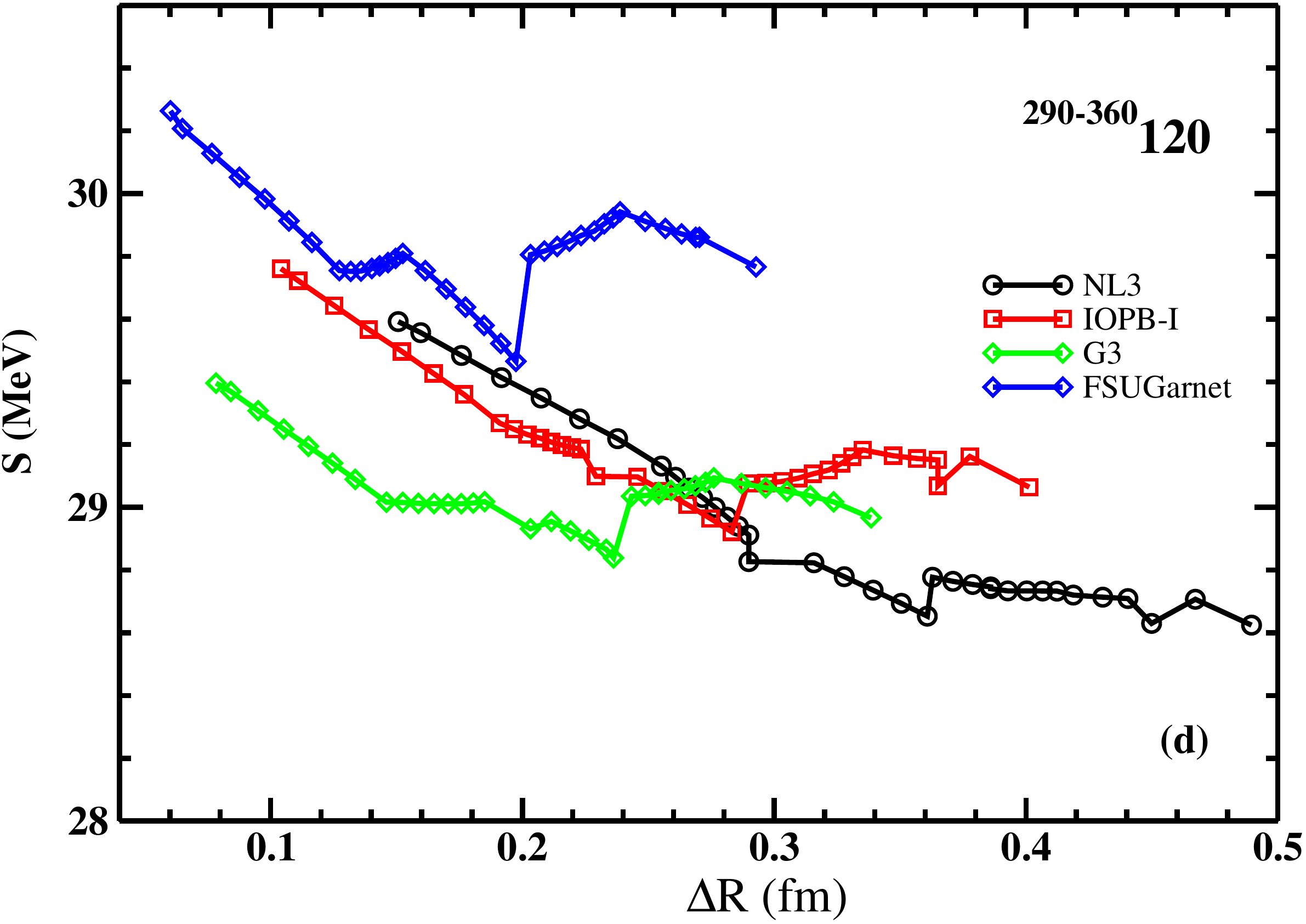}
        \caption{A correlation of the neutron skin-thickness with the symmetry energy is
shown for the isotopes of $Ca$, $Sn$, $Pb$,  and $Z = 120$ nuclei as the representative cases corresponding to NL3,
IOPB-I, G3, and FSUGarnet parameter sets. }
        \label{drS}
\end{figure*}

\section{Correlation of Skin-thickness with the Symmetry Energy}
\label{correlation}
The skin-thickness is found to be linearly correlated with the surface properties of isotopic series of nuclei 
, except for some kinks, which correspond to the magic/semi-magic nuclei of thta isotopic chain \cite{gai11,bhu18,anto,gai12}. 
Here, we present the correlation between 
the symmetry energy and the neutron skin-thickness for the isotopic series of $Ca$, $Sn$, $Pb$, and $Z = 120$ nuclei 
for the NL3, IOPB-I, G3, and FSUGarnet parameter sets. It is remarked in Refs. \cite{iopb1} and shown 
in Table \ref{table4.3} that stiffer EoS of nuclear matter predicts larger neutron skin-thickness of 
nuclei. Among the chosen parameter sets, NL3 is the stiffest which predicts larger skin-thickness, while 
FSUGarnet as being softer estimate smaller skin-thickness. On the other hand, it has been shown in Fig. 
\ref{spkn} that the symmetry energy is maximum at the neutron magic number of an isotope. 

Fig. \ref{drS} shows the correlation of the symmetry energy with the neutron skin-thickness of nuclei. The 
panels (a), (b), (c), and (d) of the figure represent the correlation for the isotopic series of $Ca$, $Sn$, $Pb$, 
and $Z = 120$, respectively. It is clear from the figure that the skin-thicknesses of the nuclei are larger 
corresponding to the NL3 parameter set and smaller for the FSUGarnet set. It can be noticed from the figure that 
with some exceptions the symmetry energy 
predicted by FSUGarnet is higher compared to the rest of the parameter sets. The peaks in the symmetry energy 
curves (in Fig. \ref{drS}) correspond to the magic or semi-magic neutron numbers. The symmetry energy 
decreases with varying neutron numbers in either direction of the magic/semi-magic number. It implies that for  
exotic nuclei (nuclei lie at the drip line) less amount of energy is required to convert one proton to 
neutron or vice-versa, depending on the neutron-proton asymmetry. The behavior of the symmetry energy with 
skin-thickness is undermined for a few cases. In the case of $Ca$, $Sn$, and $Pb$, the symmetry energy curve is 
almost linear before and after the peaks. Further improvement in the results can be obtained by solving the 
field equations on an axially deformed basis. 

\section{Conclusions}
\label{summary4}
We have studied the effective surface properties like the symmetry energy, neutron pressure, 
and the symmetry energy curvature for the isotopic series of $O$, $Ca$, $Ni$, $Zr$, $Sn$, $Pb$, and $Z = 120$ nuclei within 
the coherent density fluctuation model. We have used the spherically symmetric effective field theory 
motivated relativistic mean-field model to study the ground-state bulk properties of nuclei with the recent 
parameter sets like IOPB-I, FSUGarnet, and G3. The calculated results are compared with the predictions by 
the well known NL3 parameter set and found in good agreement. The densities of nuclei calculated within the E-RMF 
formalism are used as the inputs to the coherent density fluctuation model to obtain the weight functions 
for the isotopes. The symmetry energy, neutron pressure, and symmetry energy curvature of infinite nuclear 
matter are calculated within the BEDF model which are further folded with the 
weight function to find the corresponding quantities of finite nuclei. 
The FSUGarnet parameter set predicts 
a large value of the symmetry energy while the smaller symmetry energy values are for the G3 set with some exceptions. We found 
a larger value of the skin-thickness for the force parameter that corresponds to the stiffer EoS and vice-versa. 
We also found a few mass-dependence peaks in the symmetry energy curve corresponding to the neutron 
magic/semi-magic number. Observing the nature of the symmetry energy over the isotopic chain, we predict a 
few neutron magic numbers in the neutron-rich exotic nuclei including superheavy. The 
transparent signature of magicity is diluted for a few cases over the isotopic chain of $Pb$ and $Z = 120$ 
nuclei. Similar behavior is also observed for the neutron pressure and symmetry energy curvature for 
these isotopes. Concurrently, the present calculations tentatively reveal a way to calculate the effective 
surface properties of unknown drip line nuclei including superheavy. More detailed studies are disclosed by 
considering deformation into account. The calculated quantities are important for the structural properties of finite 
nuclei and may be useful for the synthesis of neutron-rich or superheavy nuclei. These effective surface properties 
can also be used to constrain an EoS of the nuclear matter and consequently nucleosynthesis processes.  


\chapter{Temperature-Dependent Symmetry Energy of Finite Nuclei and Nuclear Matter}{\label{chap5}}  

\rule\linewidth{.5ex}

\section{Introduction} 

As disussed in Chapter \ref{chap1}, \ref{chap3}, and \ref{chap4}, 
the symmetry energy is nearly equal to the energy cost used to convert symmetric nuclear matter to asymmetric matter. 
Isospin asymmetry in nuclear matter arises due to differences in protons and neutrons densities and masses. It is remarked in Ref. 
\cite{iopb1} that the density type isospin asymmetry is described by $\rho-$ meson (isovector-vector) 
exchange and the mass type asymmetry by $\delta-$ meson (isovector-scalar) exchange. 
The findings of S. J. Lee $et al.,$ show that the surface symmetry energy term 
is more sensitive to temperature than the volume energy term \cite{lee10}. Furthermore, the 
dependence of the symmetry energy on density and temperature have a crucial role in explaining various 
phenomena in heavy-ion collision, supernovae explosions, the liquid-gas phase transition of 
energy coefficient is sensitive to temperature as well as to the force parameters \cite{aqjpg}. 
asymmetric nuclear matter, and mapping the location of the neutron drip line in the nuclear landscape 
\cite{lattimer07,baran05,baron85}. 
Furthermore, the internal configuration of a nuclear system (especially, 
of a neutron-rich nucleus) such as its distribution of nucleons, interaction strengths, and the 
nucleon dynamics influence the neutron pressure and the observables related to it.

In Chapter \ref{chap3}, we found that the symmetry 
energy coefficient is sensitive to temperature as well as to the force parameters. 
The importance of the symmetry energy and its sensitivity to temperature and density have motivated 
us to study it for neutron-rich thermally fissile nuclei. In this chapter, we 
study the symmetry energy, neutron pressure, and symmetry energy curvature of $^{234,236,250}U$ and 
$^{240}Pu$ at finite temperature. The 
motivation behind choosing these nuclei has already been stated, i.e., the thermally fissile nature 
of the nuclei and their importance for energy production. 
Several studies have been carried out to investigate structural properties and reaction dynamics
of the thermally fissile nuclei and the isotopic series of actinide nuclei
\cite{bh15,seinthil17,bharat17,je18,sw18}. In Chapter \ref{chap3}, we studied the ground
and excited-state bulk properties of $^{234,236}U$, $^{240}Pu$, $^{244-262}Th$ and $^{246-264}U$ nuclei. 
Apart from the other bulk properties of a nucleus, it is also fruitful to know the
properties, such as the symmetry energy coefficient, neutron pressure, and symmetry energy curvature,
which can be helpful in synthesizing neutron-rich super heavy and exotic nuclei. 


Various approaches to calculating the nuclear symmetry energy along with its significance 
have been discussed in Chapter \ref{chap3} and Chapter \ref{chap4}.
In this chapter, we have used the local density approach to calculate the symmetry energy and the observables related to it. 
A brief methodology of the local density method is discussed in the upcoming section. 
Within a Thomas-Fermi model, the density of a nucleus is calculated by subtracting the
density profile of a gas phase from that of the liquid-gas phase \cite{jn64,sks75,bka89,Zhang14}. 
This density (obtained through a subtraction procedure) is then used to calculate the rest of 
the properties of a nucleus. In the present analysis, we calculate the densities of the nuclei 
along with some of the ground and excited-state properties within the TRMF model by considering 
particle number conservation. The properties of symmetric nuclear matter are also obtained 
within the TRMF formalism, which are further used to calculate the corresponding quantities of 
finite nuclei.



\section{Local Density Approximation (LDA)}
{\label {LDA}}
The effective bulk properties of a nucleus can be found by using the LDA once its density profile 
is known. In the LDA, the symmetry energy coefficient $S (T)$ can be defined as \cite{bkaprl,sam07,De12}, 
\begin{eqnarray}
S(T)\left(\frac{N-Z}{A}\right)^2= \frac{1}{A}\int \rho(r) S^{NM}[\rho(r),T] \alpha^2(r) d{\bf r},  
\label{seqn}
\end{eqnarray}
where, $S^{NM} [\rho(r),T]$ is the symmetry energy coefficient of infinite symmetric nuclear matter  
at finite temperature (T), and at the local density $\rho(r)$ of a nucleus and $\alpha(r)$ is the 
isospin asymmetry parameter defined earlier. Similar expressions as in Eq. (\ref{seqn}) can be used to 
find the neutron pressure, and symmetry energy curvature by replacing $S^{NM}[\rho(r),T]$ with the 
corresponding nuclear matter quantities at the local density of a nucleus. The $S^{NM}[\rho(r),T]$ 
is found from Eq. (\ref{slda}), where $\rho$ is the density of a nucleus $\rho(r)$. Similarly,  
$P^{NM}[\rho(r),T]$ and $K^{NM}[\rho(r),T]$ at the density of a nucleus are found by using Eqs. 
(\ref{lsym}) and (\ref{ksym}), respectively. As stated earlier, the density distribution for a nucleus 
is calculated within the TRMF (Hartree approximation) for a given parameter set where the equations of motion are solved 
self-consistently \cite{aqjpg,furnstahl97}. As a result, the density from the Hartree approximation 
is purely quantal one with a well-defined surface. There are thus no corrections to be included externally 
for the surface, as would be the case for semi-classical model 
\cite{cent93,wigner32,kirk33,brack85,cent91,sam08}. Generally, semi-classical methods such as the extended Thomas-Fermi
(ETF) method \cite{brack85} and the relativistic extended Thomas-Fermi (RETF) formalism \cite{cent93} are based on the Wigner-Kirkwood (WK)
$\hbar$ expansion of density matrix \cite{wigner32,kirk33} of a nucleus. In this case, the calculated density of a nucleus 
does not have a well-defined surface, which is the expected error (qualitative) in assuming the LDA for the 
symmetry energy of a finite nucleus. To sort out the deficiencies of the Thomas-Fermi approximations, 
and thus, to provide a more accurate description of the nuclear surface, at least $\hbar^2$ order gradient 
corrections coming from inhomogeneity and non-local effects have to be included in the energy density functional \cite{cent93}. 
It is worth mentioning that the density obtained here from the Hartree approximation can be directly used 
in Eq. (\ref{seqn}) for further calculations \cite{aqjpg,furnstahl97,sam08,bkaprl} without adding surface corrections externally.   

In general, the symmetry energy coefficient of a finite nucleus is a bulk property that mainly depends 
on the isospin asymmetry of the nucleus. The symmetry energy of a nucleus can be written in terms of 
volume and surface symmetry energy coefficients in order to study its approximate mass dependence 
[see Eq. (2) of Ref. \cite{bkaprl}]. The surface term is important in determining the symmetry energy 
of light mass nuclei and become small for heavy and super-heavy nuclei since the surface symmetry 
energy coefficient is proportional to $A^{-1/3}$, where $A$ is the mass number \cite{bkaprl,nikolov}. 
The volume symmetry energy is almost independent of the shape degrees of freedom of a nucleus 
\cite{nikolov}. In Ref. \cite{hong15}, one can find the relative change in the symmetry energy with very 
large deformation ($\beta_2 \approx 0.6$) is around 0.4 MeV. It is also mentioned in Ref. \cite{hong15} 
that the effect of deformation on the symmetry energy decreases with respect to the mass number. Here 
we study the Uranium and Plutonium isotopes, which are super-heavy in nature, with ground-state 
deformations of magnitude $\beta_2 \approx 0.2$. Hence the effects of deformation on the symmetry energy 
are very small and for the sake of computational simplicity, we have taken the monopole term (spherical 
equivalent) of the density distribution without changing the volume of a nucleus in the present analysis. 
The decomposition of the density in terms of even values of the multipole index 
$\lambda$ \cite{moya91} is given as,
\begin{eqnarray}
\rho(r_\bot,z)=\sum_\lambda \rho_\lambda (r) P_\lambda (\cos \theta),
\label{spher.}
\end{eqnarray}
where $P_\lambda$ is a Legendre polynomial and r is the radial variable. 


\section{Ground and Excited-State Properties of the Nuclei} 
{\label {finite}}
The main aim of this work is to study the symmetry 
energy coefficient $S$, neutron pressure $P$, and symmetry energy curvature $K_{sym}$ of neutron-rich 
thermally fissile nuclei at finite temperature. In this work, we have taken $^{250}U$ as a 
representative case of the neutron-rich thermally fissile nuclei \cite{satpathy}. $^{234,236}U$ and 
$^{240}Pu$ are also studied because of the importance in the fission of the known thermally fissile 
$^{233,235}U$ and $^{239}Pu$ nuclei. Before proceeding to the  
observables described above, we have calculated the ground and excited-state properties of these nuclei. 
For the calculations, we have used the FSUGarnet \cite{chai15}, IOPB-I \cite{iopb1}, and NL3 \cite{lala97} 
parameter sets within the TRMF model. Among these, the NL3 \cite{lala97} set is one 
of the best-known and most used RMF parameter sets and describes the properties of nuclei remarkably 
well over the nuclear chart. The FSUGarnet \cite{chai15} and IOPB-I \cite{iopb1} are more recent 
parameter sets with the advantage that their equations of state are softer than that of the 
NL3 parameter set. 
In the relativistic mean-field model, the field equations are solved self-consistently by taking an initial estimate 
of the deformation $\beta_0$ \cite{lala97,chai15,iopb1,bhu18}. The mesonic and Fermionic fields are 
expanded in terms of the deformed harmonic oscillator basis. The numbers of major shells for the 
fermionic and bosonic fields (NOF and NOB, respectively) have been taken as 14 and 20. At these 
values of NOF and NOB, we obtain a converged solution in this mass region. 

The binding energy per particle (B/A), charge radius ($R_{c}$), and deformation parameter 
($\beta_2$) of the nuclei $^{208}Pb$, $^{234,236,250}U$ and $^{240}Pu$ at finite temperature (T) 
with the FSUGarnet \cite{chai15}, IOPB-I \cite{iopb1}, and NL3 \cite{lala97} parameter sets are 
shown in Table \ref{tab3} with the available experimental data \cite{audi12,Angeli2013}. The 
calculated values corresponding to all parameter sets are in good agreement with each other. These 
results are comparable to the corresponding experimental data at $T=0$ MeV. All 
of the calculations show good agreement with the binding energy but slightly underestimate the 
deformation at T = 0. The IOPB-I parameter set furnishes changes in the radii slightly longer than 
those of the other sets at all values of the temperature. The binding energies obtained with the 
NL3 parameter set are slightly smaller than those of the other sets for T $>$ 0. The binding 
energy and quadrupole deformation of the nuclei decrease with T. The temperature at which the 
deformation of a nucleus becomes zero is known as the critical temperature (T$_c$), as discussed in Chapter \ref{chap3}. 
For the IOPB-I set, T$_c$ value is obtained at a lower temperature than for the other sets.   


\renewcommand{\baselinestretch}{1.0}
\begin{table*}
\caption{The calculated binding energy per particle (B/A) (MeV), charge radius ($R_c$) (fm),  
and deformation parameter $\beta_2$ of the nuclei $^{234,236,250}U$ and $^{240}Pu$ at finite 
temperature T (MeV) are tabulated and compared with the available experimental data 
\cite{audi12,Angeli2013}.}
\renewcommand{\tabcolsep}{0.12cm}
\renewcommand{\arraystretch}{1.3}
\begin{tabular}{ccccccccccccccccccc}
\hline
\hline
\multicolumn{1}{c}{T}&
\multicolumn{3}{c}{$^{234}U$}&
\multicolumn{4}{c}{$^{236}U$}&
\multicolumn{4}{c}{$^{240}Pu$}&
\multicolumn{4}{c}{$^{250}U$}&
\multicolumn{2}{c}{Parameter}\\
 & B/A & R$_{c}$ & $\beta_2$ && B/A & R$_{c}$ & $\beta_2$&& B/A & R$_{c}$ & $\beta_2$&& B/A & R$_{c}$ & $\beta_2$
&&\\
\hline
0     & 7.60 & 5.84 & 0.20 && 7.57 & 5.86 & 0.22 && 7.56 & 5.91 & 0.24 && 7.41 & 5.95 & 0.22 &&FSUGar. \\
      & 7.61 & 5.88 & 0.20 && 7.59 & 5.90 & 0.22 && 7.57 & 5.95 & 0.25 && 7.43 & 5.99 & 0.23 &&IOPB-I\\
      & 7.60 & 5.84 & 0.24 && 7.58 & 5.86 & 0.25 && 7.55 & 5.90 & 0.27 && 7.42 & 5.94 & 0.22 &&NL3 \\
      & 7.60 & 5.83 & 0.27 && 7.59 & 5.84 & 0.27 && 7.56 & 5.87 & 0.29 &&  -   &  -   &  -   &&Exp.\\
\\
1     & 7.55 & 5.83 & 0.16 && 7.54 & 5.85 & 0.19 && 7.51 & 5.90 & 0.23 && 7.37 & 5.94 & 0.20 &&FSUGar. \\
      & 7.56 & 5.85 & 0.02 && 7.55 & 5.89 & 0.20 && 7.52 & 5.94 & 0.23 && 7.39 & 5.99 & 0.22 &&IOPB-I\\
      & 7.51 & 5.84 & 0.22 && 7.50 & 5.85 & 0.24 && 7.47 & 5.90 & 0.25 && 7.33 & 5.94 & 0.21 &&NL3 \\
\\
2     & 7.35 & 5.82 & 0.02 && 7.32 & 5.83 & 0.02 && 7.29 & 5.87 & 0.04 && 7.15 & 5.92 & 0.05 &&FSUGar. \\
      & 7.35 & 5.86 & 0.01 && 7.33 & 5.88 & 0.02 && 7.29 & 5.91 & 0.03 && 7.16 & 5.96 & 0.03 &&IOPB-I\\
      & 7.28 & 5.81 & 0.00 && 7.26 & 5.82 & 0.00 && 7.22 & 5.86 & 0.00 && 7.09 & 5.91 & 0.00 &&NL3 \\
\\
3     & 6.95 & 5.86 & 0.02 && 6.94 & 5.87 & 0.02 && 6.90 & 5.90 & 0.02 && 6.75 & 5.95 & 0.02 &&FSUGar. \\
      & 6.94 & 5.90 & 0.01 && 6.93 & 5.91 & 0.02 && 6.89 & 5.95 & 0.02 && 6.75 & 6.00 & 0.01 &&IOPB-I\\
      & 6.87 & 5.84 & 0.00 && 6.85 & 5.86 & 0.00 && 6.82 & 5.91 & 0.00 && 6.68 & 5.94 & 0.00 &&NL3 \\
\\
\hline
\hline
\end{tabular}
\label{tab3}
\end{table*}
\renewcommand{\baselinestretch}{1.5}
\begin{figure}[!b]
\includegraphics[width=1.0\columnwidth]{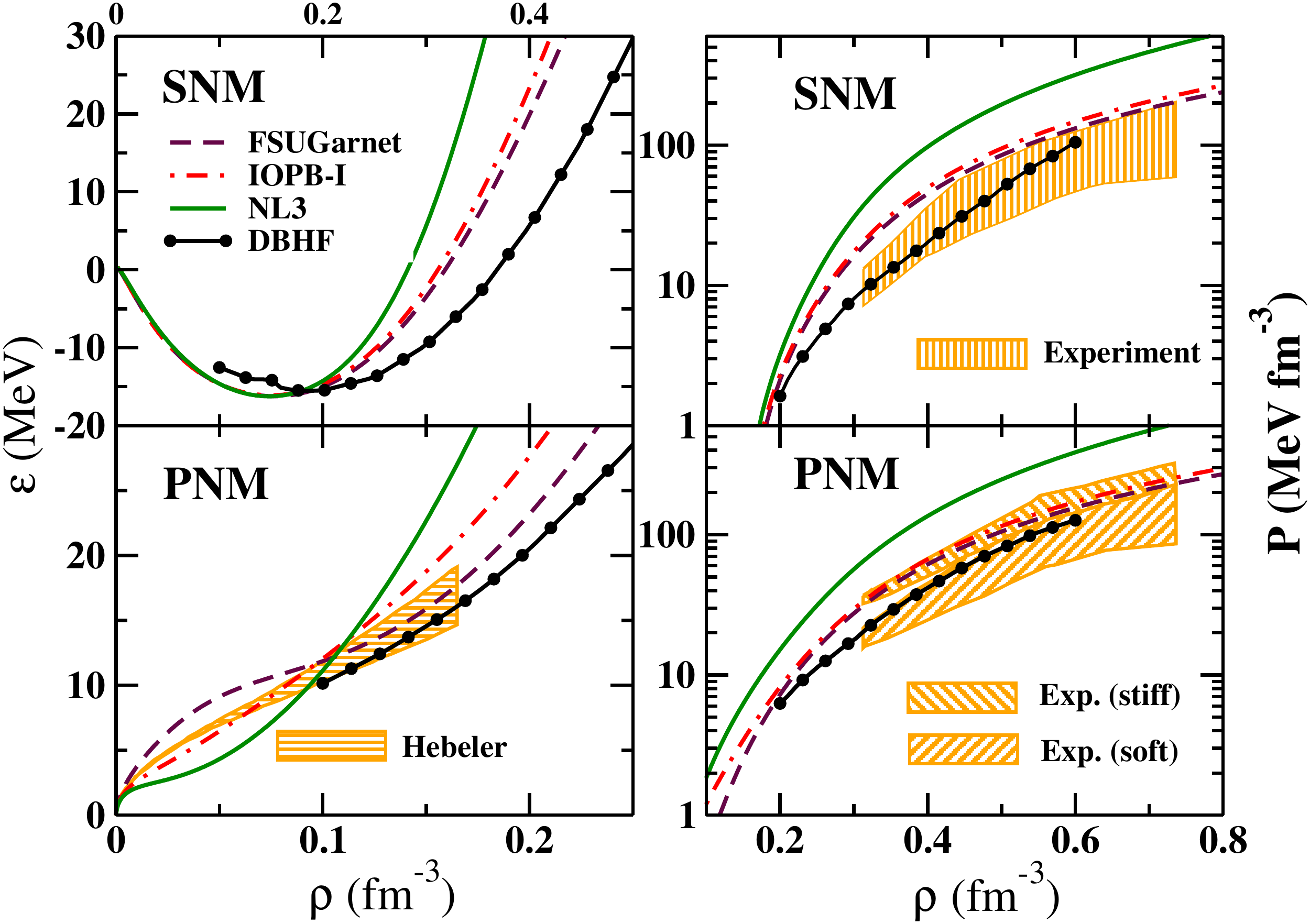}
\caption{The energy density and pressure for SNM and PNM at zero temperature with FSUGarnet, IOPB-I, and NL3 parameter sets. The calculated results are compared with the results by Hebeler \cite{hebe13}, the DBHF equation of 
state \cite{dbhf,baldo}, and the available experimental data \cite{daniel02}.}
\label{esrho}
\end{figure}

\section{Temperature-Dependent Symmetry Energy of Nuclear Matter} {\label {eos}}
In order to study the surface properties of the nuclei at finite T, we will use the 
corresponding temperature-dependent quantities of nuclear matter at the local density of 
the nuclei in LDA. First, we reproduce the equation of states of symmetric 
nuclear matter (SNM) and pure neutron matter (PNM) at zero temperature, as have been 
presented in Ref. \cite{iopb1}. The energy density and pressure of SNM and PNM for the 
FSUGarnet (maroon dashed curve), IOPB-I (red dot-dashed curve), and NL3 (green bold curve) parameters sets are shown 
in Fig. \ref{esrho}. The upper-left panel of the figure represents the energy density of 
symmetric nuclear matter. It is clear from this panel of the figure that the FSUGarnet 
and IOPB-I EOSs are softer than that of the NL3 set. FSUGarnet is the softest among the chosen 
parameter sets. The similar nature of the EOS of PNM is shown in the lower-left panel. 
The shaded region is the range for the EOS obtained by Hebeler {\it et. al.}, 
\cite{hebe13}. It can be seen in the figure (Fig. \ref{esrho}) that the energy densities 
corresponding to NL3 and FSUGarnet (it is even softer) do not pass through the low-density 
region of pure neutron matter. Here, the EOS of IOPB-I passes comparatively close to the 
low as well as the high-density region of the experimental band and through it in the 
intermediate region of the density, while that of FSUGarnet passes only through the high-density region. 
The black curve (in the left panels) represents the DBHF data \cite{dbhf}.

The right panel of the figure shows the pressure of SNM (upper-right) and PNM (lower-right) 
for the same parameter sets. The black curves represent the BHF results of 
Baldo et al., \cite{baldo} and the shaded region is the experimentally consistent 
range of values \cite{daniel02}. In the case of SNM, both the FSUGarnet and IOPB-I satisfy the 
experimental limits, while for PNM they only pass through the upper boundary of the experimental 
soft EOS. In both the cases of SNM and PNM, the NL3 is far from the experimental bounds. The 
softer nature of the EOSs of FSUGarnet and IOPB-I is attributed to the cross-coupling of 
$\omega-$ and $\rho-$ mesons \cite{iopb1}. This is one of the reasons for choosing these 
parameter sets in this work. The values of SNM properties such as the energy density 
($\mathcal{E}$), symmetry energy  $J^{NM}$, symmetry energy curvature ($K^{NM}_{sym}$), 
slope parameter of the symmetry energy ($L^{NM}$) and $Q^{NM}_{sym}$ at the saturation density 
$\rho_0$ are given in Table \ref{force1}. 

\begin{figure}[!b]
\includegraphics[width=1.0\columnwidth]{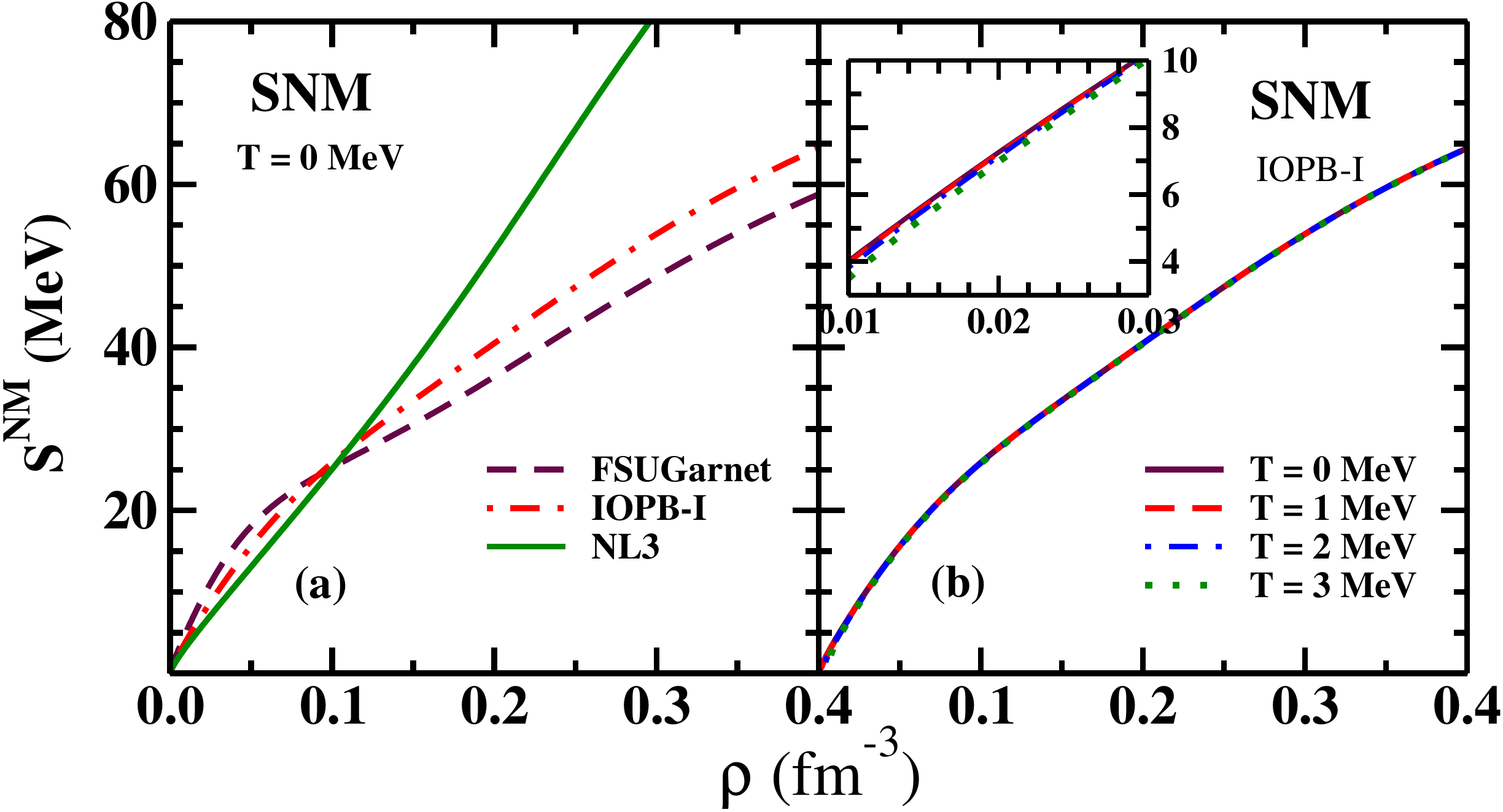}
\caption{(a) The symmetry energy ($S^{NM}$) of infinite SNM at zero temperature
for the FSUGarnet, IOPB-I, and NL3 parameter sets. (b) The same quantity at the finite temperature
corresponding to IOPB-I. The zoomed part shows $S^{NM}$ at low density.}
\label{srhoT}
\end{figure}

The symmetry energy of SNM is shown in Fig. \ref{srhoT}. The left panel of the figure 
represent the symmetry energy for the FSUGarnet (maroon dashed curve), IOPB-I (red dot-dashed curve), and NL3 
(green bold curve) parameter sets at T = 0. The parameter sets produce different symmetry energy 
curves, clearly visible in the figure. We show the symmetry energy of SNM at finite T for the 
IOPB-I parameter set as a representative case in the right panel of Fig. \ref{srhoT}. The 
effects of temperature on the nuclear matter can be observed at higher values 
of T. Here, we have calculated the symmetry energy for T = 0, 1, 2, and 3 MeV. The symmetry 
energy $S^{NM}$ at these temperatures is almost the same. The minute difference at low density 
can be seen in the zoomed part of the figure. This minute difference causes a significant 
change in the value of $S(T)$ of a nucleus at finite T. The quantities involved in the expansion 
of the symmetry energy at the local density of a nucleus [Eq. (\ref{slda})] are $J^{NM}$, $L^{NM}$, 
and $K^{NM}_{sym}$ at the saturation point of nuclear matter. These quantities are not considered 
as constant values throughout the calculation of the temperature-dependent symmetry energy. Rather, 
they slightly change due to the temperature. Thus, in the calculation of the symmetry energy of 
nuclear matter at the local density of a nucleus, we have taken the temperature-dependent saturation 
point to get numerically consistent results. Similarly, all the quantities at the saturation point 
in Eqs. (\ref{slda}) are evaluated at finite T.

\begin{figure}[!b]
\includegraphics[width=1.0\columnwidth]{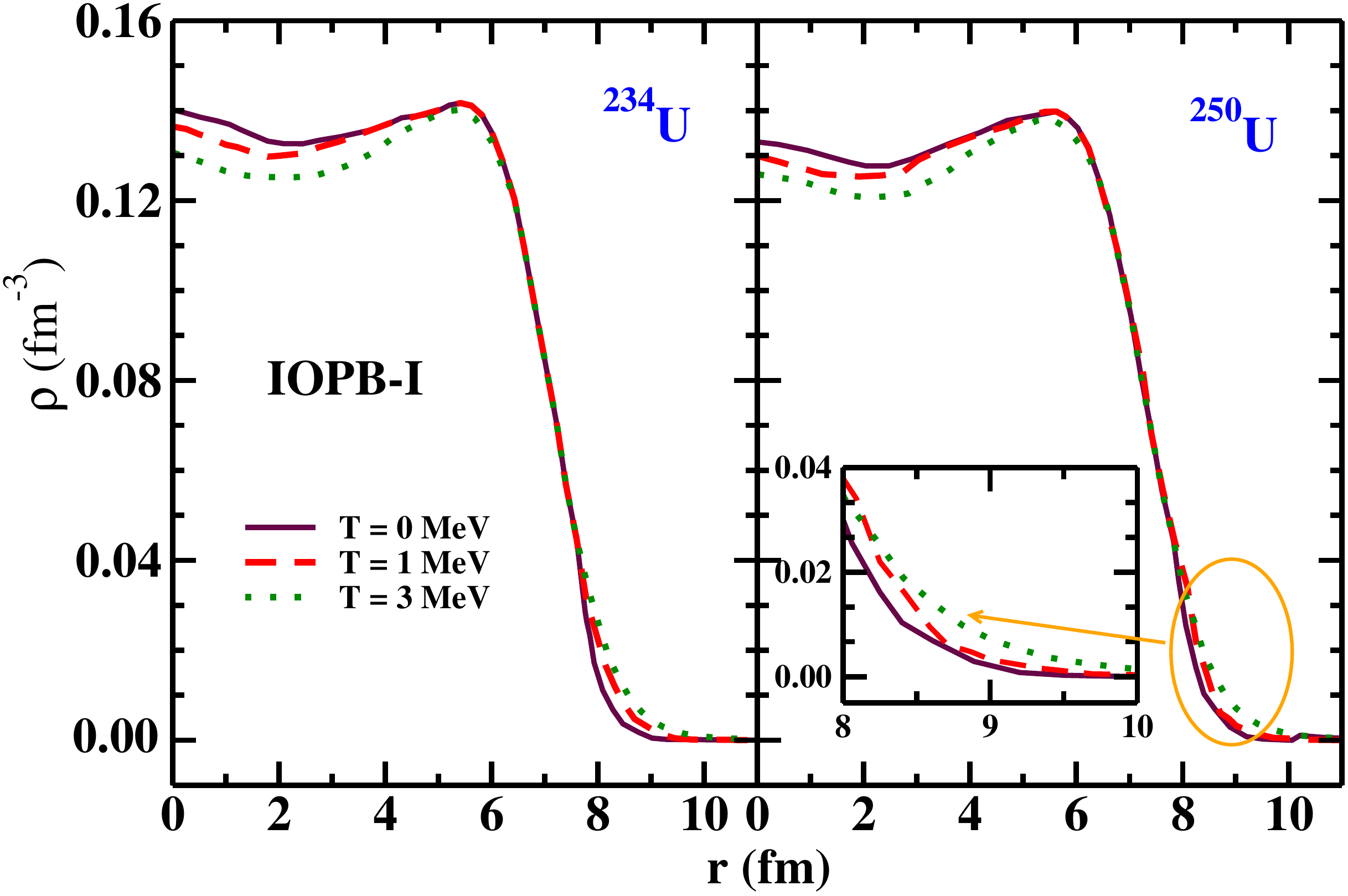}
\caption{The densities of the nuclei $^{234}U$ (left panel) and $^{250}U$ 
(right panel) corresponding to the IOPB-I set at finite T. The zoomed part shows the effect 
of T on the density at the surface of the nuclei.}
\label{density}
\end{figure}
\begin{figure}[!b]
\includegraphics[width=1.0\columnwidth]{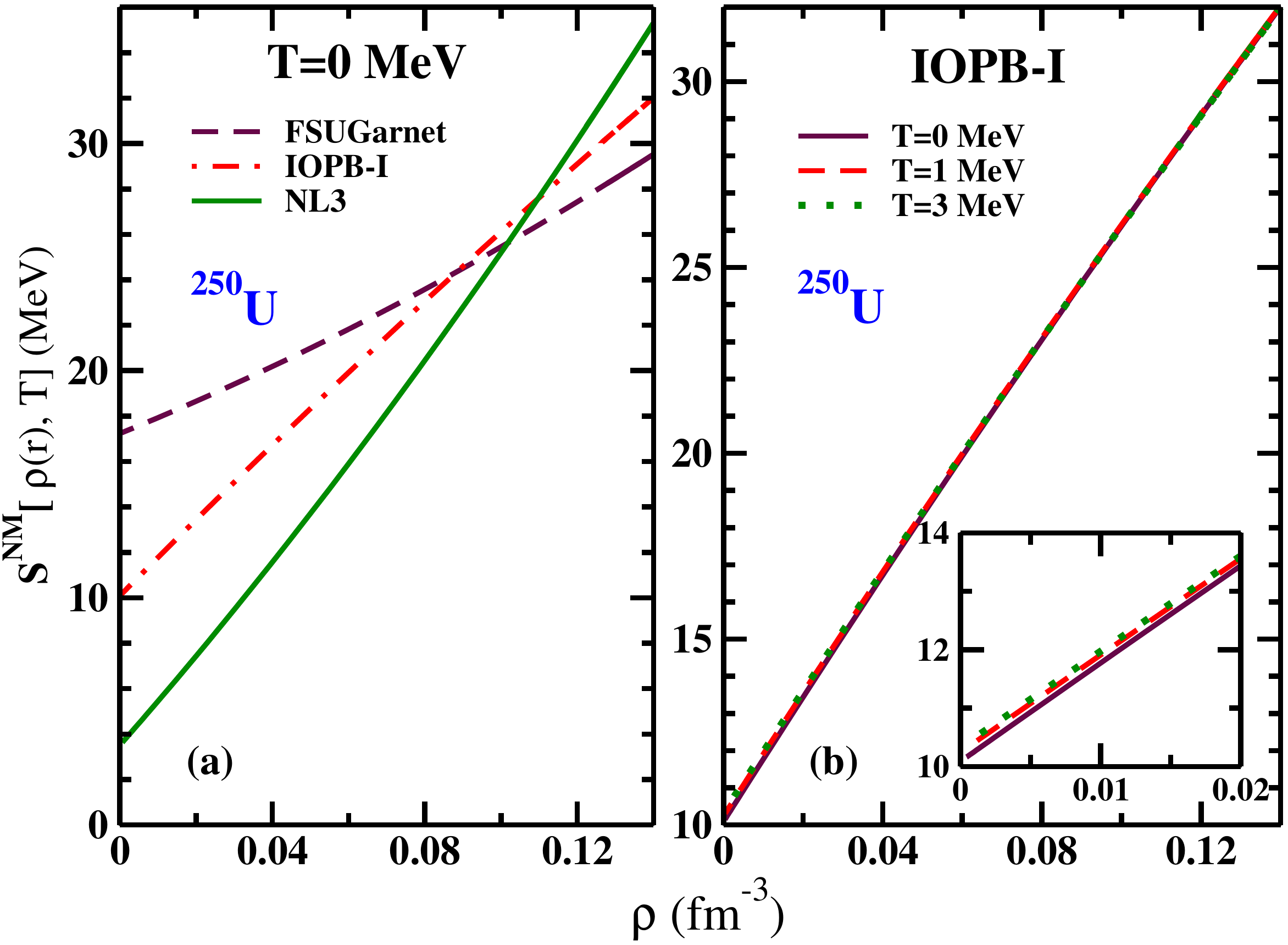}
\caption{The symmetry energy of nuclear matter $S^{NM}[\rho(r),T]$ at the local 
density of $^{250}U$ (a) at T=0 MeV corresponding to the FSUGarnet, IOPB-I, and NL3 parameter sets, 
and (b) at finite T corresponding to IOPB-I set. The zoomed part of panel (b) shows the 
effect of T at low density.}
\label{250rhoS}
\end{figure}

\section{The Symmetry Energy at the Local Density of a Nucleus}
{\label{denS}}
The spherical equivalent densities of the deformed nuclei $^{234,250}U$ at finite T obtained 
within the TRMF model with IOPB-I parameter set as a representative case are shown in Fig. 
\ref{density}. The color code is described in the legends (legend box). The effect of T on the 
densities can be observed from the figure. At finite temperature, the random motion of the 
nucleons is increased. As a result, their density distribution is changed. The central density 
of the nuclei is found to decrease with the increase of T. Consequently, the nucleons are pushed 
towards the surface and hence, the surface density of nuclei has a slight enhancement with T. 
The effect of temperature on the size of a nucleus can also be observed from Table 
\ref{tab3}, where it is found to be grown. The zoomed part of the figure shows the surface 
part of the density. The calculated densities are used to obtain the effective surface 
properties of the nuclei through Eq. (\ref{seqn}).  

The symmetry energy of nuclear matter at the local density of $^{250}U$ as a representative 
case is shown in Fig. \ref{250rhoS}. The left panel of the figure exhibits 
the symmetry energy at zero temperature with the three parameter sets, while, the right panel 
shows the same at finite T corresponding to the IOPB-I set. It is clear from the figure that 
the symmetry energy is very sensitive to the choice of the parameter set. The symmetry 
energy corresponding to FSUGarnet is larger while that corresponding to NL3 is smaller. This 
trend is reversed at higher values of the density $i.e., \approx 0.10$ fm$^{-3}$. The symmetry 
energy has a weak temperature dependence in the range of T considered. The zoomed part of the 
right panel shows this small effect of T at low nuclear density. This dependence is similar to that 
shown in Fig. \ref{srhoT}.  

\begin{figure}[!b]
\includegraphics[width=1.0\columnwidth]{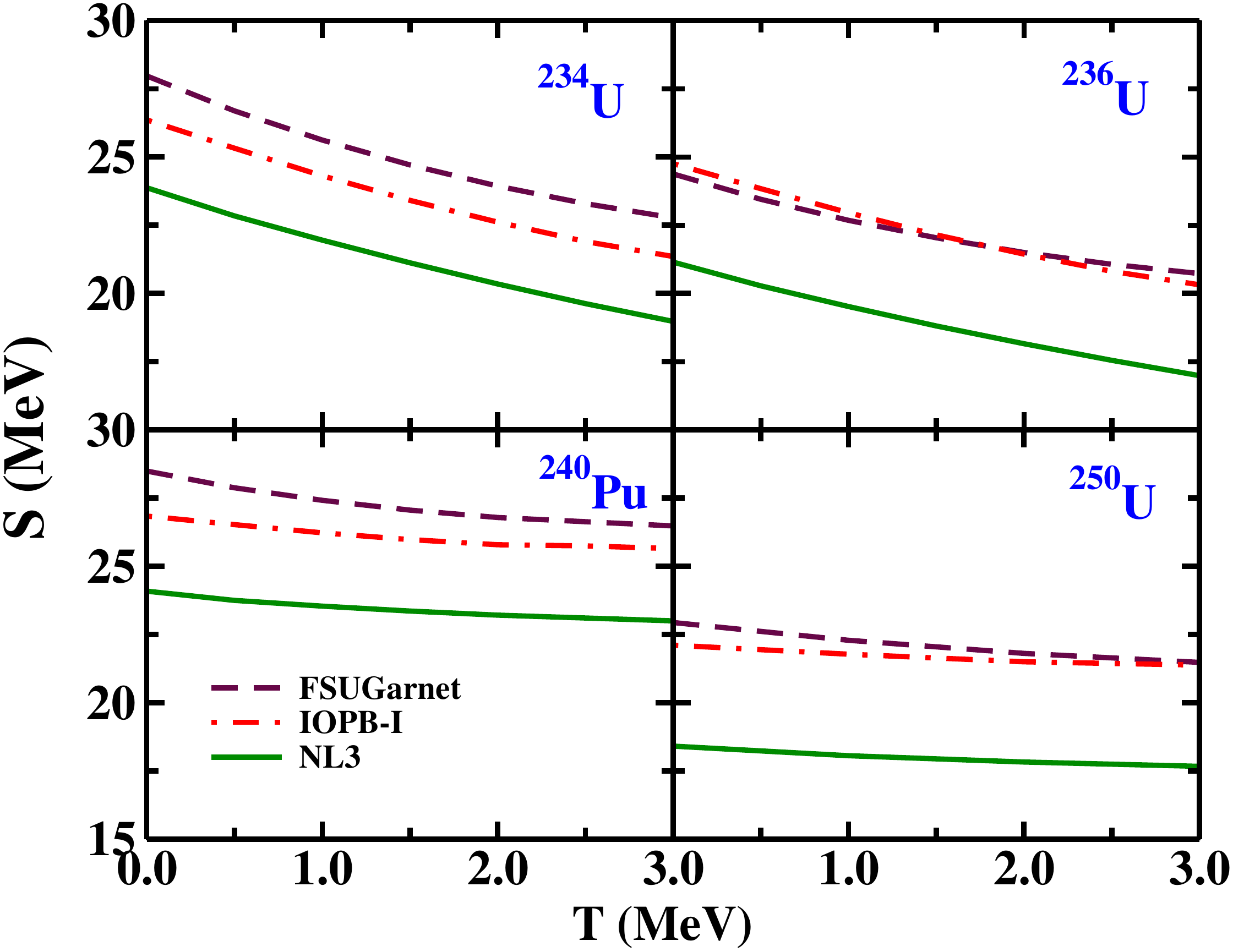}
\caption{The effective symmetry energy coefficient ($S$) of $^{234,236,250}U$ 
and $^{240}Pu$ at finite.}
\label{symmetry}
\end{figure}
\begin{figure}[!b]
\includegraphics[width=1.0\columnwidth]{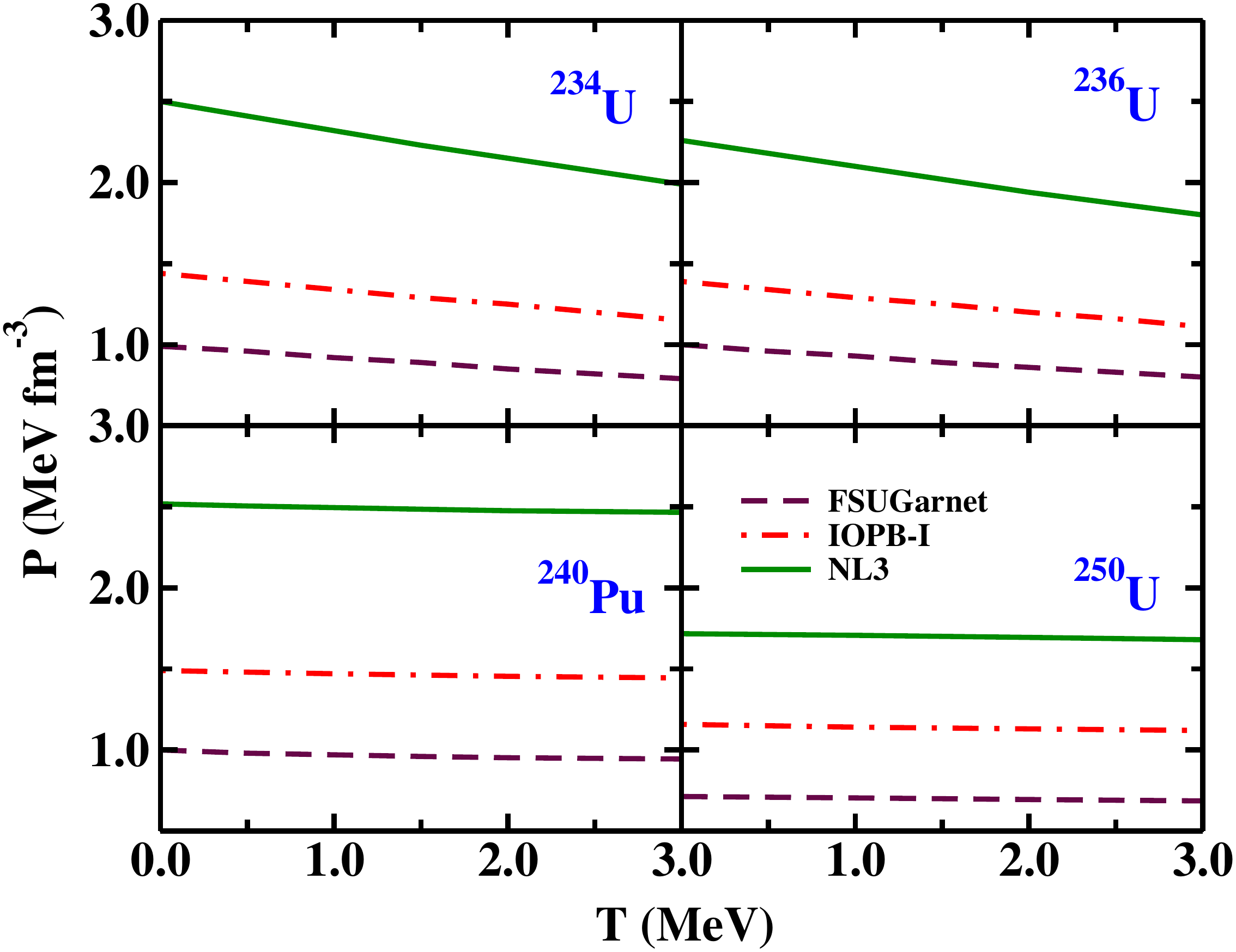}
\caption{The effective neutron pressure ($P$) of the nuclei $^{234,236,250}U$ 
and $^{240}Pu$ at finite T for the chosen parameter sets.}
\label{pressure}
\end{figure}
\begin{figure}[!b]
\includegraphics[width=1.0\columnwidth]{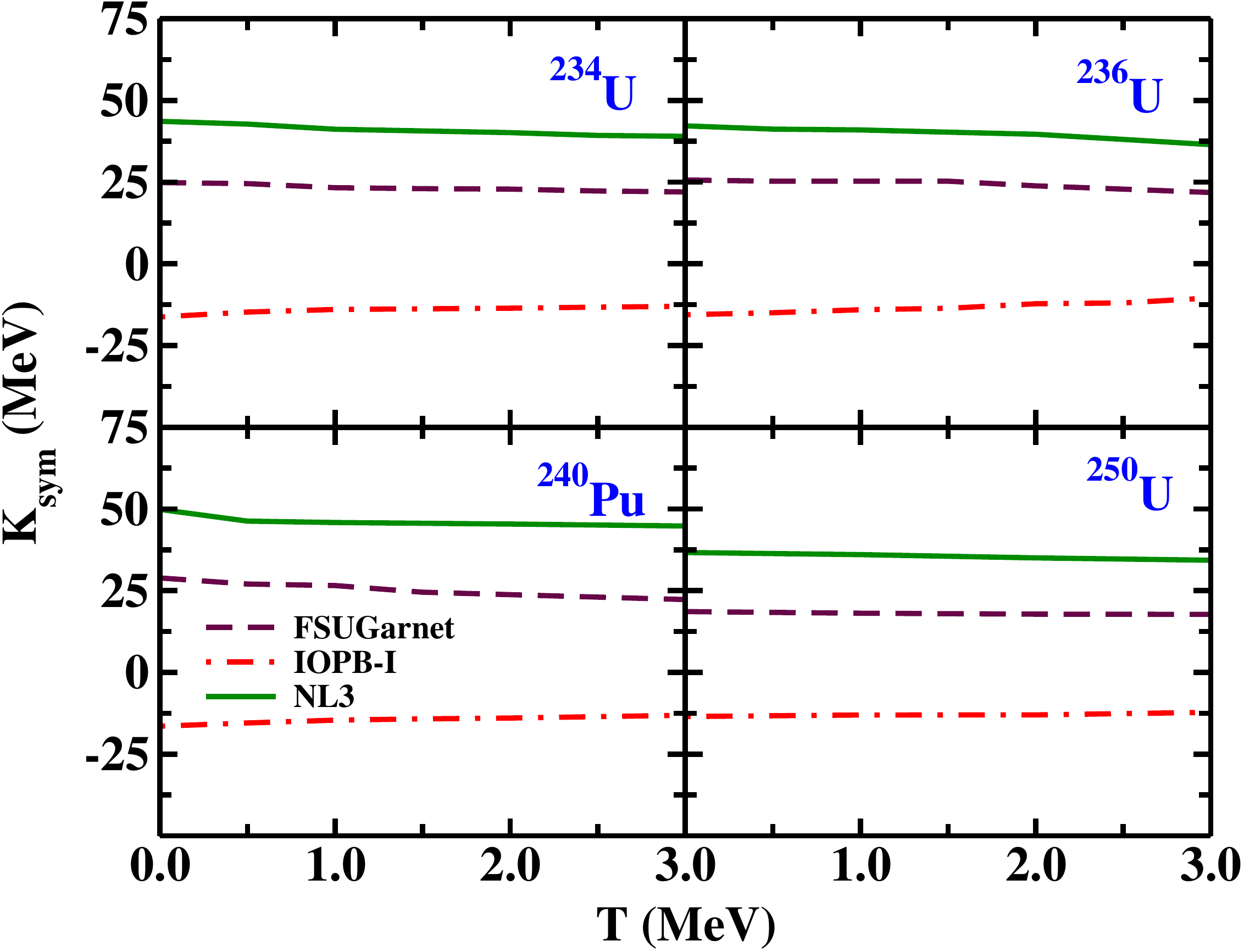}
\caption{The symmetry energy curvature ($K_{sym}$) of the nuclei $^{234,236,250}U$ and 
$^{240}Pu$ at finite T.}
\label{compress}
\end{figure}

\section{The Symmetry Energy, Neutron Pressure and Symmetry Energy Curvature of Nuclei at Finite Temperature}
{\label{surface}}
Fig. \ref{symmetry} shows the temperature-dependent effective symmetry energy coefficient $S$ of 
the nuclei $^{234,236,250}U$ and $^{240}Pu$ for the FSUGarnet (maroon dashed curve), IOPB-I (red dot-dashed curve) and 
NL3 (green bold curve) parameter sets. The values of $S$ in all the cases decreases monotonically with 
T. The results corresponding to the IOPB-I set are intermediate to those corresponding to FSUGarnet 
and NL3. As mentioned above, the FSUGarnet set is the softest among the chosen sets while NL3 is the 
stiffest (see Fig. \ref{esrho}). Thus, the properties of the nuclei and nuclear matter predicted by 
IOPB-I lie in between those predicted by FSUGarnet and NL3 \cite{iopb1,aqjpg}. In the case of 
$^{236,250}U$, the symmetry energy curve corresponding to IOPB-I is almost equal to that of FSUGarnet. 
Among the four nuclei, the $S(T)$ coefficient for $^{250}U$ is the smallest due to its large isospin 
asymmetry. A lower value of the symmetry energy enhances the rate of conversion of protons to neutrons 
through electron capture \cite{Steiner05,janka07}. The low value of $S(T)$ for $^{250}U$ means that 
the conversion of asymmetric to symmetric matter requires a smaller amount of energy. Similarly, the 
lowering of the values of symmetry energy coefficient at higher T implies that less energy is needed 
to convert a neutron to proton or vice versa. It can be remarked, here, that the rate of $\beta-$ 
decay will increase with T ($i.e.,$ in the excited-state of a nucleus).      

The effects of T on the effective neutron pressure and symmetry energy curvature are shown in Fig. 
\ref{pressure} and \ref{compress}, respectively. The trend of the curves of $P$ and $K_{sym}$ is 
similar to that observed in the case of $S$. But, the order of the curves corresponding to the 
parameter sets is different. Here, FSUGarnet predicts smaller values of the pressure compared to 
the IOPB-I and NL3 sets. It implies that the softer the EOS, the smaller the pressure of the system 
is. The $^{250}U$ nucleus has smaller $P$ and $K_{sym}$ values than the other nuclei. This behavior 
can also be attributed to the large isospin asymmetry of the nucleus. The neutron pressure decreases 
with T because the volume of a nucleus expands ($i.e.,$ the radius increases as seen in Table 
\ref{tab3}) as T increases. The rate of emission of neutrons from a nucleus also increases with T 
\cite{zhu14}, which leads to a decrease in the pressure. The values of $K_{sym}$ corresponding to the 
IOPB-I set are negative which is consistent with the negative value of the symmetry energy curvature 
at the saturation density of nuclear matter (see Table \ref{force1}). 

\begin{figure}[!b]
\includegraphics[width=1.0\columnwidth]{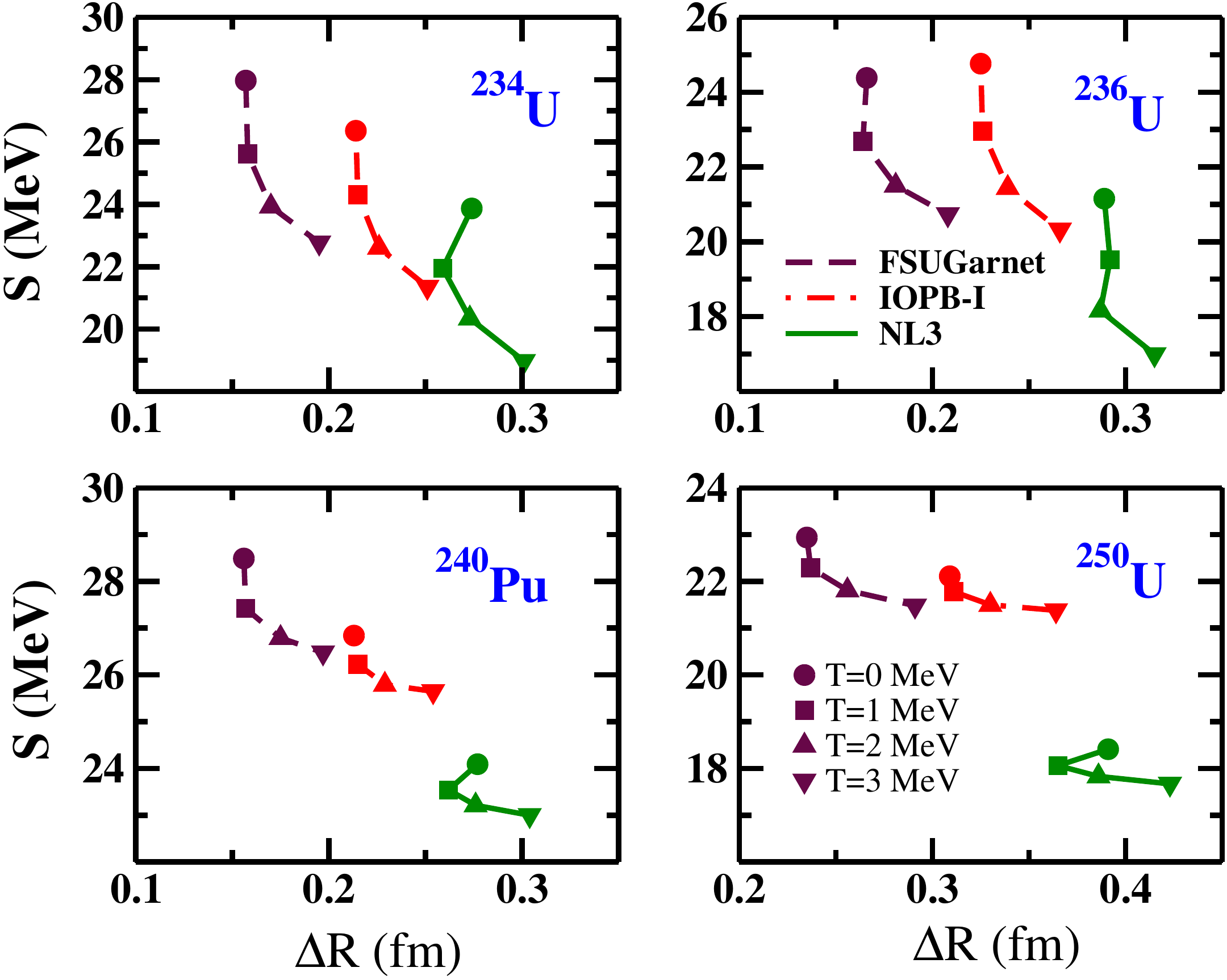}
\caption{The variation of the symmetry energy coefficient with the neutron skin-thickness 
for the FSUGarnet, IOPB-I, and NL3 parameter sets. The circle, square, triangle up, and 
triangle-down symbols correspond to the values at temperature 0, 1, 2, and 3 MeV.}
\label{Sdr}
\end{figure}
\begin{figure}[!b]
\includegraphics[width=1.0\columnwidth]{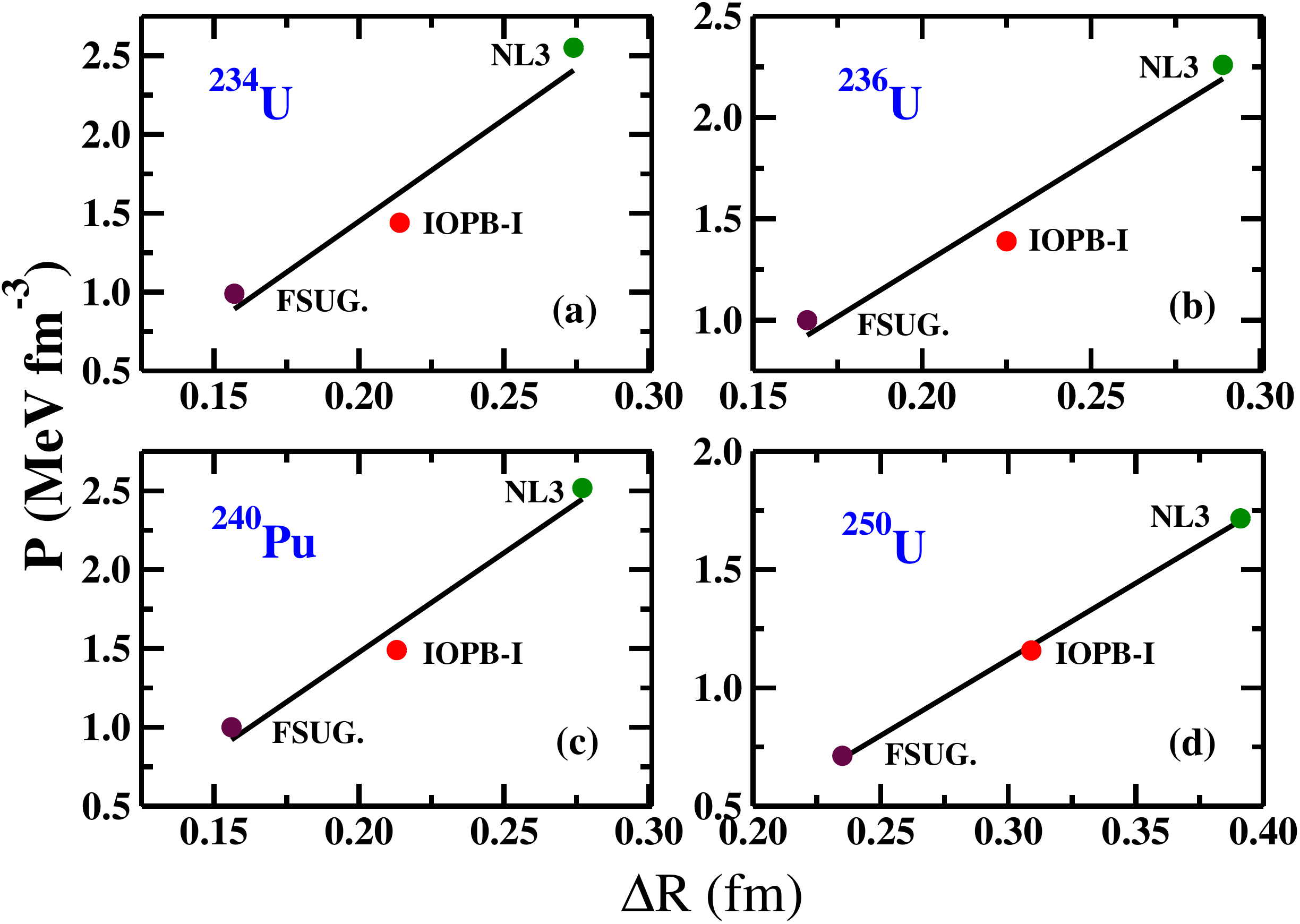}
\caption{The effective neutron pressure $P$ for (a) $^{234}U$, (b) $^{236}U$, 
(c) $^{240}Pu$, and (d) $^{250}U$, as a function of the neutron skin-thickness $\Delta R$ for the three sets of parameter sets; FSUGarnet, IOPB-I, and NL3. The lines in all panels represent a linear fit.}
\label{drPT00}
\end{figure}
\begin{figure}[!b]
\includegraphics[width=1.0\columnwidth]{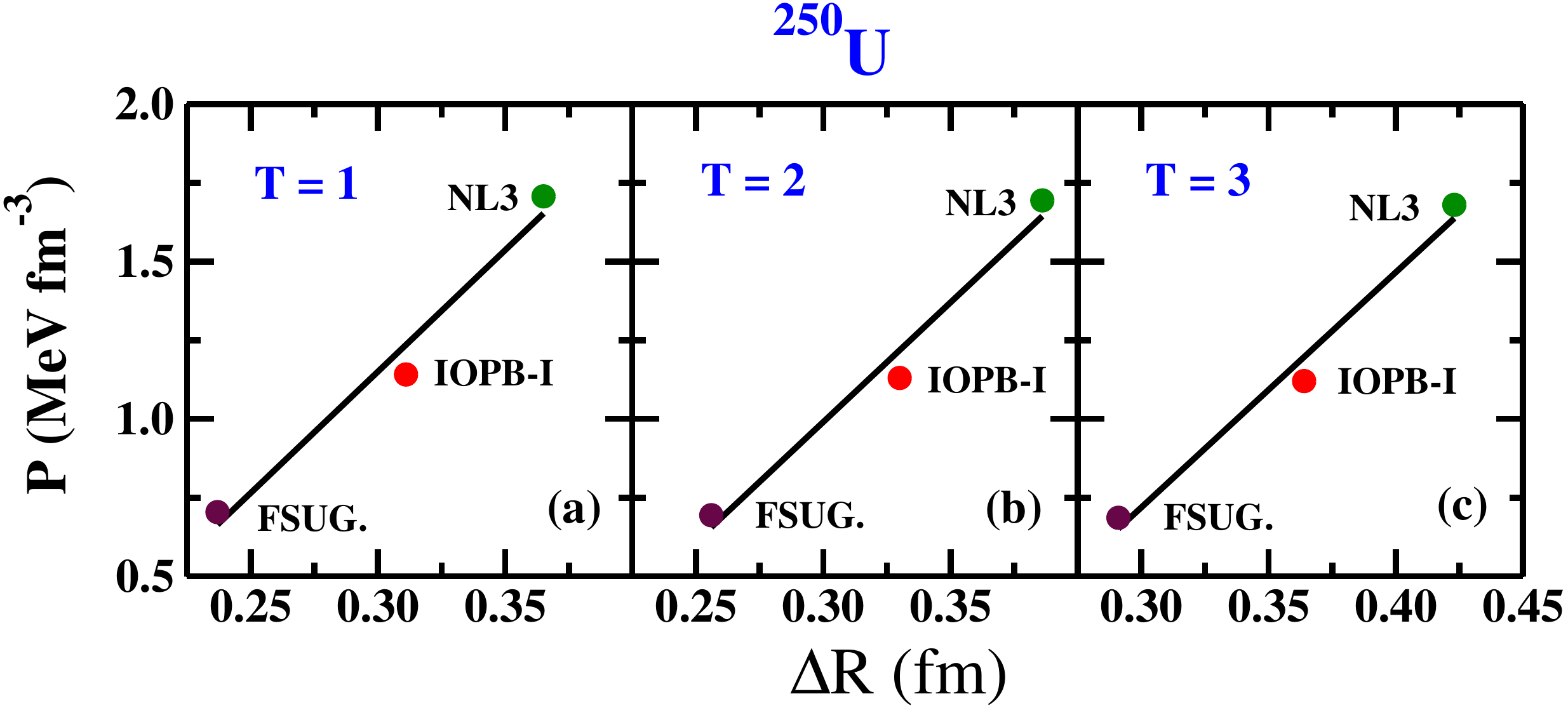}
\caption{The effective neutron pressure $P$ as a function of the neutron skin-thickness $\Delta R$ for $^{250}U$ at temperatures $T=1,2,3$ (MeV) (panels a, b, and c, respectively) 
for the three sets of parameter sets; FSUGarnet, IOPB-I, and NL3. The lines in all panels 
represent a linear fit.}
\label{drP250}
\end{figure}

The skin-thickness has been shown to be correlated with the effective surface properties in Refs. 
\cite{anto,anto80,gai11,bhu18,gai12} over particular isotopic chains. It depends linearly on the 
surface properties of a nucleus with a kink at a magic/semi-magic nucleus of an isotopic chain 
\cite{anto,anto80,gai11,bhu18,gai12}. Although, we have not considered a whole isotopic chain of 
nuclei, we have found a correlation among the skin-thickness, symmetry energy, and the neutron 
pressure due to the variation in these quantities at finite T. Fig. \ref{Sdr} shows the correlation 
between the effective symmetry energy coefficient and the skin-thickness of the nuclei at finite T. 
The skin-thickness of the nuclei grows with the 
temperature \cite{aqjpg}. We have seen in Fig. \ref{symmetry} that the symmetry energy coefficient 
decreases with T. This implies that the symmetry energy coefficient should decrease with the increase 
of skin-thickness at finite T. The same tendency is observed in Fig. \ref{Sdr}. It is clear from the 
figure that NL3 predicts a larger neutron skin-thickness, with lower values of symmetry energy coefficient 
compared to the other parameter sets. This decrease is monotonic in nature. In going to heavier nuclei, 
the curves become flattered, which means that slight changes in the symmetry energy coefficient lead to 
large changes in the skin-thickness and vice versa.

The slope parameter ($L-$ coefficient), and thus the pressure [see Eq. (\ref{lsym})] have been shown to 
be strongly correlated with the neutron skin-thickness of $^{208}$Pb \cite{brown,rj02,cent09,rocaprl11} 
and the radius of a neutron star. This motivates us to examine the correlation between the pressure of 
the finite nuclei and their neutron skin-thickness. In Fig. \ref{drPT00}, we show a linear correlation 
between the neutron pressure and the skin-thickness of finite nuclei at T = 0 MeV for all four nuclei. 
The panels (a), (b), (c), and (d) of the figure contain the data of $P$ versus $\Delta R$ for the nuclei 
$^{234,236}U$, $^{240}Pu$ and $^{250}U$, respectively with FSUGarnet, IOPB-I, and NL3 parameter sets. 
The skin-thickness of the nuclei is increased with the parameter sets (going from softer to stiffer). 
One observes in Fig. \ref{esrho} that the stiffer the EOS, the larger be the pressure for nuclear matter. 
A similar trend is obtained for finite nuclei. The neutron pressure is greater for the NL3 set, while it 
is a minimum for the parameter set FSUGarnet. Here, by fixing the temperature, the effects on pressure 
may be studied by going from one force parameter to the others. The lines in the graph are the linear 
fitted curves (as, $y=a x+b$) with the values of $a$ and $b$ (31.715, -2.523), (23.429, -1.644), 
(31.421, -2.399), and (19.177, -2.484) for $^{234,236}U$, $^{240}Pu$ and $^{250}U$, respectively. This 
correlation is also seen when the temperature of the nuclei is increased (see, Fig. \ref{drP250}).  
 
\section{Conclusions}{\label{summary}}
We have studied the effective bulk symmetry energy properties of $^{234,236,250}U$ and $^{240}Pu$ nuclei 
within the TRMF model. We have calculated the densities of the nuclei along with ground and excited-state 
bulk properties at finite T within the axially deformed TRMF model. The nuclear matter EOS for SNM and 
PNM along with the symmetry energy and its related observables are also estimated within TRMF formalism. 
We have used recently determined force parameter sets i.e., FSUGarnet and IOPB-I and compared the results 
with the widely accepted NL3 interaction. The nuclear matter properties are calculated at the local 
density of the nuclei, which are further used in the LDA to calculate the corresponding properties of the 
nuclei.  
We have observed a minute effect of the temperature on the nuclear matter symmetry energy, which 
causes a significant change in the symmetry energy and related quantities of a nucleus according to its 
density profile. However, the effect of temperature on the calculated properties is almost the same irrespective 
of the parameter sets. The symmetry energy, neutron pressure, and symmetry energy curvature decrease with 
the increase of T, while the skin-thickness increases. These properties are found to be smaller for $^{250}U$ 
due to its large isospin asymmetry, which enhances the rate of electron capture. The correlation among the 
calculated properties of the nuclei is found at finite T for each parameter set. The symmetry energy 
coefficient is found to vary inversely with the pressure. It is found that the softer the EOS is, the larger 
the symmetry energy coefficient and the smaller the neutron pressure of the nuclei are. The neutron pressure 
is linearly correlated with $\Delta R$ at finite T as calculated with all three parameter sets. 



\chapter{Neutron Star in the Presence of Dark Matter}{\label{chap6}}

\rule\linewidth{.5ex}

\section{Introduction}
\label{intro6}

Currently, there is a plethora of modern observations that support and confirm the existence of DM on a wide range of scales.
Many DM candidates have been proposed and studied over the years by cosmologists and particle physicists alike in an effort to constrain its properties. For a nice list of the existing DM candidates see e.g. \cite{taoso}, and for DM searches go through Refs. \cite{DM3,jennifer,DM5}. Despite that, the origin and nature of DM still
remain unknown. Indeed, the determination of the type of elementary particles that play the role of DM in the Universe is one of the current challenges of Particle Physics and modern Cosmology.
WIMPs, which are thermal relics from the Big-Bang, are perhaps the most popular DM candidates. Initially, when the Universe was very hot, WIMPs were in thermal equilibrium with their surrounding particles. As the Universe expands and cools down, at a certain temperature which depends on the
precise values of the mass of the DM particle and its couplings to the Standard Model (SM) particles, WIMPs decouple from the thermal bath, and its abundance freezes out. After freezing out, they can no longer annihilate, and their density is the same since then comprising the observed DM abundance of the Universe \cite{carlos}.

If WIMPs (let us have in mind the lightest neutralino in supersymmetric models, although the analysis and the obtained results still hold true for any fermionic WIMP) are the main candidates for DM, they will cluster gravitationally with stars, and also form a background density in the Universe. In Ref. \cite{smith90}, it was remarked that our own galaxy, the Milky Way, contains a large amount of DM. This raises the hope of
detecting relic WIMPs directly, by scattering experiments in Earth-based detectors. The interaction of the DM particle with nuclei through elastic \cite{carlos} or inelastic scattering \cite{ellis88,stark95} is being studied in various
laboratories worldwide. As mentioned in Chapter \ref{chap1}, some of the direct detection experiments are: 
the DArk MAtter (DAMA) experiment \cite{dama1,dama2}, Cryogenic Dark Matter Search (CDMS) experiment \cite{cdms}, 
EDELWEISS experiment \cite{edel}, IGEX \cite{igex}, ZEPLIN \cite{zeplin},
 GErmanium DEtectors in ONe cryostat (GEDEON) \cite{morales}, CRESST \cite{cresst}, GErmanium in liquid NItrogen Underground Setup
(GENIUS) \cite{genius}, and LHC. Furthermore, Fermi-LAT, GAMMA-400, IceCube, Kamiokande, and AMS-02 are some of the indirect DM detection experiments \cite{jennifer}.

Despite the null results of other experiments \cite{akerib}, which only put an upper limit on the nucleon-DM particle, the DAMA/LIBRA collaboration, located underground at the Laboratori Nazionali del Gran Sasso in Italy, has been reporting for many years an annual modulation caused by the variation of the velocity of the detector relative to the galactic DM halo as the Earth orbits the Sun. In particular, the final model-independent results of phase 1 were published in 2013 \cite{phase1}, while last year they published the first model-independent results of phase 2 \cite{phase2} collecting data from 6 annual cycles. In the second upgraded phase, which started at the end of 2010, the two main improvements in comparison with the first phase are the doubled exposure as well as the lower energy threshold from $2~keV$ to $1~keV$. If the signal reported by the DAMA/LIBRA collaboration is interpreted as WIMP DM, it gives rise to the $\sim 10~GeV$ WIMP hypothesis with a spin-independent (SI) DM-nucleon at $\sim 10^{-40}~cm^2$, see Fig.~1 of \cite{theor1,theor2}. To be more precise

\begin{eqnarray}
3 \times 10^{-41} cm^{2} \lesssim \sigma_p^{SI} \lesssim 5 \times 10^{-39} cm^{2}
\label{crossrange}
\end{eqnarray}

while the range of the mass of the DM particle is
\begin{eqnarray}
3 GeV \lesssim M_\chi \lesssim 8 GeV .
\label{mdm}
\end{eqnarray}
Various studies have taken these results into account \cite{foot,feng,bott,avigno}.
In the WIMP scenario, a one-to-one relation is seen between the SI direct detection rate and DM relic density if its elastic scattering on nuclei occurs dominantly through Higgs exchange \cite{sarah}.
The SI direct detection cross-section of elastic scattering of DM ($\chi$) with nuclei is given by \cite{sarah}
\begin{eqnarray}
\sigma(\chi N \rightarrow \chi N) = \frac{y^2}{\pi} \frac{\mu^2_r}{v^2 M_h^4} f^2 M^2 , 
\label{sicross}
\end{eqnarray}
where $v=246~GeV$ is the Higgs vacuum expectation value, $M,M_h$ are the nucleon mass and the Higgs mass, respectively. The variables $y$,
$f M/v$ are the Yukawa couplings for DM interaction with Higgs boson and the interaction of Higgs particle with
the nucleons, respectively. Finally, $\mu_r= \frac{M M_\chi}{M+M_\chi}$ is the reduced mass of a nucleon-DM particle system. The unknown parameters entering Eq. \ref{sicross} may be fixed as follows: First, the mass of the DM particle as well as the DM-nucleon system are taken to be the ones suggested by the aforementioned results of the DAMA/LIBRA collaboration. Then, for a given mass of the Higgs boson the Yukawa coupling $y$ can be determined.

Neutron stars (NSs) are considered to be unique cosmic laboratories to test non-standard physics. A brief introduction of NS is 
given in one of the paragraphs of Chapter \ref{chap1}.  
The NS properties are predicted by an equation of state. On the other hand, by knowing the experimental values of the NS observables such as 
mass, radius, tidal deformability, and moment of inertia, its equation of state can be constrained. Since the exact EOS, 
describing the NS interior is not known yet, we assume the possibility of DM inside the NS core. 
The consequences of DM inside NS have been discussed in the literature \cite{raj,gould,goldman,kouvaris08,kouvaris10,lava10,guver, 
ellis18,ellisplb}. These discussions include the effect of charged massive DM particle on NSs \cite{gould},
trapped WIMPs inside NSs \cite{goldman}, DM annihilation and its effect on NSs \cite{kouvaris08,kouvaris10,lava10},
or the collapse of an NS due to accretion of non-annihilating DM \cite{guver} etc.
The possible effects of DM cores on certain properties of NS have been studied in \cite{ellisplb,ellis18} assuming
different nuclear EOSs {as well as} different fractions of DM. In particular, first in \cite{ellisplb} using a mechanical
model the authors found that the DM cores may produce a supplementary peak in the characteristic gravitational
wave spectrum of neutron star mergers, and then in \cite{ellis18} they investigated the impact of Fermionic asymmetric DM
as well as bosonic self-interacting DM on mass-to-radius profile, maximum mass and tidal deformability $\Lambda$ of NSs.
In \cite{nelson}, the authors studied the properties of NS considering the DM halo and constrained the DM parameters using the GW170817 data.
It was pointed out in \cite{panta} that the mass-radius relation of an NS can be affected in the presence of DM inside the object. The authors of \cite{panta,tuhin} considered, respectively, the Walecka model \cite{wal74} and the NL3 \cite{lala97} EOSs, within the framework of relativistic mean-field (RMF) theory,
for the nucleonic part and fermionic DM inside the neutron star with additional self-coupling of Standard Model Higgs boson in Ref. \cite{panta}.

In this work, we have investigated for the first time the effects of DM inside an NS adopting the $\sim 10~GeV$ WIMP hypothesis as suggested by the results of the DAMA/LIBRA collaboration, which can be realized e.g. in the framework of the
Next-to-Minimal Supersymmetric Standard Model (NMSSM).
The EOSs of nuclear matter can be generated within the framework of an effective theory with
appropriate degrees of freedom. In particular, for example, perturbative QCD \cite{aleksi,tyler}, non-relativistic Skyrme type
density functional theory \cite{tond}, and relativistic mean field approach \cite{wal74} are used to predict EOSs of nuclear matter.
We have considered the effective-field theory motivated relativistic mean-field
model (E-RMF) to generate the EOS of NS by considering the IOPB-I \cite{iopb1}, G3 \cite{G3}, and NL3 \cite{lala97} parameter sets.
Recently as relativistic microscopic approaches, the E-RMF model is used widely to predict the properties of finite nuclei as well as nuclear matter.
Here, in addition to the mass-radius relations, and the tidal deformability, we have analyzed the effects of DM on the moment
of inertia of an NS.
Our work differs from other similar works in two respects, namely i) we have considered more EOSs (especially G3 within E-RMF) for hadronic matter, and ii) we have studied the impact of DM on more NS observables.

\section{Equation of State of Neutron Star in the Presence of Dark Matter}{\label{EOS-NS}}

The Lagrangian density for DM-nucleon interaction through the exchange of Higgs bosons $h$ is given by \cite{tuhin}

\begin{eqnarray}
{\cal{L}} & = & {\cal{L}}_{had.} + \bar \chi \left[ i \gamma^\mu \partial_\mu - M_\chi + y h \right] \chi + 
              \frac{1}{2}\partial_\mu h \partial^\mu h  
\nonumber\\
& &
- \frac{1}{2} M_h^2 h^2 + f \frac{M_n}{v} \bar \varphi h \varphi , 
\label{lag-tot}
\end{eqnarray}

where ${\cal{L}}_{had.}$ is the Lagrangian density for pure hadronic matter, which is discussed in Section \ref{nuclear-matter} of Chapter 2. 
The wave functions $\chi$ and $\varphi$ correspond to
DM particle and nucleon, respectively. We have not considered the higher-order terms of the Higgs scalar potential ($i.e.,$ $h^3$ and $h^4$) since in the mean-field theory approximation these terms are negligible \cite{panta}.
The factor $f$ parameterizes the Higgs-nucleon coupling, and a complete expression for $f$ can be found in \cite{singlet3}.
Following the lattice computations \cite{lattice1,lattice2,lattice3,lattice4}, we shall consider the central value $f=0.3$ in agreement with \cite{singlet3}. For the DM sector, we shall assume a mass range and an SI DM-nucleon cross-section suggested by the DAMA results \cite{dama1,dama2}. It is easy to verify that if we take the Higgs mass to be 125 GeV, the Yukawa coupling $y$ computed using Eq. \ref{sicross} lies in the non-perturbative regime. Therefore we have to assume a light Higgs boson with a mass $M_h=40~GeV$, so that $y < 1$.
The authors of \cite{john10,basic1} have shown that such a scenario can be realized in the framework of the NMSSM in agreement with the rest of the experimental constraints.

Before we continue with our discussion, perhaps it should be useful to briefly mention here the basic features of the NMSSM \cite{nmssm1,nmssm2,nmssm3} (for a review see \cite{nmssm4}). It is a simple extension of the MSSM in which a singlet supermultiplet is added, and it is characterized by the following properties: i) It preserves the nice properties of the MSSM, i.e. it solves the hierarchy problem while at the same time
it provides us with an excellent DM candidate, ii) it solves the $\mu$ problem \cite{muproblem}, and iii) there is a rich Higgs sector with 2 Higgs bosons more in comparison with MSSM. In particular, if some of the Higgs bosons have a significant singlino component they can be light, $M_H \leq 70~GeV$, without any contradiction to current experimental constraints \cite{teixeira}. As a matter of fact, it has been shown that in NMSSM it is possible to obtain a DM-nucleon SI cross-section as high as the one indicated by the DAMA/LIBRA results, precisely due to the exchange of light Higgs bosons, which cannot be achieved in the MSSM \cite{john10,basic1}.

Moreover, one may briefly summarize the current status of DM in SUSY models taking into account LHC searches as follows: Supersymmetric models have been under siege after the Higgs boson discovery \cite{higgs1,higgs2} combined with the lack of any signal for sparticles \cite{null1,null2}, pushing the SUSY spectrum in the multi-TeV region \cite{siege1,siege2}. In natural SUSY with low fine-tuning electroweak symmetry breaking, the lightest neutralino is higgsino-like with a mass at (100-300)~GeV \cite{baer1}, which has been excluded as a single
DM candidate \cite{baer2}, as its abundance is lower than the WMAP/PLANCK measured
value by a factor of 10-15 \cite{baer3}. A mixed axion/higgsino dark
matter scenario has emerged, in which the axion is the dominant DM component in the bulk
of the parameter space \cite{baer4}. On the other hand, in the framework of the NMSSM a
light singlino with a mass lower than $60~GeV$ is still a viable DM candidate in a few
regions of the allowed parameter space \cite{Ref1}, while the lightest CP-even Higgs
boson, which is predominantly a singlet, can be as light as $48~GeV$ \cite{Ref2}.

Solving the full Lagrangian density (Eq. \ref{lag-tot}) by the variational principle, and taking care of all the mean-field and INM approximations \cite{tuhin,iopb1}, the energy density and pressure are given by

\begin{eqnarray}
{\cal{E}} & = &  {\cal{E}}_{had.} + \frac{2}{(2\pi)^{3}}\int_0^{k_F^{DM}} d^{3}k \sqrt{k^2 + (M_\chi^\star)^2 } 
+ \frac{1}{2}M_h^2 h_0^2 .
\label{etot}
\end{eqnarray}

\begin{eqnarray}
P & = &  P_{had.} + \frac{2}{3(2\pi)^{3}}\int_0^{k_F^{DM}} \frac{d^{3}k k^2} {\sqrt{k^2 + (M_\chi^\star)^2}} 
- \frac{1}{2}M_h^2 h_0^2 .
\label{ptot}
\end{eqnarray}

The Fermi momentum of DM particles ($k_f^{DM}$) is taken to be constant throughout the calculation with the value fixed at 0.06 GeV, although in Refs. \cite{tuhin,panta} the authors have considered values within a certain range.
The effective mass of the nucleon ($M^\ast$) is modified due to the interaction with the Higgs boson. The new effective mass of the nucleon $M^\star$ and the effective mass of the DM particle $M_\chi^\star$ are given by,

\begin{eqnarray}
M_i^\star &=& M_i + g_\sigma \sigma -\tau_3 g_\delta \delta - \frac{f M_n}{v}h_0, 
\nonumber\\
M_\chi^\star &=& M_\chi -y h_0.
\label{effmass}
\end{eqnarray}

So far the discussion on the EOS has been for INM. Now we present in the discussion to follow how to obtain the EOS for NS.
In a neutron star, the Fermi momentum of neutrons and protons are different due to the different number densities of these particles.
For the stability of NSs, the $\beta -$ equilibrium condition is imposed, which is given by,

\begin{eqnarray}
\mu_n &=& \mu_p +\mu_e,   \nonumber \\
\mu_e &=& \mu_\mu.
\label{bequi}
\end{eqnarray}

where, $\mu_n$, $\mu_p$, $\mu_e$, and $\mu_\mu$ are the chemical potentials of neutrons, protons, electrons, and muons, respectively. The muon comes into play when the chemical potential of the electrons reaches the muon rest mass
and maintains the charge of NS as follows

\begin{eqnarray}
\rho_p = \rho_e +\rho_\mu. 
\label{charge}
\end{eqnarray}

The chemical potentials $\mu_n$, $\mu_p$, $\mu_e$, and $\mu_\mu$ are given by,

\begin{eqnarray}
\mu_n &=& g_\omega \omega_0 + g_\rho \rho_0+ \sqrt{k_n^2+ (M_n^\star)^2},
\label{mun}
\end{eqnarray}

\begin{eqnarray}
\mu_p &=&  g_\omega \omega_0 - g_\rho \rho_0+ \sqrt{k_p^2+ (M_p^\star)^2},
\label{mup}
\end{eqnarray}

\begin{eqnarray}
\mu_e &=& \sqrt{k_e^2+ m_e^2},
\label{mue}
\end{eqnarray}

\begin{eqnarray}
\mu_\mu &=& \sqrt{k_\mu^2+ m_\mu^2}.
\label{mumu}
\end{eqnarray}

The particle fraction is determined by the self-consistent solution of Eq. \ref{bequi} and Eq. \ref{charge}
for a given baryon density. The total energy density and pressure for $\beta-$ stable NS are given by,

\begin{eqnarray}
{\cal {E}}_{NS} &=& {\cal {E}} + {\cal {E}}_{l} \nonumber \\
P_{NS} &=& P + P_l .
\label{eos-nstar}
\end{eqnarray}

Where,
\begin{eqnarray}
{\cal {E}}_{l} &=& \sum_{l=e,\mu}\frac{2}{(2\pi)^{3}}\int_0^{k_l} d^{3}k \sqrt{k^2 + m_l^2 },
\label{elep}
\end{eqnarray}

and
\begin{eqnarray}
P_{l} &=& \sum_{l=e,\mu}\frac{2}{3(2\pi)^{3}}\int_0^{k_l} \frac{d^{3}k k^2} {\sqrt{k^2 + m_l^2}}
\label{plep}
\end{eqnarray}

are the energy density and pressure for leptons ($e$ and $\mu$).
The EOSs of NS (Eq. \ref{eos-nstar}) are used as the input (with the representation ${\cal {E}}_{NS}\equiv {\cal {E}}$ and $P_{NS}\equiv p$)
to TOV equations to find the NS observables.

\section{Introducing Equations for Calculating NS Observables}{\label {tov-tidal}}
The structural properties of NS, such as the mass-to-radius profile, the tidal deformability, the moment of inertia are studied in this work.
Given an EOS it is straightforward to calculate the mass and radius of the NS by using the TOV equations \cite{tov}.
For slowly rotating objects, we make as usual for the line element the following ansatz \cite{molnvik}

\begin{eqnarray}
ds^2 &=& - e^{\nu(r)} dt^2 + e^{\lambda(r)} dr^2 + r^2 (d\theta^2+\sin^2\theta d\phi^2)
-2 \omega(r) (r sin \theta)^2 dt d \phi 
\label{metric}
\end{eqnarray}

The TOV equations are given by,
\begin{equation}
e^{\lambda(r)} = \left(1- \frac{2 m}{r}\right)^{-1},
\label{lambda}
\end{equation}

\begin{equation}
\frac{d\nu}{dr}= 2 \: \frac{m + 4 \pi p r^3}{r (r-2 m)}, 
\label{nu}
\end{equation}

\begin{equation}
\frac{dp}{dr} = - \frac{({\cal{E}}+p)(m+4\pi r^3p)}{r (r-2m)}, 
\label{pres}
\end{equation}

\begin{equation}
\frac{dm}{dr} = 4 \pi r^2 {\cal{E}}. 
\label{mass}
\end{equation}

The moment of inertia (MI) of NSs is obtained by solving the TOV equations along with the equation including the rotational frequency (given below). For a slowly rotating NS, the MI is given by \cite{hartle67,lattimer01},

\begin{eqnarray}
I=\frac{8\pi}{3} \int_0^R r^4 ({\cal {E}}+p) e^{(\lambda-\nu/2)} \frac {\bar \omega}{\Omega} dr, 
\label {mi}
\end{eqnarray}
where $\Omega$ and $\bar \omega(r) \equiv \Omega - \omega(r)$ are the angular velocity and the rotational drag function, respectively, for a
uniformly rotating NS. The rotational drag function $\bar \omega$ meets the boundary condition,

\begin{equation}
\bar \omega (r=R)=1-\frac{2I}{R^3}, \; \; \; \; \; \frac{d\bar\omega}{dr}|_{r=0}=0 
\label{bound.}
\end{equation}

The quantity $\frac{\bar \omega}{\Omega}$, evolve in Eq. \ref{mi}, is the dimensionless frequency satisfying the equation
\begin{eqnarray}
\frac{d}{dr} \left(r^4 j \frac{d\bar \omega}{dr}\right) = -4 r^3 \bar \omega \frac{dj}{dr}, 
\label{unitlessw.}
\end{eqnarray}
with $ j= e^{-(\lambda + \nu )/2}$.

The tidal deformability of an NS is one of the most important measurable physical quantities.
It characterizes the degree of deformation of NS due to the tidal field of its companion in BNS.
During the last stage of NS binary, each component star of a binary system develops a mass quadrupole due to
the tidal gravitational field of the partner NS.
The tidal deformability for $l=2$ quadrupolar perturbations is defined to be,

\begin{eqnarray}
\lambda_2 = \frac{2}{3}k_2 R^5 ,
\label{lam2}
\end{eqnarray}
where $R$ is the NS radius, and $k_2$ is the tidal love number which depends on the stellar structure.
The $k_2$ is calculated using the expression \cite{tanja},

\begin{eqnarray}
k_2 = \frac{8C^5}{5}(1-2C)^2 (2(1-C)+(2C-1)y_R) \times \nonumber \\
 \{ 4C^3 \left(13-11y_R+2C^2(1+y_R)+C(-2+3y_R) \right) \nonumber \\
 + 2C \left( 6-3y_R+3C (5y_R-8) \right) +3 (1-2C)^2 \nonumber \\
 \times (2+2C(y_R-1)-y_R)\log (1-2C) \}^{-1}, 
\label{k2}
\end{eqnarray}

where $C=M/R$ is the compactness of the NS, and $y_R=y(R)$ is obtained by solving the following
differential equation

\begin{eqnarray}
r \frac{dy}{dr} + y^2 + y F(r) + r^2 Q(r)=0
\label{yeqn}
\end{eqnarray}
with

\begin{eqnarray}
F(r) &=& \frac{r-4\pi r^3 ({\cal{E}}-p)}{r-2M},  \nonumber \\
Q(r) &=& \frac{4\pi r \left(5 {\cal {E}}+9 p + \frac{{\cal{E}}+p}{\partial p/ \partial {\cal{E}}} -\frac{6}{4\pi r^2}\right)}{r-2M} 
\nonumber \\
&& -4\left[\frac{M + 4\pi r^3 p}{r^2(1-2M/r)}\right]^2, 
\end{eqnarray}
along with the TOV equations (\ref{lambda}, \ref{nu},\ref{pres}, and \ref{mass}) with the appropriate boundary conditions, as given in \cite{tuhin}.
After solving these equations and obtaining the values $M$, $R$, $k_2$ $etc.$, one can compute the dimensionless tidal polarizability as: $\Lambda = 2/3 k_2 C^{-5}$.


\section{Effect of Dark Matter on the EOS of Neutron Star} {\label {eos-dm}}

The main goal of this work is to investigate the possible effects of DM on the properties of NSs, which depend entirely on the nature of the EOS of the object. It should be noted here that contrary to
\cite{ellis18,ellisplb}, in the present work the DM particles are everywhere inside the NSs and so they do not form a compact core. Consequently, additional peaks in post-merger power spectral density (PSD), as studied in \cite{ellisplb}, most likely will not be produced. We have obtained the EOSs within the E-RMF model using recently developed parameter sets, such as IOPB-I \cite{iopb1} and G3 \cite{G3}.
The merit of these parameter sets is that they pass through or relatively close to the low as well as high-density region \cite{iopb1}. The EOS corresponding to the G3 set is softer than the one corresponding to the IOPB-I parameter set \cite{iopb1}. For comparison reasons, we also consider the NL3 parameter set \cite{lala97}, which is one of the best known and widely used set. The EOS corresponding to the NL3 set is the stiffest among the chosen parameter sets. Table \ref{force1} shows values of the coefficients of Eqs. \ref{eq20}, \ref{eq21} for the parameters sets considered here. To obtain the EOS of an NS in the presence of DM, we need the values of the parameters for the DM part in the full Lagrangian (Eq. \ref{lag-tot}).
The values of the quantities involved in Eq. \ref{lag-tot} have already been mentioned in Sub-section \ref{EOS-NS} except
for $M_{\chi}$  and the Yukawa coupling $y$. It has been stated that the values of $y$ are obtained using Eq. \ref{sicross}, varying the mass
of the DM particle in the range specified in Eq. \ref{mdm}.

In Figure \ref{eos}, the EOSs of NS corresponding to the G3, IOPB-I, and NL3 parameter sets are shown for two different values of DM particle mass i.e., $M_\chi=4$ GeV (dashed curve) and $M_\chi=8$ GeV (dash-dotted curve).
For comparison reasons, the EOSs without DM (bold curves) are also shown for all the parameter sets.
The grey and yellow shaded regions represent the 50\% and 90\% confidence level of EOS, obtained from GW data \cite{abbott18}.
This was done using the spectral EOS parameterization with the condition that the EOS must support at least a $1.97~M_{\odot}$ star \cite{abbott18}.
In this way, the pressure posterior band \cite{abbott18} shrinks about three times from the prior pressure \cite{abbott17} (not shown here). It is remarked in \cite{abbott18} that posterior EOS becomes softer than the prior EOS. The vertical lines (blue lines) represent the nuclear saturation density and twice its value. These densities are assumed to almost correlate with bulk macroscopic properties of NS \cite{ozel16}. The pressure at twice of the nuclear saturation density is measured to be $21.88^{+16.88}_{-10.62}$ MeV-fm$^{-3}$ \cite{abbott18}.
We see that the presence of DM inside NS softens the EOS. A higher mass of the DM particle has a stronger impact in softening the EOS. The shift of the EOSs curves with the increase of the mass of the DM particle can be easily seen. It can be observed from the figure that G3 and IOPB-I EOSs with and without DM pass through the 90\% credible limit of the experimental band (posterior EOS) at and around the nuclear saturation density $\rho_{nucl}$. These EOSs pass through the 50\%
confidence level too at slightly larger energy density than $\rho_{nucl}$, while at larger nuclear density, these EOSs become
softer than the shaded band. However, NL3 EOSs satisfy the 90\% as well as 50\% confidence level of posterior EOSs only at a very large value of the energy density.
The effects of DM on the EOSs are consistent with what was obtained in \cite{panta,tuhin}, where the authors fixed the mass of the DM particle and varied the wave number of the DM particle. In \cite{panta,tuhin}, the effects of DM were larger
due to the fact that there the SM Higgs boson was the mediator and the mass of the DM particle was taken as 200 GeV.
In this work, however, we have considered lighter DM particle
and Higgs bosons, as was mentioned before, which can accommodate for scattering results consistent with the DAMA/LIBRA experiment. It is important to mention that the EOS becomes stiffer when considering the DM haloes around NS \cite{nelson}.
In this case, an enhancement in structural properties is reported in \cite{nelson}.

\begin{figure}
        \includegraphics[width=1.0\columnwidth]{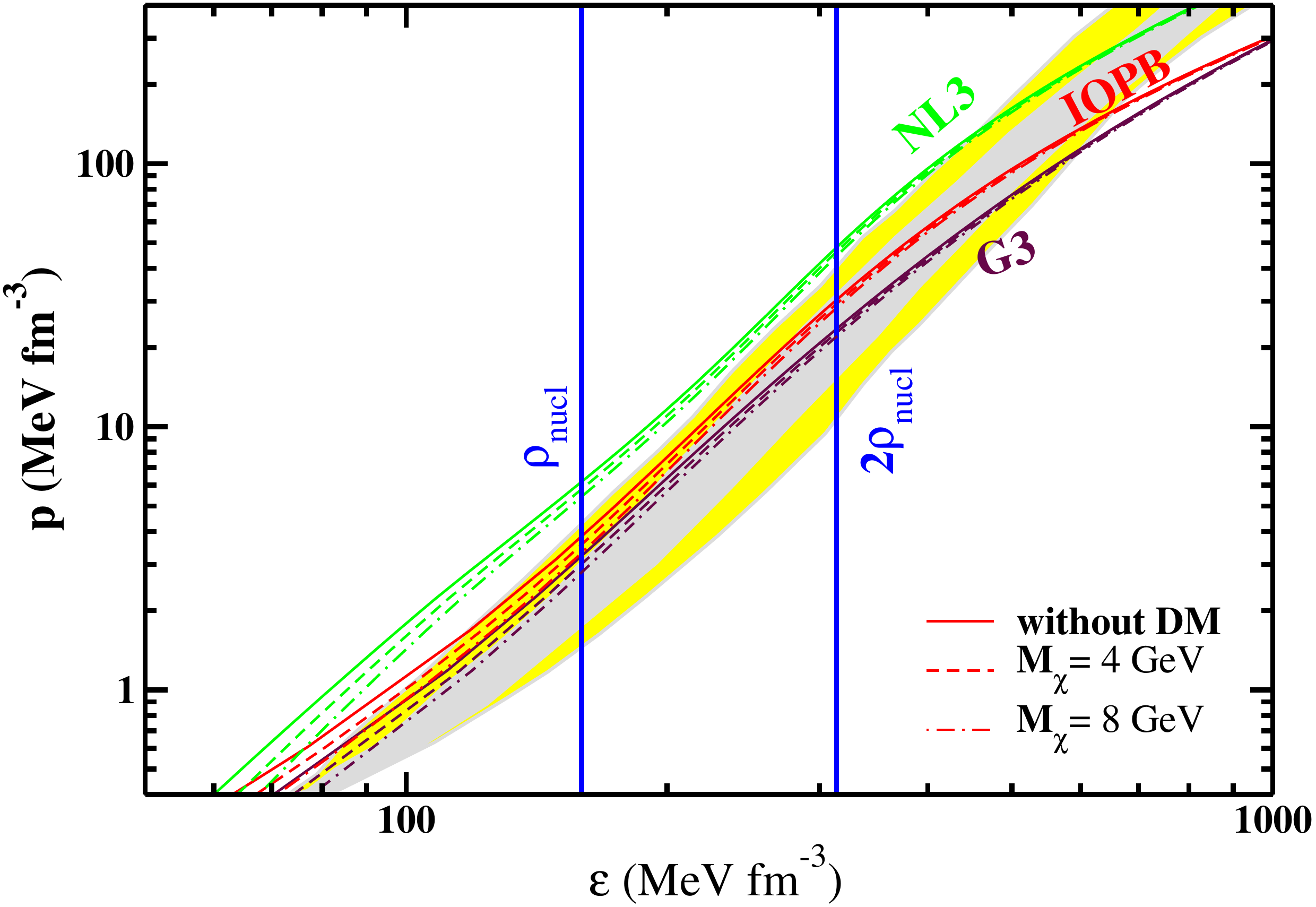}
        \caption{The EOSs of NSs in the presence of DM, corresponding to G3, IOPB-I, and NL3 parameter sets.
The bold lines represent the EOSs without considering DM. The dashed and dash-dotted lines represent the
EOSs in the presence of DM with a WIMP mass $M_\chi=4$ GeV and  $M_\chi=8$ GeV, respectively. The grey (yellow)
shaded region correspond to the 50\% (90\%) posterior credible limit from the GW data \cite{abbott18}.}
        \label{eos}
\end{figure}

\begin{figure}
        \includegraphics[width=1.0\columnwidth]{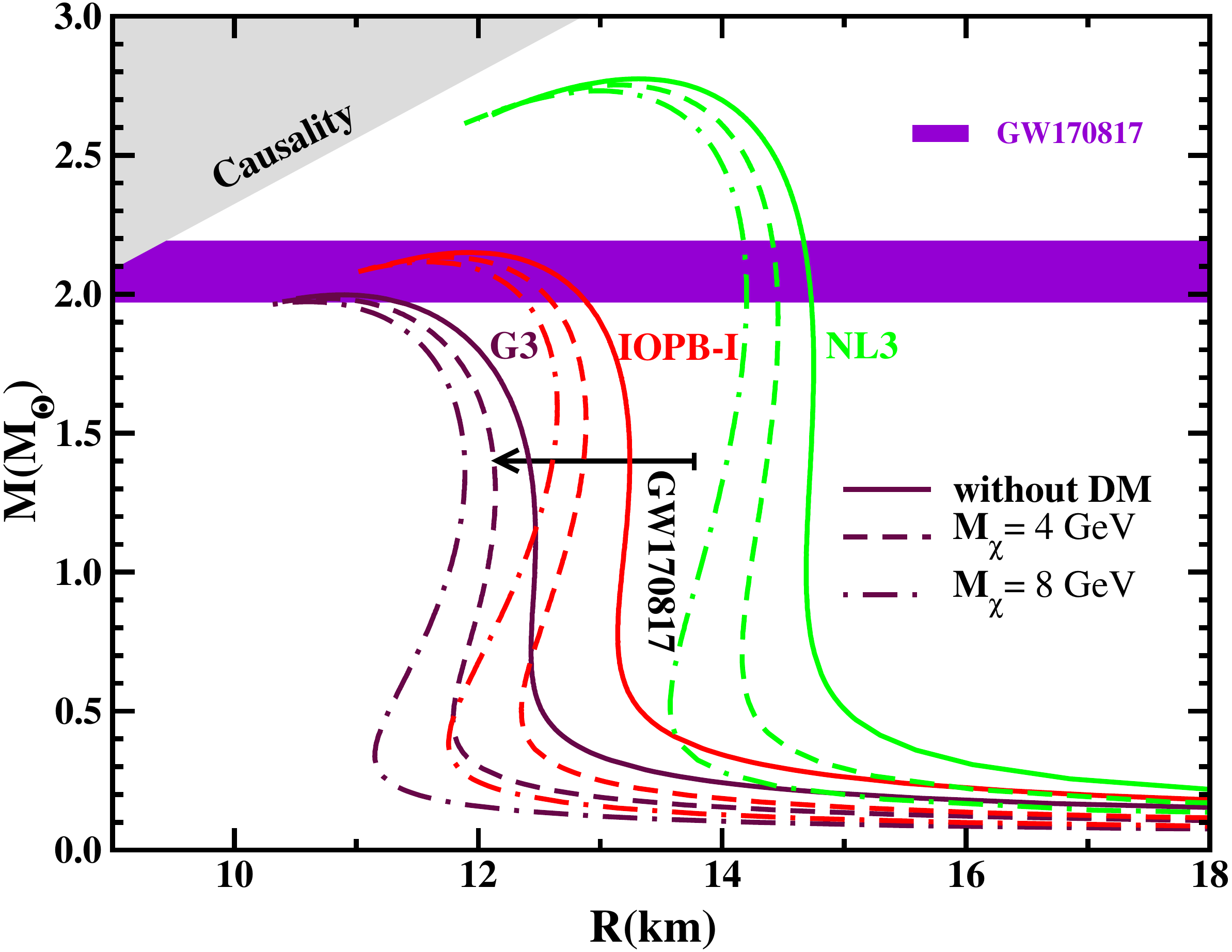}
        \caption{The mass-radius profile for NSs in the presence of DM corresponding to the
 IOPB-I \cite{iopb1}, G3 \cite{G3}, and NL3 \cite{lala97} parameter sets. The recent constraints
on the mass \cite{rezzo18} and radii \cite{fattoyev18} of NS are also shown. The grey shaded region shows the causality region \cite{lattimer07}.}
        \label{mr}
\end{figure}

\section{Neutron Star Observables}{\label {observables}}

The mass-radius profile for an NS is presented in Fig. \ref{mr} using the EOSs as shown in Fig. \ref{eos}. The violet band represents the maximum mass range for a non-rotating NS \cite{marg17,rezzo18,antoni13}.
This band also satisfies the precisely measured mass of NS, such as PSR J0348+0432 with mass $(2.01 \pm 0.04) M_\odot$ \cite{antoni13}.
These results imply that the theoretically predicted masses of NSs should reach the limit $\sim 2.0~M_\odot$.
The black arrow represents the radius at the canonical mass of NS \cite{fattoyev18} with the maximum value $R_{1.4} \leq 13.76$~km.
As anticipated, the mass-radius (MR) profiles are shifted downwards in the presence of DM inside an NS. The small effect of the DM on EOS
produces a significant shift of the MR profile to the left with a slightly lower highest mass. The bold, dashed, and dash-dotted lines correspond to the same representation
throughout this work, as mentioned for the EOS figure (Fig. \ref{eos}). The NL3 set, being the stiffest among the considered parameter sets, predicts large mass and radius. The G3 and IOPB-I EOSs with DM predict the maximum masses of NS that satisfy
the mass range constrained in \cite{rezzo18} from the prior GW data \cite{abbott17}, while GW170817 data rule out the NL3 EOSs.
On considering the large value of DM wave number, the EOSs for the NL3 set can be significantly reduced to satisfy the GW170817
mass range. In the figure, the lowering in the maximum mass of NS for the EOSs with DM is small. The effects of DM are more important for masses below the highest mass. In other words, the radius is more reduced at a fixed mass other than the maximum mass.
Among the three parameter sets considered here, the IOPB-I EOSs with and without DM satisfy the radius range at canonical mass constrained by the event GW170817 \cite{fattoyev18}.

\begin{figure}
        \includegraphics[width=1.0\columnwidth]{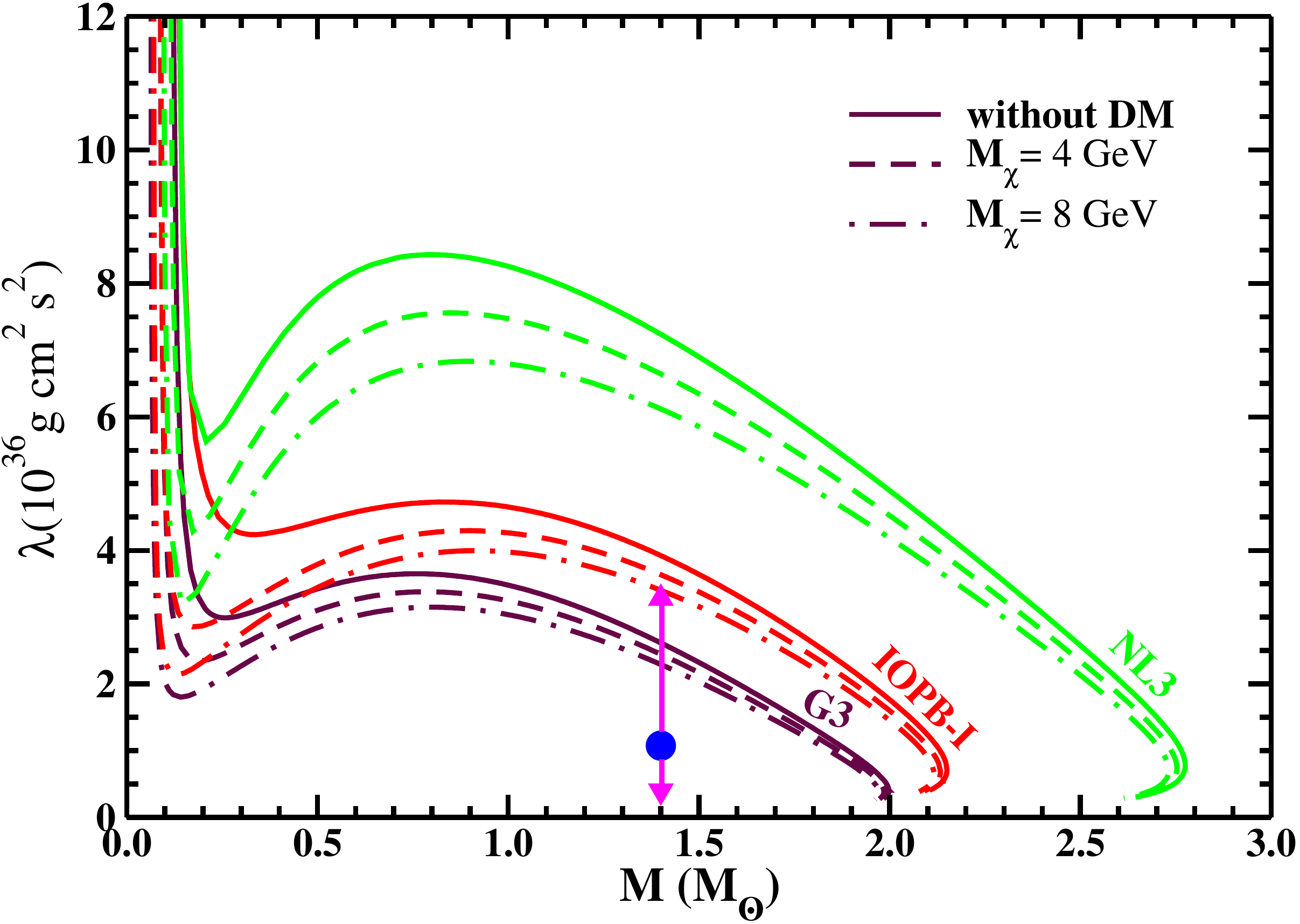}
        \caption{Tidal deformability $\lambda_2$ as a function of NS mass corresponding to the IOPB-I, G3,
and NL3 EOSs. The dashed and dash-dotted lines represent the EOSs in the presence of DM with the neutralino mass
$M_\chi=4$ GeV and  $M_\chi=8$ GeV, respectively. The blue circle with the arrow bar represent the $\lambda_2$ value
at 1.4M$_\odot$ NS mass obtained from GW data \cite{abbott18}.}
        \label{lm}
\end{figure} 

\begin{figure}
        \includegraphics[width=1.0\columnwidth]{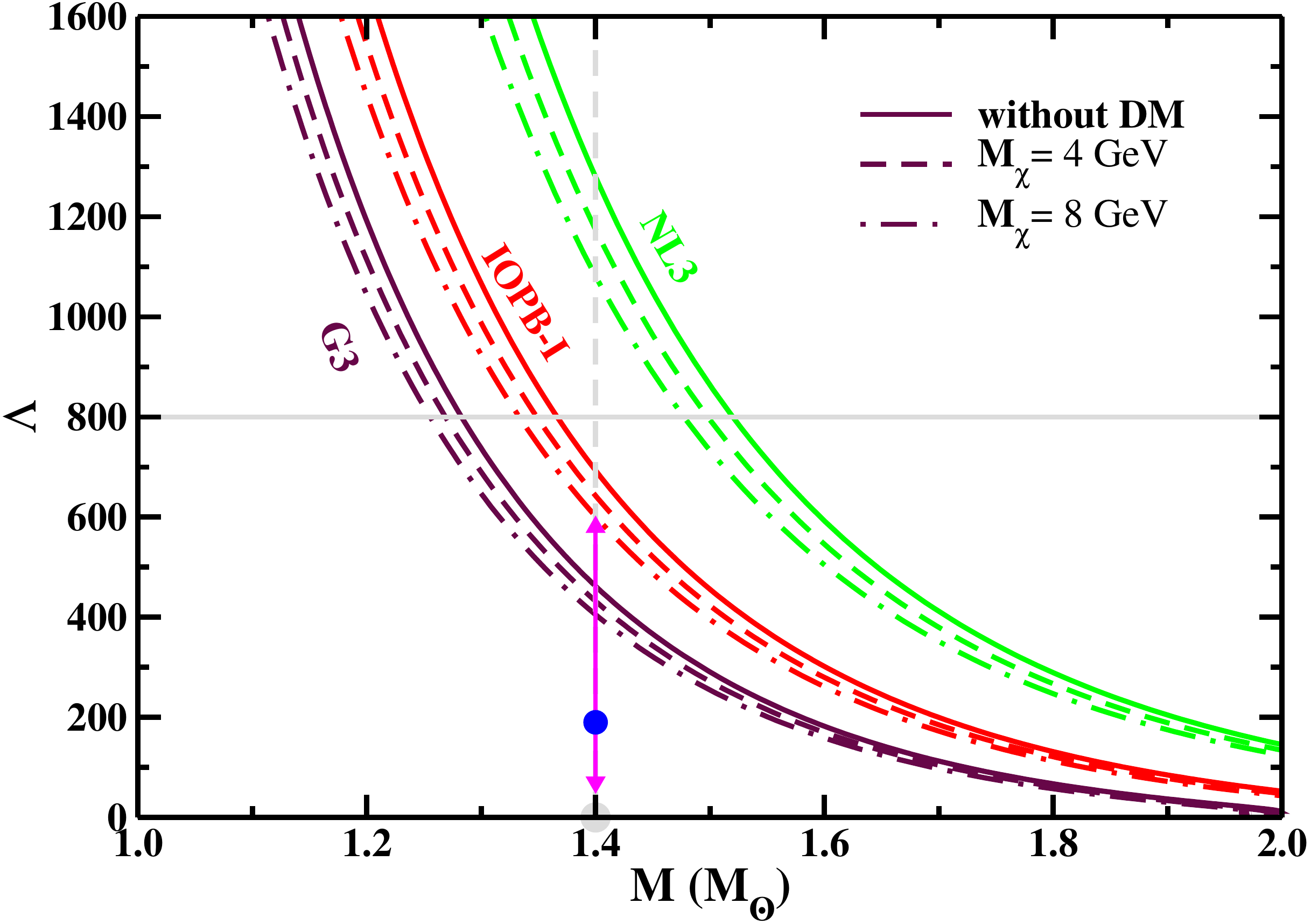}
        \caption{The dimensionless tidal deformability parameter ($\Lambda$) as a function of NS mass
corresponding to the IOPB-I, G3, and NL3 EOSs. The dashed grey line represent the canonical mass of NS.
However, the bold grey line shows the upper limit of $\Lambda$ value from GW170817 data \cite{abbott17}.
The blue circle with the error bar represent the $\Lambda_{1.4}$ value for posterior GW170817 data \cite{abbott18}. }
        \label{LM2}
\end{figure}

The tidal deformability of NS depends on its mass quadrupole, which is developed due to the tidal gravitational field of another component of NS binaries, as discussed above. It quantifies mainly the surface part of NS.
We have calculated the tidal polarizability for $l=2$ perturbation, i.e., $\lambda_2$. Recently, tidal deformability was discussed
for the GW170817 data \cite{abbott17}. It is clear from its definition (Eq. \ref{lam2}) that $\lambda_2$ depends on the radius of a star and on its tidal love number $k_2$, which describes the internal structure of NS.
As the radius of NS increases, $\lambda_2$ values grow and the surface becomes more deform. It simply means that soft EOSs predict
less value for $\lambda_2$. In Figure \ref{lm}, we plot $\lambda_2$ for the chosen EOSs with and without DM. The blue circle with the arrow bar (error bar) represents the $\lambda_2$ of an NS at the mass 1.4M$_\odot$ corresponding
to the $\hat{\Lambda}_{1.4} = 190^{+390}_{-120}$, which is constrained from the GW data \cite{abbott18} at 90\% confidence level.
The NL3 set predicts large values for the tidal deformability and hence large deformation. The $\lambda_2$ corresponding to the NL3 EOSs,
even in the presence of DM, does not pass through the experimental range at the canonical mass.
On the other hand, $\lambda_2$ curves for the G3 EOSs lie within the observationally allowed region. However,
the IOPB-I EOS at neutralino mass $M_\chi=8$~GeV predicts a $\lambda_2$ value that just satisfies the upper range of
the experimental $\lambda_2$ value. The shift of the curves in the presence of DM can easily be noticed from the figure. The significant changes in $\lambda_2$ due to the DM occur at the canonical mass of NS. We also show the dimensionless tidal deformability ($\Lambda$) of a single NS in Figure \ref{LM2}. The blue circle represents $\Lambda$ at the canonical mass of NS from GW170817 posterior data \cite{abbott18}, the numerical value of which is given above.

\begin{figure}
        \includegraphics[width=1.0\columnwidth]{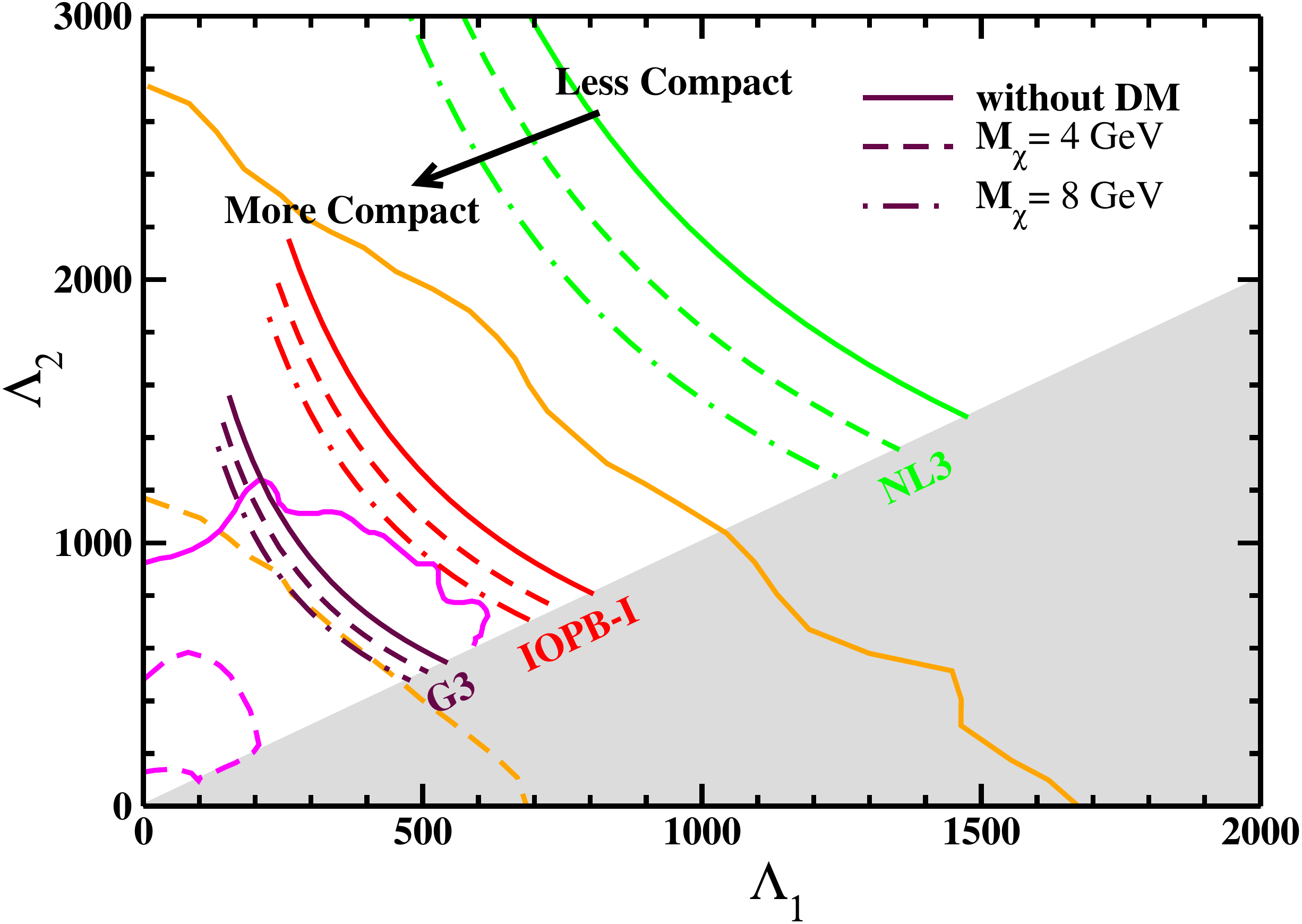}
        \caption{Dimensionless tidal deformability $\Lambda$ generated by using IOPB-I, G3, and NL3 EOSs with DM and
without DM inside the NS. The calculated values are compared with 50\% (dashed) as well as 90\% (bold) probability contour for
the case of low spin, $|\chi| \leq 0.05$, and represented by Orange and Magenta color for, respectively, prior \cite{abbott17}
and posterior \cite{abbott18} GW170817 data. }
        \label{tidal}
\end{figure}

In Fig. \ref{tidal} we display the dimensionless tidal deformabilities $\Lambda_1 $ and $\Lambda_2 $ of a binary NS
corresponding to the G3, IOPB-I, and NL3 parameter sets. Here, we consider the tidal deformability constraint from
GW170817 observation on EOSs of NS in the presence of DM. The individual dimensionless tidal deformabilities
$\Lambda_1 $ and $\Lambda_2 $ correspond to high mass $m_1$ and low mass $m_2$ of BNS. We vary the mass $m_1$
in the range $1.365 < m_1/M_\odot < 1.60$, and determine the range of $m_2$ by fixing the chirp mass as ${\cal{M}}_c = 1.188~M_{\odot}$.
It can be seen in the figure that the G3 and IOPB-I sets are in excellent agreement with the $90\%$ (bold line)
probability contour of prior GW170817, shown by orange curves \cite{abbott17}.
We also show the recently re-analyzed results of GW170817 data in magenta color \cite{abbott18}. The figure shows that only the curves
corresponding to the G3 EOS with and without DM lies within the 90\% confidence level allowed region of prior as well as
posterior GW170817 data.
The shaded part (grey color) in the figure marks the $\Lambda_2 < \Lambda_1$ region that is naturally excluded for a common realistic EOS \cite{abbott18}. The analysis of \cite{abbott18} suggests that soft EOSs, which predict lower values for $\Lambda$ are favored over stiffer EOSs.
For the EOSs corresponding to DM admixed NSs, the curves are shifted to the left and predict lower values for $\hat{\Lambda}$ corresponding to less compact NSs. The NL3 EOSs lie outside the 90\% confidence lever region (bold line) of prior (orange)
as well as posterior (magenta) analysis.
On adjusting the parameters of the DM Lagrangian for an EOS, the curve can be shifted even more to the left. That way, the parameters of the DM Lagrangian can be optimized satisfying the GW170817 constraints.

\begin{figure}
        \includegraphics[width=1.0\columnwidth]{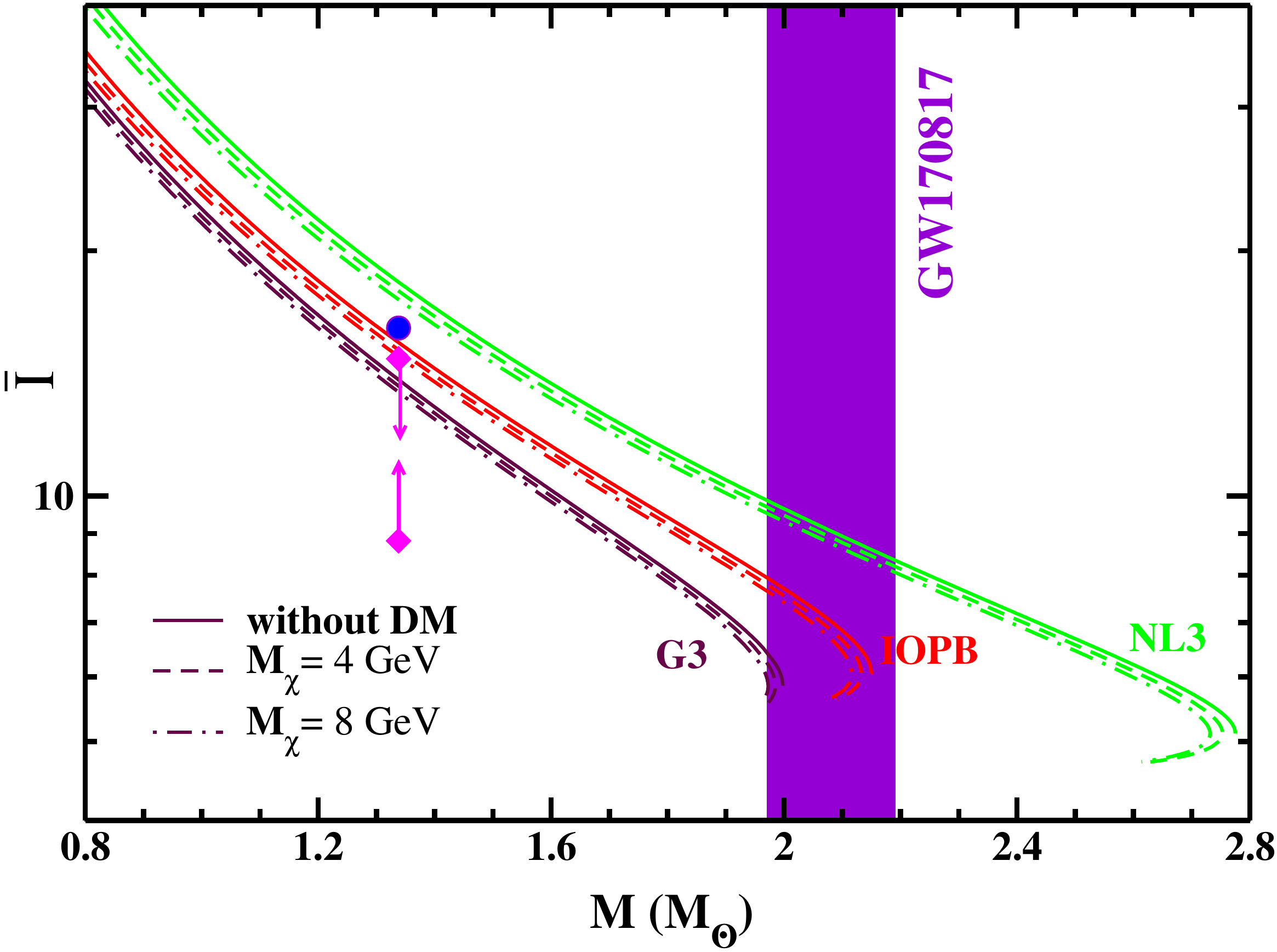}
        \caption{The dimensionless moment of inertia $\bar I$ as a function of NS mass for EOSs shown in Fig. \ref{eos}.
The overlaid arrows represent the constraints on MI of PSR J0737-3039A set \cite{bharatAPJ}
from the analysis of GW170817 data \cite{abbott18}.
The circle shows the upper bound from the minimal-assumption analysis of Ref. \cite{abbott17}.
From a reference point of view, the mass range of NS, constrained from GW data \cite{rezzo18}, is also shown by the violet band.}
        \label{moi}
\end{figure}

The moment of inertia (MI) of NSs strongly depends on the structure of the object. For a slowly rotating NS, the moment of inertia is computed solving Eqs. \ref{mi}, \ref{bound.}, and \ref{unitlessw.} together with the TOV equations. It is one of the most important macroscopic quantities that can be used to constrain the EOS of NSs. The MI of the binary pulsar, PSR J0737-3039, is expected to be determined within $\sim 10\%$ accuracy by measuring its angular momentum \cite{lyne04,kramer09,lattimer05}. The mass distribution of an NS, the final stage of the BNS merger, and $r-$ process nucleosynthesis are determined by the EOS, and therefore a precise measurement of the MI, tidal deformability, etc are very important. In Fig. \ref{moi} we plot the dimensionless MI, which decreases with the mass of an NS. Stiffer EOSs predict a larger MI for a given mass of an NS. In the presence of DM, the soft nature of EOSs generates a lower MI. The overlaid arrows in the figure indicate the MI of PSR J0737-3039, constrained by the analysis of GW170817 \cite{abbott18}, while the circle represents the upper bound on MI from minimal assumption analysis \cite{abbott17}.

\section{Conclusions}{\label{summary}}

We have analyzed the effect of DM on the properties of NSs.
We have considered a generic WIMP of fermionic nature (one may have in mind the lightest neutralino), which is trapped inside the object. The WIMPs interact with the baryonic matter through the exchange of light Higgs bosons. We have adopted the $\sim 10~GeV$ WIMP hypothesis, as suggested by the DAMA/LIBRA results. As it has been shown in the literature that such a scenario can be realized within NMSSM, one can have in mind the lightest CP-even eigenstate of the NMSSM as the mediator Higgs boson. The EOSs of NSs are generated using the E-RMF Lagrangian density including the interaction Lagrangian density of DM with the baryonic matter, and applying the $\beta-$ equilibrium and charge neutrality conditions.
Within E-RMF, we have used the recent parameter sets, such as G3 and IOPB-I, along with the older and widely accepted NL3 set. 
Out of the three EOSs considered here, G3 is the softest one and predicts relatively small values of NS observables in agreement
with the GW170817 results. We have observed that the presence of DM in NS softens the EOS, which results in lowering the values of NS observables, such as mass, radius, tidal deformability, and even moment of inertia. We have imposed the constraints
from GW170807 on the mass, $\lambda_2$ values, dimensionless tidal deformability $\Lambda$ and MI of NSs. The effects of DM are small at the maximum values of NS mass, while its impact is more significant in the mass region other than the maximum mass.








 

\chapter{Summary and Conclusions}{\label{chap8}}  
\rule\linewidth{.5ex}

The work presented in this thesis has touched various areas of contemporary 
research in nuclear structure, infinite nuclear matter, neutron star, and dark matter. 
In this chapter, we summarize and conclude the most important outcomes obtained in this thesis work. 

\begin{enumerate}
	\item After a brief introduction of the various phenomenon of finite nuclei, infinite nuclear matter, neutron star, and dark matter 
in Chapter \ref{chap1}, we present an overview of relativistic mean-field theory along with the derivations of field equations in Chapter \ref{chap2}. 
Temperature-dependent equation of state of nuclear matter, its key parameters, and the formalism for BCS and quasi-BCS pairing are 
discussed in Chapter \ref{chap2}. In the same chapter, we have also explained the reason to consider the G3, IOPB, FSUGarnet parameters within 
the extended version of RMF theory, known as effective field theory motivated relativistic mean-field models. 
	\item For the first time we have applied the recently proposed FSUGarnet and IOPB-I parameter sets
of RMF formalism to deform nuclei at finite temperature.
	\item The temperature-dependent bulk properties of $^{234,236}U$, $^{240}Pu$, $^{244-262}Th$, and $^{246-264}U$ nuclei 
are presented in Chapter \ref{chap3}. The findings of this study are the following:
	\begin{itemize}
		\item The quadrupole and hexadecapole deformation parameters $\beta_2$ and $\beta_4$
approach to zero at the critical temperature $T_c$, which is about 2 MeV for the neutron-rich thermally fissile nuclei
irrespective of the force parameters.
		\item All the analysis, such as the $S_{2n}$ trends, $S^2\sim E^*$ relation, $\epsilon\sim T$
(single-particle energy with temperature), $\beta_2$ and $\beta_4$, shape transition, and the vanishing of shell
correction consistently agree with the transition temperature of $T_c\sim 2$ MeV.
		\item The inverse level density parameter is found to be quite sensitive to the temperature and interactions at low T.
		\item In spite of similar predictions of FSUGarnet and IOPB-I models, the IOPB-I predicts
moderate neutron skin-thickness as compared to the NL3 and FSUGarnet sets. 
These results conclude that the fission yield with asymmetric fragments will be more
in the case of the NL3 set than the FSUGarnet and then with IOPB-I.
		\item We have also studied the symmetry energy coefficient with temperature as well as mass number.  The $a_{sym}$
increases with both T and A. The rate of increase of $a_{sym}$ with T for an isotope is slower as compared
to the rate of increase of asymmetry energy coefficient in an isotopic chain with mass number A.
	\end{itemize}
	\item After studying the bulk properties of finite nuclei, we move to study the temperature-independent symmetry energy of 
finite nuclei and nuclear matter.
	\item The effective surface properties such as the symmetry energy, neutron pressure, and the symmetry energy curvature 
are calculated for the isotopic series of $O$, $Ca$, $Ni$, $Zr$, $Sn$, $Pb$, and $Z = 120$ nuclei within 
the coherent density fluctuation model. The densities of the isotopes are evaluated within the spherically symmetric 
E-RMF model with G3, FSUGarnet, and IOPB-I sets and used in the input of the CDFM. The results are presented in Chapter \ref{chap4}. 
	\item The symmetry energy, neutron pressure, and symmetry energy curvature of infinite nuclear
matter are calculated within the Bruckner energy density functional model which are further folded with the
weight function to find the corresponding quantities of finite nuclei.
	\item The FSUGarnet parameter set predicts the 
large value of the symmetry energy while the smaller symmetry energy values are for the G3 set
with some exceptions.
	\item We found a larger value of the skin-thickness for the force parameter that corresponds to the stiffer EoS and vice-versa.
	\item Observing the nature of the symmetry energy over the isotopic chain, we predict a
few neutron magic numbers in the neutron-rich exotic nuclei including superheavy. The
transparent signature of magicity is diluted for a few cases over the isotopic chain of $Pb$ and $Z = 120$
nuclei.
	\item Similar behavior is also observed for the neutron pressure and symmetry energy curvature for
these isotopes.
	\item The calculated quantities are important for the structural properties of finite
nuclei and may be useful for the synthesis of neutron-rich or superheavy nuclei. These effective surface properties
can also be used to constrain an EoS of the nuclear matter and consequently nucleosynthesis processes.
	\item Following the work of Chapter \ref{chap3} where temperature-dependent bulk properties of neutron-rich thermally fissile 
nuclei are presented, we study the temperature-dependent symmetry energy, neutron pressure, and symmetry energy curvature of 
neutron-rich thermally fissile nuclei in Chapter \ref{chap5}. 
	\item The densities of the nuclei along with their ground and excited state bulk properties at finite T and the 
symmetry energy and related quantities of nuclear matter are estimated within TRMF formalism. 
	\item The nuclear matter properties are calculated at the local density of the nuclei, which are further 
used in the LDA to calculate the corresponding properties of the nuclei.
 	\item  The symmetry energy, neutron pressure, and symmetry energy curvature decrease with
the increase of T, while the skin-thickness increases. These properties are found to be smaller for $^{250}U$
due to its large isospin asymmetry, which enhances the rate of electron capture.
	\item The correlation among the calculated properties of the nuclei is found at finite T for each parameter set. The symmetry energy
coefficient is found to vary inversely with the pressure. It is found that the softer the EOS is, the larger
the symmetry energy coefficient and the smaller the neutron pressure of the nuclei are. The neutron pressure
is linearly correlated with $\Delta R$ at finite T as calculated with all three parameter sets.
	\item In this thesis work, we have also tried to constrain the parameters of one of the mysterious objects, known as the dark matter with the help of GW170817 data.
	\item The effect of DM on the properties of NSs is discussed in Chapter \ref{chap6}, where an algorithm of the equation of state of 
hadronic matter of NS in the presence of dark matter is presented. 
	\item We assume that WIMPs interact with baryonic matter through the exchange of light Higgs bosons, considered within NMSSM. 
We have adopted the $\sim 10~GeV$ WIMP hypothesis, as suggested by the DAMA/LIBRA results.
	\item Out of the NL3, G3, and IOPB-I EOSs within the E-RMF model, the G3 is the softest one and predicts relatively 
small values of NS observables in agreement with the GW170817 results. 
	\item We have observed that the presence of DM in NS softens the EOS, which results in lowering the values of NS observables, such as mass, radius, tidal deformability, and even moment of inertia. The effects of DM are small at the maximum values of NS mass, while its impact is more significant in the mass region other than the maximum mass.
	\item  We have imposed the constraints from GW170807 on the mass, $\lambda_2$ values, 
dimensionless tidal deformability $\Lambda$ and MI of NSs.
\end{enumerate}


\begin{appendix}

\chapter{Solution of RMF Equations for Spherically Symmetric Case}{\label{append1}}

Since spherical nuclei preserve rotational symmetry, the spherical coordinates are used to solve RMF equations of motion. 

\begin{eqnarray}
x=rsin{\theta}cos{\phi}, \;\;\;\;y=rsin{\theta}sin{\phi}, \;\;\;\;z=rcos{\theta} 
\end{eqnarray}

The nucleon densities and meson fields depends only on radial coordinate {\it r}. The 
spinor for a nucleon $i$ with angular momentum $j_i$, its projection $m_i$, parity  $n_i$, and the isospin projection
$t_i$ (= ±1/2 for neutrons and protons) is expressed as 

\begin{equation}
{\Psi_i}(\vec{r},s,t)=\left( \begin{array}{c} {f_i}(r){\Phi_{{l_i}{j_i}{m_i}}}({\theta},\phi,s) \\ i{g_i}(r){\Phi_{{\tilde{l_i}}{j_i}{m_i}}}({\theta},\phi,s) \end{array} \right){\chi_{t_i}}(t).
\label{aaa1}
\end{equation}

Where,
$l_i$ and ${\tilde{l_i}}$ are, respectively, the orbital angular momentum corresponding to large and small spinor components 
which are defined as:
\begin{eqnarray}
l_i= {j_i}\pm{\frac{1}{2}}, \;\;\;{\tilde{l}_i= {j_i}\mp{\frac{1}{2}}}, \;\;\;{\pi}=(-1)^{{j_i}\pm{\frac{1}{2}}}, \;\;\;
{\kappa}={\pm}{j_i}\pm{\frac{1}{2}}.
\end{eqnarray}

Moreover, ${\chi_{t_i}}$ represents the isospin wave function and ${\Phi_{{l_i}{j_i}{m_i}}}$, having {\it ljm} quantum numbers,
 denotes the two dimensional that is expressed as

\begin{equation}
\Phi_{ljm}(\theta,\phi,s)={[{\chi_{(1/2)}}(s){\otimes}{Y_l}(\theta,\phi)]}_{jm}
\end{equation}

Using the spinor as expressed in Eq. \ref{aaa1} in Dirac equation for nucleons, 
one gets a set of coupled ordinary differential equations in r for large and small
components of Dirac spinors.

\begin{eqnarray}
(M^{*}(r)+V(r)){f_i}(r)+({\partial_r}-{\frac{{\kappa_i}-1}{r}}){g_i}(r)&=&{\epsilon_i}{f_i}(r)\\
-({\partial_r}+{\frac{{\kappa_i}+1}{r}}){f_i}(r)-(M^{*}(r)-V(r)){g_i}(r)&=&{\epsilon_i}{g_i}(r).
\end{eqnarray}

In the above equations, $M^{*}(r)$ represents the effective mass of a nucleon and V(r) is the potential, defined as:
\begin{equation}
V(\vec{r})={g_\omega}\omega+{g_\rho}{\tau_3}\rho+e{A_0}+{\Sigma_0}^R.
\end{equation}

The term ${\Sigma_0}^R$ in the above equation is the rearrangement contribution to vector self-energy that is defined below as: 
\begin{equation}
{\Sigma_0}^R={\frac{\partial{g_\sigma}}{\partial\rho_v}}{\rho_s}\sigma
+{\frac{\partial{g_\omega}}{\partial\rho_v}}{\rho_v}\omega+
{\frac{\partial{g_\rho}}{\partial\rho_v}}{\rho_{tv}}\rho.
\end{equation}

The radial functions ${R_{nl}}(r,b_0)$ of a spherical harmonic oscillator potential having frequency $\hbar{\omega_0}$ and 
oscillator length $b_0=\sqrt{\hbar/m{\omega_0}}$ is used to expand the small and large components of Dirac spinors, as expressed below:

\begin{eqnarray}
{f_i}(r)&=&{{\Sigma_{n=0}}^{n_{max}}}{{f_n}^{(i)}}{R_{nl_i}}(r,b_0),\\
{g_i}(r)&=&{{\Sigma_{n=0}}^{\tilde{n}_{max}}}{{g_{\tilde{n}}^{(i)}}{R_{{\tilde{n}}{\tilde{l}_i}}}}(r,b_0)
\end{eqnarray}

with

\begin{equation}
R_{nl}(r,b_0)={{b_0}^{-3/2}}{R_{nl}}(\xi)={{b_0}^{-3/2}}{{\cal N}_{nl}}{\xi^l}{L_n}^{l+1/2}(\xi^2){e^{-{\xi^2}/2}},
\end{equation}

where $n_{max}$ and $\tilde{n}_{max}$ are determined, respectively, by using the expressions $N_{max}=2n_{max}+l_{max}$ and 
$\tilde{N}_{max}=2\tilde{n}_{max}+\tilde{l}_{max}$ having relation $\tilde{N}_{max}=N_{max}+1$. 
The $\xi=r/{b_0}$ is the radial distance in units of oscillator length $b_0$. 
The ${{L_n}^m}(\xi^2)$ represents the associated Laguerre polynomial that is defined in Ref.~\cite{va73}.
${\cal N}_{nl}$ is the normalization factor which is defined as:

\begin{equation}
{\cal N}_{nl}={(2n!/(l+n+1/2)!)}^{1/2}
\end{equation}

with $n=0,1,2,........$ as the number of radial nodes.

\section{\label{sec:level1}The Dirac Hamiltonian}
The $|\alpha\rangle=|nljm\rangle$ are the basis states. In case of the spherical symmetry, the matrix elements of Dirac Hamiltonian are:
\begin{eqnarray}
{\cal {A}}_{{\alpha}{\alpha'}}&=&{\int_{0}^{\infty}}d\xi{R_{nl}}(\chi){R_{n'l}}(\xi)[M^{*}({b_0}{\xi})+V({b_0}{\xi})],\\
{{\cal C}_{{\tilde{\alpha}}{\tilde{\alpha'}}}}&=&{{\int_0}^{\infty}}{d_\xi}{R_{n\tilde{l}}}(x){R_{n'\tilde{l}}}(\xi)[M^{*}({b_0}{\xi})-V({b_0}{\xi})],\\
{{\cal B}_{{\tilde{\alpha}}{\alpha'}}}&=&{\cal N}_{n\tilde{l}}{\cal N}_{n'\tilde{l}}{{\int_0}^{\infty}}{d_\xi}
{e^{-\xi^2}}{\xi^{2l}}{{L_n}^{\tilde{l}+1/2}}{{L_{n'}}^{\tilde{l}+1/2}}{\times}(2n'+l+1+\kappa-\xi^2).
\end{eqnarray}
\section{\label{sec:level2}The Coulomb Interaction}

The potential for protons has contribution from coulomb direct field which is expressed as: 
\begin{equation}
{V_C}(\vec{r})={e^2}{\int}{d^3}{r'}{\frac{{\rho_p}(\vec{r'})}{|\vec{r}-\vec{r'}|}}.
\end{equation}

The above integration has a logarithmic singularity that is eliminated by using the identity~\cite{2} as:
\begin{equation}
{\Delta_{r'}}{|\vec{r}-\vec{r'}|}=\frac{2}{|\vec{r}-\vec{r'}|},
\end{equation}

After elimination of the logarithmic singularity and doing integration by parts yields



\begin{equation}
{V_C}(\vec{r})=\pi{e^2}{{\int_0}^{\infty}}{dr'}{r'^2}(3r+\frac{r'^2}{r}){\frac{{d^2}\rho_p{(r')}}{dr'^2}}.
\end{equation}

\section{\label{sec:level3}Klein-Gordon Equations}
The equation
\begin{equation}
({-\frac{\partial^2}{\partial{r^2}}}-{\frac{2}{r}}{\frac{\partial}{\partial{r}}}+{m_\phi}^2){\phi(r)}={s_\phi}(r).
\end{equation}
is the Helmholtz equation for field $\phi$ where ($\phi=\sigma, \omega, $and $\rho$), the solution of which is obtained by expanding in 
the basis states as:
\begin{eqnarray}
\phi(r)={{\Sigma_{n=0}}^{n_b}}{\phi_n}{R_{n0}}(r,b_0),;\;\;\;\;{s_\phi}(r)={{\Sigma_{n=0}}^{n_b}}{{s_n}^{\phi}}{R_{n0}}(r,b_0),
\end{eqnarray}
 with $n_b$ is determined by cutt-off parameter (${N_B}={2n_b}$). A set of inhomogeneous linear equations,
\begin{equation}
\sum_{n'}^{n_b}{{\cal H}_{nn'}}{\phi_{n'}}={{s_n}^{\phi}},\\
\end{equation}

with the matrix
\begin{eqnarray}
{{\cal H}_{nn'}}=-{{b_0}^{-2}}{\delta_{nn'}}(2n+3/2)
+{{b_0}^{-2}}{\delta_{nn'+1}}{\sqrt{(n+1)(n+3/2)}}\nonumber\\
+{{b_0}^{-2}}{\delta_{n'n+1}} {\sqrt{(n'+1)(n'+3/2)}}
\end{eqnarray}
is solved by inversion.
\chapter{Solution of RMF Equations for Axially Symmetric Case}{\label{append2}}
In the case of axially deformed nuclei, the $z-$ component of total angular momentum, i.e, $j_z$ is conserved. The 
quantum number $\Omega_i$ is 
\begin{equation}
{\Psi_i}(\vec{r},s,t)=\left( \begin{array}{c} {{f_i}^+}({r_\perp},z)e^{i{\Lambda_-}\phi} \\ {{f_i}^-}({r_\perp},z)e^{i{\Lambda_+}\phi}\\
{i{g_i}^+}({r_\perp},z)e^{i{\Lambda_-}\phi}\\{i{g_i}^-}({r_\perp},z)e^{i{\Lambda_+}\phi} \end{array} \right){\chi_{t_i}}(t),
\end{equation}

where $\Lambda_\pm={\Omega_i}\pm1/2$, and ($r_\perp,z,\phi$) are the cylindrical coordinates
\begin{eqnarray}
x={r_\perp}cos{\phi},\;\;\;\;y={r_\perp}sin{\phi},\;\;\;\;z=z
\end{eqnarray}

The axially symmetric harmonic oscillator potential is expanded as:
\begin{equation}
V_{osc}(z,r_\perp)={\frac{1}{2}}m{{\omega_z}^2}{z^2}+{\frac{1}{2}}m{{\omega_\perp}^2}{{r_\perp}^2}
\end{equation}

Two oscillator frequencies, conserving the volume of the oscillator are expressed as 

\begin{eqnarray}
\hbar{\omega_z}=\hbar{\omega_0}{e^{-{\sqrt{\frac{5}{4\pi}}}\beta_0}},\;\;\;\;
\hbar{\omega_\perp}=\hbar{\omega_0}{e^{{\frac{1}{2}}{\sqrt{\frac{5}{4\pi}}}\beta_0}}
\end{eqnarray}

The corresponding oscillator length parameters; 
\begin{eqnarray}
{b_z}=\sqrt{\frac{\hbar}{m\omega_z}}, \;\;\;\;{b_\perp}=\sqrt{\frac{\hbar}{m{\omega_\perp}}}
\end{eqnarray}

and due to the conservation of volume ${{b^2}_\perp}{b_z}={b_0}^3$, the basis is now defined in terms of the constants
$\hbar\omega_0$ and $\beta_0$.

The eigen function are labelled as

\begin{equation}
|\alpha\rangle=|{n_z}{n_r}\Lambda{m_s}\rangle,
\end{equation}

where
$n_z$: number of nodes in z-direction, \\
$n_r$: number of nodes in ${r_\perp}$-direction,\\
$\Lambda$: the z-axis projection of orbital angular momentum, and \\
$m_s$: the z-axis projection of spin.

Now, using dimensionless variables

\begin{eqnarray}
\xi=z/{b_z}, \;\;\;\;\eta={{r_\perp}^2}/{{b_\perp}^2},
\end{eqnarray}

The eigen function of harmonic oscillator is defined as 
\begin{equation}
{\Phi_\alpha}(\vec{r},s)={\phi_{n_z}}(z,b_z){{\phi^\Lambda}_{n_r}}(r_\perp,b_\perp)
{\frac{e^{i\Lambda\phi}}{\sqrt{2\pi}}}\chi(s),
\end{equation}

where

\begin{eqnarray}
{\phi_{n_z}}(z,b_z)&=&{{b_z}^{-1/2}}{\phi_{n_z}}(\xi)={{b_z}^{-1/2}}{{\cal N}_{n_z}}{{\cal H}_{n_z}}(\xi){e^{-{\xi^2}/2}},\\
{{\phi^\Lambda}_{n_r}}(r_\perp,b_\perp)&=&{{b_\perp}^{-1}}{{{\cal N}^\Lambda}_{n_r}}{\sqrt{2}}{\eta^{|\Lambda|/2}}
{{L^{|\Lambda|}}_{n_r}}(\eta){e^{-\xi/2}}.
\end{eqnarray}

${H_{n_z}}(\xi)$ and ${{L_{n_r}}^{|\Lambda|}}(\eta)$ denote the Hermite and associated Laguerre polynomials, respectively.
The normalization factors are

\begin{eqnarray}
{\cal N}_{n_z}=({\sqrt\pi}{2^{n_z}}{n_z}!)^{-1/2}\;\;\;\;{{{\cal N}^\Lambda}_{n_r}}=({n_r}!/({n_r}+|\Lambda|)!)^{1/2}.
\end{eqnarray}

\section{\label{sec:level1}The Dirac Hamiltonian}

For the case of axially symmetric nuclei, the matrix elements of the Dirac Hamiltonian are:   

\begin{eqnarray}
\left( \begin{array}{c} {\cal A}_{\alpha{\alpha'}} \\ {\cal C}_{\alpha{\alpha'}} \end{array} \right)=
{\delta_{\Lambda{\Lambda'}}}{\delta_{{m_s}{{m_s}'}}}{{\int_{-\infty}}^{\infty}}{d\xi}{\phi_{n_z}}(\xi){\phi_{n'_z}}(\xi)
\times{{\int_0}^{\infty}}{d\eta}{{\phi_{n_r}}^{\Lambda}}(\eta){{\phi_{n'_r}}^{\Lambda}}(\eta) \nonumber \\
\times[M^{*}({b_z}\xi,{b_\perp}\sqrt{\eta}){\pm}V({b_z}\xi,{b_\perp}\sqrt{\eta})]
\end{eqnarray}

\begin{eqnarray}
{{\cal B}_{\alpha{\alpha'}}}
&=&{\delta_{\Lambda{\tilde\Lambda}}}{\delta_{{m_s}{\tilde{m_s}}}}{\delta_{{n_r}{\tilde{n_r}}}}(-1)^{1/2-{m_s}}{\frac{1}{b_z}}\nonumber \\
&&\times\left({\delta_{{n_z}{\tilde{n_z}-1}}}{\sqrt{\frac{\tilde{n_z}}{2}}}-{\delta_{{n_z}{\tilde{n_z}+1}}}{\sqrt{\frac{\tilde{n_z}}{2}}}\right) \nonumber \\
&&+{\delta_{\Lambda{\tilde{\Lambda}+1}}}{\delta_{{m_s}{\tilde{m_s}-1}}}{\delta_{{n_z}{\tilde{n_z}}}}
\frac{{{{\cal N}^\Lambda}_{n_r}}{{{\cal N}^{\tilde\Lambda}}_{\tilde{n_r}}}}{b_\perp}\nonumber \\
&&{\times}{{\int_0}^\infty}{d\eta}{e^{-\eta}}{\eta^{\Lambda-1/2}}{{L_{n_r}}^{\Lambda}}(\eta)
\left({{\tilde{L}_{\tilde{n}_r}}^{\tilde\Lambda}}(\eta)-{\Lambda}{L_{\tilde{n}_r}^{\tilde\Lambda}}(\eta)\right) \nonumber \\
&&+{\delta_{\Lambda{\tilde{\Lambda}-1}}}{\delta_{{m_s}{\tilde{m_s}+1}}}{\delta_{{n_z}{\tilde{n_z}}}}
\frac{{{{\cal N}^\Lambda}_{n_r}}{{{\cal N}^{\tilde\Lambda}}_{\tilde{n_r}}}}{b_\perp}\nonumber \\
&&{\times}{{\int_0}^\infty}{d\eta}{e^{-\eta}}{\eta^{\Lambda-1/2}}{{L_{n_r}}^{\Lambda}}(\eta)
\left({{\tilde{L}_{\tilde{n}_r}}^{\tilde\Lambda}}(\eta)+{\Lambda}{L_{\tilde{n}_r}^{\tilde\Lambda}}(\eta)\right).
\end{eqnarray}
\section{Coulomb Interaction}
In case of axially deformed nuclei, the potential for protons include the direct coulomb field as
\begin{equation}
{V_C}(\vec{r})={e^2}{\int}{d^3}{r'}{\frac{{\rho_p}(\vec{r'})}{|\vec{r}-\vec{r'}|}}.
\end{equation}

And in the same way as of spherical symmetry, the logarithmic singularity is eliminated as:
\begin{equation}
{\Delta_{r'}}{|\vec{r}-\vec{r'}|}=\frac{2}{|\vec{r}-\vec{r'}|},
\end{equation}



After integrating by parts over the azimuthal angle $\phi$, one gets the expression

\begin{equation}
{V_C}(r_\perp,z)=2{e^2}{{{\int_0}^{\infty}}{r'_\perp}d{r'_\perp}}{{\int_{-\infty}}^{\infty}}{dz'}d(r_\perp,z)
{\times}E(\frac{4{r_\perp}{r'_\perp}}{d(r_\perp,z)}){\Delta{\rho_p}(r'_\perp,z')},
\end{equation}

with $d(r_\perp,z)=\sqrt{{(z-z')}^2+{({r_\perp}+{r'_\perp})}^2}$. The complete elliptical
integral of second kind is approximated by standard polynomial formula~\cite{va73}.

\section{\label{sec:level3}Klein-Gordon Equations}
Again the Helmholtz equations for meson fields ($\phi=\sigma,\omega,$and $\rho$) 
\begin{equation}
({-\frac{\partial^2}{\partial{{r_\perp}^2}}}-{\frac{2}{r_\perp}}{\frac{\partial}{\partial{r_\perp}}}+{m_\phi}^2){\phi(r_\perp,z)}={s_\phi}(r_\perp,z)
\end{equation}

are expanded in the basis states as:

\begin{equation}
\phi(z,r_\perp)={\Sigma_{{n_z}{n_r}}^{N_B}}{\phi_{{n_z}{n_r}}}{\phi_{n_z}}(z,b_z){{\phi_{n_r}}^0}(r_\perp,b_\perp)
\end{equation}

Finally set of inhomogeneous linear equations is obtained,

\begin{equation}
{\Sigma_{{n'_z}{n'_r}}^{N_B}}{{\cal H}_{{n_z}{n_r}{n'_z}{n'_r}}}{\phi_{{n'_z}{n'_r}}}={s^\phi}_{{n_z}{n_r}},
\end{equation}

with the matrix elements

\begin{eqnarray}
{{\cal H}_{{n_z}{n_r}{n'_z}{n'_r}}}&=&{\delta_{{n_r}{n'_r}}}{\delta_{{n_z}{n'_z}}}\left({\frac{1}{{b^2}_z}}({n_z}+\frac{1}{2}\right)+
{\frac{1}{{b^2}_\perp}}(2{n_r}+1)+{{m_\phi}^2})\\ \nonumber
&&-{\frac{1}{2{b^2}_z}}{\delta_{{n_r}{n'_r}}}\left(\sqrt{({n_z}+1){n'_z}}{\delta_{{n_z}{{n'_z}-2}}}
+\sqrt{({n'_z}+1){n_z}}{\delta_{{n_z}{{n'_z}+2}}}\right)\\ \nonumber
&&+{\frac{1}{{b^2}_\perp}}{\delta_{{n_z}{n'_z}}}\left({n'_r}
{\delta_{{n_r}{{n'_r}-1}}}+{n_r}{\delta_{{n_r}{{n'_r}+1}}}\right).
\end{eqnarray}
The Hamiltonian equation is solved by inversion.

\end{appendix}

  \backmatter
 \markboth{}{}


\cleardoublepage
\phantomsection

\end{document}